\documentclass[aps,pra,superscriptaddress]{revtex4}
\usepackage[intlimits]{amsmath}
\usepackage{amsfonts}
\usepackage{psfrag}
\usepackage{comment}
\usepackage{subfigure}
\usepackage{enumitem}
\usepackage[usenames]{color}
\usepackage{braket}

\advance\voffset by -1cm
\advance\hoffset by -0.5cm
\newcommand\be            {\begin{equation}}
\newcommand\ee            {\end{equation}}
\newcommand\bes           {\begin{subequations}}
\newcommand\esu           {\end{subequations}}

\newcommand\labl[1]       {\label{#1}\ee}

\newcommand\sgn {\operatorname{sgn}}
\newcommand\Tr {\operatorname{Tr}}

\newcommand{\twovec}[2]{\left(\begin{array}{c}
#1\\#2
\end{array}
\right)}

\newcommand{\twovecT}[2]{\left(\begin{array}{cc}
#1 & #2
\end{array}
\right)}

\newcommand{\TT}{\mathcal{T}}
\newcommand{\ud}{\mathrm d}

\newcommand{\bigx}[1]{\bBigg@{#1}}

\newcommand\eps           {\varepsilon}
\newcommand\fii           {\varphi}

\newcommand\mc            {\mathcal}

\newcommand\p            {\partial}

\newcommand\no[1]{{\,:\!#1\!:\,}}

\def\3pt#1#2#3{{\langle{#1}\vert{#2}\vert{#3}\rangle}}

\newcommand\doi[2]        {\href{http://dx.doi.org/#1}{#2}}

\newcommand{\EQ}{\begin{equation}}
\newcommand{\EN}{\end{equation}}
\usepackage{epsfig}
\usepackage{color}
\usepackage{psfrag}
\usepackage{amsmath}
\usepackage{graphicx}
\usepackage{amsfonts}
\usepackage{amssymb}
\graphicspath{{Figures/}}
\begin{document}
\bibliographystyle{plainnat}

\title{{\Large {\bf Equilibration Properties of Classical Integrable Field Theories }}}

\author{Andrea De Luca} 
\affiliation{LPTMS, CNRS, Univ. Paris-Sud, Universit\'{e} Paris-Saclay, 91405 Orsay, France}
\author{Giuseppe Mussardo}
\affiliation{SISSA and INFN, Sezione di Trieste, via Bonomea 265, I-34136, 
Trieste, Italy}
\affiliation{International Centre for Theoretical Physics (ICTP), 
I-34151, Trieste, Italy}

\begin{abstract}
\noindent
We study the equilibration properties of classical integrable field theories at a finite energy density, with a time evolution that starts from initial conditions far from equilibrium. These classical field theories may be regarded as quantum field theories in the regime of high occupation numbers. This observation permits to recover the classical quantities from the quantum ones by taking a proper $\hbar \rightarrow 0$ limit. 
In particular, the time averages of the classical theories can be expressed in terms of a suitable version of the LeClair-Mussardo formula relative to the Generalized Gibbs Ensemble. For the purposes of handling time averages, our approach provides a solution of the problem of the {\em infinite gap solutions} of the Inverse Scattering Method.

\vspace{3mm}
\noindent
Pacs numbers: 11.10.St, 11.15.Kc, 11.30.Pb

\end{abstract}
\maketitle


\section{Introduction}
Recent advances in ultra-cold atom systems and other quantum devices \cite{Weiss1,Weiss2,Weiss3} have revitalized the study of long standing issues in statistical physics, such as the approach to equilibrium of an extended system subjected to some non trivial initial conditions. Can we understand the microscopic laws that drive the system asymptotically to equilibrium? What are the time scales involved in this process? Do all macroscopic systems equilibrate? How to calculate the expectation values of various observables when time goes to infinity? These and other related questions have recently received a lot of attention from various groups and much progress has been made (see, for instance, refs. \cite{CC1,CC2,quenches1,quenches2,quenches3,quenches4,Rigol,2011_Pozsgay_JSTAT_P01011,MC,CK,CE,Cauxreview,FEreview,CEF,Sotiriadis,prethermalization,BertiniSchurichEssler,FM,GMPRL13,GGE_problems,completeGGE,EMP}), especially thanks to the cross-fertilization of new theoretical tools, efficient numerical methods and important inputs coming from the experiments. 

In the quantum setting, a framework sufficiently general to formulate the study of systems out of equilibrium goes under the name of {\em quantum quench} \cite{CC1,CC2}: an isolated system is prepared in the ground state of a Hamiltonian $H_{0} = H(\lambda_0)$, where $\lambda_0$ is some controllable parameter, function of the experimental knobs. At $t_0 = 0$, such a parameter is abruptly switched to a different value, resulting in a unitary time evolution of the system under the new Hamiltonian $H = H(\lambda)$. Since the only role played by $H_0$ consists of preparing the system in an initial state $\mid \Psi_0\rangle$ that is not an eigenstate of the post-quench Hamiltonian $H(\lambda)$, we can free ourselves of considering from now on $H_0$ and simply formulate the out-of-equilibrium problem as the time evolution of a system subjected to some boundary condition encoded in the 
initial state $\ket{\Psi_0}$. It is important to stress that in all the situations we are interested in this paper, the time evolution is purely Hamiltonian, i.e. there is neither coupling to external bath nor dissipative terms. In other words, we are interested in understanding the approach to equilibrium of an extended system subjected only to its own interactions. Still, recently it has been shown how the interactions with quantum reservoirs can be reproduced by choosing inhomogeneous initial conditions~\cite{inho1, inho2, inho3, inho4}. Indeed, in this paper, we will not assume translation invariance.

In the quantum case, one of the most important problems is to predict the Quantum Dynamical Average (QDA) of local observables ${\cal O}$ of these systems defined as   
\be 
\langle {\cal O}\rangle_{QDA} \,\equiv \, \lim_{t \rightarrow \infty} \frac{1}{t}\, \int_0^t \, dt \,  
\, {\langle \Psi_0 |\cal O}(t)| \Psi_0\rangle \,\,\,. 
\label{QuantumDynamicalAverage}
\ee

Here, we are interested in the equilibration aspects of local quantities in the context of a purely (1+1) {\em classical} relativistic invariant field theory made of a scalar field $\phi(x,t)$. In particular, 
we focus on the Classical Dynamical Average (CDA) of local functions $F[\phi(x,t)]$ of the scalar field $\phi(x,t)$ defined as 
\be 
\langle F[\phi] \rangle_{CDA} \,\equiv\,\lim_{t \rightarrow\infty} \frac{1}{t} \,\int_{0}^t dt\, \left[\frac{1}{L} \int_0^L F[\phi(x,t)] dx\right]\,\,\,,
\label{ClassicalDynamicalAverage}
\ee
where the time and space dependence of the observable $F[\phi(x,t)]$ is induced by the field $\phi(x,t)$, which evolves according to the equation of motion and its initial boundary conditions, as discussed in more details below. $L$ is the size of the interval on which the theory is defined. 

Although the two averages (\ref{QuantumDynamicalAverage}) and (\ref{ClassicalDynamicalAverage}) seem to be rooted on two different grounds, 
one the aims of this paper is to show their deep connection, as it can be intuited from these considerations  
\begin{enumerate}
\item 
it is well known that the classical field theories can be regarded as quantum field theory in the limit in which the Planck constant goes to zero, $\hbar \rightarrow 0$. This is also the regime in which the mode occupations of the systems are very high, and the effective description of the quantum dynamics can be indeed well captured by the classical equation of motion; 
\item on the other side, the purely classical field theories and the relative equations of motion can be regarded as the formalism that only describes the dynamics of a particular set of {\em matrix elements} of the quantum field theory, i.e. those where all the observables are sandwiched between the coherent states of the system. 
\end{enumerate}

\noindent
As we are going to see, each of these points of view has it own benefits and, altogether, they help in clarifying several aspects of the equilibration 
process that occur both at the classical and quantum levels. In particular, it is worth stressing that a key feature of this paper is the change of perspective with respect to the traditional studies in classical integrable systems: indeed, our goal is to show how to compute the time averages of classical quantities, i.e. those expressed in eq~(\ref{ClassicalDynamicalAverage}), not using the formulas coming from the classical Inverse Scattering Method \cite{Faddev,Novikov,Ablowitz} but taking instead full advantage of the known solution of the quantum problem! As recalled below, for the quantum problem the solution is provided by the LeClair-Mussardo formula \cite{LM} and its out-of-equilibrium generalization established by one of the authors \cite{GMPRL13}. The advantage of the approach based on the out-of-equilibrium LeClair-Mussardo formula is that we can get around the very difficult problem of handling the almost intractable infinite gap solution provided by the classical Inverse Scattering Method.  

At a more technical level, the classical limit of a quantum theory is achieved by firstly restoring $\hbar$ in the path integral and then sending $\hbar \rightarrow 0$. The relevant quantities selected in this way correspond to the {\em tree-level} Feynman diagrams and, as shown in the text, they match with the classical solutions of the equation of motion for the elementary field $\phi(x)$ and all its various powers $:\phi^k(x):$ which give rise to the infinite tower of composite operators. Term by term, these classical solutions can be regarded as the {\em classical} limit of the quantum matrix elements (Form Factors) of the corresponding operator: the check of this statement is particularly efficient in the case of an integrable QFT, where the exact computation of these matrix elements can be done thanks to the known expressions coming from the Form Factor technique \cite{KW, Smirnov,Luky,KM}. 

In this paper we consider bosonic massive Quantum Field Theories (QFT) in (1+1) dimensions and their classical counterparts. Let's briefly present here the relevant formulas relative to the field equilibration in the quantum case, keeping in mind though that the main goal of this paper is precisely to transpose to the classical context the quantum identities shown in Eqs.\,(\ref{BasicIdentity}) and (\ref{EnsembleAverageCharge}) given below, and in particular the one encoded in the formula (\ref{EnsembleAverageMus23}).   

To start with, an aspect particularly relevant for the equilibration dynamics of quantum field theories is whether the field theory is integrable or non-integrable. While a generic non-integrable QFT possesses the Hamiltonian $H$ and the momentum $P$ as the only conserved quantities, an integrable QFT is instead supported by an infinite number of conserved charges $Q_i$ which can be local or non-local and of course also include the Hamiltonian.  For this reason, the dynamics of integrable models is strongly constrained \cite{zamzam} (see also \cite{GMUSSARDO} and references therein). To fix the ideas, in the following we will choose as prototype of non-integrable field theory the $\Phi^4$ Landau-Ginzburg (LG) QFT, with Lagrangian and Hamiltonian densities given by   
\begin{eqnarray}
{\mathcal L} &\,=\,& \frac{1}{2} (\partial_\mu \phi)^2 - \frac{m^2}{2} \phi^2 - \frac{g}{4!} \phi^4 \,\,\,,
\label{Lagrangianphi4}
\\
{\mathcal H} &\,=\,& \frac{1}{2} \Pi^2 + \frac{1}{2} \phi_x^2 + \frac{m^2}{2} \phi^2 + \frac{g}{4!} \phi^4 \,\,\,,
\label{Hamiltonianphi4}
\end{eqnarray}
($\Pi(x,t) \equiv \phi_t(x,t) $),  
while our prototype of an integrable QFT will be the Sinh-Gordon model, with Lagrangian and Hamiltonian densities 
\begin{eqnarray}
{\mathcal L} &\,=\,& \frac{1}{2} (\partial_\mu \phi)^2 - \frac{m^2}{g^2} (\cosh(g\phi) -1)  \,\,\,,
\label{LagrangianShG}
\\
{\mathcal H} & \,=\, &\frac{1}{2} \Pi^2 + \frac{1}{2} \phi_x^2 + \frac{m^2}{g^2} (\cosh(g\phi) -1) \,\,\,. 
\label{HamiltonianShG}
\end{eqnarray}
Both theories share a $Z_2$ symmetry and have only one massive excitation, with no further bound states. They are sufficiently 
rich but at the same time sufficiently simple to address many topics of non-equilibrium physics in the most direct way. 

\vspace{1mm}
{\bf Non-integrable QFT}. In the quantum case, for the equilibration process of a generic non-integrable QFT one expects that, as time goes by, the interactions present in the Hamiltonian give rise to non-trivial scattering processes among different numbers of particles, with a  consequent mixing of the modes (alias, the occupation number of particles of given momentum). Consequently the systems will asymptotically reach a situation of {\em thermal} equilibrium, with the temperature $T=\beta^{-1}$ fixed in terms of the energy $E$ of the initial state. As well known, this is in a nutshell the basis of the Ergodic Hypothesis: translated in formula, this implies the equality between the Quantum Dynamical Average and the Gibbs Ensemble Average (GE) 
\be 
\langle {\cal O}\rangle_{QDA} \,=\,\langle {\cal O}\rangle_{GE}\,\,\,.
\label{ergodic}
\ee 
where the Gibbs Ensemble Average is defined as 
\begin{eqnarray} 
\langle {\cal O} \rangle_{GE} &\,\equiv &\, Z^{-1} \,  {\rm Tr} \,\left({\cal O} e^{-\beta H}\right) \,=\, 
Z^{-1}  \, \sum_{n=0}^\infty \langle n | {\cal O} | n \rangle \, e^{-\beta E_n} \,\,\,, \label{EnsembleAverage1}
\\
&& Z = {\rm Tr} \, e^{-\beta H} \,=\,\sum_{n=0}^\infty e^{-\beta E_n} \nonumber \,\,\,,  
\label{GibbsPartitionFunction}
\end{eqnarray}
with the vectors $| n \rangle$ chosen to be eigenvectors of the Hamiltonian with eigenvalues $E_n$. 
In a QFT context, notice that the diagonal matrix elements $\langle n | {\cal O}  | n\rangle$ on the particle states are divergent quantities which need to be properly defined in order to give meaning to the ensemble average. 

\vspace{1mm}
{\bf Integrable QFT}. For an integrable quantum field theory, the situation is however quite different. In this case it is easy to see (and we will give some examples below) that the time average of the observables does not generally coincide with its Gibbs Ensemble Average. In other words,  these systems violate the ergodicity property, yet the time averages may be recovered by employing a Generalized Gibbs Ensemble (GGE) which involves higher conserved charges -- a scenario originally advocated in \cite{Rigol} and further studied in a series of papers, among which \cite{2011_Pozsgay_JSTAT_P01011,MC,CK,CE,CEF,Sotiriadis,prethermalization,BertiniSchurichEssler,FM,GMPRL13,GGE_problems,completeGGE,EMP}.  
Namely, for an Integrable QFT one expects that it will hold a Generalized Ergodic Hypothesis, expressed by the following equality between the Quantum Dynamical Average and the Generalized Gibbs Ensemble (GGE) Average  
\be 
\langle {\cal O}\rangle_{QDA} \,=\,\langle {\cal O}\rangle_{GGE} \,\,\,, 
\label{BasicIdentity}
\ee 
where 
\begin{eqnarray} 
\langle {\cal O} \rangle_{GGE} & \,= & \, Z^{-1} \,  {\rm Tr} \,\left({\cal O} e^{-\sum_i \beta_i Q_i}\right)
\,=\, Z^{-1} \, \sum_{n=0}^\infty \langle n | {\cal O}  | n\rangle\, e^{-\sum_i \beta_i q_i(n)} \,\,\,, 
\label{EnsembleAverageCharge} \\
&& Z \,=\, {\rm Tr}\, e^{-\sum_i \beta_i Q_i} \,=\,\sum_{n=0}^\infty e^{-\sum_i \beta_i q_i(n)}\,\,\,. \nonumber  
\label{EnsembleAverageCharge2}
\end{eqnarray}
In writing eq\,(\ref{EnsembleAverageCharge}) we have assumed that the basis of vectors $| n \rangle$ are 
made of eigenvectors of all the charges $Q_i$, with eigenvalues $q_i(n)$. As before, in a QFT context
one faces the divergence of the ensemble average for the divergence of the diagonal matrix elements $\langle n | {\cal O}  | n\rangle$ 
on the particle states. 

\vspace{1mm}
{\bf LeClair-Mussardo formula out of equilibrium}.The formula that cures the divergences of the original expression 
(\ref{EnsembleAverageCharge2}) and implemented the GGE Average has been established in \cite{GMPRL13}:  it is very similar 
to the LeClair-Mussardo (LM) formula \cite{LM}, originally established for the pure thermal case, and reads 
\be
\langle {\cal O}\rangle_{GGE}\,\equiv\, 
\sum_{n=0}^\infty\frac1{n!}
\int_{-\infty}^\infty 
\left(\prod_{i=1}^n\frac{\ud\theta_i}{2\pi} f(\theta_i) 
\right) 
\3pt{\overleftarrow{\theta_n}}{\mc O(0)}{\overrightarrow{\theta_n}}_\text{conn}\,,   
\label{EnsembleAverageMus23} 
\ee
This is the formula we are going to use in this paper and all our further considerations will crucially depend upon it. Such a formula employs the 
following quantities: 
\begin{enumerate}
\item the function $f(\theta)$, that is the filling fraction of the states. This function can be obtained in terms of the so-called {\em pseudo-energy} $\epsilon(\theta)$ that satisfies the non-linear integral Bethe Ansatz equation. In the thermal case, this function is obtained self-consistently by 
the solution of the non-linear integral equations driven by the temperature $T$; in the out of equilibrium case, instead, the Bethe Ansatz equations are driven by the initial state and the corresponding pseudo-energy solution $\epsilon(\theta)$ contains at once all information about the conserved charges entering the Generalized Gibbs Ensemble (\ref{EnsembleAverageCharge2}). 
\item the so-called {\em connected Form Factors} $\3pt{\overleftarrow{\theta_n}}{\mc O(0)}{\overrightarrow{\theta_n}}_\text{conn}$: these are functions of the various rapidities $\theta_i$ of the particles, and are obtained as special limit of the matrix elements (Form Factors) of the operator ${\mc O}$. The limit is defined in such a way to ensure that the diagonal Form Factors, computed in the kinematic configurations where particles in the bra and ket vectors have exactly the same rapidities, are actually finite expressions. 
\end{enumerate}

\vspace{1mm}
{\bf Plan of the paper}. The main idea behind this paper is to compute the Classical Dynamical Average of quantities $F[\phi(x,t)]$, which are functions of the classical field $\phi(x,t)$, by using the identity (\ref{BasicIdentity}) and the definition of the GGE average which is implemented by the formula (\ref{EnsembleAverageMus23}).  In other words, we would like to make sense in the classical context of the out-of-equilibrium LM formula given in  eq~(\ref{EnsembleAverageMus23}). To do so, we will need to implement:  
\begin{itemize}
\item  
the classical limit of the connected Form Factors of the observables $F[\phi(x,t)]$. The striking fact is that, in classical field theory, there is of course neither the notion of particles nor of matrix elements of operators between such particle states! Despite this puzzling aspect, we will see that it will be nevertheless possible to make sense of a notion as {\em classical Form Factors} and to compute these expressions. 
\item the classical notion of "filling fraction" $f(\theta)$ which enters eq\,(\ref{EnsembleAverageMus23}). We will see that this function can be numerically determined in terms of the transfer matrix coming from the Inverse Scattering Method of the classical theories, and it is entirely fixed in terms of the initial boundary conditions $\phi_0(x,t)$ and $\partial_t \phi_0(x,t)$ for the classical dynamics. 
\end{itemize} 
As we will shown in the following, the definition and the computation of the classical Form Factors and the classical filling fraction will bring us in an interesting journey through several fields of theoretical physics, among which: Exact Scattering Theory, Form Factors, Bethe Ansatz, Transfer Matrix Method and Semi-classical Approach. Given the broad range of topics faced in our study, the paper is naturally divided in three Parts which coherently address different aspects of the classical and quantum field theories.   
 
\vspace{1mm}

{\bf Part A}, made of Sections \ref{Sectionnonint} and \ref{transfermatrixsection}, concerns with all relevant aspects of a generic classical field theory. In this Part we discuss issues as the numerical implementation of the equation of motion, the virial theorem, the existence of different time scales present in the non-linear dynamics of the field equations and the useful Transfer Matrix formalism for computing the thermal averages of local observables. We also comment on the obstructions for directly implementing the GGE Average in classical field theory. 

\vspace{1mm}
{\bf Part B}, which includes Sections \ref{classicallimit} up to Section \ref{summarynextsteps}, deals with the formalism of Quantum Field Theories and the $\hbar \rightarrow 0$ limit which defines the corresponding classical field theories. In this Part, we discuss the exact expressions of the S-matrix and the Form Factors, the classical limit of these quantities, the formalism of the coherent states, the Generalized Bethe Ansatz equations, the equivalence between the fermionic and bosonic formulations of the Bethe Ansatz, and the classical version of the Bethe Ansatz equations once we take the limit $\hbar \rightarrow 0$. 

\vspace{1mm}
{\bf Part C}, which includes Sections \ref{lightconeSH} up to Section \ref{conclusions}, concerns with the problem of determining the root density $\rho^{(r)}(\theta)$ in the classical field theory. For this aim, we discuss the action-angle structure of the classical field theory. The analysis carried on in this Part has several by-products, which helps enlightening the richness of the subject: we introduce, for instance, the monodromy matrix and we show how it can be numerically determined, we present the computation of the higher conserved charges both in the light-cone or the laboratory frames, we also present the Inverse Scattering Method at a {\em finite energy density} on a cylinder geometry and its relation with (classical) Bethe Ansatz equations. The important output of Part C is the identification of the algorithmic steps which lead to the computation of the filling fraction $f(\theta)$ for classical field theories. 

\vspace{1mm}
The paper also contains several appendices: Appendix A collects the main definitions of Fourier expansions of the field, Appendix B deals with the ground state energy of the modified Mathieu equation expressed as solution of a classical Bethe Ansatz integral equations, Appendix C contains the light-cone formalism, Appendix D gathers the explicit expressions of the lowest conserved charges of the Sinh-Gordon model.

\newpage

\begin{center}
{\Large {\bf PART A}}
\end{center}


\section{Time evolution of classical fields} \label{Sectionnonint}
In Part A of the paper we discuss the time evolution of classical fields, we introduce the main quantities of interest and we analyse their properties. Further, we discuss the time scale(s) involved in the process of equilibration. All these results refer to equations of motion of purely classical field theories. In this context, an important source of information is provided by a paper by Boyanovski et al. \cite{BDD}, where one can find detailed results about the $\Phi^4$ LG theory but also several discussions about general aspects of the time evolution of classical fields in (1+1) dimensions. 
For non-relativistic field theory, the reader may consult the classical work by Fermi, Pasta and Ulam \cite{FPU} or a recent report written by various authors on the status of the Fermi-Pasta-Ulam problem \cite{FPUreport}.

\subsection{Classical equation of motion} 
Since we are interested in the equilibration process that takes place in systems at {\em finite energy density}, in the following we will study the dynamics of the (1+1) dimensional field theories on a cylinder geometry of width $L$, in the limit in which $L \rightarrow \infty$, with $E/L$ finite, where $E$ is the energy of the system. In more details, the time variable $t$ takes all positive values $t \geq 0$, while the space coordinate $x$ spans the interval $0 \leq x \leq L$, where the field variable $\phi(x,t)$ satisfies at all times the periodic boundary conditions 
\be 
\phi(x+L,t) = \phi(x,t) \,\,\,.
\label{periodicboundaryconditionsalways}
\ee
Some relevant formulas of various field expansions are collected in Appendix A. For both our prototype models of interest, the $\Phi^4$ Landau-Ginzburg and the Sinh-Gordon models, the Lagrangian density has the general form   
\be 
{\mathcal L} \,=\,\frac{1}{2} (\partial_\mu \phi) - m^2 V(g \phi)  
\,\,\,, \label{generalL}
\ee  
where $m$ is a mass parameter and $g$ the coupling constant. In both models the canonical momentum of the field $\phi$ is given by  $\Pi = \frac{\partial \phi}{\partial x_0}$ and the Hamiltonian is expressed as 
\be 
H[\Pi,\phi] \,=\,\int_0^{L} dx_1 \left[\frac{1}{2} \Pi^2 + \frac{1}{2} \left(\frac{\partial\phi}{\partial x_1}\right)^2 + m^2 V(g \phi) \right] \,\,\,.
\label{generalH} 
\ee
As well known, at the classical level the dependence from the mass and the coupling constant of the theory can be reabsorbed in a proper definition of the field variable and the coordinates:  with the rescaling of both the coordinates and the fields as
\begin{eqnarray}
& t\equiv m x^{0} 
\,\,\,\,\,\,\,\,\,\,
&,
\,\,\,\,\,\,\,\,\,\,
x \equiv m x^1\,\,\,;  \nonumber \\   
& \varphi(x,t) = g \phi(x^0,x^1)  
\,\,\,\,\,\,\,\,\,\,
&,
\,\,\,\,\,\,\,\,\,\,
\pi(x,t) = \varphi_t(x,t)  \label{rescalingfield}
\end{eqnarray}
we end up in the dimensionless Lagrangian density $\hat {\mathcal L}$ 
\be 
{\mathcal L} \,=\,\frac{m^2}{g^2} \hat {\mathcal L} 
\,\,\,\,\,\,\,\,\,\,\,\,
,
\,\,\,\,\,\,\,\,\,\,\,\,
\hat {\mathcal L} = \frac{1}{2} (\partial_{\mu} \varphi)^2 - V(\varphi) \,\,\,.
\ee
The Hamiltonian can be also written in dimensionless form as 
\begin{eqnarray} 
H(\Pi,\phi) &\,=\,& \frac{m}{g^2} \,\hat H(\pi,\varphi)  \,\,\,,\\
\hat H(\pi,\varphi) &\,= & \, \int_0^L dx \left[\frac{1}{2} \pi^2 + \frac{1}{2} \varphi_x^2 + V(\varphi) \right] 
\equiv {\mathcal T}[\pi] + {\mathcal W}(\varphi) \,\,\,.
\label{adimensionalHamiltonian}
\end{eqnarray}
Note that, in addition to the purely potential term $V(\varphi)$, the function ${\mathcal W}(\varphi)$ also contains the square of the space-derivative of the field $\varphi$, a fact that will be important in computing thermal averages of observables which depend only on the field $\varphi$ (see Section \ref{transfermatrixsection}). Together with $H$, another conserved quantity which is always present in our theories is the total momentum $P$ of the field 
\be
P \,=\,\frac{m}{g^2} \,\hat P(\pi,\varphi) 
\,\,\,\,\,\,\,\,\,\,\,\,\,\,
,
\,\,\,\,\,\,\,\,\,\,\,\,\,\,
\hat P(\pi,\varphi) = - \int_0^{L} dx \,\pi(x,t) \,\varphi_x(x,t) \,\,\,.
\ee

In the following we are concerned with the time evolution of the field $\varphi(x,t)$ given by the (dimensionless) equation of motion 
\be 
\varphi_{tt} - \varphi_{xx} + \frac{d V}{d \varphi} \,=\,0 \,\,\,,
\label{dimensionlesseqmt}
\ee  
subjected to the initial conditions 
\be 
\begin{array}{lcl}
\varphi(x,0)  &\,= &\,  f(x) \\
\varphi_t(x,0) & \,=\, & g(x) \,\,\,, 
\end{array}
\ee
where $f(x)$ and $g(x)$ are two assigned functions, both periodic with period $L$ in order to enforce the condition (\ref{periodicboundaryconditionsalways}). Given their periodicity, they can be expanded as 
\be
f(x)  \,=\, A \, \sum_{n=1}^{N_m} c_n \,\cos\left( \frac{2 \pi n}{L} x + 2\pi \gamma_n\right) \,\,\,\,\,\,\,,\,\,\,\,\,\,
g(x)  \,=\,  B \, \sum_{n=1}^{N_m} d_n \,\cos\left( \frac{2 \pi n}{L} x + 2\pi \delta_n\right) \,\,\,. \label{randomfunctionn}
\ee
In order to study a generic evolution of the field, the ideal choice is to take both $f(x)$ and $g(x)$ as random functions, a condition that can be easily implemented by allowing the parameters $c_n$, $d_n$, $\gamma_n$ and $\delta_n$ to be random variables distributed according to certain 
probability distribution, as for instance uniformly  
distributed in the interval $[-1,1]$.  Substituting these initial values of the field and its time derivative into the Hamiltonian (\ref{adimensionalHamiltonian}), the overall constants $A$ and $B$ can be then used to tune the energy density $\hat E = E/L$ to any desired value. Note that, choosing $B =0$, we have identically $P=0$. Finally, varying the integer $N_m$, one has the possibility to start from field configurations with different spreads of the mode number occupations. For what concerns the equilibration process in non-integrable theories, the most interesting situation occurs with those initial conditions in which only few low-energy (infrared) modes are occupied: this because one is mostly interested to observe the energy cascade towards the ultraviolet modes and the consequent thermalization of the system \cite{BDD,FPU}. For integrable models, instead, the mode occupations associated to the action variables are essentially frozen during all the time evolution and, in the absence of such a mixing of the modes, it is then not particularly relevant to concentrate the attention only on field configurations peaked in the infrared region.    

For obtaining a numerical solution of the equation of motion (\ref{dimensionlesseqmt}), one strategy is to discretize both the space and time intervals in units of $a_x$ and $a_t$, with $r\equiv a_t/a_x$. Each space-time point $(x,t)$ is then identified by two integers $(x,t) =  (n \, a_x, m \, a_t) \equiv (n,m)$. Approximating the second derivatives as 
\begin{eqnarray*}
&& \varphi_{tt}(x,t) \simeq \frac{\varphi(n,m+1)+\varphi(n,m-1) - 2 \varphi(n,m)}{a_t^2}  
\,\,\,,\\ 
&& \varphi_{xx}(x,t) \simeq \frac{\varphi(n+1,m)+\varphi(n-1,m) - 2 \varphi(n,m)}{a_x^2}\,\,\,,
\label{derappr}
\end{eqnarray*}
the equation of motion (\ref{dimensionlesseqmt}) then becomes the recursive relation
\begin{eqnarray}
\varphi(n,m+1) & \,=\,&  r^2 \left[\varphi(n+1,m)+\varphi(n-1,m)\right] - \varphi(n,m-1) + \\
&  & \,\,\, + 2 \,\varphi(n,m) \,(1-r^2) - a_t^2 \, \frac{dV}{d\varphi}(n,m) \,\,\,.\nonumber 
\label{recursiveeqm}
\end{eqnarray}
Therefore, assigning at $t=0$ the initial values of the field $\varphi(x,t)$ and its time derivative $\varphi_t(x,t)$  \hspace{1mm} ---  i.e. fixing 
the values of the field on the two first rows $(n,0)$ and $(n,1)$  ---  \hspace{1mm} one can recursively determine the field at any later time and at any point. To ensure enough stability of this algorithm for sufficiently large number of time steps, in addition to take $a_t $ and $a_x$ sufficiently small,  it is also necessary to take $r < 1$.

\subsection{Local observables and their averages}
\begin{figure}[t]
\centering
$\begin{array}{cc}
\includegraphics[width=0.5\textwidth]{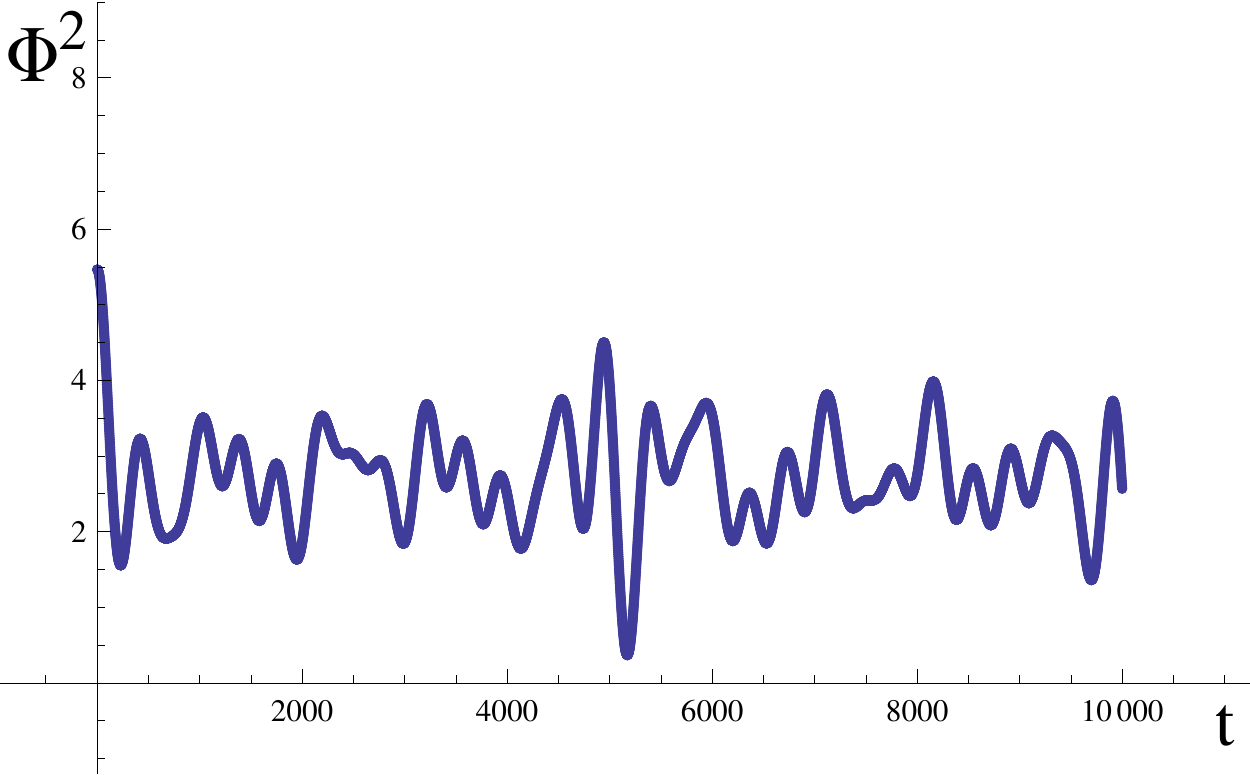} &
\end{array}$
\caption{Plot of $\phi^2$ at a given space point x as function of time for a typical initial state of the Landau-Ginzburg theory 
at $E/L =50$. }
\label{pippo}
\end{figure}

\begin{figure}[b]
\centering
$\begin{array}{cc}
\includegraphics[width=0.5\textwidth]{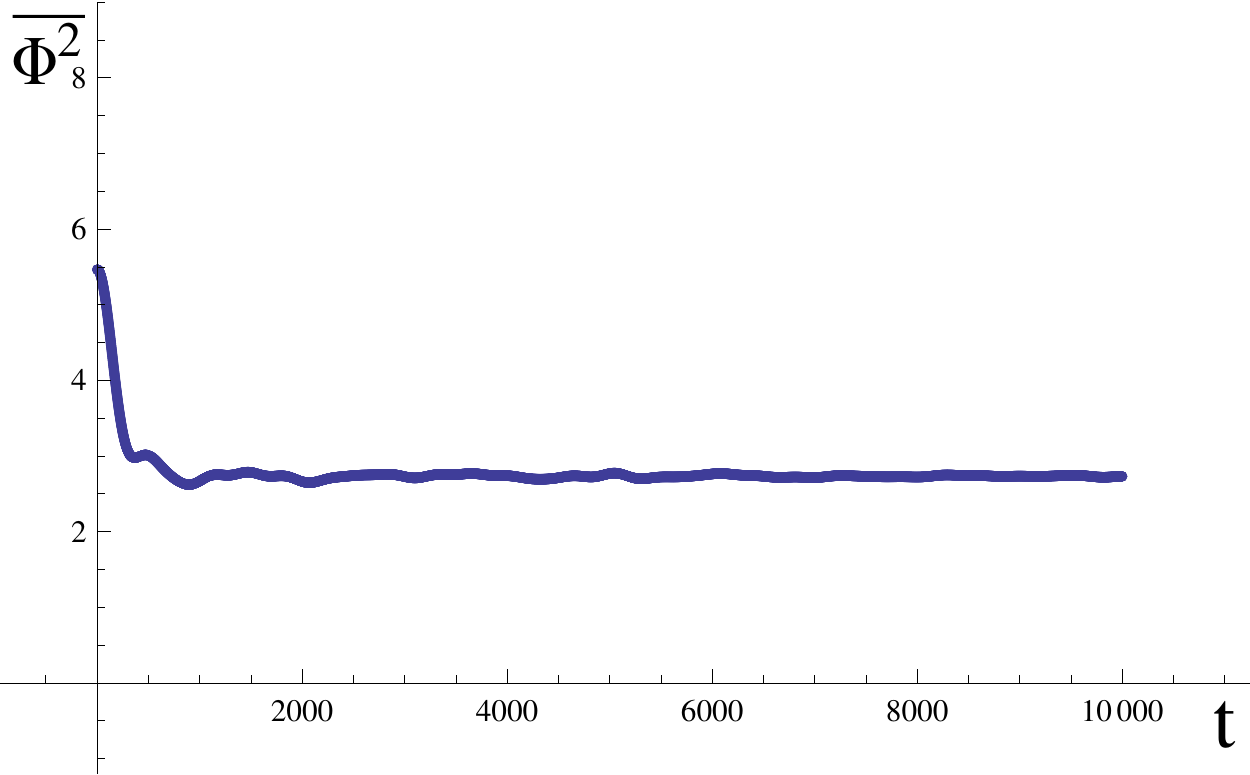} &
\end{array}$
\caption{Plot of the space and time averages of $\phi^2$ as function of time for a typical initial state of the Landau-Ginzburg theory 
at $E/L =50$. }
\label{pippoave}
\end{figure}

Following the time evolution of the field by solving (say numerically) the equation of motion, one can focus the attention on the time average of local quantities which do not trivially vanish for symmetry reasons. As a typical set of operators one can take for instance (in the quantum context it is necessary to consider their normal order version)
\begin{itemize}
\item the family of the even powers of the elementary field, ${\cal O}_k(x,t) \equiv \phi^{2k}(x,t)$; 
\item the family of the $Z_2$ even combination of vertex operators ${\cal V}_\alpha(x,t) \equiv \cosh [\alpha \phi(x,t)]$; 
\item the derivative operators $\phi_t^2(x,t)$ and $\phi_x^2(x,t)$; 
\item the trace of the stress-energy tensor $\Theta(x,t)$ that, in the two theories, takes the form 
\be 
\Theta(x,t) \,=\, 
\left\{
\begin{array}{lll}
\frac{m^2}{2} \phi^2 + \frac{g}{4!} \phi^4 & , & \phi^4 \,\mbox {LG theory} \\
& & \\
\frac{m^2}{g^2} \left[\cosh(g \phi) -1 \right] & , & \mbox{ShG theory} 
\end{array}
\right. 
\label{tracestressenergytensor}
\ee 
\end{itemize}
In classical systems the time dependence of any local observable ${\cal F}[\phi(x,t)]$ is induced by the time dependence of 
the elementary field $\phi(x,t)$ by a simple substitution. The field, as well as all other observables, presents then point-wise persistent fluctuations which do not vanish also at $t \rightarrow \infty$, as shown for instance in Figure \ref{pippo} for the observable $\phi^2(x,t)$. 
A procedure to smooth out these fluctuations and select a set of quantities which are less sensitive to the microscopic 
degrees of freedom of the model consists of a sequence of two types of averages enforced on the local observables: 
\begin{description}
\item[The spatial average] 
\be 
{\cal F}[\phi(x,t)] \rightarrow \overline {\cal F}(t) \equiv \frac{1}{L} \int_0^L dx \, {\cal F}[\phi(x,t)] \,\,\,;
\ee 
\item[The time average] 
\be 
\langle {\cal F} \rangle(t) \,=\, \frac{1}{t} \, \int_0^t dt \, \overline{ {\cal F}(t)}	 \,\,\,.
\ee
\end{description}
In the limit $t\rightarrow \infty$, we have that $\langle {\cal F} \rangle(t) \rightarrow \langle {\cal F} \rangle_{CDA}$. Performing, for instance, these two averages on the observable $\phi^2(x,t)$ of the LG model, we obtain the plot shown in Figure \ref{pippoave}, where the asymptotic constant value of this curve corresponds to the asymptotic average value of the observable. 

\begin{figure}[t]
\centering
$\begin{array}{cc}
\includegraphics[width=0.5\textwidth]{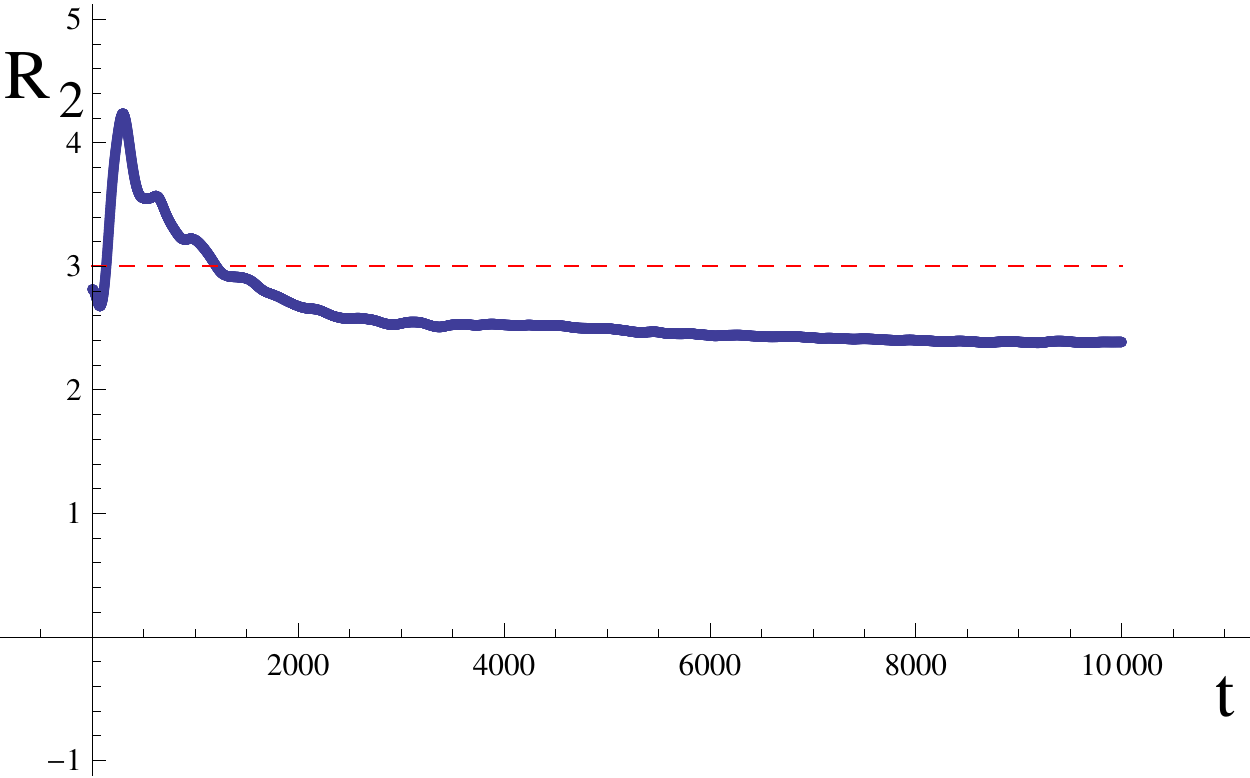} &
\end{array}$
\caption{Plot of the universal ratio $R_2$ as function of time for the Landau-Ginzburg theory 
at $E/L =1000$. The red dashed line corresponds to the asymptotic free value of $R_2$.}
\label{ratiofig}
\end{figure}

In the following it is useful to consider some {\em Universal Ratios}, such as 
\be 
R_k \,=\,\frac{\langle \phi^{2k} \rangle_{CDA}}{(\langle \phi^2 \rangle_{CDA})^k} 
\,\,\,\,\,\,\,\,\,
;
\,\,\,\,\,\,\,\,\,
{\cal R}(\alpha,\beta) \,=\,\frac{\log \langle \cosh [\alpha \phi(x,t)] \rangle_{CDA}}{\log \langle \cosh [\beta \phi(x,t)]\rangle_{CDA}}
\,\,\,.
\label{UniversalRatios}
\ee
For a free bosonic theory (see Section \ref{Sectionfree}), these universal ratios assume the values 
\be 
R_k \,=\, (2 k - 1)!! 
\,\,\,\,\,\,\,\,\,
;
\,\,\,\,\,\,\,\,\,
{\cal R}(\alpha,\beta) \,=\,\left(\frac{\alpha}{\beta}\right)^2 
\,\,\,,
\ee
and therefore any observed violation of the universal ratios from these values can be attributed to the interaction present in the theory, as in the example of LG theory shown in Figure \ref{ratiofig}. 

\subsection{Virial theorem}
One can get a series of identities involving the CDA of local fields by employing the virial theorem \cite{BDD}. These identities only depend on the equation of motion satisfied by the field, therefore their validity does not rely on the integrable or non integrable nature of the theory. In classical mechanics, the virial theorem simply expresses the fact that the time average of quantities which are total time derivatives of bounded functions vanishes. Consider, for instance, the quantity 
\be 
I(t) \,=\,\frac{1}{2} \int_0^L dx \,\phi^2(x,t) \,\,\,, 
\label{startingpoint} 
\ee
where the field $\phi(x,t)$ satisfies periodic boundary conditions. 
Taking its second derivative with respect to time, one has 
\be 
\ddot{I} \,= \, 
\int_0^L dx \left\{ \phi(x,t) \ddot{\phi}(x,t) + \left[\dot{\phi}(x,t)\right]^2 \right\} \,\,\,.
\label{idoubledot}
\ee
If now one uses the equation of motion (\ref{dimensionlesseqmt}) and an integration by part, we can express the quantity 
above as 
\be 
\ddot{I} \,= \, 
\int_0^L dx \left\{ \left[\dot{\phi}(x,t)\right]^2 - [\phi_x(x,t)]^2 - \phi(x,t) \frac{d V}{d \phi} \right\} \,\,\,.
\label{idoubledotgeneral}
\ee
Taking now the time average of left and right sides of this expression, with $\langle \ddot{I}\rangle_{CDA} =0$, and using the 
translation invariance of the EV of the local fields, we arrive to the identity
\be
\langle [\dot{\phi}]^2 \rangle_{CDA} = \langle \left[{\phi_x}\right]^2 \rangle_{CDA} + \langle \phi \, \frac{d V}{d \phi} \rangle_{CDA} .
\label{virial4}
\ee
If we now consider the energy of the theory 
\be 
E \,=\,\int_0^L dx \left[\frac{1}{2} \dot\phi^2 + \frac{1}{2} \phi_x^2 + V(\phi) \right]\,\,\,,
\ee
and its time average, using eq~(\ref{virialphi4}) it is easy to see that  we can equivalently express the energy density 
at equilibrium as 
\be 
\frac{E}{L} \,=\,\langle \dot\phi^2\rangle_{CDA} +  \langle V - \phi \, \frac{d V}{d \phi}\, \rangle_{CDA} \,\,\,.
\label{endenvir}
\ee
Let's now work out explicitly these formulas for the $\Phi^4$ LG and the Sinh-Gordon models. 
\begin{itemize}
\item  {\bf $\Phi^4$ theory}. In this case eq.\,(\ref{virial4}) becomes 
\be
\langle [\dot{\phi}]^2 \rangle_{CDA} \,=\, \langle \left[{\phi_x}\right]^2 \rangle_{CDA} + m^2 \langle \phi^2 \rangle_{CDA} 
+ \frac{g}{3!} \langle \phi^4 \rangle_{CDA} \,\,\,.
\label{virialphi4}
\ee
while eq.\,(\ref{endenvir}) reads 
\be 
\frac{E}{L} \,=\,\langle \dot\phi^2\rangle_{CDA} - \frac{g}{4!} \langle \phi^4 \rangle_{CDA} \,\,\,.
\ee
\item {\bf Sinh-Gordon theory}. For this theory eq.\,(\ref{virialphi4}) becomes 
\be 
\langle [\dot{\phi}]^2 \rangle_{CDA} = \langle \left[{\phi_x}\right]^2 \rangle_{CDA} + \frac{m^2}{g} \langle \phi\,\sinh g\phi \rangle_{CDA} \,\,\,.
\label{virialShG}
\ee
while the energy density can be expressed as  
\be 
\frac{E}{L} \,=\,\langle \dot\phi^2\rangle_{CDA} - \frac{m^2}{2 g} \langle \phi \sinh g\phi \rangle_{CDA} + \frac{m^2}{g^2} 
\langle (\cosh g\phi -1) \rangle_{CDA} \,\,\,.
\ee
\end{itemize}
It is evident that, varying the quantity given in eq.~(\ref{startingpoint}) as the starting point of the procedure, 
the virial theorem permits to derive many other identities for the EV's.

\subsection{Thermalization time scales}
\begin{figure}[t]
\centering
$\begin{array}{cc}
\includegraphics[width=0.69\textwidth]{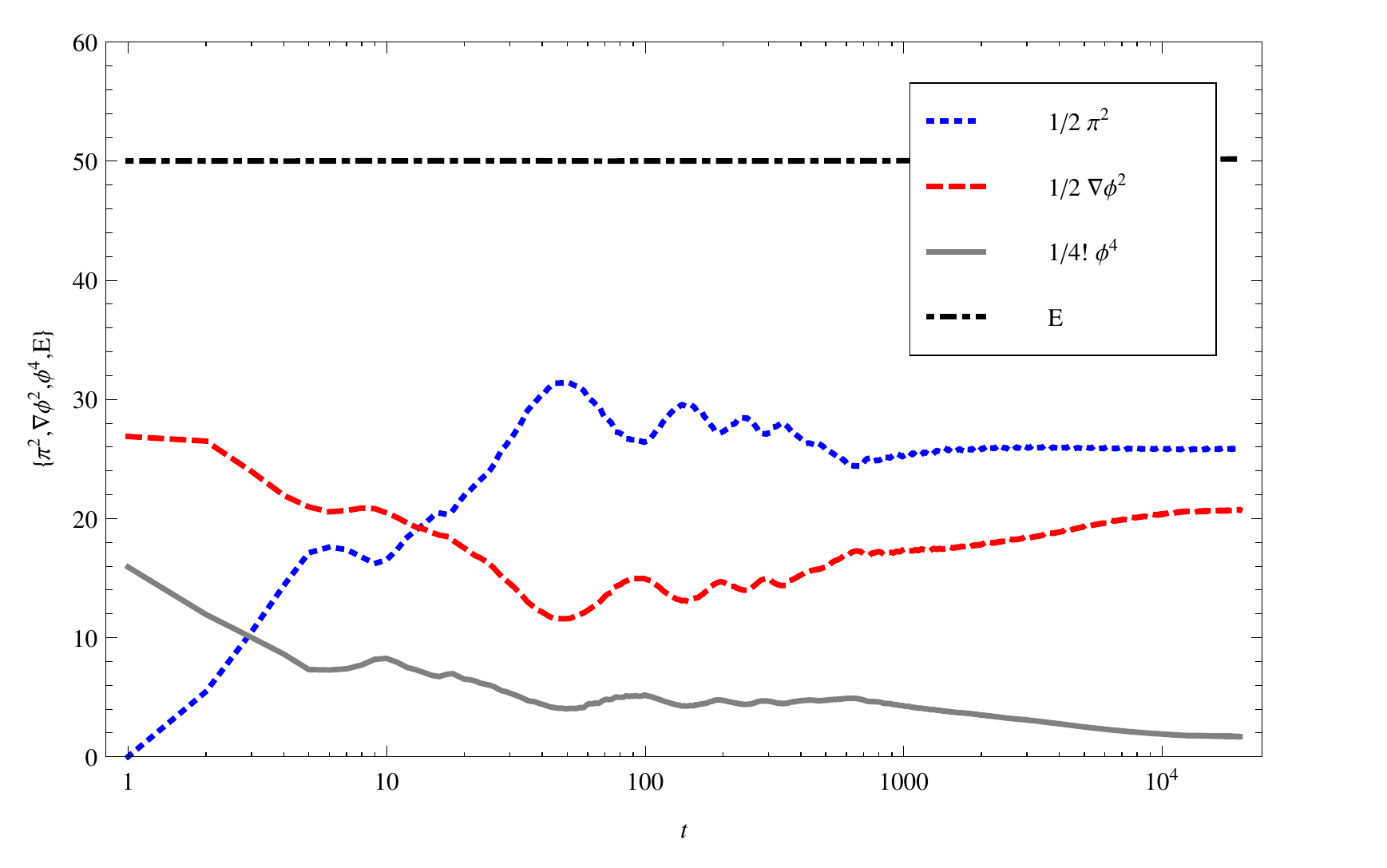} 
\end{array}$
\caption{Time average of several observables in $\Phi^4$ LG theory. The time axis is in logarithmic scale.}
\label{evolutionLLGG}
\end{figure}

Let's consider an initial configuration of the field where only the soft Fourier modes are occupied, i.e. $|\tilde \phi(k,t)| > 0$ only for $k<k_0$ where
\be
\label{phimodes}
\tilde \phi(k,t) = \frac{1}{\sqrt{2\pi}} \int_0^L dx e^{-i k x} \phi(x,t) \;. 
\ee
On the basis of the equipartition principle, we know that, if the system satisfies an ergodic dynamics, after some time all the other modes will be occupied. It is clear that the exchange among the different modes is due to the interaction and, as time goes by, there is a flow of energy from the potential to the kinetic term -- a mechanism that is captured by the virial theorem. If the system asymptotically thermalized and the equilibrium state were described by the Gibbs ensemble measure $e^{-\beta H}$ (for some value of $\beta$ fixed by the initial conditions), for very large $k$, where the mass and interaction terms are negligible, only the space derivative in the Hamiltonian will matter: it follows that $|\tilde \phi(k,t)|^2 \propto k^{-2}$ for large $k$. This power-law decay in the Fourier-transform implies however a non-smooth behaviour for $\phi(x,t)$. Since the time evolution of the field will keep it smooth at any finite time $t$, this implies that a complete thermalization can occur in a classical field theory only in an {\em infinite} amount of time \cite{fucito1982approach}. From a practical point of view one can argue however about the possibility to distinguish three different time regimes \cite{BDD}: 
\begin{itemize}
 \item a small time-scale $t_S$, during which the already occupied modes start to mix, producing a quasi-thermal state, restricted though only to this subset of modes. Roughly speaking, $t_S$ marks the time scale at which the interaction and the gradient terms are of the same order of magnitude. 
\item an intermediate time-scale $t_I$, at which the gradient terms are much larger than all non-linear local terms present in the 
Hamiltonian. This time scale sets the cross-over of two regimes of the dynamics, originally dominated by the interaction terms while later by gradient terms. 
 \item and, finally,  a long time-scale $t_L$ that may be considered as the actual time-scale in which equilibration takes place through a mechanism that we call the \textit{drop-phenomenon}. In this latest stage, higher and higher modes get activated although very slowly, i.e. {\em drop by drop}. This happens simply because the non-linear term in the Hamiltonian has become much smaller than the gradient terms: hence, in this regime, 
 the equation of motion becomes almost the one of a free theory and the interactions are unable to efficiently transfer energy among the modes. This explains why the mixing of the modes becomes slower at larger times, particularly at high $k$ where the interaction is less and less relevant. 
\end{itemize}
Looking at Fig. \ref{evolutionLLGG}, which refers to the time averages of kinetic, gradient and potential terms in the 
$\Phi^4$ LG theory at energy density $E/L =50$, one can spot the short time scale $t_S$ as the point where the potential 
term crosses the kinetic term, i.e. $t_S \simeq\,10^{0.5} $; the intermediate time scale $t_I$ as the point where the curves change their 
concavity, heading to the final stage of equilibration, i.e. $t_I \simeq 10^3$, and the long time scale $t_L \simeq 10^5$ as the point where the quantities reach their asymptotic equilibrium values.   

\begin{figure}[t]
\centering
$\begin{array}{cc}
\includegraphics[width=0.4\textwidth]{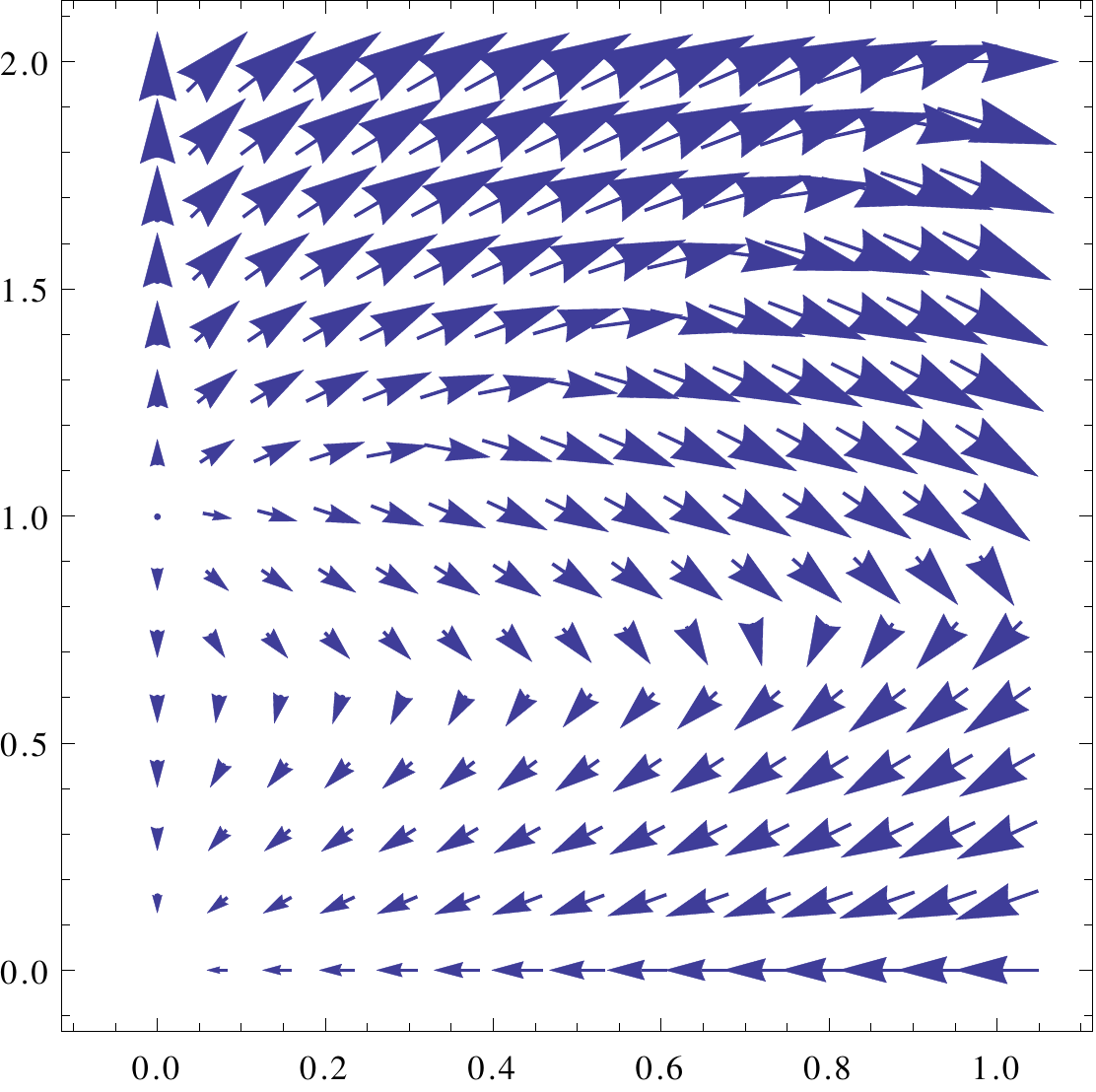} &
\includegraphics[width=0.4\textwidth]{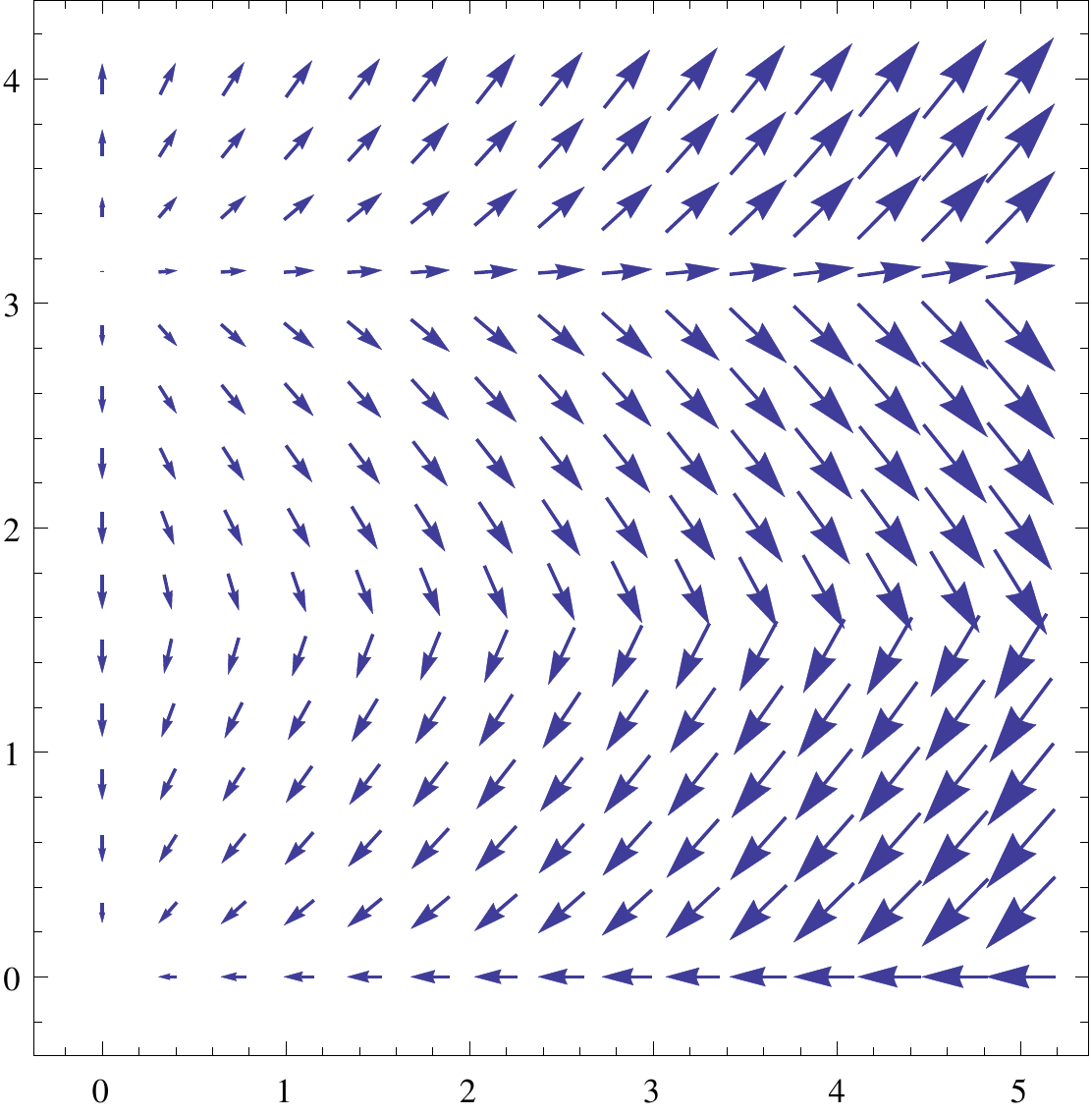}\\
\end{array}$
\caption{Force field, neglecting the space dependence, for Landau-Ginzburg (left) and Sinh-Gordon (right) at $g=m=1$.}
\label{fucitofield}
\end{figure}

It is possible to provide a qualitative estimation of the small and intermediate time scales $t_S$ and $t_I$ following an argument originally presented in \cite{fucito1982approach,bassetti1984complex}. We know that the field $\phi(x,t)$ has to develop singularities when $t\to \infty$. By promoting $x$ to a complex variable, we can see that it happens by the moving of simple poles in the complex $x$ plane toward the real axis, as $t \to \infty$. It follows from \eqref{phimodes} that 
\be
\label{largestPole}
|\tilde \phi(k,t)|^2 \simeq e^{- 2 k y_S (t) } \; ,
\ee
where $y_S(t)$ is the imaginary part of the pole of $\phi(x,t)$ closest to the real axis. 
We now estimate the value of $y_S(t)$. Let's suppose that for small $t$, the second spatial derivative can be neglected in the equation of motion so that, separating real and imaginary parts, we have  
\be
\label{forcefield}
\left\{ \begin{array}{ll}
         \ddot\phi_R & =  -\Re[\partial_{\phi} V(\phi)]\,\,\,;\\
         \ddot\phi_I & = -\Im[\partial_{\phi} V(\phi)]\,\,\,.
        \end{array}\right.
\ee
The corresponding two-dimensional force fields for the $\Phi^4$ LG model and the Sinh-Gordon model are shown in Fig.~\ref{fucitofield}.
It is useful to analyse the two cases in details. 

\vspace{1mm}
{\bf LG theory $\Phi^4$ LG}.  
It is easy to check from eq.~(\ref{forcefield}) and Fig.~\ref{fucitofield} (left) that there is an unstable point on the imaginary axis 
\be
\phi(t=0) \,=\, i \phi_{\mbox{\tiny u}} \,\equiv \,i\, \sqrt\frac{ 6 m^2 }{g} \,\,\,. 
\ee
Therefore, for any $\phi_0 > \phi_u $, 
the time evolution will bring the field at infinity in a finite amount of time
\be
\label{timeLG}
t_{\infty} \,=\, \int_{\phi_0}^\infty (2 V (i \phi) - 2 V(i \phi_0) )^{-\frac 1 2} d\phi \stackrel{\phi_0\gg1}{\simeq}  \sqrt{\frac{12}{g}} \phi_0^{-1}
  \ee
If we then assume to start with a plane-wave initial condition 
$$\phi(z,0) = A \cos(k_0 z)\,,$$ where $z = x + i y \in \mathbb{C}$, 
we expect that, at time $t$, a divergence will appear for all $z$ such that $\Im(\phi(z,0)) \simeq \phi_0 $ in (\ref{timeLG}). 
Setting $z = x + i y$, we have the approximate estimation for the imaginary part of the pole closest to the real axis
$y_S(t)$ in (\ref{largestPole}) 
\be
\label{smalltLG}
 y_S(t) \simeq  -\frac{\log \frac{\sqrt{g} A t}{\sqrt {12}}}{k_0}\,\,\,. 
\ee
This mechanism permits the appearance of divergences if the imaginary part of the initial condition is sufficiently big. 
Otherwise, the field would remain in principle always bounded.  However, in this case, it is crucial the role played by the Laplacian in the equation of motion: it produces fluctuations that are capable of overcoming the barrier $\phi_0 > \phi_u$. Neglecting the interaction, one can estimate this time with the fluctuation in the Gaussian theory, obtaining
\be
\label{intermtLG}
 y_S(t) = m (g \log t k_0^2 A^2)^{-\frac{1}{2}} \,\,\,.
\ee
These two expressions (\ref{smalltLG}) and (\ref{intermtLG}) show the behaviour of the tail of the mode distribution, respectively, for small and intermediate times $t$. The logarithmic dependence on the time $t$ in eq.~(\ref{smalltLG}) is in any case a manifestation of the slow dynamics of the mixing among the modes.

For the $\Phi^4$ LG theory, using an extensive numerical analysis, Boyanovski et al. \cite{BDD} were also able to extract the 
dependence of the long-time scale of thermalization from the energy density and the lattice spacing $a$ 
\be 
t \simeq \left(\frac{\pi}{a}\right)^\alpha \,\left(\frac{E}{L}\right)^{-1} \,\,\,,
\label{longtimethermalization}
\ee
where $0.21 < \alpha < 0.25$ according to the value of $E/L$. 

\vspace{1mm}
{\bf Sinh-Gordon theory}
Even though we know that such an integrable theory never thermalizes, it is anyhow interesting to evaluate its short-time scale.  
The analysis proceeds as in the previous case: taking into account the force field plotted in Fig.~\ref{forcefield} (right), we see 
that for every initial condition of the form $\phi(t=0) = \phi_0 + i \pi$, for arbitrary real $\phi_0$, the field will reach infinite in a finite amount of time given by
\be
\label{timeSG}
t_{\infty} \stackrel{\phi_0\gg1}{\simeq} \frac{\sqrt{2}e^{- \frac{g \phi_0}{2}}}{m}
\ee
As before, taking a plane wave as initial condition, we obtain the estimation
\be
\label{smalltSG}
 y_S(t) \simeq  \frac{1}{k_0} \ln \ln \frac{\sqrt 2}{ m t}
\ee
This expression shows the extremely slow evolution of the field with time, even compared with eq.~(\ref{smalltLG}). Moreover, in this case there is no activation mechanism: no threshold appears in the force field. We remark that this effect can in principle be even stronger once all the details of the exact dynamics are take into account. All these facts show, from a different perspective, the peculiarities of integrable models. As we will argue later for the Sinh-Gordon model, its action variables coming from the Inverse Scattering Transform appear as a smooth deformation of the free ones so that such a dramatic slowing down of the mode mixing finds a natural explanation in the integrable structure of the model. 

For integrable models, the quasi-thermal state reached in the short-time scale has to be promoted to equilibrium state. On one side, this will represent an advantage from the numerical point of view, since the simulation time can be considerably reduced to a much smaller value. On the other side, it poses a technical challenge because the corresponding {\em equilibrium thermodynamics} will have to take into account the constraints set by the initial conditions. A framework to deal with this problem will be developed in Section \ref{GBAEQUATIONS}. In the meanwhile, it is now instructive to see how, in a generic non-integrable model, one can deal with the computation of the thermal values alone. 

\section{The transfer matrix approach for thermal equilibrium}\label{transfermatrixsection}

Assuming that the dynamics will bring the classical system to an equilibrium steady state at $t \rightarrow \infty$, it becomes important to compare the expectation values of the observables got from the time evolution with their values computed by employing an equilibrium ensemble. For non-integrable models such ensemble is provided by the Gibbs Ensemble while for integrable models is given by the Generalized Gibbs Ensemble. In this section we discuss the formalism of the Gibbs Ensemble for classical field theories and we show that its implementation is particularly simple in view to a mapping to a one-dimensional Quantum Mechanics problem via the Transfer Matrix technique. Such a mapping is however not suitable for implementing the Generalized Gibbs Ensemble and this clearly shows the technical difficulty of handling the asymptotic values of various observables in classical integrable models.  

Here we discuss the formalism of the Gibbs ensemble for both non-integrable and integrable models: for integrable models it is equally important to develop the formalism of the canonical ensemble in order to be able to prove or disprove their asymptotic thermalization by comparing the Dynamical Averages of the observables with the Gibbs Ensemble averages.  Let's discuss then this equilibrium thermal formalism. 
\subsection{Thermal averages}  
In the Gibbs ensemble, the thermal averages of the various observables ${\mathcal O}(\pi,\varphi)$ is expressed by a path integral that involves the Hamiltonian   
\be 
\langle {\mathcal O}(\pi,\varphi) \rangle_T \,=\,\frac{
\int {\mathcal D} \varphi \,{\mathcal D}\pi 
\, {\mathcal O}(\pi,\varphi)\,
e^{-\beta H} }
{\int {\mathcal D} \varphi \,{\mathcal D}\pi 
\,e^{-\beta H} }\,\,\,.
\label{thermalpathintegral}
\ee
Notice that, due to the rescaling (\ref{rescalingfield}) of the fields, the variable $\beta$ entering eq. (\ref{thermalpathintegral}) defines an {\em effective} temperature $T$, which is related to the {\em physical} temperature $T_p$ by  
\be
T \,=\, \frac{1}{\beta} \equiv \frac{g^2}{m} T_p \,\,\,.
\label{relationTT}
\ee
Thanks to this relation, fixed the physical temperature $T_p$, one can reach the low-temperature limit $T \ll 1$ also making the coupling constant smaller, i.e. probing the field theory in its weak-coupling regime. Similarly, at fixed $T_p$, the high-temperature limit $T \gg 1$ is equivalent to the strong coupling regime of the theory. The value of the temperature can be fixed in terms of the energy density of the theory, as shown in eq.~(\ref{temperaturefixed}) discussed below. 

For observables that separately depend on $\pi$ and $\varphi$, i.e. ${\mathcal O}(\pi,\varphi) =  {\mathcal O}_1(\pi) \,{\mathcal O}_2(\varphi)$,  the thermal average factorizes 
\be 
\langle {\mathcal O}(\pi,\varphi) \rangle_T \,=\,
\langle {\mathcal O}_1(\pi) \rangle_T \, 
\langle {\mathcal O}_2(\varphi) \rangle_T 
\ee
with 
\begin{eqnarray}
&& \langle {\mathcal O}_1(\pi) \rangle_T \,=\, 
\frac{
\int {\mathcal D} \pi 
\, {\mathcal O}_1(\pi) \,
e^{-\beta {\mathcal T}[\pi] }
}
{\int {\mathcal D}\pi 
\,e^{-\beta {\mathcal T}[\pi] }
}\,\,\,; \label{piaverage}\\
&& \langle {\mathcal O}_2(\varphi) \rangle_T \,=\, 
\frac{
\int {\mathcal D} \varphi 
\, {\mathcal O}_2(\varphi) \,
e^{-\beta {\mathcal W}[\varphi] }
}
{\int {\mathcal D}\varphi 
\,e^{-\beta {\mathcal W}[\varphi] }
}\,\,\,. \label{psiaverage}
\end{eqnarray}
For the observable ${\mathcal O}_1(\pi)$ which depends only on the momentum $\pi$, the path integral reduces to a multiple integral with Gaussian measure. On the other hand, for the observables ${\mathcal O}_2(\varphi)$ which depend only on the field $\varphi$, the corresponding thermal average can be computed in terms of the Transfer Matrix method \cite{Scalapino} or, equivalently, in terms of quantum mechanics formalism, as discussed in the Section \ref{TMQM} below.  

\subsection{Partition function of free theory and determination of the temperature}\label{classicalfreepartitionfunction}
Particularly instructive is the computation of the thermal expectation value of $\pi^2(x)$, because in the continuum this quantity is divergent. It is then necessary to discretize the field theory on a lattice with lattice spacing $a$, with $L = N a$. The coordinates $x$'s will be identified by the corresponding integer $j$, $x \rightarrow j a$. At this stage it is also not very difficult to perform the computation of the full partition function of the free theory with Hamiltonian 
\be 
H = \frac{1}{2} \,\int_0^L dx\,\left[\pi^2(x) + (\phi_x)^2 + m^2 \phi^2 \right] \,\,\,.
\label{freehamiltonian}
\ee
This computation will turn out useful for a later comparison with the classical limit of the quantum partition function done in 
Section \ref{PARTITIONFUNCTIONQUANTUMFREE}. 
The discretization of such an Hamiltonian turns out to be 
\be 
H_d \,=\, \frac{1}{2} a \sum_{j=1}^N \left[
\frac{\pi_j^2}{a^2} + \frac{(\phi_j - \phi_{j-1})^2}{a^2} + m^2 \phi_j^2 \right] 
\,\,\,. 
\ee
Notice that the discretized form of the momentum $\pi$ needs to have a factor $a^{-1}$ in order to preserve the equation of motion. Indeed, with the expression above we have 
\begin{eqnarray*}
\dot{\phi}_i & = & \frac{\delta H_d}{\delta \pi_i} \,=\, \frac{\pi_i}{a} \,\,\,, \\
\dot{\pi}_i & = & \frac{\delta H_d}{\delta \phi_i} \,=\, a \left[\frac{(\phi_{i+1} - 2 \phi_i + \phi_{i-1})}{a^2} - m^2 \phi_i\right] \,\,\,,
\end{eqnarray*} 
and combining the two equations we get the correct discretized version of the equation of motion 
\[
\ddot{\phi_i} \,= \, \frac{(\phi_{i+1} - 2 \phi_i + \phi_{i-1})}{a^2} - m^2 \phi_i \,\,\,,
\] 
where the first term on the right hand side is easily identified with the discretized version of $\phi_{xx}$. With the discretization adopted, the partition function is given by the multi-integral 
\be 
Z^{(0)}(\beta) \,=\,\int \prod_{i=1}^N \left(\frac{d\pi_i \,d\phi_i}{h}\right) \,e^{-\beta H_d} \,\equiv\, e^{-\beta F}\,\,\,.
\label{discretizedpartition}
\ee
Notice that, in order to get a dimensionless expression for $Z(\beta)$, we have introduced the Planck constant $h$ for the phase-space of each degree of freedom. The partition function factorizes as 
\be
Z^{(0)}(\beta) \,=\, h^{-N} \,Z_{\pi}^{(0)} \, Z_{\phi}^{(0)}\,\,\,.
\label{factorizationzzz}
\ee 
Let's first compute $Z_\pi^{(0)}$  
\be
Z_\pi^{(0)} \,=\, \int {\mathcal D}\pi e^{-\frac{\beta}{2} \int_0^L dx\,\pi^2(x)}  
= \int \prod_{i=1}^N \left(d\pi_i \right)\, e^{-\frac{\beta}{2a} \sum_{i=1}^N \pi_i^2}
 =  \left(\frac{2 \pi a}{\beta}\right)^{N/2} \,\,\,.
\ee 
and then $Z_\phi^{(0)}$ 
\begin{eqnarray}
Z_\phi^{(0)} & = & \int {\mathcal D}\phi \,e^{-\frac{\beta}{2}\int_0^L dx \,\left[(\phi_x)^2 + m^2 \phi^2 \right]} = \int \prod_{i=1}^N d\phi_i \,e^{-\frac{\beta}{2}\,a\,\sum_{j=1}^N \left[
\frac{(\phi_j - \phi_{j-1})^2}{a^2} + m^2 \phi_j^2 \right] } 
\end{eqnarray}
This is also a Gaussian integral which can be easily computed by going to Fourier space 
\be 
\phi_n \,=\,\frac{1}{\sqrt{N}}\,\sum_{m=0}^{N-1} \hat\phi_m e^{i k_m a n} \,\,\,,
\ee 
where $k_m = \frac{2 \pi m}{N a}$ ($ m = - N/2,\ldots N/2$). Hence 
\begin{eqnarray}
Z_\phi^{(0)} & = & \int \prod_{i=1}^N d\hat\phi_m \,e^{-\frac{\beta}{2}\,a\,\sum_{m=1}^N \left[
\omega^2_m |\hat\phi_m|^2 \right] } = \left(\frac{2\pi}{\beta a}\right)^{N/2} \,
\prod_{m=-N/2}^{N/2} \frac{1}{\omega_m} \,\,\,,
\end{eqnarray} 
where 
\be 
\omega_m^2 \,=\,m^2 + \left(\frac{\sin\frac{1}{2} k_m a}{\frac{1}{2} a}\right)^2 \,
\,\,. 
\ee
Combining $Z_\pi^{(0)}$ and $Z_\phi^{(0)}$ into eq.~(\ref{factorizationzzz}), in the limit $N\rightarrow \infty$ the free energy $F(\beta)$ is given by 
\be 
\beta F(\beta) \,=\, \frac{L}{2\pi} \int_{-\pi/a}^{\pi/a} dk \, \log[\beta \hbar (m^2 + k^2)^{1/2}] \,\,\,.
\label{freefreeenergy}
\ee
The free energy is an extensive quantity in $L$ but is divergent in the limit $a \rightarrow 0$ for the infinite number of degrees of freedom of the field theory 
\be
\beta F(\beta) \,\simeq \, 
L \left[a^{-1} \,\log(\hbar \beta a^{-1}) -a^{-1} + \frac{m}{2} + {\mathcal O}(a) \right] \,\,\,.  
\label{freefreeenergyazero}
\ee  
The factorization of the partition function into a piece that involves only the momentum and another piece that involves the 
field also holds in an interacting theory. This allows us to compute in general the average of $\pi^2(x)$ 
\be 
\int_0^L dx\, \langle \pi^2(x) \rangle_T \rightarrow 
\frac{1}{a}\,\sum_{i=1}^N \langle \pi_i^2\rangle_T \,=\,
\frac{N}{\beta} =\frac{L}{a} T   \,\,\,, 
\ee
so that 
\be 
\langle \pi^2 \rangle_T \,=\,\frac{T}{a} \,\,\,.
\label{VEpi}
\ee
This expectation value has two features: firstly, it provides a direct measure of the temperature; secondly, it presents an explicit dependence on the cut-off $a$. Since in the continuum $\langle \pi(x) \pi(y) \rangle_T = T \delta(x-y)$, the expression given above then appears as the regularized form of the $\delta(0)$ divergence. eq. (\ref{VEpi}), together with the virial theorem relation (\ref{endenvir}), can be used to fix the value of the temperature in terms of the energy density: in fact, given that  
\be 
\frac{E}{L} \,=\,\langle \pi^2\rangle_{T} +  \langle V - \phi \, \frac{d V}{d \phi}\, \rangle_{T} \,\,\,, 
\ee
and $L = N a$, we have 
\be 
T \,=\, \frac{E}{N} - a  \langle V - \phi \, \frac{d V}{d \phi}\, \rangle_{T}\,\,\,.
\label{temperaturefixed}
\ee
In particular, if in the limit $a\rightarrow 0$ the second term on the right-hand side vanishes, the temperature becomes equal to the energy per degree of freedom.   

\vspace{3mm}

\subsection{Transfer Matrix and Quantum Mechanics} \label{TMQM}
Let's now go back to the problem of performing the path integral for the observables which depend only on the field $\phi(x)$. The path integral (\ref{psiaverage}) involves the weight 
\be
{\mathcal W}(\phi) \,=\,\int_0^L \left[\frac{1}{2} \phi_x^2 + V(\phi)\right] \,\,\,, 
\ee
and functions $\phi(x)$ which take the same value at the ends of the interval $[0,L]$, $\phi(0) = \phi(L)$. One can then interpret the coordinate $x$ as {\em euclidean time} $\tau$ and therefore convert the path integral in the euclidean time interval $0 \leq \tau \leq L$ into a quantum trace \cite{Scalapino,BDD} 
\be 
\langle {\mathcal O}(\phi) \rangle_T \,=\, 
\frac{
\int {\mathcal D} \phi 
\, {\mathcal O}(\phi) \,
e^{-\beta {\mathcal W}[\phi] }
}
{\int {\mathcal D}\phi 
\,e^{-\beta {\mathcal W}[\phi] }
}\,=\,\frac{
{\rm Tr} \left[ e^{-L \hat H} {\mathcal T}\{{\mathcal O}(\phi)\} \right]}
{{\rm Tr} \left[ e^{-L \hat H}\right]}\,\,\,,
\label{quantumtracee}
\ee
where the quantum Hamiltonian $\tilde H$ is given by 
\be
\tilde H \,=\, \frac{p^2}{2\beta} + \beta\, V(q) \,\,\,. 
\label{quantumHamiltoniantransfer}
\ee
Here $q \equiv \phi$, while $p$ is the conjugate momentum of this coordinate variable, with canonical commutation relation $[q,p] = i$. In such a formalism, the original classical field $\phi(x)$ is promoted to be a quantum operator, subjected to the time evolution of the Heisenberg representation $\phi(x) = e^{x \tilde H} \phi(0) e^{-x \tilde H}$.  The operator $ {\mathcal T}$ in eq. (\ref{quantumtracee}) implements the time ordering along $x$, i.e. 
\[
 {\mathcal T}\{\phi(x_1) \phi(x_2)\} = 
 \theta(x_1 - x_2) \phi(x_1) \phi(x_2) + 
 \theta(x_2-x_1) \phi(x_2) \phi(x_1) \,\,\,.
 \]
The temperature $T =1/\beta$ enters the quantum Hamiltonian $\tilde H$ as a parameter and therefore both its eigenvalues ${\mathcal E}_n(T)$ and the corresponding eigenfunctions $\chi_n(q,T)$ depend on $T$. These eigenfunctions satisfy the orthogonality conditions  
$$
\int_{-\infty}^{+\infty} 
dq \,\chi_n(q,T)\,\chi_m(q,T) \,=\,\delta_{n,m} \,\,.
$$
For local observables, the translation invariance of the theory implies that the thermal average is independent from the point $x$, $\langle {\mathcal O}(\phi(x)) \rangle_T \,=\,\langle {\mathcal O}(\phi(0)) \rangle_T$. In this formalism the thermal expectation value of such observables ${\mathcal O}(\phi)$ is expressed by 
\be 
\langle {\mathcal O}(\phi) \rangle_T \,=\,
\frac{1}{Z(T)} \sum_{n=0}^{\infty} {\mathcal O}_{nn}(T) e^{-L {\mathcal E}_n(T)} \,\,\,, 
\label{spectraltrace}
\ee
where 
\be
{\mathcal O}_{nn}(T) \,=\,\int_{-\infty}^{+\infty} 
dq \, {\mathcal O}(q) \,\chi_n(q,T)\,\chi_n(q,T) \,\,\,
\ee
\be
Z(T) \,=\,\sum_{n=0}^{\infty} e^{-L {\mathcal E}_n(T)}\,\,\,.
\ee
If there is a finite gap $\Delta = ({\mathcal E}_1(T) - {\mathcal E}_0(T))$ in the spectrum of $\tilde H$ for any value of $T$, in the limit $L \rightarrow \infty$ the thermal expectation value of the local observables ${\mathcal O}(\phi)$ simply reduces to the matrix element of the operator ${\mathcal O}(q)$ on the ground state up to exponentially small terms 
\be 
\langle {\mathcal O}(\phi) \rangle_T \,\simeq\,
{\mathcal O}_{00}(T) + o(e^{-L \Delta})
\,\,\,\,\,\,\,\,\,\,\,
, 
\,\,\,\,\,\,\,\,\,\,\,
L \gg 1 
\label{spectraltrace00}
\ee
Examples of this formula will be give below for various observables and various theories. 

\subsection{Thermal expectation values of the $\Phi^4$ theory}
The formalism of the transfer matrix of the $\Phi^4$ theory has been extensively studied in \cite{BDD} and here we just report the main findings. First of all, making the canonical transformation 
$
q \rightarrow \frac{1}{\sqrt{T}} q 
$, 
$p \rightarrow \sqrt{T} p 
$, 
the adimensional version of the quantum Hamiltonian (\ref{quantumHamiltoniantransfer}) can be written as  
\be
\tilde H \,=\, \frac{1}{2} (p^2 + q^2) + \frac{T}{4!} \, q^4 \,\,\,. 
\label{quantumHamiltoniantransferphi4}
\ee
It is useful to discuss the low and the high temperature limits. 

\vspace{1mm}
\noindent
{\bf Low-temperature.} In limit $T \rightarrow 0$, the thermal expectation values are determined by the vacuum expectation values of the quantum harmonic oscillator and therefore we have 
\be 
\langle \varphi^2 \rangle_T \,=\,\frac{T}{2} 
\,\,\,\,\,\,\,\,\,
,
\,\,\,\,\,\,\,\,\,
\langle \varphi^{2k} \rangle_T \,=\,\left(\frac{T}{2}\right)^k \, (2 k-1) !!  
\ee
and therefore for the Universal Ratios $R_k$ we have 
\be 
R_k \,=\, (2 k -1)!! \,\,\,.
\ee
In this limit it is also simple to compute the EV of the $\cosh(a \phi)$ operators 
\be
\langle \cosh(a \varphi) \rangle_T \,=\,\exp\left[\left(\frac{\alpha}{2}\right)^2 T \right]\,\,\,,
\label{vertexoperatorVEfree}
\ee
and therefore for the Universal Ratio ${\cal R}(\alpha,\beta)$ we have 
\be 
{\cal R}(\alpha,\beta) \,=\,\left(\frac{\alpha}{\beta}\right)^2 \,\,\,.
\ee 

\vspace{1mm}
\noindent
{\bf High-temperature.} In the limit $T\rightarrow \infty$, it is convenient to rescale $q$ and $p$ according to the canonical transformation 
$ 
q \rightarrow \left(\frac{T}{12}\right)^{1/6}\, q 
$
,
$p \rightarrow \left(\frac{12}{T}\right)^{1/6} \, p 
$, 
so that the quantum Hamiltonian (\ref{quantumHamiltoniantransferphi4}) becomes
\be 
\tilde H \,=\,\frac{1}{2} \left[p^2 + q^4 + \left(\frac{12}{T}\right)^{2/3} \, q^2 \right] \,\,\,,
\ee
i.e. the Hamiltonian of a quartic oscillator perturbed by a quadratic term. In particular, solving numerically the Schr\"odinger 
equation for the quartic oscillator in the limit $T \rightarrow \infty$, one finds \cite{BDD} 
\be 
\lim_{T \rightarrow \infty} \frac{\langle \varphi^2 \rangle_T}{T^{2/3}} \,=\,0.456119... 
\,\,\,\,\,\,\,\,\,
,
\,\,\,\,\,\,\,\,\,
\lim_{T \rightarrow \infty} \frac{\langle \varphi^4 \rangle_T}{T^{4/3}} \,=\,0.56104... 
\,\,\,\,\,\,\,\,\,
\ee
and for the Universal Ration $R_2$ we have 
\be 
\lim_{T\rightarrow \infty} R_2 \,=\, 2.69673...
\ee

\subsection{Thermal expectation values of the Sinh-Gordon theory}

Although the asymptotic values of local operators of the Sinh-Gordon theory are expected to follow from a Generalized Gibbs Ensemble, it is nevertheless worth deriving their thermal values. For two reasons: (i) to compare these thermal values with the ones obtained by the time average and observing, in general, their discrepancy; (ii) to get the asymptotic values of local operators if the initial state is properly chosen to be thermal.  

\begin{figure}[t]
\centering
$\begin{array}{cc}
\includegraphics[width=0.5\textwidth]{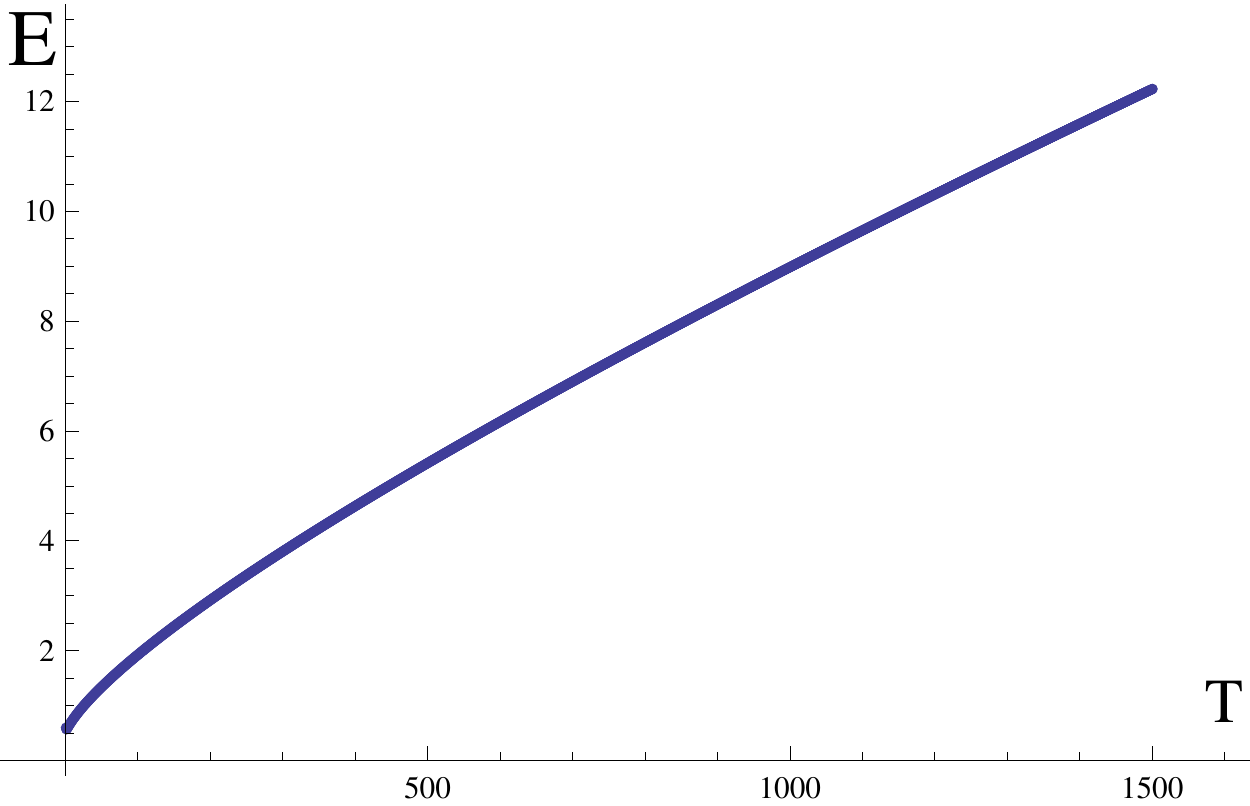} &
\end{array}$
\caption{$\tilde {\mathcal E}_0(T)$ as function of the temperature $T$, as obtained by the variational method.}
\label{hh}
\end{figure}

The adimensional version of the quantum Hamiltonian associated to the transfer matrix is given in this case by 
\be 
H\,=\,\frac{T}{2} p^2 + \frac{1}{T} (\cosh q -1) \,\,\,, 
\label{quantumHamiltonianShG}
\ee
and the corresponding Schr\"odinger equation is the modified Mathieu equation. The exact determination of the ground state energy ${\mathcal E}_0(T)$ of this Schr\"odinger equation is presented in Appendix B by using the Bethe Ansatz techniques developed in Section \ref{GBAEQUATIONS}. However a good approximation of the ground state energy ${\mathcal E}_0(T)$ and the ground state wave-function of the Hamiltonian 
(\ref{quantumHamiltonianShG}) can be obtained by using as a variational wave function  
\be
\psi_a(q) \,=\,\frac{1}{2 \sqrt{K_0(2 a)}} \,\exp{\left(-a \cosh\frac{q}{2}\right)} \,\,\,,
\label{trialwfSh}
\ee
where $a$ is a parameter. The functional form of $\psi_a(q)$ has been chosen to match the behaviour of the exact solution as $|q| \rightarrow \infty$. In the following $K_\nu(x)$ denotes the modified Bessel function of order $\nu$. We can easily compute 
\be
H(a) \equiv \langle \psi_a | H | \psi \rangle_a =
\frac{T}{16} \frac{a K_1(2 a)}{K_0(2 a)} + \frac{1}{T} \left(\frac{K_2(2 a)}{K_0(2 a)}-1\right) 
\,\,\,.
\ee
\begin{figure}[t]
\centering
$\begin{array}{cc}
\includegraphics[width=0.5\textwidth]{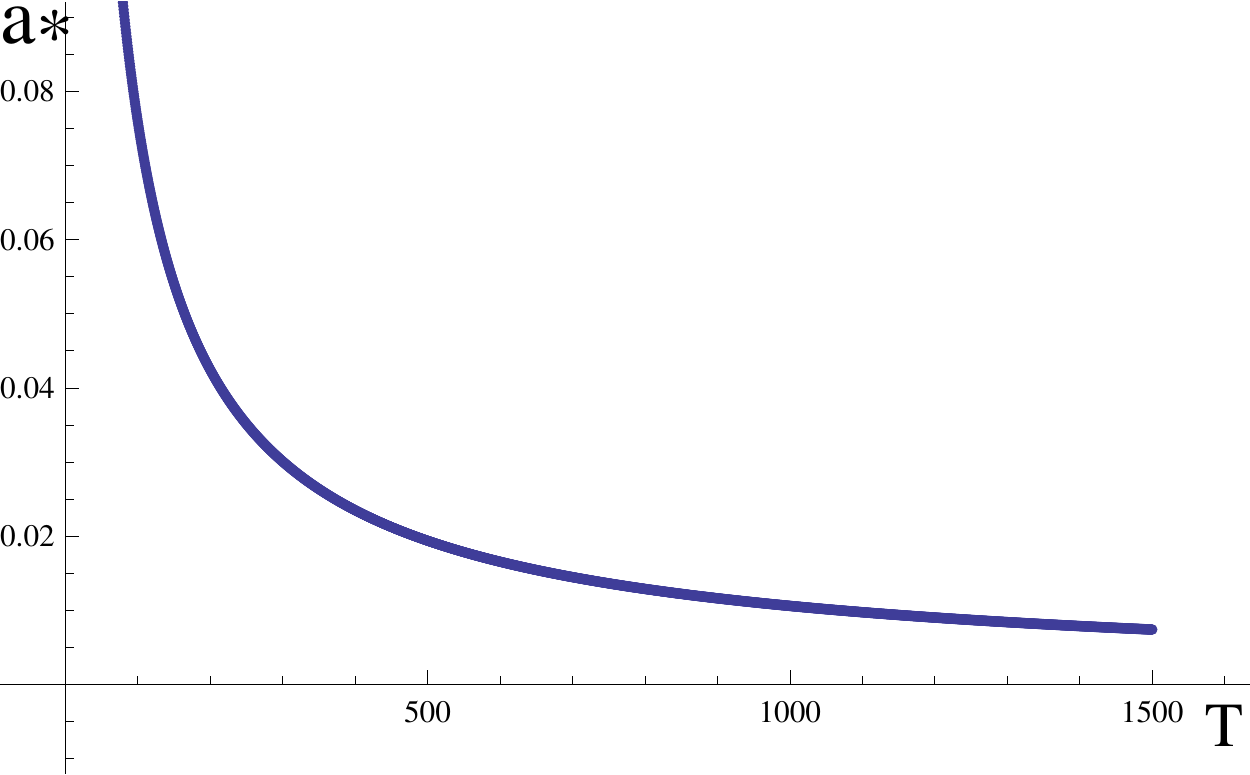} &
\end{array}$
\caption{ $a^\star$ as a function of the temperature $T$, obtained by the condition $\partial_a H(a)=0$.}
\label{aT}
\end{figure}
The value $a^\star(T)$ which minimizes $H(a)$ is determined by the condition $\partial_a H(a) = 0$ and is plotted in Figure \ref{aT}.  
Substituting this function $a^\star(T)$ into $H(a)$, we get an estimate of the ground state energy ${\mathcal E}_0(T)$
\be
{\mathcal E}_0(T) \leq \tilde {\mathcal E}_0 = H(a^\star(T)) \,\,\,.
\ee
The plot of $\tilde {\mathcal E}_0(T)$ is shown in Figure \ref{hh}. This function starts at $T=0$ from $1/2$ (the ground state of the harmonic oscillator) and, for $T\rightarrow\infty$, asymptotically grows linearly with logarithmic correction,  ${\mathcal E}_0(T) \simeq C\, T/\log T$, with $C \simeq 0.05...$.  
 
\begin{figure}[b]
\centering
$\begin{array}{cc}
\includegraphics[width=0.5\textwidth]{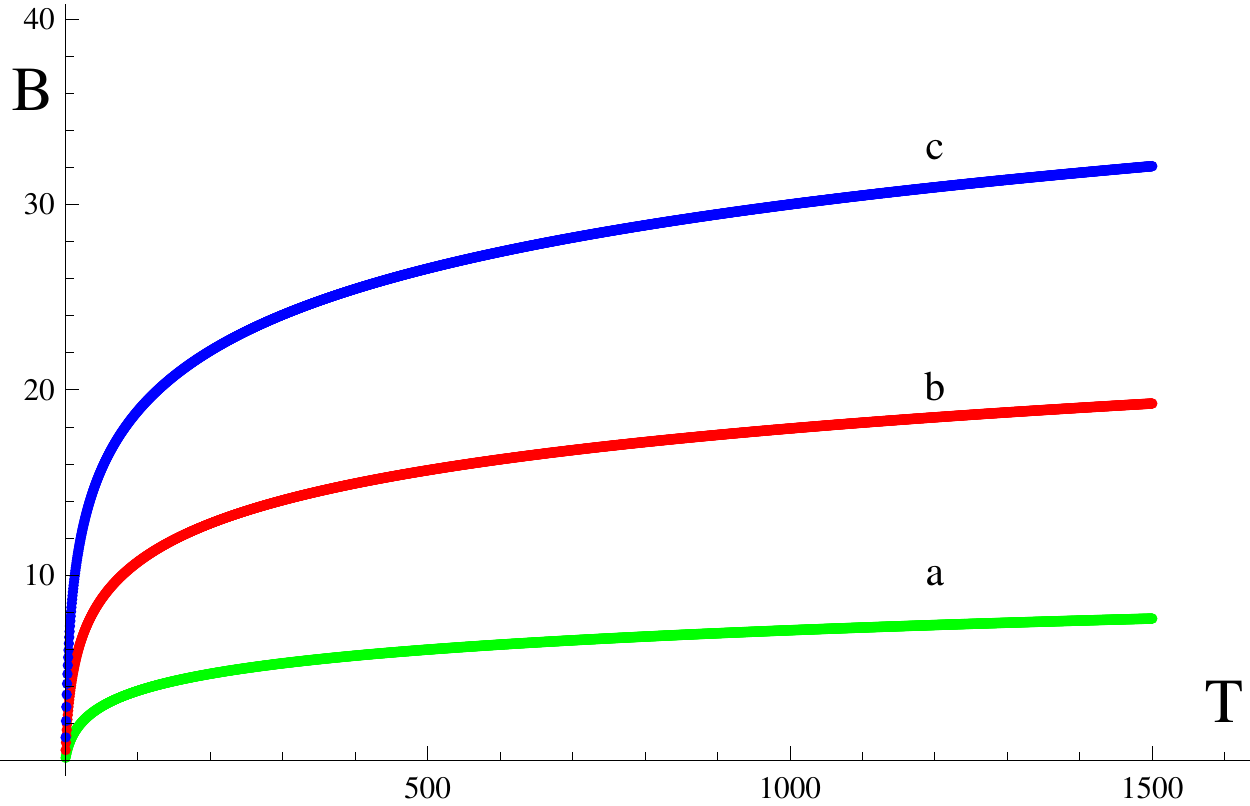} &
\end{array}$
\caption{$B\equiv \log {\mathcal V}(\alpha,T)$ as a function of the temperature $T$ for $\alpha = 1$ (curve $a$), $\alpha = 2$ (curve $b$) and 
$\alpha = 3$ (curve $c$). }
\label{bg}
\end{figure}

Using the variational wave-function (\ref{trialwfSh}) we can now easily compute the thermal expectation values of the vertex operator 
\be
{\mathcal V}(\alpha,T) \equiv \langle \cosh \alpha \varphi \rangle = \int_{-\infty}^{+\infty} \cosh(\alpha q) \psi_a^2(q) dq \,=\,
\frac{K_{2 \alpha}(2 a^\star(T))}{K_0(2 a^\star(T))} \,\,\,.
\label{thermalcosh}
\ee
At low temperature, using the divergent behaviour of $a^\star(T)$ at $T \rightarrow 0$ and the asymptotic values of the Bessel functions, ${\mathcal V}f(\alpha,T)$ reduces to the free theory results (\ref{vertexoperatorVEfree}) 
\be
{\mathcal V}(\alpha,T) \simeq \left(1+ \frac{T \alpha^2}{4} \right) \simeq e^{T \alpha^2/4} \,\,\,.
\ee
At high-temperature, $a^\star(T)$ goes to zero and the expectation value is given by the 
short distance expansion of the Bessel functions 
\be 
{\mathcal V}(\alpha,T) \simeq -\frac{1}{2} \frac{\alpha(2 \alpha)}{a^{\star 2 \beta}(T)} \,\left(\frac{1}{\log a^\star(T) - \gamma}\right)
\,\,\,,
\ee
where $\gamma$ is the Euler-Mascheroni constant. The plot of $\log {\mathcal V}(\alpha,T)$ as  function of $T$ for various values of $\alpha$ is shown in Figures \ref{bg}.

\subsection{Difficulties to implement the Generalized Gibbs Ensemble in classical field theories}

The Transfer Matrix approach proves to be a very efficient tool to compute the thermal expectation values of local operators in the Gibbs Ensemble. At the basis of this approach there is the interpretation of the term ${\cal W}(\varphi)$ in the Hamiltonian as Euclidean action of a one-dimensional particle, with the original spatial gradient term $\left(\frac{\partial \varphi}{\partial x}\right)^2$ interpreted as the kinetic term for the one-dimensional problem. This has allowed us to adopt the operator formalism of Quantum Mechanics and to convert the original path integral problem (\ref{quantumtracee}) into the problem of finding the spectrum of the Schr\"odinger operator (\ref{quantumHamiltoniantransfer}), in particular its ground state energy and wave function. 

Unfortunately this route cannot be followed to compute Expectation Values of local operators in classical integrable theories by using the Generalized Gibbs Ensemble. There are indeed a series of difficulties in handling an expression as 
\be 
\langle {\mathcal O}(\pi,\varphi) \rangle_{GGE} \,=\,\frac{
\int {\mathcal D} \varphi \,{\mathcal D}\pi 
\, {\mathcal O}(\pi,\varphi)\,
e^{-\beta_i {\mathcal Q}_i} }
{\int {\mathcal D} \varphi \,{\mathcal D}\pi 
\,e^{-\beta_i {\mathcal Q}_i} }\,\,\,, 
\label{GGEpathintegral}
\ee
obtained by substituting in (\ref{thermalpathintegral}) $\beta H$ $\rightarrow $ $\beta_i {\mathcal Q}_i$ (sum on the index $i$ is understood), where 
${\mathcal Q}_i$ are the (infinite) number of local conserved charges. The {\em cahier de doleances} include: 
\begin{itemize}
\item first of all, the determination of all the Lagrange multiplier $\beta_i$ of the conserved charges $Q_i$. In principle, they can be fixed by computing the Expectation Values of the conserved charges $Q_i$, therefore solving the infinite dimensional system of equations for the $\beta_i$'s given by  
\be 
\langle {\mathcal Q}_k(\pi,\varphi) \rangle_{GGE} \,=\,\frac{
\int {\mathcal D} \varphi \,{\mathcal D}\pi 
\, {\mathcal Q}_k(\pi,\varphi)\,
e^{-\beta_i Q_i} }
{\int {\mathcal D} \varphi \,{\mathcal D}\pi 
\,e^{-\beta_i Q_i} }\,\,\,.
\label{systempathintegral}
\ee 
However, even assuming to be able to compute the path integrals on the right hand side of these expressions, the solution of these infinite number of transcendental equations is far from being obvious. 
\item secondly, the computation of the path integrals (\ref{GGEpathintegral}) at a given value of the $\beta_i$'s. If one employs the local conserved charges, as shown in Section \ref{ConservedChargesShG} and in Appendix \ref{CCShGexplicit}, these quantities are given by integrals of local densities made of higher partial derivatives both in $x$ and $t$. Moreover, these higher derivative terms are also mixed up with local expressions in $\varphi$, preventing any obvious "one-dimensional " interpretation of these terms as it was the case for the path integral in the Gibbs ensemble. 

The higher degrees of these partial derivatives is also an obstacle in setting up, even numerically, a Transfer Matrix approach because, once discretized, they coupled together sites arbitrarily far away one from the other. This feature therefore spoils the meaning itself of the Transfer Matrix which typically couples only sites separated by one or two lattice spacings).

\item thirdly, the absence of a "trace formula". Let's assume that one is able to express all the terms in the conserved charges ${\mathcal Q}_i$ which contain partial derivative w.r.t. the time $t$ in terms of partial derivative in $x$ by using the equation of motion (an operation that is however far from obvious and probably even false). Let's call the resulting expressions $\tilde {\mathcal Q}_i$.  Then one may conceive the idea to pose the path integral (\ref{GGEpathintegral}) equal to the "quantum trace" of a suitable operator $\hat {\mathcal G}$, namely
  \be 
\langle {\mathcal O}(\varphi) \rangle_GGE\,=\, 
\frac{
\int {\mathcal D} \varphi 
\, {\mathcal O}(\varphi) \,
e^{-\beta_i \tilde {\mathcal Q}_i }
}
{\int {\mathcal D}\varphi 
\,e^{-\beta {\mathcal W}[\varphi] }
}\,=\,\frac{
{\rm Tr} \left[ e^{-L \hat {\mathcal G}} {\mathcal T}\{{\mathcal O}(\varphi)\} \right]}
{{\rm Tr} \left[ e^{-L \hat {\mathcal G}}\right]}
\label{quantumtraceeG}
\ee
Differently from the Gibbs Ensemble, where the corresponding operator $\hat {\mathcal G}$ is linear and a second order 
derivative operator, in this case the operator $\hat {\mathcal G}$ is a non-linear and infinite order differential operator, for which 
there is no consolidate mathematical literature which ensures the completeness of its spectrum, the monotonic 
increasing of its eigenvalues and even the existence of its "ground state" eigenvalue and eigenfunction. 

In other words, establishing the validity of an identity as the one given in (\ref{quantumtraceeG}) is a very interesting 
problem in the subject of classical integrable models but, presently, this is an impracticable route. This forces us to follow another 
approach for computing the Generalized Gibbs Ensemble averages, the one based on Integrable Quantum Field Theory. 
 
\end{itemize}

\newpage

\begin{center}
{\Large {\bf PART B}}
\end{center}

\vspace{1mm}

This part of the paper, which includes Section \ref{classicallimit} till Section \ref{summarynextsteps}, concerns with two basic quantities of Integrable Quantum Field Theory: 
\begin{itemize}
\item the connected Form Factors of local operators $\3pt{\overleftarrow{\theta_n}}{\mc O(0)}{\overrightarrow{\theta_n}}_\text{conn}$; 
\item the filling fraction $f(\theta)$ of the particle states of the out-of-equilibrium system. 
\end{itemize}
We recall that these quantities enter the LM formula of the GGE average, eq.\.(\ref{EnsembleAverageMus23}), and allow us to compute the Quantum Dynamical Average according to the basic identity (\ref{BasicIdentity}). Hence, the aims of the next Sections are twofold: (a) firstly, to define all quantities entering the GGE average given in eq.\,(\ref{EnsembleAverageMus23}); (b) secondly, to understand how to extend their definition to the classical case, in order to set an identity similar to eq.\,(\ref{BasicIdentity}) for classical integrable field theories. For this reason we are interested in studying quantum field theory properties and their limit when $\hbar \rightarrow 0$.  

\section{The classical limit of quantum fields}\label{classicallimit} 
It is well known that the classical field theory can provide useful insights on the (non-perturbative) structure of a quantum field theory: this is the case, for instance, of soliton solutions of classical field equations \cite{Rajaraman}. But it is also true the vice-versa, alias a quantum field theory may allow us to have access, in a proper limit, to classical quantities. What makes the difference between the classical and quantum theories is of course the presence of $\hbar$ which enters the commutation relations of conjugate variables. One then expects that taking the limit $\hbar \rightarrow 0$ of a quantum theory should provide a way to recover the classical results. Although there may be certain subtleties going on in this limit (see, for instance, \cite{Brodsky}), the standard implementation of this procedure turns out to be useful for our future purposes. The first thing to do is to restore the presence of $\hbar$ and consider the path integral formulation of a bosonic theory 
\be 
Z\,=\,\int {\mathcal D}\phi \exp\left\{\frac{i}{\hbar} {\mathcal S}[\phi] \right\} \,\,\,.
\label{partitionfunctionh}
\ee
In the limit $\hbar \rightarrow 0$, the rapidly varying phase ${\mathcal S}/\hbar$ selects field configurations for which the action ${\mathcal S}$ is stationary, i.e. the solutions of the classical equation of motion. But there is more than that, because a systematic expansion in $\hbar$ allows us to organize differently the perturbative solution of a quantum field theory: as well known, it provides an expansion in the {\em number of loops} of the Feynman diagrams \cite{Amit}. For a Lagrangian made of a quadratic part ${\mathcal L}_0 = \frac{1}{2} (\partial_{\mu} \phi)^2 - m^2 \phi^2$ and interaction term $V(\phi)$, the path integral in the presence of an external current $J(x)$ can be written as 
\be 
Z\{ J \} \,=\, {\mathcal N} \, \exp\left\{i \hbar^{-1} \int dx \, V\left(\frac{\delta}{\delta J}\right) \right\} \, 
\exp\left[\frac{i}{2} J(x) (\hbar G_0(x-y)) J(y) dx dy \right] \,\,\,.
\label{partfunctcurrent}
\ee
where 
\be 
G_0(x-y)\,=\,- \langle x |\, (\Box + m^2)^{-1} \,| y\rangle \,=\,\int \frac{dk}{2 \pi} \frac{e^{-i k (x-y)}}{k^2 - m^2 + i \epsilon }\,\,\,
\ee
is the propagator. Hence, with respect to the case when $\hbar$ is put equal to 1, the changes consist in multiplying each interaction vertex by $\hbar^{-1}$ and every internal line (given by the propagator) by $\hbar$. In this way, any previous Feynman diagram of the $n$th order in perturbation theory which involves $E$ external points, $I$ internal lines and $n_r$ vertices of $r$ legs, gets multiplied by $\hbar^{I-n_r}$. But a simple combinatoric argument relates $I$ to the number of loops $L$ of the diagram, 
\be 
L = I - (n_r-1)
\label{topologicalrelation}
\ee
and therefore the original diagram gets multiplied by $\hbar^{L-1}$. This argument shows that in the $\hbar \rightarrow 0$ limit the leading contributions come only from tree level diagrams, which can be regarded as those terms coming from the iterative perturbative solution of the classical non-linear equation of motion, as we are going to show below. 

\vspace{3mm}
\noindent
{\bf Role of $\hbar$}. It is important to comment more on the role of $\hbar$ in the classical field theory and the computation we are going to present. In the pure classical formalism there is of course no trace of $\hbar$. On the other hand, from the arguments given above, it is clear that the role of $\hbar$ simply consists in selecting the "classical" terms present in the full quantum theory. So, even though $\hbar$ will finally disappear from the classical expressions by getting reabsorbed into the definition of the various quantities, a rule of thumb to simply get rid of $\hbar$ from the final classical expression is the following: 

\vspace{1mm}
{\bf Rule of thumb:} {\em Select the classical quantities by initially restoring $\hbar$ and then considering the limit $\hbar \rightarrow 0$, expanding 
correspondingly the quantum expressions. Once the classical quantities are identified and extracted in this way, take $\hbar =1$ in the remaining expressions.}  

\subsection{Tree Level Diagrams for the Elementary Field}
As examples of our perturbative considerations, let's analyse the $\Phi^4$ Landau-Ginzburg theory and the Sinh-Gordon model. 

\vspace{0.5cm}
\noindent
{\bf $\Phi^4$ Landau-Ginzburg theory}. Let's consider the purely {\em classical} equation of motion of this theory  
\be 
(\Box + m^2) \phi \,=\,- \,\frac{g}{3!} \,\phi^3 \,\,\,, 
\label{classicalphi44}
\ee
and let's look for its solution in terms of a series expansion 
\be
\phi(x) \,=\,\sum_{n=0}^\infty  g^n \, \phi_n \,\,\,.
\ee
Substituting into (\ref{classicalphi44}) and matching the powers in $g$, we get the iterative equations for $\phi_n$ 
\be 
(\Box + m^2) \phi_n \,=\,-\,\sum_{\substack{k,l,m \\ k+l+m+1=n}} \phi_k \phi_l \phi_m \,\,\,, 
\ee
i.e. 
\begin{eqnarray}
(\Box + m^2) \phi_0 & = & 0 \,\,\,,\nonumber \\
(\Box + m^2) \phi_1 & = & -\,\frac{1}{3!} \,\phi_0^3 \,\,\,,\\
(\Box + m^2) \phi_2 & = & -\,\frac{1}{2}\, \phi_0^2 \phi_1 \,\,\,,\nonumber \\
(\Box + m^2) \phi_3 & = & -\,\frac{1}{2}\, (\phi_0^2 \phi_2 + \phi_0 \phi_1^2) \,\,\,,\nonumber \\
\cdots \cdots \,\,\, &  & \,\,\, \cdots \cdots \nonumber 
\end{eqnarray}
\begin{figure}[t]
\centering
$\begin{array}{cc}
\includegraphics[width=0.24\textwidth]{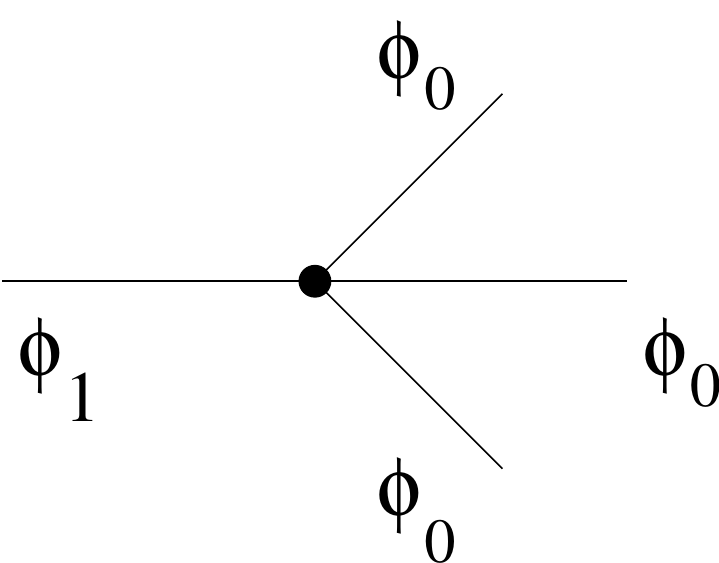} \,\,\,\,\, & \,\,\,\,\,\,
\includegraphics[width=0.3\textwidth]{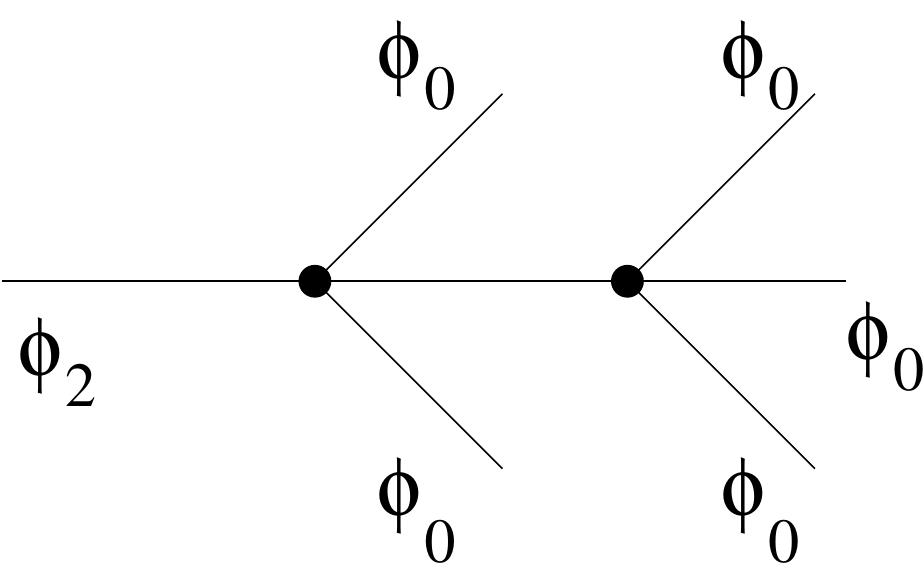}\\
\end{array}$
\caption{Perturbative expansion of the classical solution of the $\Phi^4$ LG, on the left the term $\phi_1(x)$ and on the right the term $\phi_2(x)$.}
\label{treelevel}
\end{figure}
The solutions of these equations can be given in terms of $\phi_0$ (expressed by the usual Fourier series of the free theory) and the 
inverse operator of $(\Box + m^2)$, i.e. the propagator $G_0(x-y)$. The first terms are 
\begin{eqnarray} 
\phi_1(x) & = & \frac{1}{3!}\,\int \,dy \,G(x-y) \, \phi_0^3(y) \,\,\,, \\
\phi_2(x) & = & \frac{1}{2}\, \int \,dy \,G(x-y) \,\phi_0^2(y) \,\phi_1(y) \,=\,\frac{1}{12} \int \,dy \,\int \,dz \,G(x-y) \,\phi_0^2 \,G(y-z) \,\phi_0^3(z) \,\,\, \nonumber 
\end{eqnarray}
and longer expressions for the higher terms. These expressions can be graphically expressed as in Figure \ref{treelevel} and they are clearly in correspondence with the tree level diagrams of the quantum theory. Moreover, as shown in Section \ref{sec:sg}, regarding $\phi_0(x)$ as the field that creates particle excitation at the position $x$, the various terms $\phi_n(x)$ may be consider as the classical limit of the Form Factor of the operator $\phi(x)$ on the asymptotic states, namely 
\be 
\phi_n \longrightarrow F_{classical}(p_1,\cdots,p_{2n+1}) \,=\,\langle 0 \mid \phi(x) \mid p_1,\cdots, p_{2n +1} \rangle \,\,\,.
\ee 

\vspace{0.5cm}
\noindent
{\bf Sinh-Gordon theory}. The same analysis can be also repeated for the Sinh-Gordon theory by firstly expanding in series of $g$ the equation of motion 
\be 
\Box \phi + \frac{m^2}{g} \sinh g \phi \,=\,
\Box \phi + m^2 \phi + m^2 \left(\frac{g^2}{3!} \phi^3 + \frac{g^4}{5!} \phi^5 + \cdots \right) \,=\, 0 \,\,\,,
\label{classicaleqsmotionShG}
\ee 
and then looking for a solution as a series expansion 
\be 
\phi(x) \,=\,\sum_{n=0}^{\infty} g^{2n} \,\phi_n(x) \,\,\,.
\ee
While the first differential equations satisfied by the $\phi_n$'s are similar to $\phi^4$ theory, the presence of the additional vertices of the Sinh-Gordon model sensibly alters those of higher order  
\begin{eqnarray}
(\Box + m^2) \phi_0 & = & 0 \nonumber \,\,\,, \\
(\Box + m^2) \phi_1 & = & -\,\frac{1}{3!} \,\phi_0^3 \,\,\,, \nonumber \\
(\Box + m^2) \phi_2 & = & -\,\frac{1}{2}\, \phi_0^2 \phi_1 - \frac{1}{5!} \phi_0^5  \,\,\,, \\
(\Box + m^2) \phi_3 & = & -\,\frac{1}{4!}\, \phi_0^4 \phi_1 - \frac{1}{2} (\phi_0 \phi_1^2 + \phi_0^2 \phi_2) -
\frac{1}{7!}\,\phi_0^7 \,\,\,, \nonumber \\
\cdots \cdots &  & \cdots \cdots \nonumber 
\end{eqnarray}
The solution of these equations follows the same scheme as in the $\phi^4$ theory and it is expressed in terms of the propagator $G_0(x-y)$ and the field $\phi_0(x)$. 
\begin{figure}[b]
\centering
$\begin{array}{c}
\includegraphics[width=0.7\textwidth]{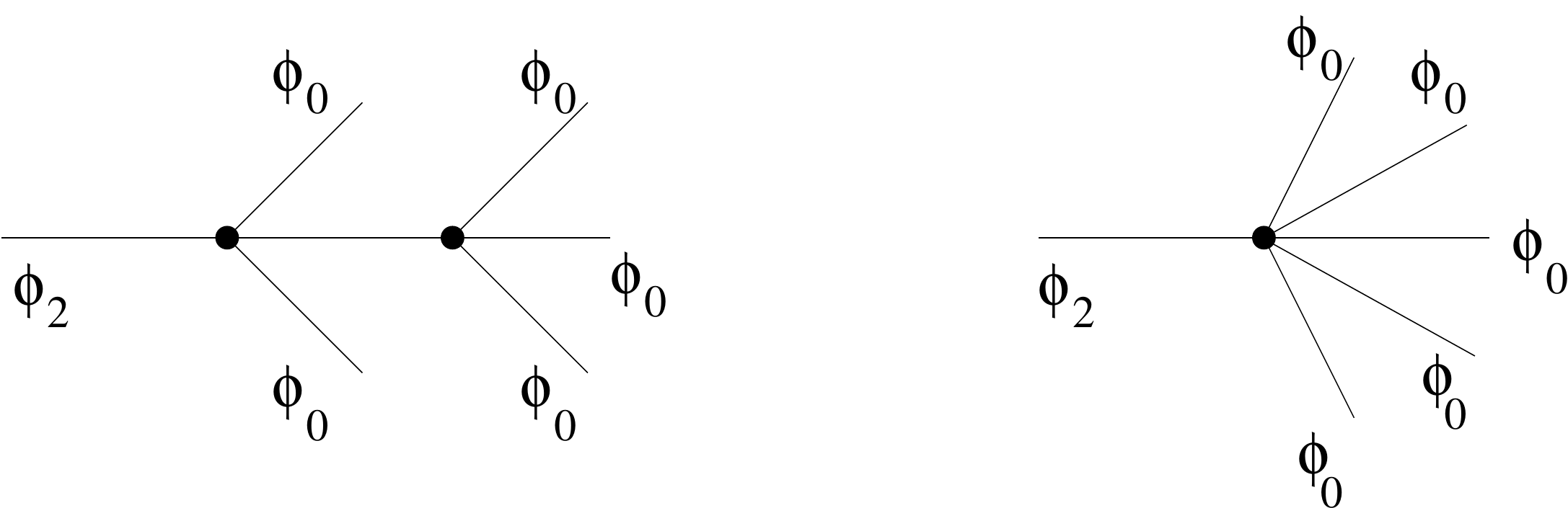}
\end{array}$
\caption{Graphs relative to the second perturbative term $\phi_2(x)$ of the classical solution of the Sinh-Gordon model.}
\label{phi2SH}
\end{figure}
Also in this case there is a correspondence between the classical solution and the tree level diagrams of the quantum Sinh-Gordon theory relative to the elementary field $\phi(x)$. The various terms $\phi_n(x)$ can be regarded as the classical limit of the Form Factors of the field $\phi(x)$ on the asymptotic states
\be 
\phi_n \longrightarrow F_{classical}(p_1,\cdots,p_{2n+1}) \,=\,\langle 0 \mid \phi(x) \mid p_1,\cdots, p_{2n +1} \rangle \,\,\,.
\ee 
For the field $\phi_2(x)$, for example, with respect to the LG $\phi^4$, we have an additional term and both graphs, with the relative combinatoric factors, enter the expression of the classical Form Factor $F_{cl}(p_1,\ldots,p_5)$, as shown in Fig.\ref{phi2SH}.

\subsection{Tree Level Diagrams for the Composite Fields}\label{ClassicalCompositeFields}
\begin{figure}[t]
\centering
$\begin{array}{c}
\includegraphics[width=0.7\textwidth]{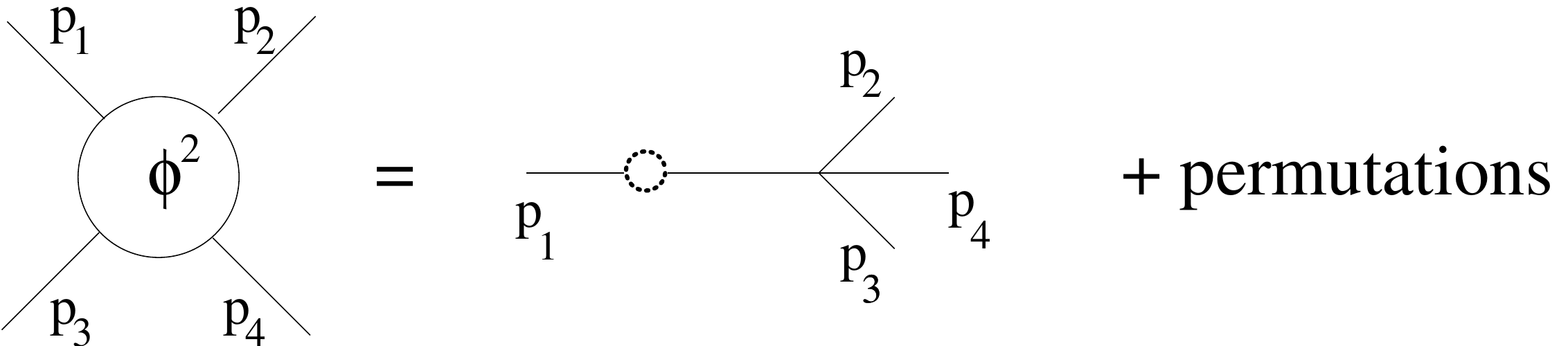}
\end{array}$
\caption{Graphs relative to the classical 4-particle Form Factor of the composite operator $\phi^2(x)$ of the Sinh-Gordon model.}
\label{FF24SH}
\end{figure}

The perturbative classical solution for the elementary field $\phi(x)$ can be further exploited to find explicit expression for the composite fields ${\mathcal O}(x) =F[\phi(x)]$, where we consider here only analytic functions $F[z]$. The thing to do is to substitute in $F[\phi(x)]$ the power series expansion of $\phi(x)$ and then expand the function $F[\phi(x)$ in power of the coupling constant. So, for instance, for the composite operators $\phi^2(x)$ and $\phi^4(x)$ of, say, the Sinh-Gordon model we have
\begin{eqnarray} 
\phi^2(x)   \, \equiv\, \sum_{n=0}^{\infty} g^{2n}\,\phi^{(2)}_n  &\,=\,& 
\phi_0^2 + g^2 \,(2 \phi_0 \phi_1) + g^4 
\,(\phi_1^2 + 2 \phi_0 \phi_2) + \\
& + & g^6 \, (2 \phi_1 \phi_2 + 2 \phi_0 \phi_3) + g^8 \, (\phi_2^2 + 2 \phi_1 \phi_3 + 2 \phi_0 \phi_4) 
+ \cdots \nonumber 
\end{eqnarray}
\begin{eqnarray} 
\phi^4(x)  \,\equiv\, \sum_{n=0}^{\infty} g^{2n}\,\phi^{(4)}_n & \,=\,& \phi_0^4 + g^2 \,(4 \phi_0^3 \phi_1) 
+ g^4 
\,(6 \phi_0^2 \phi_1^2 + 4 \phi_0^3 \phi_2) + \nonumber \\
& + & g^6 \, (4 \phi_0 \phi_1^3 + 12 \phi_0^2 \phi_1 \phi_2 + 
4 \phi_0^3 \phi_3) + \\
& + & g^8 \, (\phi_1^4 + 12 \phi_0 \phi_1^2 \phi_2 + 6 \phi_0^2 \phi_2^2 
+12 \phi_0^2 \phi_1 \phi_3 + 4 \phi_0^3 \phi_4) 
+ \cdots \nonumber 
\end{eqnarray}
Analogous expressions hold for the LG theory $\phi^4$ replacing $g^2 \rightarrow g$. Each perturbative term $\phi_n^{(k)}(x)$ of the composite operator $\phi^k(x)$ can then be expressed in terms of the tree level diagrams corresponding to the perturbative terms $\phi_n(x)$ of the elementary field $\phi(x)$. The graphs contributing to $\phi_n^{(k)}(x)$ can be considered as those entering the classical limit of the Form Factor of the composite operator $\phi^k(x)$ on the asymptotic particles, whose number is given by the number of fields $\phi_0$ entering the final expression. For instance, at the tree level, the classical Form Factor of the operator $\phi^2(x)$ on four particle states is given by the $g^2$ term of the expansion of this operator. Its graphical form is given in Figure \ref{FF24SH} 
and its analytic expression (up to normalization factor) is 
\begin{eqnarray}
F^{\phi^2}_{classical}(p_1,\ldots,p_4) & \,=\,& 
\frac{1}{(p_1+p_2+p_3)^2 - m^2 + i \epsilon} + 
\frac{1}{(p_1+p_2+p_4)^2 - m^2 + i \epsilon} 
\label{classicalFF24}\\ 
&+& \frac{1}{(p_1+p_3+p_4)^2 - m^2 + i \epsilon} + 
\frac{1}{(p_2+p_3+p_4)^2 - m^2 + i \epsilon} \,\,\,.\nonumber
\end{eqnarray}

As a matter of fact, one can set up a Feynman diagram analysis that allows us to determine the number of vertices entering the tree level diagrams of the Form Factors of the composite operator. In order to do so, one needs to absorb differently the $\hbar$ dependence of the path integral (\ref{partitionfunctionh}). Let's discuss how this method works for our two prototype theories. 

\vspace{3mm}
\noindent
{\bf $\Phi^4$ Landau-Ginzburg theory}. Given the Lagrangian density of this theory 
\be 
{\mathcal L} \,=\,\frac{1}{2} (\partial_{\mu} \phi)^2 - \frac{m^2}{2} \phi^2 - \frac{g}{4!} \phi^4 \,\,\,.
\ee 
there is a way of absorbing the $\hbar$ present in the path integral that consists in introducing a new field $\psi = \hbar^{-1/2}  \phi$ and a new coupling constant $\hat g = g \hbar^2$. Once this is done, consider now the matrix elements of the composite field $: \psi^k(x):$ on the asymptotic particle states of the theory created by the field $\psi$
\be 
\langle 0 | : \psi^{k}(x) : | A_1 \ldots A_m \rangle \,\,\,.
\ee 
In this theory, these matrix elements are different from zero only when $m$ and $k$ have the same parity. Leaving out their momentum dependence, the diagrams which contribute to these matrix elements consists of graph of external legs $E = m$, related to the number of vertices $n_4$ (with four legs coming from the interaction) and to the number of internal lines $I$ by the relation 
\be 
E + 2 I \,=\, 4 \, n_4 + k \,\,\,. 
\ee
In this respect the composite operator $:\phi^k(x):$ can be considered as an extra vertex of $k$ legs, so that the previous relation (\ref{topologicalrelation}) gets modified as $L = I - n_4$.  In the $\hbar \rightarrow 0$ limit only those diagrams with zero loops survive: therefore, with $L=0$, we get  $I = n_4 $ and therefore the previous equation becomes 
\be 
E  \,=\, 2 n_4 +k \,\,\,, 
\label{diophantic1}
\ee 
i.e. there is a diophantine equation that fixes the number of vertices $n_4$ entering the matrix elements of the composite operator $\phi^k(x)$ in the $\hbar \rightarrow 0$ limit. Consider for instance the matrix element 
\be 
\langle 0 | :\psi^6 : | A_1 \ldots A_{2 m} \rangle \,\,\,. 
\ee 
In this case $E = 2m$, $k=6$ and eq. (\ref{diophantic1}) becomes 
\be 
n_4 \,=\,  m - 3 \,\,\,. 
\label{ruleee}
\ee 

\vspace{3mm} 
\noindent 
{\bf Sinh-Gordon theory}. Given the Lagrangian density of this theory  
\be 
{\mathcal L} \,=\,\frac{1}{2} (\partial_{\mu} \phi)^2 - \frac{m^2}{g^2} \left(\cosh(g \phi) -1\right) 
\,\,\,, 
\ee 
also in this case we can re-absorb the $\hbar$ present in the path integral by introducing a new field $\psi = \hbar^{-1/2}  \phi$ and a new coupling constant $\hat g = \hbar^{1/2} g$. The difference with respect to the previous case is that now we have interaction vertices with arbitrarily large number of even legs $\psi^{2 r}$ and coupling constants $\frac{1}{(2 r)!} \, g^{2 (r-1)} $. Consider the matrix elements of the composite field $: \psi^k(x):$ on the asymptotic particle states of the theory created by the field $\psi$
\be 
\langle 0 | : \psi^{k}(x) : | A_1 \ldots A_m \rangle \,\,\,. 
\ee
\begin{figure}[b]
\centering
$\begin{array}{c}
\includegraphics[width=0.7\textwidth]{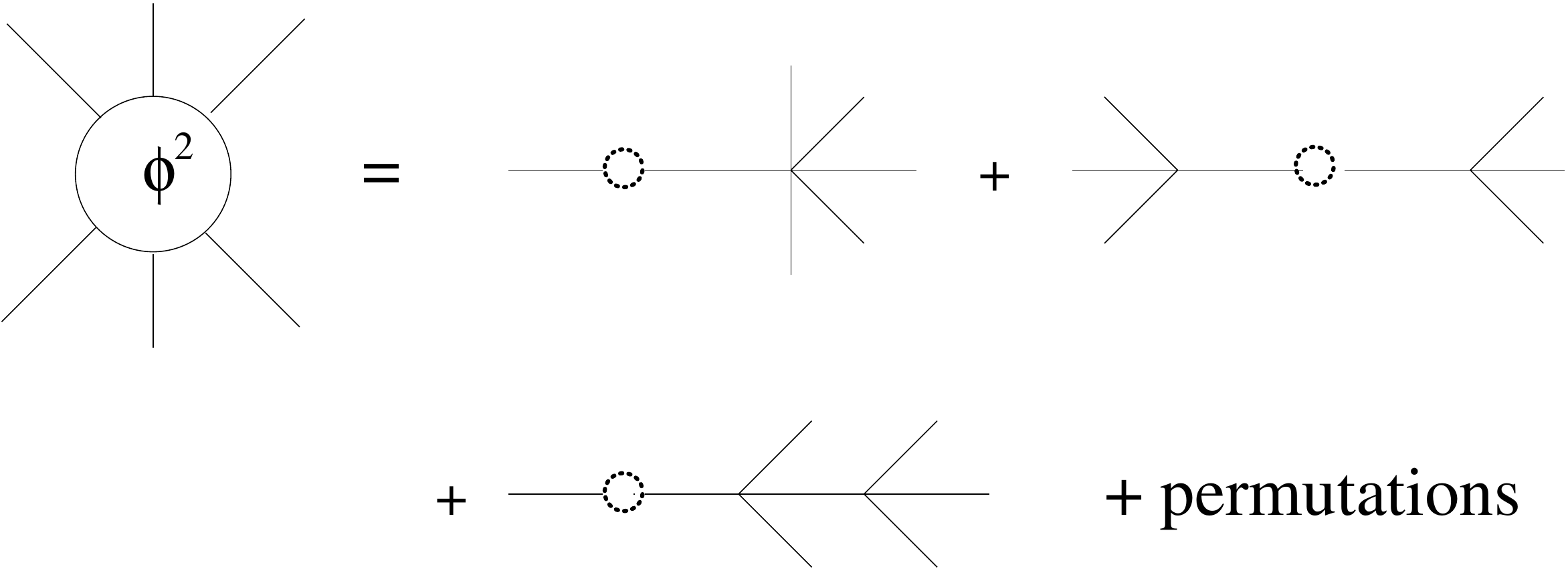}
\end{array}$
\caption{Graphs relative to the classical 6-particle Form Factor of the composite operator $\phi^2(x)$ of the Sinh-Gordon model.
The permutations are with respect to the external momenta.}
\label{phi26SH}
\end{figure} 
\noindent
Repeating the same argument given above, but keeping track of the presence of $n_{2 r}$ vertices with $2 r$ legs, we arrive to the diophantine equation  
\be 
E - k  \,=\, \sum_r 2(r-1) n_{2 r}   \,\,\,.
\label{diophantic2}
\ee
Consider, for instance, $k=2$ and $E=m=4$, so that $E- k = 2$ and the above equation becomes 
\be 
2 \,=\, 2 \, n_4 + 4 \, n_6 + 6 \, n_8 + 8 \, n_{10} + 10 \, n_{12} + \cdots 
\ee
Hence $n_4 =1$ while $n_{2l} = 0$ for $l \geq 2$, a solution that corresponds to the graphs of Figure \ref{phi2SH}. 
 
As another example, consider instead $k=2$ and $E=m=6$, so that $E- k = 4$ and the above equation becomes 
\be 
4 \,=\, 2 \, n_4 + 4 \, n_6 + 6 \, n_8 + 8 \, n_{10} + 10 \, n_{12} + \cdots 
\ee
Hence $n_{2l} = 0$ for $l \geq 4$, while solution are found for 
\begin{eqnarray}
&& n_4 \,=\, 2 \,\,\,\, , \,\,\,\,  n_6 \,=\, 0 \\
&& n_4 \,=\, 0 \,\,\,\, , \,\,\,\, n_6 \,=\, 1 \, \nonumber 
\end{eqnarray}
The corresponding graphs at tree level of the matrix elements of $\phi^2(x)$ on $6$-particle states is given in Figure \ref{phi26SH}.

More insights on the classical limit of quantum field theories will come from the coherent state formulation of the free theory, that is discussed in Section {\ref{Sectionfree}.

\section{Prototype of integrable field theory: the free field} \label{Sectionfree}
As discussed in the Introduction of this paper, integrable models admits a canonical change of variables that brings them into action-angle coordinates in which the time-evolution consists of multi-frequency oscillators. In order to enlighten many of the results for the interacting integrable models given below, it is instructive to work out in certain detail the free field case, in which all computations can be carried out explicitly. At the conceptual level, interacting and free integrable models are very similar, their difference being in the canonical transformation that brings them into action-angle variables: in the free case this is just Fourier transform, while in the interacting case is the Inverse Scattering Transform. 

In the following we address a series of issues that will appear later in the interacting case, so that the results of this section can serve as a guideline for more complicated situations.     

\subsection{Action-angle variables}
For the free theory, the classical and quantum cases can be treated simultaneously. One has only to care that classical expressions concerning higher powers of the fields or other composite operators have to be substituted in the quantum case by proper normal order quantities. Given this rule, in the following we will adopt for simplicity the notation of the classical case. So, the Lagrangian of the free (Klein-Gordon) theory is given by 
\be 
{\mathcal L} \,=\,\frac{1}{2} (\partial_{\mu} \phi)^2 - \frac{m^2}{2} \phi^2 \,\,\,,
\ee 
while its Hamiltonian reads 
\be 
H \,=\,\frac{1}{2}\,\int \left[ \Pi^2 +  \phi_x^2 + m^2 \phi^2\right] \, dx \,\,\,,
\label{freeHam}
\ee
where $\Pi(x,t) = \frac{\delta {\mathcal L}}{\delta \phi_t} = \phi_t(x,t)$. The momentum carried by the field is given by 
\be 
P \,=\, - \int \,\Pi \,\phi_x \, dx \,\,\,.
\label{momentum}
\ee 
Such a theory can be considered as Hamiltonian system with coordinates and momenta given by $\phi(x,t)$ and $\Pi(x,t)$ respectively. Their Poisson bracket (for the classical case) and their commutator (for the quantum case) are 
\be
\{\phi(x,t), \Pi(y,t) \} = \delta(x-y) 
\,\,\,\,\,\,\,\,\,\,\,\,\,\,\,\,
,
\,\,\,\,\,\,\,\,\,\,\,\,\,\,\,\,
[ \phi(x,t), \Pi(y,t) ] \,=\, i \delta(x-y) 
\,\,\, .
\label{Poissonbracket1}
\ee
The equation of motion 
\be 
(\Box + m^2) \,\phi(x,t) \,=\, 0 \,\,\,
\label{klein-gordon}
\ee 
has a solution that, in the infinite volume, can be expressed as
\be 
\phi(x,t)\,=\,\int_{-\infty}^{\infty} \frac{dk}{2 \pi} \frac{1}{\sqrt{2 \omega(k)}} \,
\left[A(k) \,e^{-i \omega t + i k x} + A^{\dagger}(k) \,e^{i \omega t - i k x} \right] \,\,\,,
\label{exactsolutionfreetheory}
\ee
where $\omega(k) = \sqrt{m^2 + k^2}$. The modes $A(k)$ and $A^{\dagger}(k)$ are fixed in terms of the boundary conditions at $t=0$ of the field and its time derivative. Using Eqs.(\ref{Poissonbracket1}) and (\ref{exactsolutionfreetheory}), one can derive the Poisson bracket (for the classical case) and the commutator (for the quantum case) of the variables $A(k)$ and $A^{\dagger}$   
\be 
\{ A^{\dagger}(k),A(q)\} \,=\,2 \pi i \delta(k-q) 
\,\,\,\,\,\,\,\,\,\,\,\,\,\,\,\,
,
\,\,\,\,\,\,\,\,\,\,\,\,\,\,\,\,
[ A(q), A^{\dagger}(k) ] \,=\, 2 \pi \delta(q-k) 
\,\,\,.
\ee
Substituting the solution of the equation of motion into the Hamiltonian, one has 
\be 
H \,=\,\int \frac{dk}{2\pi} \omega(k) \,|A(k)|^2 \, dk \,\,\,. 
\ee
Therefore the free theory admits an action-angle variable formulation in terms of $P(k)$ and $Q(k)$, with $0 \leq Q(k) < 2 \pi$, and Poisson bracket $\{Q(k), P(q)\} = \delta(k-q)$, with the role of $P(k)$ and $Q(k)$ played by 
\be
P(k) \,=\,|A(k)|^2 
\,\,\,\,\,\,\,\,\,\,\,\,\,\,\,\,
,
\,\,\,\,\,\,\,\,\,\,\,\,\,\,\,\,
Q(k) = {\rm arg} \,A(k) 
\,\,\,.
\ee
The time evolution of $A(k)$ and $A^{\dagger}(k)$ are then given by $A(k,t) = A(k) e^{-i \omega(k) t}$ and $A^{\dagger}(k,t) \,=\, A^{\dagger}(k) e^{i \omega(k) t}$. The action variables $P(k)$'s are the mode occupations of the field which, in the quantum case, express the number of bosonic particles of momentum $k$.  Clearly $\dot P(k)=0$ for any $k$ and therefore we have an infinite number of conserved quantities during all the time evolution. Since these expressions are given in the momentum space, they are non-local. The theory also admits denumerable {\em local} conservation laws, as shown in the next section. For the role of local and non-local charges in Quantum Field Theories, we refer the reader to the reference \cite{EMP}. 

\subsection{Conserved Charges}
The Klein-Gordon theory admits an infinite number of local conservation laws that can be easily derived using the light-cone coordinates $\tau$ and $\sigma$ defined by 
\be 
t \,=\,\frac{1}{2} (\tau - \sigma) 
\,\,\,\,\,\,\,\,\,\,\,
,
\,\,\,\,\,\,\,\,\,\,\,
x \,=\,\frac{1}{2} (\tau + \sigma) 
\,\,\,.
\label{lightconevariable}
\ee
In the light-cone variables the equation of motion becomes 
\be 
\phi_{\sigma\tau} \,=\,m^2 \,\phi \,\,\,.
\ee
Taking $\tau$ as "time" variable,  we have the infinite chain of conservation laws coming from the equation of motion 
($n = 1,2, \cdots $)
\begin{eqnarray} 
\frac{\partial}{\partial \tau} \frac{1}{2} \phi^2_{n \sigma} & \,=\, & 
m^2 \,\frac{\partial}{\partial \sigma} \frac{1}{2} \phi^2_{(n-1) \sigma} \label{ligthconeconservationlawKG} \\
\frac{\partial}{\partial \sigma} \frac{1}{2} \phi^2_{n \tau} & \,=\, & 
m^2 \,\frac{\partial}{\partial \tau} \frac{1}{2} \phi^2_{(n-1) \tau} \nonumber  
\end{eqnarray}
where $\phi_{n \sigma} = \frac{\partial^n \phi}{\partial \sigma^n}$ and analogously for $\phi_{n \tau}$. The equations above are the 
general form 
\be
\partial_{\tau} A = \partial_{\sigma} B \,\,\,,
\label{lc1}
\ee
and therefore, going back to the original laboratory coordinates $(x, t)$, they can be expressed in terms of the continuity equation 
\be 
\partial_t (A+B) \,=\,\partial_x (B-A) \,\,\,,
\label{lc2}  
\ee
so that the conserved charges are $Q = \int dx (A+B)$. For the Klein-Gordon we have then the following set of conserved charges 
\begin{eqnarray}
& Q_n & \,= \,  \int dx \left[\frac{1}{2} \phi^2_{n \sigma} + \frac{m^2}{2} \phi^2_{(n-1) \sigma} \right] \,\,\,\\
& Q_{-n} & \,= \,  \int dx \left[\frac{1}{2} \phi^2_{n \tau} + \frac{m^2}{2} \phi^2_{(n-1) \tau} \right] \,\,\,\nonumber 
\end{eqnarray}
Taking the sum and the difference of these quantities, we can define the even and odd conserved charges 
\begin{eqnarray}
{\mathcal E}_n & \,= \, & (Q_n + Q_{-n}) = 
\frac{1}{2} \int dx \left[ \phi^2_{n \sigma} + \phi^2_{n \tau} + m^2 (\phi^2_{(n-1) \sigma} + \phi^2_{(n-1) \tau}) \right] \,\,\,\\
{\mathcal O}_n & \,= \, &  (Q_n - Q_{-n}) = 
\frac{1}{2} \int dx \left[ \phi^2_{n \sigma} - \phi^2_{n \tau} + m^2 (\phi^2_{(n-1) \sigma} - \phi^2_{(n-1) \tau}) \right] \,\,\,\nonumber
\end{eqnarray} 
It is now easy to see that they can be expressed in terms of the mode occupation of the field. To do so, it is more convenient to adopt the rapidity variable and use the expansion (\ref{rapidityexpansion}), arriving to the expressions 
\begin{eqnarray}
{\mathcal E}_n & \,= \, &  
m^{2n-1} \,\int \frac{d\theta}{2\pi} \,|A(\theta)|^2 \, \cosh[(2 n-1) \theta] \,\,\,\\
{\mathcal O}_n & \,= \, & m^{2n-1} \,\int \frac{d\theta}{2\pi} \, |A(\theta)|^2 \, \sinh[(2 n-1) \theta] 
 \,\,\,\nonumber
\end{eqnarray} 
The first representatives of these expressions correspond to the energy and the momentum of the field. In the quantum field theory interpretation, the equations above imply that each particle state $| \theta \rangle$ of rapidity $\theta$ is a common eigenvectors of all these conserved quantities, with eigenvalues 
\be 
{\mathcal E}_n \,| \theta \rangle \,=\, m^{2n-1} \,\cosh[(2 n -1) \theta] \, | \theta \rangle    
\,\,\,\,\,\,\,\,\,\,\,
,
\,\,\,\,\,\,\,\,\,\,\,
{\mathcal O}_n \,| \theta \rangle \,=\, m^{2n-1} \,\sinh[(2 n -1) \theta] \, | \theta \rangle  \,\,\,.
\ee

\subsection{Partition Function} \label{PARTITIONFUNCTIONQUANTUMFREE}
The partition function of a classical free field theory 
\be 
Z_{classical}(\beta) \,=\,\int {\mathcal D}\Pi \,{\mathcal D} \phi \,
\exp \left(-\beta H[\Pi,\phi]\right) \,\equiv\,\exp\left(-\beta F(\beta) 
\right) 
\,\,\,, 
\label{partitionfunction1}
\ee
was computed in Section \ref{classicalfreepartitionfunction}, with the final expression
given by 
\be 
\beta F(\beta) \,=\, \frac{L}{2\pi} \int_{-\infty}^{\infty} dk \, \log[\hbar \beta \omega(k)]  \,\,\,,   
\label{clexpfreetp}
\ee 
where 
$
\omega(k) \,=\,(m^2+k^2)^{1/2} 
$.
It is interesting to compare this expression with the one coming from the $\hbar\rightarrow 0$ limit of the partition function of 
the quantum theory. This is expressed by the path integral in which one performs the Wick rotation $t \rightarrow - i \tau$,  
compactifies the time direction to a circle $[ 0, \hbar \beta ]$ and makes all quantities periodic, with period $\hbar \beta$, 
along this axis  
\be 
Z_{quantum}(\beta) \,=\,\int {\mathcal D}\Pi \,{\mathcal D} \phi \,
\exp \,(\hbar^{-1} \,S[\phi]) 
\,\,\,.
\label{partitionfunction2}
\ee
\be 
\hbar^{-1}\, S[\phi]\,=\, \hbar^{-1} \,\int_0^{\beta \hbar} d\tau \,\left[\int \Pi(x,\tau) \dot\phi(x,\tau) \,dx - H[\phi] \right] \,\,\,, 
\ee
where $H$ and all other quantities are classical quantities. Taking $\hbar \rightarrow 0$, it is easy to see that the first term in $S[\phi]$ reduces to 
$\Pi(x,\tau) \,d\phi(x,\tau)$ and it vanishes due to the periodic boundary condition $\phi(x,0) = \phi(x,\beta \hbar)$ for any $x$, and therefore $Z_{quantum}(\beta)$ reduces to the classical partition function (\ref{partitionfunction1}) in this limit. 

For the cylinder geometry of the quantum problem, in the thermodynamic limit $L \rightarrow \infty$, one can express the partition function as 
\be 
Z_{quantum} \,\simeq \, e^{-L E_0(\hbar \beta) /\hbar} \,\,\,,
\ee 
where $E_0(\hbar\beta)$ is the ground state energy of the quantum field with periodic boundary condition. Put $\hbar \beta = R$ and $r= m R$, such a quantity is given by 
\be
\frac{E_0(R)}{\hbar} \,=\, \,\left[\frac{m}{2} + \frac{2\pi}{R}\, \sum_{n=1}^\infty \sqrt{n^2 + \left(\frac{r}{2\pi}\right)^2 } \,\right] \,\,\,,
\ee
and needs to be regularized since it is ultraviolet divergent. This can be done by noticing that the divergence of the series is due to the large $n$ behaviour of the first two terms of the expansion 
\be
\sqrt{n^2 + \left(\frac{r}{2\pi}\right)^2} \,=\, n + \frac{1}{2} \, \left(\frac{r}{2\pi}\right)^2\,\frac{1}{n} + {\mathcal O}\left(\frac{1}{n^2}\right) 
\ee
Therefore, subtracting and adding these divergent terms we have 
\begin{eqnarray}
S(r) \,=\, \sum_{n=1}^{\infty} \sqrt{n^2 + \left(\frac{r}{2\pi}\right)^2} & \,=\, &  
\sum_{n=1}^{\infty} \left\{\sqrt{n^2 + \left(\frac{r}{2\pi}\right)^2} - n - \frac{1}{2} \, \left(\frac{r}{2\pi}\right)^2\,\frac{1}{n} \right\} \\
&+& \sum_{n=1}^{\infty} n + \frac{1}{2} \, \left(\frac{r}{2\pi}\right)^2\,\sum_{n=1}^{\infty} \frac{1}{n} \nonumber 
\end{eqnarray} 
Regularizing the divergent expressions as 
\be 
 \sum_{n=1}^{\infty} n \,=\,-\frac{1}{12} 
 \,\,\,\,\,\,\,\,\,\,\,
 ,
 \,\,\,\,\,\,\,\,\,\,\,
 \sum_{n=1}^{\infty} \frac{1}{n}\,=\,\frac{r}{2 \pi} + \gamma_E 
\ee 
we finally get 
\be 
\frac{E_0(R)}{\hbar} \,=\,\frac{1}{R} \,\left[
-\frac{\pi}{6} + \frac{r}{2} + \frac{r^2}{4\pi} \left(\log \frac{r}{4\pi} + \gamma_E -\frac{1}{2}\right) + 
\sum_{n=1}^{\infty} \left(\sqrt{(2 \pi n)^2 + r^2} - 2\pi n - \frac{r^2}{4 \pi n}  \right) \right] \,\,\,, 
\ee
a series which can be summed and put in the form
\be 
\frac{E_0(R)}{\hbar}\,=\, \int_{-\infty}^{\infty} \frac{d\theta}{2\pi}\, m \cosh\theta \,\log\left(1-e^{-mR \cosh\theta}\right) \,=\,
\frac{1}{2\pi} \int_{-\infty}^{\infty} dk \log\left(1-e^{-R \omega(k)}\right)\,\,\,.
\ee  
Substituting now $R = \hbar \beta$ and taking the limit $\hbar \rightarrow 0$, one has 
\be 
\frac{E_0(R)}{\hbar} \,\simeq\, \frac{1}{2\pi} \int_{-\infty}^{\infty} m\cosh\theta \log (\hbar m \beta\cos\theta) \frac{d\theta}{2\pi} =
\int_{-\infty}^{\infty} dk \log \left(\hbar \beta \omega(k)\right)\,\,\,,
\label{nicecheckfreeenergy}
\ee
i.e. it reproduces the classical free energy (\ref{clexpfreetp}). 

\subsection{Stationary Expectation Values and Generalized Gibbs Ensemble} 
In the free theory, given that the exact solution (\ref{exactsolutionfreetheory}) of the equation of motion is known, we can easily compute the stationary expectation values of any local function $F[\phi(x,t)]$. If the field is defined on a lattice in a finite volume $L$, with $L = N a$, the expectation values is defined by averaging over all points $x_n$ and taking the time average, i.e. 
\be
\langle F[\phi] \rangle \,=\,\lim_{t\rightarrow \infty} \frac{1}{t} \int_0^t \, dt \,\left(\frac{1}{N} \sum_{m=1}^N \, F[\phi_m(t)] \right)\,\,\,.
\label{timeaveragelattice}
\ee 
The two averages employed in this formula (on the number of points and on time) smooth out all microscopic fluctuations of the observables and give rise to the stationary value that emerges asymptotically. It is worth to have a look at the result produced by the two averages: the typical time evolution of an observable $F[\phi]$, once we have averaged only on the space lattice points, is shown on the left plot in Figure \ref{evolution}, and one can see that there are persistent fluctuations of the observable as time goes by. However, taking the time average, the curve flattens and rapidly converges to its asymptotic value. 

\begin{figure}[t]
\centering
$\begin{array}{cc}
\includegraphics[width=0.4\textwidth]{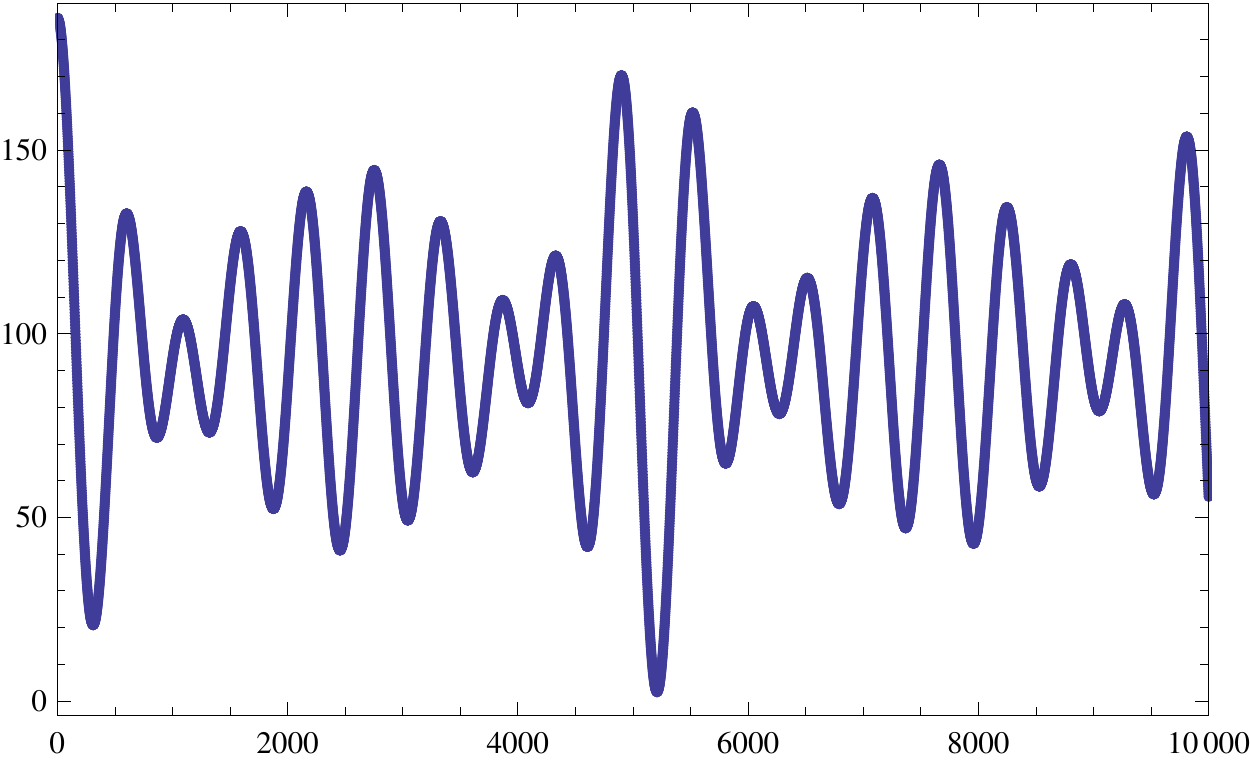} &
\includegraphics[width=0.4\textwidth]{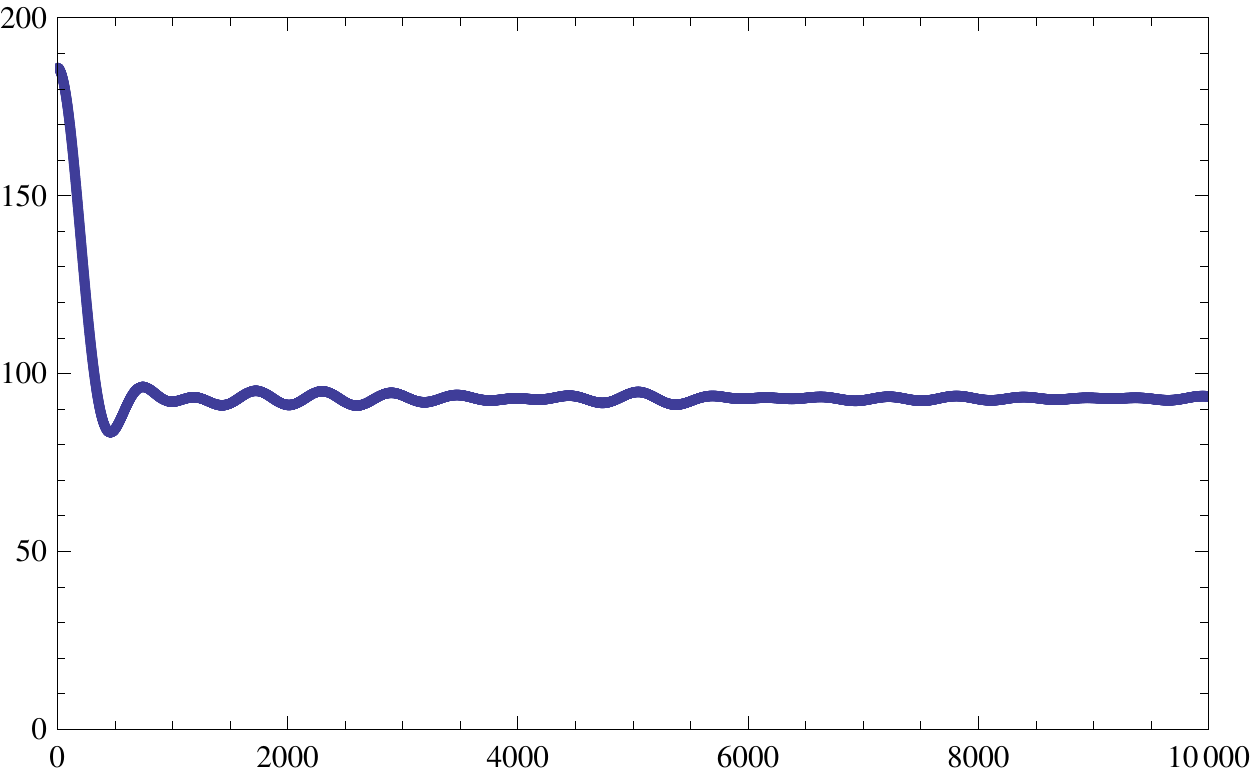}\\
\end{array}$
\caption{On the left: time evolution of $\frac{1}{N} \sum_{m=0}^N \,\phi^2_n(t)$ for $t=10^4 t_0$. On the right: 
the time average of the same observable. In both figures the horizontal line is the time measured in unit of $m^{-1}$. }
\label{evolution}
\end{figure}

For simplicity, let's concentrate our attention on the asymptotic values of powers of the field $\phi(x,t)$ for the lattice theory. It is easy to see that all odd powers $\phi^{2n +1}(x,t)$ of the field average to zero, while for the even powers we have instead 
\begin{eqnarray} 
&& \langle \phi^2 \rangle \,=\, \sum_{n=-N/2}^{N/2} \frac{|A(k_n)|^2}{\omega(k_n)} \,\equiv\, b \,\,\,, \label{vevphin}\\ 
&& \langle \phi^{2n} \rangle \,=\, (2 n -1) !! \, b^n 
\end{eqnarray}
These expressions explicitly depend on the initial condition of the occupation numbers $|A(k_n)|^2$ and therefore they cannot be derived by a Gibbs ensemble average, i.e. 
\be 
\langle \phi^{2n} \rangle \,\neq \, Z^{-1} \,\int {\mathcal D}\Pi\, 
{\mathcal D}\phi \, \phi^{2n}(x) \, e^{-\beta H} 
\ee
where $\beta$ is the conjugate variable of the conserved value of the energy 
\be
E\,=\,\sum_{n=-N/2}^{N/2} \omega(k_n) \, |A(k_n)|^2 \,\,\,.
\ee 
Indeed, keeping the energy fixed but varying the various occupation numbers, one get a rather spread set of asymptotic values for the various observables, as shown in Figure (\ref{spread}) for a representative one. The reason of such a behaviour is due to all other infinite conserved charges of the model, whose values change by changing the occupation number of the various modes. Since the motion of the field takes place on the manifold that is the intersection of the surfaces (in phase space) of constant values of the conserved charges, the asymptotic values explicitly break the ergodicity property.    

In order to derive the asymptotic values of the observable by an ensemble average, one needs to introduce a Generalized Gibbs Ensemble, which includes in addition to the energy, all other conserved charges. Since the conserved charges are expressed in terms of the mode occupation, in this case it is convenient to consider the statistical weight 
\be 
\rho \,=\,Z^{-1} \, \exp\left[ -\sum_{k} \eta_k \, |A_k|^2  \right] \,\,\,,
\ee
where 
\be 
Z\,=\, \int {\mathcal D}A_k \,{\mathcal D}A^{\dagger}_k \, \exp\left[ -\sum_{k} \eta_k \, |A_k|^2  \right] \,=\, 
\pi^n \prod_k \eta_k^{-1} \,\,\,.
\ee
The Lagrangian multipliers $\eta_k$ are fixed in terms of the initial occupation numbers, i.e. 
\be 
n^{(0)}_k \,=\,\overline{|A_k|^2} \,=\,Z^{-1} \,\int {\mathcal D}A_k \,{\mathcal D}A^{\dagger}_k \, |A_k|^2 \exp\left[ -\sum_{k} \eta_k \, |A_k|^2  \right]\,=\,\eta_k^{-1} \,\,\,.\label{lagrangianmultipliers}
\ee 
\begin{figure}[t]
\centering
$\begin{array}{cc}
\includegraphics[width=0.5\textwidth]{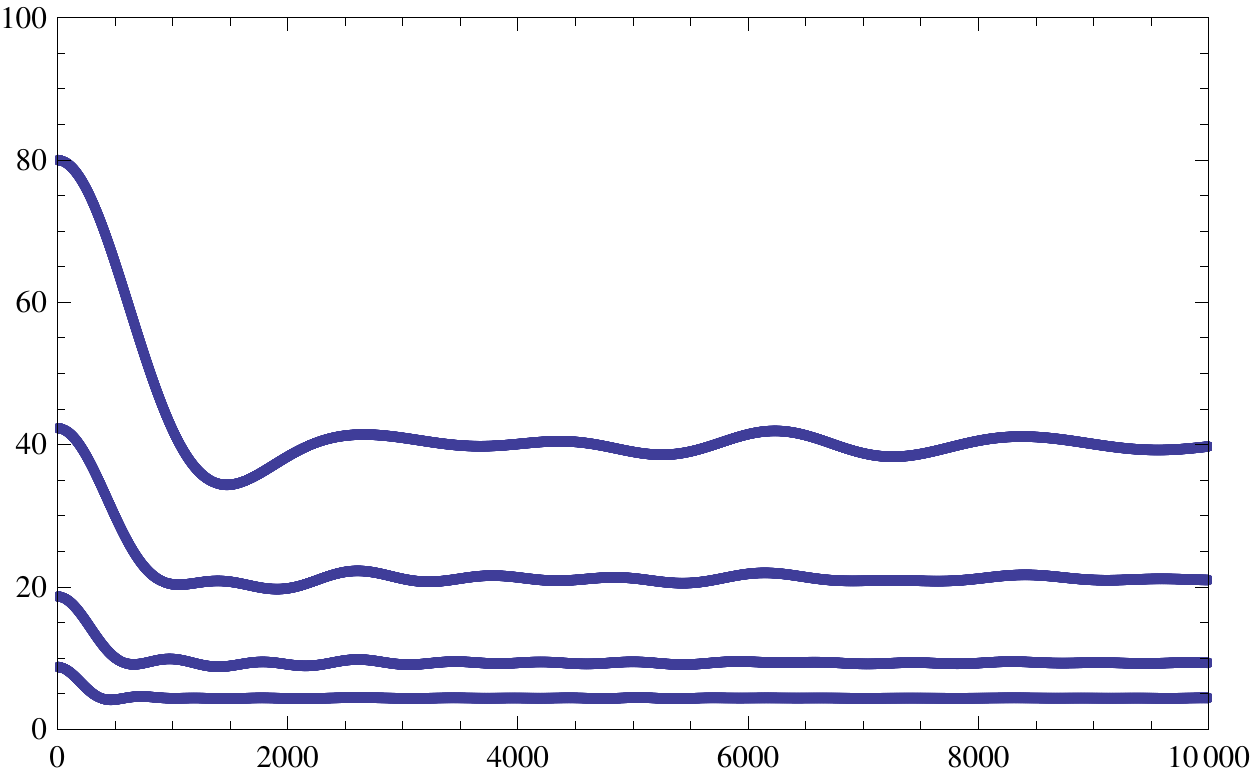} &
\end{array}$
\caption{Time average of $\phi^2$ for the same value of the energy density (in this case $E/L =100$) but different occupation numbers of the modes. 
On the horizontal line there is the time measured in unit of $m^{-1}$.}
\label{spread}
\end{figure}It is easy to see that such a statistical weight implies the following ensemble averages 
\be 
\overline{A_k A^{\dagger}_m} \,=\,\delta_{k,m} 
\,\,\,\,\,\,\,\,\,\,\,
, 
\,\,\,\,\,\,\,\,\,\,\,
\overline{A_k A_m} =  \overline{A^{\dagger}_k A^{\dagger}_m} = 0 
\,\,\,.
\ee
It is now easy to compute the generating function of all ensemble averages of the field 
\be 
\overline{\exp[ i a \phi(x)]} \,=\, \exp \left[-\frac{a^2}{2} \sum_k \frac{1}{\eta_k \omega_k}\right] 
\,=\, \exp \left[-\frac{a^2}{2} \sum_k \frac{n^{(0)}_k}{\omega_k}\right] \,\,\,. 
\label{stationargeneratingfunction}
\ee
Expanding in power series in $a$ left and right terms, one can easily recover the previous result (\ref{vevphin}) coming 
from the time-evolution averages. 

While the stationary values of various observables depend on the initial conditions on the field and its time derivative, there are 
certain universal ratios which are completely independent. This is case, for instance, for 
\be 
\frac{\langle \phi^{2n} \rangle}{\langle \phi^2 \rangle^n} \,=\, (2 n -1) !! \,\,\,\,\,\,\,, \,\,\,\,\,\,\,\,
\frac{\log \langle \cosh (\alpha \phi) \rangle}{\log \langle \cosh (\gamma \phi) \rangle} \,=\,  \left(\frac{\alpha}{\gamma}\right)^2 \,\,\,.
\label{vertex}
\ee  
It is interesting to notice that the second result coincides with the one relative to the vertex operators of Conformal Field Theory 
in the high-temperature limit \cite{LM}.   
 
\subsection{Coherent states}
Classical description of a bosonic quantum field theory is expected to emerge when there is a large occupation number of the modes, as it happens for Maxwell equations, for instance. The natural formalism to envisage such a situation is provided by the {\em coherent state approach} \cite{sudarshan} that, in particular, helps to enlighten the meaning of the classical configurations of the field and to put in the proper perspective the study of the time-evolution of the classical field. 

Consider the free bosonic quantum field $\phi(x,t)$ defined on a circle $L$, with periodic boundary condition $\phi(x+L,t) = \phi(x,t)$. 
Discretized on the lattice, this field can be expressed as  
\be
\phi_m(t)  \,=\, \,\sum_{n =0}^{N-1}  \frac{1}{\sqrt{2 \omega_n}}  \,
\left[A_n \,e^{-i \omega_n t + i k_n m } + A^{\dagger}_n \,e^{i \omega_n t - i k_n m } \right] 
\equiv   \phi_+(x,t) + \phi_-(x,t)  \,\,\,,
\ee
where $A_n$ and $A^{\dagger}_n$ satisfy the commutation relation $[A_n , A^{\dagger}_m] = \delta_{n,m}$. 
Consider now the operator 
\be 
S\,=\,{\mathcal N}\, \exp\left[\frac{1}{2} \sum_k \sqrt{2 \omega_k} \, f_k A^{\dagger}_k \right] \,\,\,,
\label{operatorS}
\ee
where $f_k$ is a set of complex numbers with $f^*_k = f_{-k}$ and ${\mathcal N}$ is the constant 
$ 
{\mathcal N} \,=\, \exp\left[-\frac{1}{2} \sum \omega_k \, |f_k|^2 \right]
$. 
Using the Baker-Haussdorf formula, it is easy  to see that 
\be 
S^{-1} \, A_q \, S \,=\, A_q + \frac{1}{2} \sqrt{2 \omega_q} \, f_q \,\,\,,
\ee 
and therefore 
\be
S^{-1} \, \phi_+(x,0) \, S \,=\, \phi_+(x,0) + \frac{1}{2}  \, f(x) \,\,\,, 
\ee
where $f(x)$ is the real function defined by the Fourier transform 
$
f(x) \,=\,\sum_q f_q \,e^{i q x} 
$. 
Let's us now define the coherent state 
\be 
| f \rangle \,=\, S\,| 0 \rangle \,\,\,, 
\ee
(with $\langle f | f \rangle = 1$) that is an eigenvector of all annihilation operators $A_q$, 
\be 
A_q \,| f \rangle \,=\, \frac{1}{2} \sqrt{2 \omega_q} f_q \,| f \rangle \,\,\,.
\ee
On the state $| f \rangle$, the number operators $N_k = A^{\dagger}_k \,A_k$ have expectation values 
\be 
\langle f | N_k | f \rangle \,=\, \frac{1}{2} \omega_k \,|f_k|^2 \,\,\,,
\label{particlenumber}
\ee
and the total number of particles contained in the state $| f \rangle$ is an extensive quantity in the thermodynamic limit  
\be 
N \,=\,\frac{1}{2} \sum_k \omega_k |f_k|^2 
\longrightarrow
 \frac{L}{2} \int_{-\pi/a}^{\pi/a} \frac{dk}{2\pi}\, \omega(k) |f(k)|^2 \,\,\,.
\ee
Using $\phi_+(x,0) | 0 \rangle = 0$, we have 
$ 
\phi_+(x,0) \, | f \rangle = \frac{1}{2} f(x) \, | f \rangle 
$ and 
$
\langle f | \phi_-(x,0) \,=\, \langle f | \frac{1}{2} f(x) 
$, 
so that 
\be 
\langle f | \phi(x,0) | f \rangle \,=\,f(x) \,\,\,.
\ee
This equation provides a direct meaning of the classical configurations of the field, i.e. they can be seen as matrix elements of the {\em quantum field} on the coherent states. This interpretation also extends to arbitrary normal ordered powers of the field and its space derivatives
\begin{eqnarray}
&& \langle f | :\phi^2(x,0): | f \rangle \,=\,f^2(x)    \nonumber \\
&& \langle f | :\phi^3(x,0): | f \rangle \,=\,f^3(x)    \nonumber\\
&& \langle f | :G[\phi(x,0)]: | f \rangle \,=\,G[f(x)]     \\
&& \langle f | :\nabla \phi(x,0): | f \rangle \,=\,\nabla f(x)    \nonumber \\
&& \langle f | :(\nabla \phi)^2(x,0): | f \rangle \,=\,(\nabla f)^2(x)    \nonumber 
\end{eqnarray}
We can also assign a time-dependence to the coherent state of the free theory by let evolving the operator $A^{\dagger}_k$ in (\ref{operatorS}) as $A^{\dagger}_k(t) = e^{i \omega_k t} A^{\dagger}(0)$. 

The picture that emerges from this formalism is particularly remarkable: we can regard the classical field configurations as a collection of particles of the quantum field, whose number in each mode is given by eq.\,(\ref{particlenumber}). In a free theory, these occupation numbers do not change during the time evolution while in an interactive theory they change according to the dynamics of the classical field. The time variation of $N_k$ can be interpreted in this case as due to particle creation and annihilation processes of the quantum theory.

\section{The Quantum Sinh--Gordon model}
\label{sec:sg}
\noindent
In this section we are going to remind the key features of the quantum theory of Sinh-Gordon model which will be important for our later developments. 
The quantum Sinh-Gordon model is an integrable relativistically invariant field theory in $1+1$ dimensions defined by the Lagrangian density
\be
\mc{L}= \frac12\left(\frac{\p\phi}{\,\p
  t}\right)^2 - V(\phi) \,\,\,, 
\ee
where $\phi=\phi(x,t)$ is a {\em real} scalar field and 
\be
  V(\phi) \,=\,\frac{m_0^2}{g^2} \,\left(\cosh g\phi -1\right) \,=\, 
  m_0^2 \,\left(\frac{1}{2} \phi^2 + \frac{g^2}{4!} \phi^4 + \frac{g^4}{6!} \phi^6 + \cdots\right) \,\,\,.
\label{potentialSHG}
\ee
$m_0$ is a mass scale, related to the physical (renormalized) mass $M$ of the particle by 
\be
m_0^2\,=\,M^2\frac{\pi\alpha}{\sin(\pi\alpha)}\,\,\,, 
\labl{eq:mu_m}
where $\alpha$ is the dimensionless renormalized coupling constant
\be
\alpha\,=\,\frac{g^2}{8\pi}\,\left(1+\frac{g^2}{8\pi}\right)^{-1}\,\simeq \frac{g^2}{8 \pi} + {\cal O}(g^4) \,\,\,.
\labl{eq:alpha}
Notice that, if we restore the presence of $\hbar$ in the theory, the coupling constant $g^2$ becomes accompanied 
by $\hbar$, i.e. 
\begin{equation}
g^2 \rightarrow g^2 \hbar \,\,\,,
\label{g2hbar}
\end{equation}
so that the perturbative expansion in $g^2$ is tantamount a semi-classical expansion in $\hbar$. 

It is convenient to express the dispersion relation of the energy and momentum of a particle excitation in terms of the rapidity variable $\theta$ as $E=M \cosh\theta$, $P=M\sinh\theta$, and label the asymptotic states as  $|\theta_1,\theta_2,\cdots,\theta_n \rangle $. The quantum integrability of the model is supported by the existence of an infinite number of conserved charges, the even ${\mathcal E}_{2n +1}$ and the odd ${\mathcal O}_{2n+1}$, which are diagonal on multi-particle states and act on them as 
\begin{eqnarray} 
{\mathcal E}_{2n+1} \,|\theta_1,\theta_2,\cdots,\theta_n \rangle \,=\,e_{2n+1} \, \sum_{k=1}^{n} \cosh[(2n+1)\theta_k] \, 
|\theta_1,\theta_2,\cdots,\theta_n \rangle \,\,\, 
\label{actioncharges}\\
{\mathcal O}_{2n+1} \,|\theta_1,\theta_2,\cdots,\theta_n \rangle \,=\,o_{2n+1} \, \sum_{k=1}^{n} \sinh[(2n+1)\theta_k] \, 
|\theta_1,\theta_2,\cdots,\theta_n \rangle \,\,\, \nonumber 
\end{eqnarray}
The real quantities $e_{2n+1}$ and $o_{2n+1}$ are the eigenvalues of the charges which, by rescaling their normalization, can be put equal to 1. The conserved charges ${\mathcal E}_0$ and ${\mathcal O}_0$ coincide respectively with the energy and the momentum. The existence of these conserved charges implies that, in the dynamics of the model, the momentum of each individual particle is conserved. Hence the $S$-matrix is elastic and factorizable in terms of the two-body $S$-matrix, given by \cite{ari}:
\be
S_{\text{ShG}}(\theta_{12},\alpha)\,=\,\frac{\sinh\theta_{12}-i\,\sin(\pi \alpha)}{\sinh\theta_{12} + i\,\sin(\pi \alpha)}\,\,\,,
\label{exactSmatrixSH}
\ee
where $\theta$ is the rapidity difference of the two particles. Notice that, for $\alpha \neq 0$, $S(0) = -1$ while if $\alpha =0$, we have instead $S(0)=1$. This discrepancy is entirely due to the interacting nature of the theory that, in the exact expression of the $S$-matrix, is captured by the 
resummation of the perturbative series. In fact, expanding $S(\theta_{12},\alpha)$ in power series of the coupling constant $g$, we start from $S =1$, with a leading correction order $g^2$  
\be
S_{\text{ShG}}(\theta_{12},\alpha) \,\simeq \,1 - i \,\frac{g^2}{4} \frac{1}{\sinh\theta_{12}} + \cdots 
\label{Sperturbative}
\ee
Taking into account that the $S$-matrix ${\mathcal S}(s)$ expressed in the usual Mandelstam variable $s=(p_1+p_2)^2$ is related to $S(\theta_{12})$ by the relation 
\be 
{\mathcal S}(s) \,=\,4 M^2 \,\sinh\theta_{12} \, S(\theta_{12}) \,\,\,,
\ee
it is easy to see that the term proportional to $g^2$ in eq.~(\ref{Sperturbative}) comes from the Feynman tree diagram relative to the $\phi^4$ vertex in (\ref{potentialSHG}), as shown in Figure \ref{pertSfig}. 
\begin{figure}[t]
\centerline{\scalebox{0.3}{\includegraphics{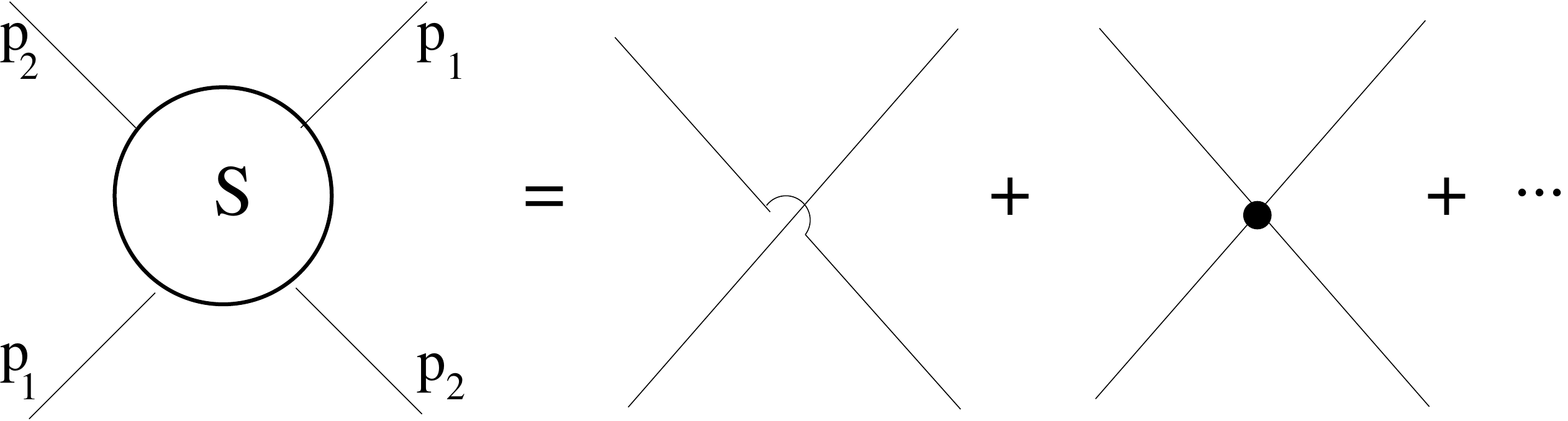}}}
\caption{First terms of the perturbative expansion of $S(\theta_{12})$.}
\label{pertSfig}
\end{figure}
One can express this fact by saying that the interacting theory ($g \neq 0$) presents "fermionic" character (i.e. $S(0) = -1$) while the perturbative expansion done with respect to the free theory presents instead "bosonic" character (i.e. $S(0) =1$).  This observation will be important in the discussion of the classical limit of the quantum theory. To this aim it is also important to anticipate some formulas relative to the phase-shift and the fermionic/bosonic nature of the $S$-matrix. 

\vspace{0.3cm}

\noindent
{\bf Phase-shifts and kernels}. The phase-shift is defined through $S(\theta) = e^{i \varsigma(\theta)}$. However, given $S(\theta)$, there is an ambiguity in choosing the branch of $-i \log S_{\text{ShG}}(\theta)$ which defines $\varsigma(\theta)$. We can in fact choose either 
\be 
\varsigma_f(\theta)\,=\,\left\{\begin{array}{lll}
-2\,\arctan\left(\sin\pi\alpha/\sinh\theta\right) - 2 \pi &,& \theta < 0 \\
-2\, \arctan\left(\sin\pi\alpha/\sinh\theta\right) &,& \theta > 0 
\end{array} \right. 
\label{phaseshiftfermionic}
\ee
called the {\em fermionic} phase-shift, or 
\be 
\begin{array}{lll}
\varsigma_b(\theta)\,=\,-2\, \arctan\left(\sin\pi\alpha/\sinh\theta\right) & , & \forall \,\theta \,\,\,
\end{array}
\label{phaseshiftbosonic}
\ee
called the {\em bosonic} phase-shift. 
\begin{figure}[t]
\centering
$\begin{array}{ccc}
\includegraphics[width=0.4\textwidth]{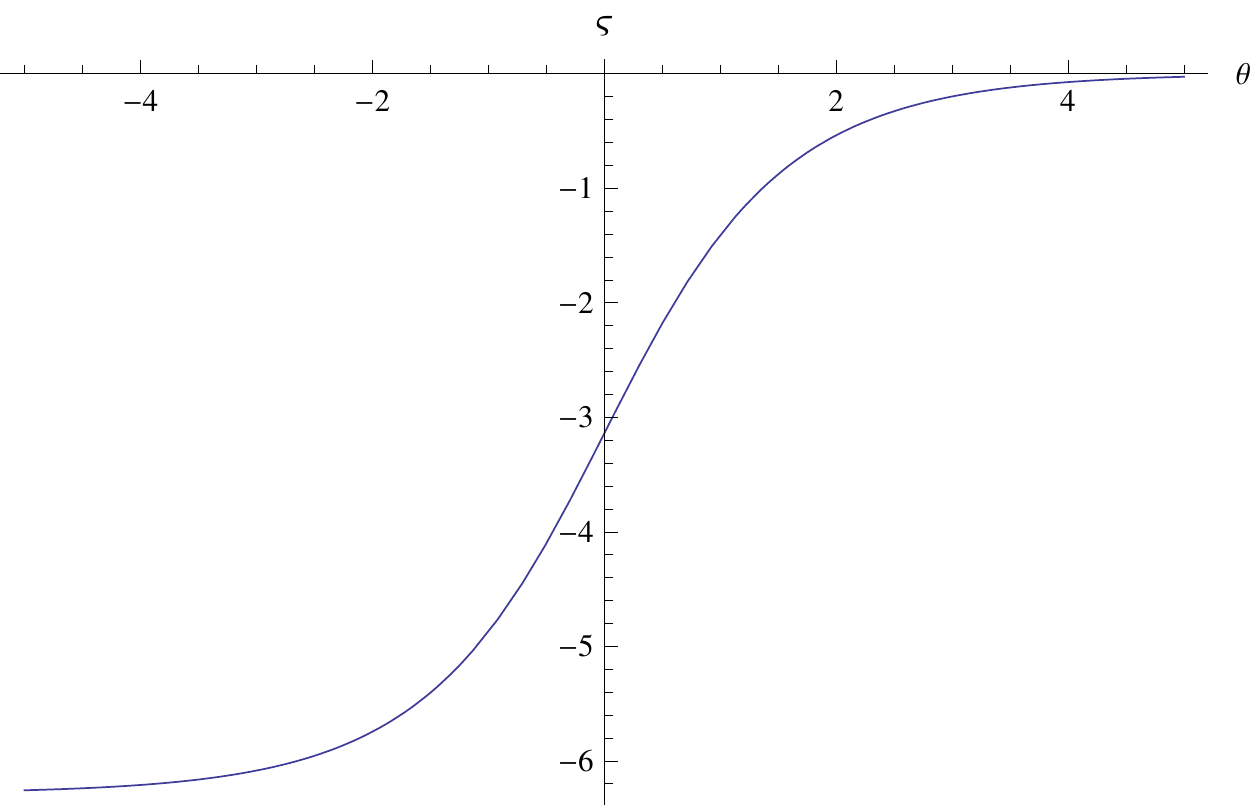} \,\,\,\,\, & & \,\,\,\,\,\,
\includegraphics[width=0.4\textwidth]{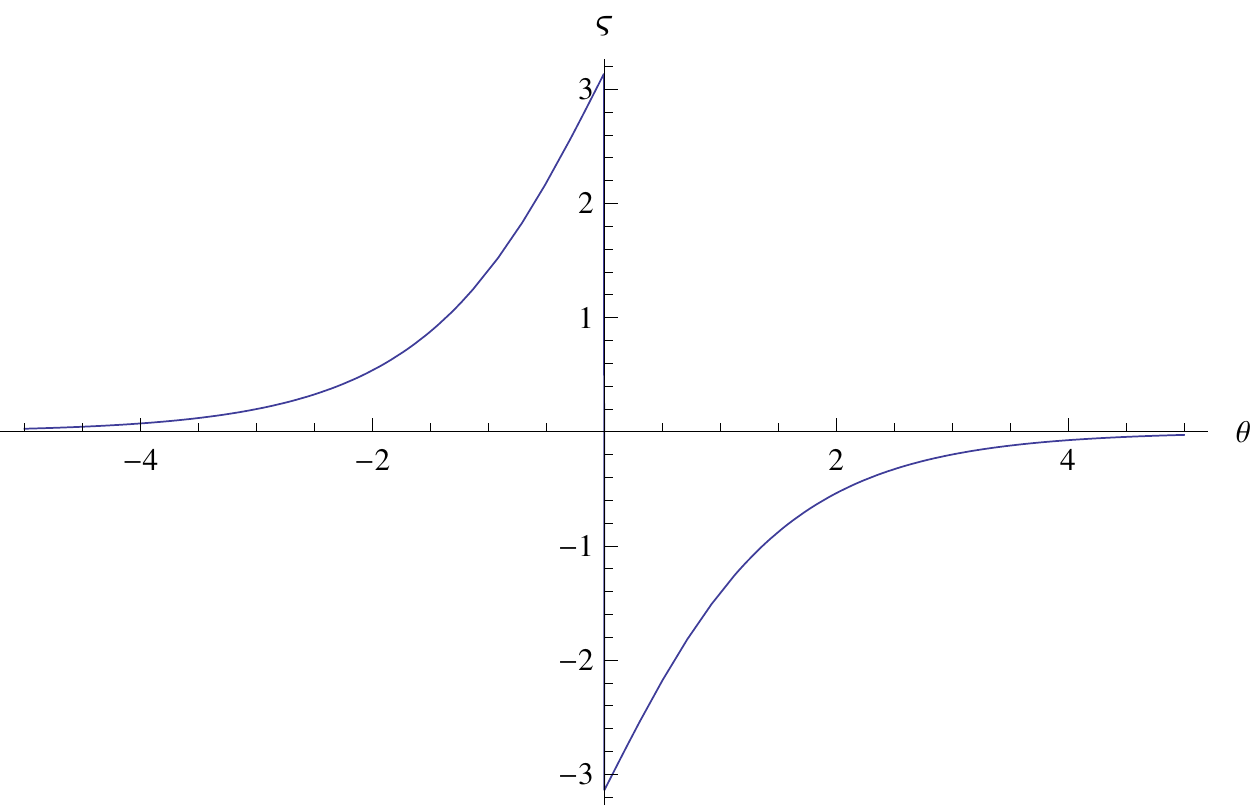}\\
\end{array}$
\caption{On the left, the fermionic phase-shift $\varsigma(\theta)$, on the right the bosonic phase-shift $\varsigma(\theta)$.}
\label{phaseshiftfig}
\end{figure}
The difference between the two functions is that, for finite $g$, $\varsigma_f(\theta)$ does not have a discontinuity along the real axis, while $\varsigma_b(\theta)$ has a jump of $-2 \pi$ at the origin (see Figure \ref{phaseshiftfig}). In the limit $g \rightarrow 0$, on the other hand, the fermionic phase shift $\varsigma_f(\theta)$ goes to the step function $-2 \pi \Theta(-\theta)$, while the bosonic $\varsigma_b(\theta)$ goes to zero.  

Associated to these two definitions of the phase-shift, we have two expressions for the kernel 
\be
\fii_{f,b}(\theta) \equiv \frac{d \varsigma_{f,b}}{d\theta}\,\,\,,
\ee 
which in the two cases are given by
\begin{eqnarray}
\fii_f(\theta)\,& \equiv & \fii_0 \,=\, \, \frac{2 \sin\pi\alpha}{\sinh^2\theta + \sin^2\pi\alpha}  \label{kernels}\\
\fii_b(\theta) & \,=\, & \fii_0 - 2 \pi \delta(\theta) \,\,\,. 
\end{eqnarray}

\vspace{1mm}
Notice that, taking into account the bosonic nature of the classical theory and restoring the presence of $\hbar$ with the substitution (\ref{g2hbar}), in the limit $\hbar\rightarrow 0$ we can define a {\em classical} phase-shift as 
\be 
\varsigma_{cl}(\theta) \,=\,-i \log\left(1 - i \,\frac{g^2 \hbar }{4} \frac{1}{\sinh\theta_{12}}\right) \simeq - \frac{g^2 \hbar }{4} \frac{1}{\sinh\theta_{12}} \,\,\,. 
\label{classicalphaseshift}
\ee

\vspace{1mm}
As well known, the two-body $S$-matrix uniquely fixes all dynamical properties of the theory, such as the matrix elements of the local fields and 
the thermodynamics, as we are going to discuss below.

\section{Form Factors of the Sinh-Gordon model and their classical limit} 
In this section we will discuss the exact Form Factors of the Sinh-Gordon model and we will argue that their leading order in the coupling constant $g$ can be considered as the matrix elements of the classical fields, i.e. they are in one-to-one correspondence with the tree level Feynman diagrams discussed in Section \ref{classicallimit}. As shown below, the Form Factors depends upon the $S$-matrix and, accordingly to the fermionic or bosonic nature of the $S$-matrix, we will have correspondingly {\em fermionic} or {\em bosonic} Form Factors, meaning that the matrix elements will be computed on a basis of particles which behave respectively as fermions or bosons. Let's anticipate that the most natural basis for the classical Form Factors is the bosonic one.   

\vspace{1mm}

Let's initially consider the matrix elements of a local and scalar operator $\mc O(x,t)$ on the asymptotic states $\mid  \theta_1,\theta_2,\cdots,\theta_n \rangle$ 
\be
\langle 0 \mid \mc O(x,t) \mid  \theta_1,\theta_2,\cdots,\theta_n \rangle \,=\, 
\exp\left[i \sum_{k=1}^n (t\, \cosh\theta_k \, - x \, \sinh\theta_k  ) \right] \, F_n^{\mc O}(\theta_1,\theta_2,\cdots, \theta_n) 
\ee
where we have used the translation operator $U= e^{-i p_{\mu} x^{\mu}}$ to extract the momenta dependence of this matrix element, so that 
\be
F_n^{\mc O}(\theta_1,\theta_2,\dots,\theta_n)\,\equiv \,\langle 0 \mid \mc O(0) \mid |\theta_1,\theta_2,\cdots,\theta_n \rangle
\,\,\,.
\ee
The function $F_n^{\mc O}$ is the $n$-particle Form Factors of this operator (see Fig.~\ref{fig:formfactor}).  
A generic matrix element of the operator ${\mc O(0,0)}$ 
\be
F_{m,n}^{\mc O}(\theta_1',\ldots,\theta_m' | \theta_1,\ldots,\theta_n) \equiv 
\langle \theta_1',\ldots,\theta_m' | {\mc O}(0) | \theta_1,\ldots,\theta_n \rangle 
\label{genericFFleftright}
\ee
can be expressed in terms of its Form Factors by using the crossing symmetry, which is implemented by an analytic continuation in the rapidity variables $\theta \rightarrow \theta + i \pi$ and the following recursive equations \cite{Smirnov}
\begin{eqnarray}
&& F_{m,n}^{\mc O}(\theta_1',\ldots,\theta_m' | \theta_1,\ldots,\theta_n) \,=\, F_{m-1,n+1}^{\mc O}(\theta_1',\ldots, | \theta_m' + i \pi,\theta_1,\ldots,\theta_n) \label{recursivecrossing}\\
&& +  2 \pi \sum_{k=1}^n \delta(\theta_m'-\theta_k) \, \left(\prod_{l=1}^{k-1} S(\theta_l-\theta_k)\right) \, 
F_{m-1,n-1}^{\mc O}(\theta_1',\ldots,\theta_{m-1}' | \theta_1,\ldots,\theta_{k-1},\theta_{k+1},\ldots, \theta_n) \nonumber \,\,\,.
\end{eqnarray} 

\begin{figure}[b]
\centerline{\scalebox{0.3}{\includegraphics{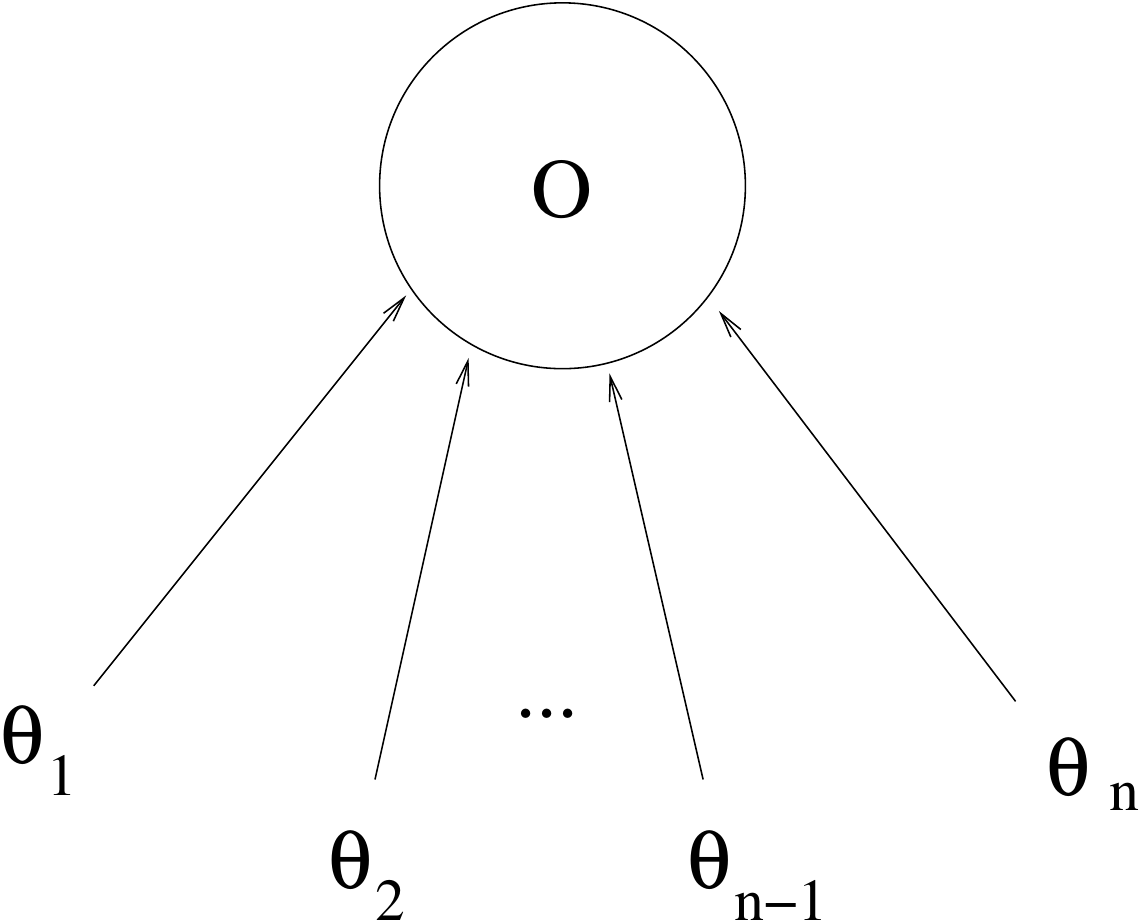}}}
\caption{Form Factor of the operator ${\mathcal O}$.}
\label{fig:formfactor}
\end{figure}

The Form Factors satisfy a series of functional and recursive equations that lead to their exact determination. The interested reader can find detailed discussion on this point in the literature \cite{KW, Smirnov, Luky, KM}. Here we briefly recall the main formulas relative to the Sinh-Gordon model which were obtained in \cite{KM}. 

\vspace{0.3cm}

\noindent
{\bf Basic properties}. For a scalar operator the Form Factors depend only on the differences of rapidities, $\theta_{ij}=\theta_i-\theta_j$. 
In order to write their exact expression one needs to introduce a series of 
quantities: 
\begin{itemize}
\item the function $F_\text{min}(\theta)$ that satisfies the equations \bes
\begin{align}
F_\text{min}(\theta)& = S(\theta)\,F_\text{min}(-\theta)\,,\\
F_\text{min}(i\pi-\theta)& = F_\text{min}(i\pi+\theta)\,,  
\end{align}
\label{eq:Fmineq}
\esu
whose solution is given by 
\begin{equation}
F_\text{min}(\theta)={\cal N}\,\exp\left\{4\int_0^\infty\frac{\ud t}t\,
\frac{\sinh\left(\frac{t}2\,\alpha\right)\sinh\left(\frac{t}2(1-\alpha)\right)}
     {\sinh(t)\cosh\left(\frac{t}2\right)}
     \,\sin^2\left(\frac{t\hat\theta}{2\pi}\right)\right\}\,,
\label{FMINSHG}
\end{equation}
where $\hat\theta = i\pi-\theta$ and ${\cal N}$ is a normalization constant, here chosen as ${\cal N}=F_\text{min}(i\pi)$.
$F_\text{min}(\theta)$ also satisfies the functional equation 
\be
F_\text{min}(i\pi+\theta)F_\text{min}(\theta)=\frac{\sinh\theta}{\sinh\theta+i \sin(\pi\alpha)}\,\,\,.
\labl{eq:Fminrel}
As already noticed for the $S$-matrix, $F_{min}(\theta)$ gets different values at $\theta=0$ according whether $g \neq 0$ or 
$g=0$: in the first case, in fact, $S(0)=-1$ and therefore the first equation in eq.~(\ref{eq:Fmineq}) implies that $F_\text{min}(\theta)$ 
vanishes at $\theta=0$ as $F_\text{min}(\theta) \simeq \theta$ while, if $g =0$ we have instead $F_\text{min}(0) =1$. Its perturbative 
expansion is given by 
\be 
F_\text{min}(\theta)  \,=\, 1 + g^2 \,f_\text{min}(\theta) + \cdots 
\ee
where 
\be 
f_\text{min}(\theta) \,=\,\frac{1}{4 \pi} \, 
\int_0^\infty dt\,
\frac{\sinh\left(\frac{t}2 \right)}
     {\sinh(t)\cosh\left(\frac{t}2\right)}
     \,\sin^2\left(\frac{t\hat\theta}{2\pi}\right)\,\,\,.
     \ee
\item The second quantities we need to introduce are the symmetric polynomials in the variables 
$x_i=e^{\theta_i}$. Such polynomials can be expressed in terms of the basis 
given by the elementary symmetric polynomials $\sigma^{(n)}_{k}$ of the $n$ variables $x_i$ 
defined by the generating functions 
\be 
\prod_{i=1}^n(x+x_i)\,=\,\sum_{k=1}^n
x^{n-k}\sigma^{(n)}_k(x_1,\dots,x_n)\,. 
\ee 
Having defined the quantities above, the final parametrization of the generic
$n$-particle Form Factor of a scalar local operator can be written as 
\be
F_n(\theta_1,\dots,\theta_n)\,=\,H_n\,Q_n(x_1,\dots,x_n)\,\prod^n_{i<j}\frac{F_\text{min}(\theta_{ij})}{x_i+x_j}\,,
\label{F_n}
\ee
where $H_n$ is a normalization factor, $x_i=e^{\theta_i}$ and $Q_n(x_1,\dots,x_n)$ is a symmetric polynomial in $x_i$. 
For a scalar operator the polynomial $Q_n$ has the total degree equal to the degree of the polynomial $\prod_{i < j}
(x_i+x_j)$ in the denominator, i.e. $n(n-1)/2$. The actual expression of the polynomials $Q_n$ can be determined by 
solving the recursive equations of the kinematic poles which occur each time that a rapidity $\theta_i$ becomes equal 
to the value $(i\pi +\theta_j)$ of another rapidity $\theta_j$. With the choice 
\be
H_{2n+1} = H_1 \left(\frac{4\sin(\pi\alpha)}{\cal N} \right)^n\,\,\,\,\,\,\,\,,\,\,\,\,\,\,\,
H_{2n} = H_2 \left(\frac{4\sin(\pi\alpha)}{\cal N} \right)^{n-1}\,
\ee
the recursive equations for the polynomials $Q_n$ entering (\ref{F_n}) are given by  
\be
(-1)^nQ_{n+2}(-x,x,x_1,\dots,x_n)=x D_n(x;x_1,\dots,x_n)Q_n(x_1,\dots,x_n)\,,
\labl{eq:rec}
where   
\be
D_n(x;x_1,\dots,x_n)=\sum_{k=1}^n\sum_{m=1,3,5,\dots}^k(-1)^{(k+1)}[m] x^{2(n-k)+m}\sigma^{(n)}_k\sigma^{(n)}_{k-m}\,.
\ee 
In this formula $\sigma_k^{(n)}$ are the elementary symmetric polynomials  while  
$
[k] \,\equiv\,\sin(k\pi\alpha)/\sin(\pi\alpha)
$. 
\end{itemize}

\vspace{0.3cm}

\noindent
{\bf Elementary Solutions and Exponential Operators}. As shown in \cite{KM}, a solution of the recursive equations (\ref{eq:rec}) is given by the class of symmetric polynomials 
\be
Q_n(k)=\det M_n(k)\,,
\ee
where $M_n(k)$ is a $(n-1)\times(n-1)$ matrix with elements
\be
\left[M_n(k)\right]_{i,j}=\sigma^{(n)}_{2i-j}[i-j+k]\,.
\ee
Notice that the form of this matrix is very close to the denominator of eq.~(\ref{F_n}) that can be expressed as determinant 
of the matrix $ [\tilde M_n]_{i,j} = \sigma^{(n)}_{2i-j}$ 
\be 
\prod_{i < j} (x_i + x_j) \,=\,\det \tilde M_n \,\,\,.
\ee 
The corresponding Form Factors can be identified as the matrix elements of a continuous family of operators identified with the exponential fields $e^{kg\phi}$ \cite{KM}. With the normalization given in this case by $H_n(k)=\left(\frac{4\sin(\pi\alpha)}{\cal N} \right)^n\,[k]$,  
the explicit form of all Form Factors of these operators is then 
\be
F_n(k) = \langle 0| e^{kg\phi}|\theta_1,\theta_2,\dots,\theta_n\rangle \,=\, 
[k]\, \left(\frac{4\sin(\pi\alpha)}{{\cal N}}\right)^{\frac{n}2}\, \det
M_n(k)\,\prod^n_{i<j}\frac{F_\text{min}(\theta_i-\theta_j)}{x_i+x_j}\,.
\labl{eq:FFexp}

\vspace{0.3cm}

\noindent
{\bf Form Factors of Normal Ordered Operators}. It is useful to express the operator content of the theory in terms of a class of particular operators, here denoted by $\no{\phi^k}$, which create $n$ particles out of the vacuum by starting only when $n \geq k$, 
i.e. they satisfy 
\begin{equation}
F^{\no{\,\phi^k\,}}_n(\theta_1,\dots,\theta_n) \,= \,
0 \,\,\,\,\,\,  \text{if} \,\,\,\,\,\, n < k \,.
\label{previous}
\end{equation} 
Their polynomial term $Q_{k}(x_1\dots,x_{k})$ for $n = k$ is equal to the polynomial $\prod_{i <
j}^{k} (x_i + x_j)$ of the denominator and they cancel each other, simply giving
\be
F^{\no{\,\phi^k\,}}_k(\theta_1,\dots,\theta_k) \,=\,  
2^{k}k!\left(\frac{\pi^2\alpha^2}{{\cal N} g^2\sin(\pi\alpha)}\right)^{\frac{k}2}\,\prod_{i<j}^k F_\text{min}(\theta_{ij})\,.
\label{eq:Fmm}
\ee
The reason of introducing these operators is that they will correspond, in the $g \rightarrow 0$ limit, to the composite operators 
of the classical level, where each power of the classical expression $\phi^k(x)$ in the perturbative expansion which always starts with 
$\phi_0^k(x)$ (see Section \ref{ClassicalCompositeFields}). In view of the kinematic recursive equations, the absence of kinematic poles in $F^{\no{\,\phi^k\,}}_k(\theta_1,\dots,\theta_k)$ obviously implies the vanishing values (\ref{previous}).  To compute the Form Factors of these operators when $n > k$, we can use the Form Factors (\ref{eq:FFexp}) of the exponential operators. Let us denote by $\tilde{\phi^m}$ the operator whose Form Factors $\tilde F^m_n$ are obtained by extracting the $\mc O(k^m)$ term in the expansion of $F_n(k)$. Due to Eqs (\ref{previous}) and (\ref{eq:Fmm}) we have
\be
\tilde F^{k}_n= F^{\no{\,\phi^{k}\,}}_n + \sum_{l=2,4,\dots}^{k-2}A^{k}_{l}\,F^{\no{\,\phi^{l}\,}}\,. 
\label{A_k_l}
\ee
This equation implies a mixing among the operators $\no{\phi^{k}}$, as discussed in \cite{KMT}: for the first even levels these mixings are explicitly given by 
\bes
\begin{align}
\tilde{\phi^2}&=\no{\phi^2}\,,\nonumber \\
\tilde{\phi^4}&=\no{\phi^4}-4\,\frac{\pi^2\alpha^2}{g^2}\no{\phi^2}\,,\\
\tilde{\phi^6}&=\no{\phi^6}-20\,\frac{\pi^2\alpha^2}{g^2}\no{\phi^4}+16\,\frac{\pi^4\alpha^4}{g^4}\no{\phi^2}\,,
\nonumber \\
\tilde{\phi^8}&=\no{\phi^8}-56\,\frac{\pi^2\alpha^2}{g^2}\no{\phi^6}+336\,\frac{\pi^4\alpha^4}{g^4}\no{\phi^4}-64\,\frac{\pi^6\alpha^6}{g^6}\no{\phi^2}\,.\nonumber 
\end{align}
\esu

\vspace{0.40mm}

\noindent
{\bf Diagonal Matrix Elements and Connected Form Factors}. For the purpose of computing averages, it is important to consider the diagonal matrix elements of local operators  
\be
F_{2n,{\rm diag}}(\theta_1,\ldots\,\theta_n | \theta_1,\ldots\,\theta_n) \,=\,
\langle \theta_1,\dots,\theta_n | {\mc O}| \theta_1,\dots,\theta_n\rangle \,\,\,. 
\label{FFFdiagonal}
\ee
This expression is formally divergent, as it can be seen by a recursive use of eq.~(\ref{recursivecrossing}). 
A way to regularize the quantity in (\ref{FFFdiagonal}) in infinite volume is to shift the singularities by the infinitesimal quantities $\eta_i$
\be 
F_{2n,{\rm diag}}(\theta_1 + \eta_1,\ldots\,\theta_n + \eta_n | \theta_1,\ldots\,\theta_n)
\,\,\,.
\ee
As shown in \cite{PozsayTakacs}, taking $\eta_i =\eta$, all equal, and sending $\eta \rightarrow 0$, the 
corresponding limit is finite and defines the functions 
\begin{eqnarray}
&& F_{2n,{\rm sym}}(\theta_1,\ldots,\theta_n) \,=\, \langle \theta_1,\dots,\theta_n | {\mc O}| \theta_1,\dots,\theta_n\rangle_\text{sym} \label{eq:symmdef}\\
&& = \left(\lim_{\eta\to0} \3pt{0}{\mc
  O}{\theta_1 + i \pi + i \eta, \dots, \theta_n +i \pi + i\eta, \theta_1,\dots,\theta_n}\right)\,.\nonumber 
\end{eqnarray}
Another way to regularise the diagonal matrix elements is to go to finite volume and the relation between the diagonal Form Factors in infinite and finite volume is given by \cite{PozsayTakacs}  
\be
\langle \theta_1,\dots,\theta_n | {\mc O}| \theta_1,\dots,\theta_n\rangle_\text{L} \,=\,\frac{1}{{\cal J}_n(\theta_1,\ldots,\theta_n)} \,
\sum_{\{\theta_+\} \bigcup \{\theta_-\}} F_{2l,{\rm sym}}(\theta_+) \, {\cal J}_{n-l}(\theta_-) \,\,\,,
\label{finitevsinfinitediagonalFF}
\ee 
where ${\cal J}_k$ is the Jacobian that will encounter later in the Bethe Ansatz formalism (see eq. (\ref{JACOBIAN})) while the sum runs on all possible bipartite partitions of the set of rapidities $\{\theta_1,\ldots,\theta_n\}$ in two disjoint sets made by $l$ and $n-l$ rapidities. 

There is, however, a third way to regularize the quantity in (\ref{FFFdiagonal}) through the so-called the Connected Form Factors, defined as \cite{balog,LM}
\begin{eqnarray}
&& F_{2n,{\rm conn}}(\theta_1,\ldots,\theta_n) \,=\, \langle \theta_1,\dots,\theta_n | {\mc O}| \theta_1,\dots,\theta_n\rangle_\text{conn} \label{eq:conndef}\\
&& = FP\left(\lim_{\eta_i\to0} \3pt{0}{\mc
  O}{\theta_1 + i \pi + i \eta_1, \dots, \theta_n +i \pi + i\eta_n,\theta_1,\dots,\theta_n}\right)\,,\nonumber 
\end{eqnarray}
where $FP$ in front of the expression means taking its finite part, i.e.\ omitting all the terms of the form $\eta_i/\eta_j$ 
and $1/\eta_i^p$ where $p$ is a positive integer. These are precisely the quantities employed in the generalized LeClair-Mussardo formula (\ref{EnsembleAverageMus23}) entering the GGE average.  

The relation between $F_{2n,{\rm sym}}$ and $F_{2n,{\rm conn}}$ was also spelled out in \cite{PozsayTakacs} and the final result can be expressed as follows: let's consider $n$ vertices labelled by the numbers $1, 2, . . . ,n$ and let $G$ be the set of the directed
graphs $G_i$ with the following properties: (a) $G_i$ is tree-like; (b) for each vertex there is at most one outgoing edge. The edge going from $i$ to $j$ is denoted as  $E_{ij}$. With these definitions, 
the function $F_{2n, { \rm sym}}(\theta_1,\ldots,\theta_n)$ can be evaluated as a sum over all graphs in $G$, where the contribution of a graph $G_i$ is given by the following two rules:
\begin{enumerate}
\item 
Let $A_i = \{a_1, a_2, . . . , a_m\}$ be the set of vertices from which there are no outgoing edges in
$G_i$. The Form Factor associated to $G_i$ is $F_{2m,{\rm conn}}(\theta_{a_1},\theta_{a_2},\ldots,\theta_{a_m})$; 
\item 
for each edge $E_{jk}$ the Form Factor above has to be multiplied by $\varphi(\theta_j-\theta_k)$, where $\varphi(\theta)$ is 
the derivative of the phase-shift $\varphi(\theta) \equiv \frac{1}{i} \frac{d \log S(\theta)}{d\theta}$ and is the kernel entering the  
Bethe Ansatz equation (see Section \ref{GBAEQUATIONS})
\end{enumerate}
So, given that $F_2(\theta)$ is a constant and $F_2 =F_{2, {\rm conn}} = F_{2, {\rm sym}}$,  
for the next few cases we have 
\begin{eqnarray*}
&& 
F_{4, {\rm sym}}(\theta_1,\theta_2) \,=\,F_{4, {\rm conn}}(\theta_1,\theta_2) + 2 \varphi(\theta_1-\theta_2) \,F_{2, {\rm conn}} 
\,\,\,;\\ [5pt]
&& 
F_{6, {\rm sym}}(\theta_1,\theta_2,\theta_3) \,=\,F_{6, {\rm conn}}(\theta_1,\theta_2,\theta_3) + 
\left[F_{4, {\rm conn}}(\theta_1,\theta_2) \left(\varphi(\theta_1-\theta_3) + \varphi(\theta_2-\theta_3)\right) + {\rm permutations}\right] \\
&& 
+ 3 F_{2, {\rm conn}} \left[\varphi(\theta_1-\theta_2) \,\varphi(\theta_1-\theta_3) + {\rm permutations} \right] 
\nonumber \,\,\,.
\end{eqnarray*}

In conclusions, thanks to eq.~(\ref{finitevsinfinitediagonalFF}) and the relation between the symmetric and the 
connected Form Factors, the regularized finite volume expression of the diagonal Form Factors can be expressed in terms of 
the connected Form Factors $F_{2n, {\rm conn}}(\theta_1,\ldots, \theta_n)$, the kernel $\varphi(\theta)$ of the Bethe Ansatz equations,  and the finite volume density of states ${\cal J}_n$.

For the connected FF of the normal ordered even powers $:\phi^{2k}:$ of the elementary field, using the functional equation (\ref{eq:Fminrel}) satisfied by $F_{\text{min}}(\theta)$, it is easy to see that  the first non-zero connected FF of these operators are given by 
\be
\langle \theta_k,\ldots,\theta_1 | :\phi^{2k}(0): | \theta_1,\ldots,\theta_k \rangle \,=\,
2^{2k} (2 k)! \left(
\frac{\pi^2\alpha^2}{{\cal N} g^2\sin(\pi\alpha)}\right)^{k}\,\prod_{i<j}^k \frac{\sinh^2\theta_{ij}}{\sinh^2\theta_{ij} + \sin^2(\pi \alpha)}
\,\,\,.
\ee
Higher-particle connected FF of these operators are more involved but they can be computed using the definition (\ref{eq:conndef}) and the exact expressions of their Form Factors.

\vspace{0.3cm}

\noindent
{\bf Form Factors of the stress-energy tensor. Fermionic and bosonic basis}.  
An important set of connected FF are those associated to the trace of the stress-energy tensor. They can be given both in the fermionic or bosonic basis. The first two representatives are given by 
\begin{eqnarray}
\langle \theta | \Theta(0) | \theta \rangle_{f,b} &\,=\, & 2 \pi m^2 \\
\langle \theta_2,\theta_1 | \Theta(0) | \theta_1,\theta_2\rangle_{f,b} &\,=\, & 4 \pi m^2 \varphi_{f,b}(\theta_1-\theta_2)\,\cosh(\theta_1-\theta_2) 
\nonumber 
\end{eqnarray}
and an inductive application of the FF residue equations leads to ($\theta_{ij} = \theta_i-\theta_j$) \cite{LM} 
\be 
\langle \theta_n\ldots \theta_1 | \Theta(0) | \theta_1 \ldots \theta_n\rangle_\text{conn} \,=\,
2\pi m^2 \,\varphi_{f,b}(\theta_{12}) \varphi_{f,b}(\theta_{23}) \ldots \varphi_{f,b}(\theta_{(n-1) \,n}) \cosh(\theta_{1n}) + {\rm permutations}
\label{CONNECTEDFFTRACE}
\ee
where the fermionic and bosonic kernels $\varphi_{f,b}(\theta)$ are given in eq.\,(\ref{kernels}).
 
\vspace{0.3cm}
\noindent
{\bf Form Factors of Classical Operators}. The exact expressions given above encode the sum of all diagrams of the perturbative series. Expanding them in the coupling constant $g$ (alias in $\hbar$) and taking just the leading order, one is expected to find then the result coming from the sum of the tree level only, i.e. the expressions for the classical operators previously recovered by solving perturbatively the classical equation of motion. This can be indeed shown explicitly for the first examples. It is important to notice that, for the classical operators, there is no mixing among them, because all coefficients $A_l^k$ vanish when $\hbar \rightarrow 0$. 
\begin{itemize}
\item Consider the Form Factors of the elementary field $\phi(x)$ obtained by the ${\mathcal O}(k)$ term in the Taylor expansion in $k$, divided by $g$. This is essentially equivalent to take $Q_n(0)$, which implies that all Form Factors with even $n$ vanish. 
The first non-zero Form Factors are 
\begin{eqnarray}
F_1^\phi &\,=\,& \frac{1}{\sqrt 2} \,\,\,, \nonumber \\
F_3^\phi & \,=\,& \frac{1}{\sqrt 2} \left(\frac{4 \sin\pi\alpha}{{\cal N}}\right) \,\sigma_3 \,\sigma_2 \, 
\prod^3_{i<j}\frac{F_\text{min}(\theta_{ij})}{x_i+x_j} \,\,\,, \\
F_5^\phi & \,=\,& \frac{1}{\sqrt 2} \left(\frac{4 \sin\pi\alpha}{{\cal N}}\right)^2 \,\sigma_5 \,\left[\sigma_2 \, \sigma_3 - 
2 \cos (\pi \alpha) \sigma_5 \right] \, 
\prod^5_{i<j}\frac{F_\text{min}(\theta_{ij})}{x_i+x_j} \,\,\,. \nonumber 
\end{eqnarray}
$F_1^\phi$ fixes the normalization of the field when it acts on one-particle state. Let's now expand the expressions above in 
powers of $g$ and keep the leading order. We have 
\begin{eqnarray}
F_1^\phi &\,=\,& \frac{1}{\sqrt 2} \,\,\,, \nonumber \\
{\cal F}_3^\phi & \,=\,& \frac{g^2}{2 \sqrt 2} \,\sigma_3 \,\sigma_2 \, 
\prod^3_{i<j}\frac{1}{x_i+x_j} \,\,\, \label{classicalFFphi} \\
{\cal F}_5^\phi & \,=\,& \frac{g^4}{4 \sqrt 2}  \,\sigma_5 \,\left[\sigma_2 \, \sigma_3 - 2 \sigma_5 \right] \, 
\prod^5_{i<j}\frac{1}{x_i+x_j} \,\,\,. \nonumber 
\end{eqnarray}
One can check that the limiting form of $F_3^\phi$ corresponds to the Feynman graph drawn on the left-side of 
Figure \ref{treelevel} while the one of $F_5^\phi$ corresponds to the sum of the Feynman graphs of Figure \ref{phi2SH}. The check 
may be tedious because, while in the Feynman diagram computation, one deals with the sum of diagrams, 
the Form Factor expression is concisely expressed, on the contrary, as a product which collects altogether all 
the Feynman diagrams.  
\item Consider now the Form Factors of the composite field $:\phi^2(x):$, whose Form Factors are given by the 
${\mc O}(k ^2)$ term in the Taylor-expansion in $k$  
\begin{eqnarray}
F_2^{:\phi^2:} &\,=\, & 
8\,\left(\frac{\pi^2\alpha^2}{{\cal N} g^2\sin(\pi\alpha)}\right)\,F_\text{min}(\theta_{12})
 \,\,\,, \\
F_4^{:\phi^2:} &\,=\,& \frac{32\pi^2 \alpha^2}{{\cal N}^2 \,g^2} \left(\sigma_1^2 \sigma_4 + \sigma_3^2\right) \,
\prod^4_{i<j}\frac{F_\text{min}(\theta_{ij})}{x_i+x_j} \nonumber \,\,\,,
\end{eqnarray}
Taking now the leading order in $g$ and dividing for $g^2$ to get the proper normalization of the $:\phi^2(x):$ operator, we arrive to the limiting expressions 
\begin{eqnarray} 
 {\cal F}_2^{:\phi^2:} &\,=\, & 1 
 \,\,\,, \\
 {\cal F}_4^{:\phi^2:} &\,=\,& \frac{g^2}{2} \left(\sigma_1^2 \sigma_4 + \sigma_3^2\right) \,
\prod^4_{i<j} \frac{1}{x_i+x_j} \label{phi24classical}\,\,\,,
\end{eqnarray} 
While the first expression gives the normalization of this operator, the second expression coincides with the sum of the tree 
Feynman diagrams drawn in Figure \ref{FF24SH}. 
\end{itemize}

\vspace{3mm}
\noindent
{\bf Classical Connected Form Factors}. In taking the leading term in the limit $g\rightarrow 0$ of the Form Factors above we have 
set $F_\text{min}(\theta_{ij}) \,=\,1$ and we have correctly reproduced the sum of the tree level diagrams of the corresponding matrix 
elements. However, as we are going to see below, $F_\text{min}(\theta_{ij})$ plays an important role when one consider the Classical Connected Form Factors. 

The first example is given by the operator $:\phi^2(x):$ and its 2-particle Classical Connected Form Factor: using (\ref{eq:conndef}) on the classical expression (\ref{phi24classical}) we find 
\be 
\langle \theta_2,\theta_1 \mid :\phi^2(0): \mid \theta_1,\theta_2 \rangle_{\text{conn}}^{\text{cl}} \,=\, \frac{g^2}{2} \,\frac{1}{\sinh^2\theta_{12}} 
\,\,\,,
\ee
which is {\em singular} when $\theta_{12} =0$. However, if we had used in the Form Factor also the term $F_{\text{min}}(\theta_{ij})$ and we had used its functional equation (\ref{eq:Fminrel}) we would have arrived instead to 
\be 
\langle \theta_2,\theta_1 \mid :\phi^2(0): \mid \theta_1,\theta_2 \rangle_{\text{conn}} \,=\, \frac{g^2}{2} \,
\frac{1}{\sinh^2\theta_{12} + \sin^2\pi\alpha} 
\,\,\,,
\ee 
an expression which is instead {\em regular} at $\theta_{12} = 0$: the regularization is provided by the term $\sin^2\pi\alpha$ in the denominator, whose origin is the infinite resummation of the contributions coming from all the loop diagrams of the theory.  Said differently, the quantum theory expression is more regular than its classical version!  

As a matter of fact, the regularization, provided by the presence of $F_\text{min}(\theta_{ij})$ in the connected Form Factors is quite a general fact in all connected Form Factors: at the practical level, this regularization can be seen as the substitution rule  
\be
\frac{1}{\sinh^2\theta_{ij}} \longrightarrow \frac{1}{\sinh^2\theta_{ij} + \sin^2\pi\alpha} \,\,\,. 
\label{substitutionrule}
\ee 
To prove it, notice that by taking the rapidity configuration of eq.~(\ref{eq:conndef}), the denominator of the Form Factors becomes 
\begin{eqnarray}
\prod_{i<j}^{2 k} \frac{1}{x_i + x_j} &\longrightarrow &\frac{1}{i^k (x_1 x_2 \cdots x_k)^{2(k-1)} (\eta_1 \eta_2 \cdot \eta_k)} \, \prod_{i < j}^k 
\frac{1}{\left(\frac{x_i}{x_j} - \frac{x_j}{x_i}\right)^2} \,=\, \label{changedenominator}
\\
& = & \frac{(-i)^k}{2^k} \, \frac{1}{(x_1 x_2 \cdots x_k)^{2(k-1)}(\eta_1 \eta_2 \cdot \eta_k)}\,
\prod_{i<j}^k \frac{1}{\sinh^2\theta_{ij}}\,\,\,.
\end{eqnarray}
On the other hand, since the symmetric polynomial which is at the numerator of the Form Factor has the same total degree 
of the one at the denominator, in the configuration of the rapidities given by eq.~(\ref{eq:conndef}) the numerator will produce a term 
$\eta_1 \eta_2 \cdots \eta_k$ that cancels the one in eq.~(\ref{changedenominator}) thus leaving a Finite Part. 
Finally, involving the product of $F_\text{min}(\theta_{ij})$ and using the functional equation (\ref{eq:Fminrel}), it is easy to see that, in the final expression, the result is equivalent to apply the substitution rule (\ref{substitutionrule}). 

\vspace{3mm}
\noindent
{\bf Summary}. The results of this section can be summarized as follows: 
\begin{itemize}
\item the exact expressions of the Form Factors of the quantum Sinh-Gordon model permit us to assign a meaning also 
to matrix elements of {\em classical operators}. 
\item these matrix elements of {\em classical operators} are nothing else but the exact resummation of the tree-level diagrams. 
From the analysis done in Section \ref{classicallimit}, we know that these diagrams are in one-to-one correspondence with the 
expressions obtained solving perturbatively the classical equation of motion.
\item the classical limit of the Form Factors are more naturally associated to a {\em bosonic} basis rather that a {\em fermionic} one, simply because the limit $\hbar \rightarrow 0$ gives rise to a bosonic $S$-matrix, with $S(0) = 1$.   
\item the connected classical Form Factors turn out to be more singular than the exact ones. This fact will force us to introduce 
a prescription when we are going to integrate them in a formula like (\ref{EnsembleAverageMus23}). 
\item analogous singular expressions appear in the classical version of the Bethe Ansatz equation, discussed in the next Section, 
and they have to be handled in the same way. 
\end{itemize}

\section{Generalized Bethe Ansatz of the Sinh-Gordon model} \label{GBAEQUATIONS}
The two-body $S$-matrix of the Sinh-Gordon model determines its thermodynamic properties, which may also include the extra conserved charges. The formalism has been developed by several authors \cite{YY,ZamTBA,FM,MC,CK}. In this section we are going to briefly recall the Generalized Bethe Ansatz (GBA) equations of the Sinh-G model and then to present the Caux-Konik formalism which allows us to find the relevant quantities needed for the out of equilibrium situation. 

\subsection{Generalized Bethe Ansatz} 
We will present the formalism of the Generalized Bethe Ansatz in two different but equivalent formulations, the {\em fermionic} and the {\em bosonic} ones, since they will be important in the sequel of this paper. 

\begin{figure}[b]
\centering
$\begin{array}{c}
\includegraphics[width=0.5\textwidth]{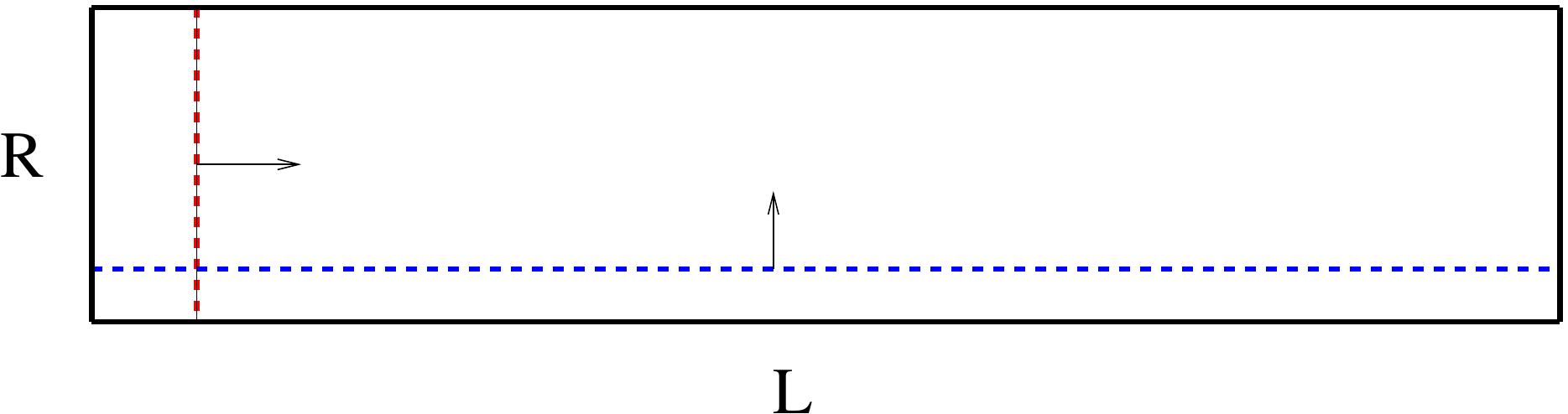} 
\\
\end{array}$
\caption{Finite geometry for the generalized partition function $Z(R,L)$.}
\label{tbafig}
\end{figure}

\vspace{0.3cm}
\noindent
{\bf L and R channels}.
The Generalized Bethe Ansatz concerns with the computation of partition function and an underlying geometrical picture of this computation can enlighten some of its important features. Let's consider our (euclidean) quantum field theory defined on rectangular geometry, as the one in Figure \ref{tbafig}, with periodic boundary conditions along both the axes $R$ and $L$. Moreover, let's take as generalized "Hamiltonian" ${\mathcal H}$ the one made of a weighted sum (with positive coefficients $\alpha_n$) of the original Hamiltonian plus all other conserved charges ${\mathcal Q}_n$ 
\be 
{\mathcal H} \,=\,\sum_{n=0}^\infty \alpha_n {\mathcal Q}_n \,\,\,,
\label{generalizedH}
\ee
where $Q_0 = H$. In the following we will restrict, for simplicity, our attention only on the even charges ${\mathcal E}_n$. In our geometry, the partition function $Z(R,L)$ can be computed in two equivalent ways \cite{ZamTBA}: 
\begin{enumerate}
\item 
in the first scheme, we take the $L$ axis as "time axis". This will be called the "L-channel" and the Hilbert space, on which act the conserved quantities $({\mathcal Q}_n)_L$ , lives along the orthogonal axis $R$. In terms of the stress-energy tensor $T_{\mu\nu}$ ($\mu,\nu=1,2$), in this channel the Hamiltonian $H_L$ is given by 
\be
H_L \,=\,\frac{1}{2\pi} \int_0^R T_{11}(y)\,dy  \,\,\,.
\label{H_L}
\ee
\item 
in the second scheme, it is the $R$ axis that plays the role of time. This will be called the "R-channel" and the 
corresponding Hilbert space, on which act the conserved quantities $({\mathcal Q}_n)_R$, lives along the orthogonal axis $L$. In this case the Hamiltonian $H_R$ is expressed in terms of the stress-energy tensor as 
\be 
H_R \,=\,\frac{1}{2\pi} \int_{-L/2}^{L/2} T_{22}(x) \, dx\,\,\,.
\label{H_R} 
\ee
\end{enumerate}
Correspondingly, we have two equivalent expression of the generalized partition function
\be
Z(R,L) \,=\,\left\{
\begin{array}{lll}
{\rm Tr}\, e^{-L {\mathcal H}_L}  &;& {\rm L-channel}\\
{\rm Tr} \, e^{-R {\mathcal H}_R} &;& {\rm R-channel} 
 \end{array}
 \right.
 \ee
In the limit $L/R \rightarrow \infty$, the first expression projects on the lowest values $E_0(R;\{\alpha_n\})$ of the generalized Hamiltonian ${\mathcal H}_L$, while the second one gives rise to the generalized free-energy $F(R;\{\alpha_n\})$ of the system
\be
Z(R,L) \,=\,\left\{
\begin{array}{lll}
e^{-L E_0(R;\{\alpha_n\})} &;& {\rm L-channel}\\
e^{-R L F(R;\{\alpha_n\})} &;& {\rm R-channel}
 \end{array}
 \right.
 \ee
with the identity 
 \be
 E_0(R;\{\alpha_n\}) \,=\,R \, F(R;\{\alpha_n\})\,\,\,.
 \label{basicidentitye0}
 \ee

\vspace{0.3cm}

\noindent
{\bf Generalized Free Energy}.
Let's now use the R-channel formulation to compute the generalized free-energy $F(R;\{\alpha_n\})$ using the Generalized Bethe Ansatz, whose starting point is the quantum integrability of the model: this implies the conservation of the number of particles and their momenta, so that we can associate to the $N$-particle configuration of the field a wave function in the Hilbert space of the $R$-channel made 
of super-position of planes waves 
\be
\psi(x_1, x_2, \ldots , x_n) \,=\, \exp\left(i \sum_j m \sinh\theta_j \,x_j \right) \,\sum_{Q \in S_N} A(Q) \Theta(x_Q) \,\,\,,
\label{BAwavefunction}
\ee
where the second sum runs on the $N!$ permutations $Q\in S_N$ of the particle coordinates and 
\[
\Theta(x_Q) \,=\, \left\{
\begin{array}{ll} 
1 & \text{if}\,\, x_{Q_1} < x_{Q_2} < \cdots \,x_{Q_N} \\
0 & \text{otherwise}
\end{array}
\right.
\]
The coefficients $A(Q)$ are determined by the $S$-matrix and for configurations $Q$ and $Q'$ which differ by the exchange of the particles $i$ and $j$ they satisfy 
\be 
A(Q) \,=\,S(\theta_i-\theta_j) \,A(Q')   \,\,\,\,
\ee
On an interval $L$ with periodic boundary conditions, the wave function must satisfy 
\be 
\psi(x_1,\ldots, x_j+L,\ldots,x_n) \,=\,\psi(x_1,\ldots, x_j,\ldots, x_n) 
\ee
for any coordinate $x_j$ and this gives rise to the $N$ equations for the rapidities $\{\theta_1,\theta_2,\ldots,\theta_n\}$ 
\be
\exp[i L\,\sinh\theta_j] \,\prod_{k\neq j} S(\theta_j - \theta_k) \,=\, 1 \,\,\,.
\label{bethe1}
\ee
Expressing the $S$-matrix in terms of the phase-shift given in eqs.\,(\ref{phaseshiftfermionic}) and (\ref{phaseshiftbosonic}), they can be written as   
\be
{\cal J}_j \,\equiv\, M L \sinh\theta_j + \sum_{k\neq j}^N \varsigma_{f,b}(\theta_j-\theta_k)\,=\, 2\pi \,N_j\,\,\,, 
\label{BETHEEQUATIONS}
\ee
where the $N_i$ are (positive or negative) integers, the so-called Bethe quantum numbers. Employing either the fermionic or the bosonic phase-shift, it gives rise to two different implementation of the Bethe Ansatz, whose equivalence we are going to prove later. 

Given a set of $N_i$'s, the $\theta_j$'s are uniquely determined. The wave function (\ref{BAwavefunction}) vanishes when any pairs of rapidities are equal. Using the fermionic phase-shift $\varsigma_{f}(\theta)$ this implies that all $N^{f}_i$ have to be different, as it happens for the quantum numbers of fermions. Using instead the bosonic phase-shift $\varsigma_{b}(\theta)$, the quantum numbers $N^{b}_i$ are allowed to be equal (as it happens for the quantum numbers of bosons), since for $\theta_k > \theta_j$, $\varsigma(\theta_j-\theta_k)$ adds an extra $-2 \pi$ in the right-hand side of the equation with respect to the fermionic case, so that $N^{b}_j$ must be greater than $N^{f}_j$ by one for each filled $k$ greater than $j$. So, for instance, with $N^{f}_i$ being a sequential integers, $N^{b}_i$ will be all equal.   

In the thermodynamic limit ($N\to\infty$, $L\to\infty$, $N/L=n=\text{fixed}$) the equations (\ref{BETHEEQUATIONS}) turn into the integral equation
\be
J(\theta)\,=\, M\sinh\theta +
\,\int_{-\infty}^\infty \mathrm{d}\theta' \,
\varsigma_{f,b}(\theta-\theta')\, \rho^{\text{(r)}}(\theta')\,\,\,,
\ee
and, by differentiating w.r.t. $\theta$, one gets
\be
\rho(\theta)\,=\, 
\frac{M}{2\pi}\cosh \theta +
\int_{-\infty}^\infty\frac{\mathrm{d}\theta'}{2\pi}\,
\fii_{f,b}(\theta - \theta') \rho^{\text{(r)}}(\theta')\,.
\labl{eq:rhos}
In the equations above
\be
\rho(\theta)\ =\, \frac1{2\pi}\frac\p{\p\theta}J(\theta)
\ee
is the density of states per unit length where the fermionic/bosonic form of the kernel $\fii_{f,b}(\theta)$ was given 
in (\ref{kernels}). The root density enters the expression of the energy and all the conserved charges: restricting the attention to the 
even conserved charges ${\mathcal E}_n$, they can be combined together into the functional per unit length 
\be 
{\cal D}[\rho^{\text{(r)}}] \,=\,\int_{-\infty}^{\infty} \,\Lambda(\theta,\{\alpha\}) \, \rho^{\text{(r)}}(\theta)\,\,\,,
\ee
where 
\be 
\Lambda\left(\theta,\left\{\alpha \right\}\right) \, = \, \sum_{n=0}^\infty \alpha_{n} \, M^{2n+1}\,\cosh[(2n+1) \theta] \,\,\,
\label{generalizedfunctional}
\ee
depends on the infinite number of variables $\alpha_{n}$ which can be considered as the conjugated variables of the conserved charges ${\mathcal E}_{2n+1}$. On the other hand, in the thermodynamic limit $L \rightarrow \infty$, there is an entropy per unit length $S(\rho,\rho^{\text{(r)}})$ due to the fact that a large number of quantum states are compatible with the densities $\rho$ and $\rho^{\text{(r)}}$. In the two cases, its expression is given by 
\begin{eqnarray}
S_f[\rho, \rho^{\text{(r)}}] \,& = & \,\int_{-\infty}^{\infty} d\theta \left[\,\rho \, \log\rho - \rho^{\text{(r)}} \, \log \rho^{\text{(r)}} - (\rho - \rho^{\text{(r)}}) 
\, \log (\rho - \rho^{\text{(r)}}) \, \right] \\
S_b[\rho,\rho^{\text{(r)}}]\,& = & \,\int_{-\infty}^{\infty} d\theta \left[\, (\rho + \rho^{\text{(r)}}) 
\, \log (\rho + \rho^{\text{(r)}}) - \rho \, \log\rho - \rho^{\text{(r)}} \, \log \rho^{\text{(r)}}  \right] 
\end{eqnarray}
To determine $\rho(\theta)$ and $\rho^{\text{(r)}}(\theta)$, one can minimise the Generalized Free-Energy, given by the 
functional 
\be
R\, F_{f,b}[\rho,\rho^{\text{(r)}}] \,\equiv\, R \, {\cal D}[\rho]  - S_{f,b}[\rho,\rho^{\text{(r)}}] \,\,\,,
\ee
with respect to the densities $\rho(\theta)$ and $\rho^\text{(r)}(\theta)$ (with the constraint (\ref{eq:rhos})). The final expressions can be expressed in terms of the pseudo-energy $\eps_{f,b}(\theta)$ defined in the two cases by  
\be 
f_{f,b}(\theta) \,\equiv\, \frac{\rho^{\text{(r)}}}{\rho} \,=\, 
\left\{ 
\begin{array}{l} 
(e^{\eps_f}+1)^{-1} \\
(e^{\eps_b} -1)^{-1}
\end{array} 
\right. 
\label{filling}
\ee
The function $f_{f,b}(\theta)$ is the filling fraction of the states respectively for the fermionic and bosonic cases. The relevant formulas for the fermionic and bosonic case are quite similar and we present them simultaneously, with the convention that the upper sign always refers to the fermionic case while the lower to the bosonic case. Moreover, in the following formulas we use for the kernel the expression $\fii(\theta)$ for both the fermionic and bosonic case, and to specialize to one or the other, one has to refer to eq.~(\ref{kernels}). 

The pseudo-energy for the two cases satisfies the integral equation 
\be
\eps(\theta) \,= \, R\,\Lambda(\theta,\{\alpha\}) \, \mp \, 
\int_{-\infty}^\infty\frac{\mathrm{d}\theta'}{2\pi}\,\fii(\theta-\theta')
\log\left(1\pm e^{-\eps(\theta')}\right)\,.
\labl{eq:TBAf}
while the final formula of the Generalized Free Energy per unit length is given by 
\be
R\,F_{f,b} \,=\,\mp 
\frac{1}{2\pi }\int_{-\infty}^\infty\mathrm{d}\theta\,M \cosh \theta\,
\log\left(1 \pm e^{-\eps(\theta)}\right) \,.
\labl{eq:GFEF}

\vspace{3mm}
\noindent
{\bf Equivalence of the bosonic and fermionic formulation}. It is now important to notice that, as in the case of Lieb-Liniger model analysed in \cite{Wadati}, also for the relativistic models the fermionic and bosonic formulation are exactly equivalent: indeed they can be mapped one onto the other by posing 
\begin{eqnarray}
&& \rho_f(\theta) - \rho^{(r)}_f(\theta) \,=\, \rho_b(\theta) \,\,\,,
\label{densitiesfermionboson} \\
&&1+e^{-\eps_f(\theta)} \,=\, (1 - e^{-\eps_b(\theta)})^{-1} \,\,\,, 
\label{mappingTBA}
\end{eqnarray}
where $\rho_f(\theta)$ and $\rho^{(r)}_f(\theta)$ denote respectively the density of states and the root density of the fermionic formulation, while 
$\rho_b(\theta)$ denotes the density of states of the bosonic formulation. By using the relation between the fermionic and the bosonic kernels given in eq.~(\ref{kernels}), it is easy to see that the fermionic version of eq.~(\ref{eq:TBAf}) transforms into the bosonic one and vice-versa. Correspondingly, the fermionic Generalized Free Energy given in (\ref{eq:GFEF}) becomes the bosonic one and vice-versa. 

Notice that the fermionic and bosonic filling fractions do {\em not} go one into the other under the transformation (\ref{mappingTBA}). We have, for instance, that under the mapping (\ref{mappingTBA}), the fermionic filling fraction is expressed as 
\be
f_f = \frac{f_b}{1+ f_b} \,=\,e^{-\eps_b} \,\,\,, 
\label{transffilling1}
\ee
and viceversa
\be
f_b = \frac{f_f}{1- f_f} \,=\,e^{-\eps_f} \,\,\,. 
\label{transffilling2}
\ee
The meaning of these transformations will become clear soon. 

\vspace{3mm}

\noindent
{\bf Expectation values of higher charges}. 
The expectation values per unit length of higher charges can be easily computed in both fermionic and bosonic cases (their Free Energy here denoted by the same symbol $F$)
\be
\langle {\mathcal E}_{n} \rangle \,=\, \frac{ \partial  F}{\partial \alpha_n} \,=\, \frac{1}{2\pi R} 
\int_{-\infty}^{\infty} M \cosh\theta \,f(\theta) \, \frac{\partial \eps}{\partial \alpha_n} \,\,\,. 
\,\,\,
\label{conscharGGEf}
\ee
The function $\frac{1}{2 \pi R} \frac{\partial \eps}{\partial \alpha_n}\equiv \epsilon^{(n)}(\theta)$ satisfies the {\em linear} integral equation 
\be
\epsilon^{(n)}(\theta) \,= \, \frac{1}{2 \pi} M^{2n+1} \,\cosh[(2n+1)\theta] \, +\, 
\int_{-\infty}^\infty\frac{\mathrm{d}\theta'}{2\pi}\,\fii(\theta-\theta')\,
f(\theta') \, \epsilon^{(n)}(\theta') \,\,\,, 
\labl{eq:TBApseudoalphanf} 
whose solution can be obtained by iteration. The linearity of this equation comes from the fact that the pseudo-energy $\eps(\theta)$ entering the filling fraction $f(\theta')$ has to be considered as a {\em given} function, i.e. the function obtained by the solution of equation (\ref{eq:TBAf}). 
Notice that the density of roots $\rho^{(r)}(\theta)$ is related to the density of states $\rho(\theta)$ by means of the filling fraction
\be 
\rho^{(r)} \,=\,f(\theta)  \,\rho(\theta) \,\,\,.
\label{rhorrho}
\ee 
Substituting this relation into eq.~(\ref{eq:rhos}) and comparing with eq.~(\ref{eq:TBApseudoalphanf}), 
one can see that the density of state can be expressed as  
\be 
\rho(\theta) \,=\,\frac{1}{2\pi R} \, \frac{\partial\eps}{\partial \alpha_0} \equiv \epsilon^{(0)}(\theta)\,\,\,.
\label{rhoepsilon0}
\ee
Using eq.~(\ref{eq:TBApseudoalphanf}) it is easy to see that the following identity holds
\be
\int_{-\infty}^{\infty} M \cosh\theta \,f(\theta) \, \epsilon^{(n)}(\theta) \,=\,
\int_{-\infty}^{\infty} M^{2n+1} \cosh[(2n+1)\theta] \,f(\theta) \, \epsilon^{(0)}(\theta)\,\,\,,
\ee
and therefore, using Eqs.(\ref{rhorrho}) and (\ref{rhoepsilon0}), the expectation value of the 
conserved charges coincides with its original definition   
\be
\langle {\mathcal E}_{n} \rangle \,=\,  
\int_{-\infty}^{\infty} M^{2n+1} \cosh[(2n+1)\theta] \, \rho^{(r)}(\theta) \,\,\,.
\label{check}
\ee

\vspace{3mm}
\noindent
{\bf Expectation value of the trace of the stress-energy tensor}. The Generalized Bethe Ansatz allows us to compute the 
expectation value of the trace of the stress-energy tensor, given by $\langle \Theta\rangle = \langle T_{11} + T_{22} \rangle$. 
This will be an important formula for setting up the formalism of computing the expectation values of other local operators. 
Let's first doing the computation in the fermionic basis. 
 
The computation of $\langle T_{22} \rangle$ is straightforward, since its integral gives the total energy in the $R$-channel, 
see eq.~(\ref{H_R}): using the translation invariance, its expectation value per unit length is then given by 
\be 
\langle T_{22} \rangle \,=\,2\pi  M \int_{-\infty}^{\infty} M \cosh\theta \rho^{(r)}(\theta) \,=\,
2\pi M \int_{-\infty}^{\infty} \cosh\theta \,f_f(\theta)  \,\epsilon_0(\theta) \,\,\,.
\ee
To compute $\langle T_{11}\rangle$ one needs to exploit instead the $L$-channel: in this channel, as shown in eq.~(\ref{H_L}), its integral provides the total energy in the $L$-channel. If we denote this energy by $e_0(R)$, using the translation invariance, we have 
\be
\langle T_{22} \rangle \,=\,2\pi \frac{e_0(R)}{R} \,\,\,.
\ee
The problem is now to extract $e_0(R)$ from the expression $E_0(R,\{\alpha_n\})$ which involves all the conserved charges. This is easily done by using, first of all, the equation (\ref{basicidentitye0}) and then, integrating by part the corresponding expression $R F(R)$, so that  
\be 
\frac{E_0(R,\{\alpha_n\})}{R} \,=\, - \frac{1}{2 \pi R} \int_{-\infty}^{\infty} M \sinh\theta \,f_f(\theta)\,  
\frac{\partial \epsilon_f}{\partial \theta} \,\,\,.
\label{partialtheta}
\ee
The key point is that the quantity $\frac{1}{2\pi R} \frac{\partial \epsilon}{\partial \theta}\equiv \epsilon_\theta$ satisfies the 
{\em linear} integral equation 
\be 
\epsilon_\theta \,=\,\frac{1}{2\pi} \sum_{n=0}^{\infty} \alpha_n M^{2n+1} (2 n+1)\,\sinh[(2 n +1) \theta)] +  
\int_{-\infty}^\infty\frac{\mathrm{d}\theta'}{2\pi}\,\fii_f(\theta-\theta')\,
f_f(\theta')\, \epsilon_\theta(\theta')\,\,\,. 
\ee
Being a linear equation, its solution is given by the {\em sum} of solutions (with weights $\alpha_n$) associated to the each individual equation 
\be  
\epsilon_{n,\theta} \,=\,\frac{1}{2\pi}  M^{2n+1} (2 n+1)\,\sinh[(2 n +1) \theta)] +  
\int_{-\infty}^\infty\frac{\mathrm{d}\theta'}{2\pi}\,\fii_f(\theta-\theta')\,
f_f(\theta')\, \epsilon_{n,\theta}(\theta')\,\,\,.
\ee
Clearly the quantity $e_0(R)$ we are looking for is given by the contribution coming from the solution $\epsilon_{0,\theta}$, so that 
\be 
\langle T_{11}\rangle \,=\,2\pi \frac{e_0(R)}{R} \,=\,- 2\pi \int_{-\infty}^{\infty} M \sinh\theta f_f(\theta) 
\epsilon_{0,\theta}(\theta) \,\,\,.
\label{t11lchannel}
\ee
Hence, for the expectation value of the trace we have 
\be
\langle \Theta \rangle \,=\, 2\pi \int_{-\infty}^{\infty} M f_f(\theta) \left[
\cosh\theta \,\epsilon_0(\theta) - \sinh\theta\,\epsilon_{0,\theta}(\theta)\right] 
\,\,\,.
\label{tobecompared}
\ee
Using the linear integral equations satisfied by $\epsilon_0(\theta)$ and $\epsilon_{0,\theta}(\theta)$, the iterative expression of this 
formula is given by 
\begin{eqnarray} 
&&\langle \Theta \rangle \,=\, 2\pi M^2 \left[ \int_{-\infty}^{\infty}  f_f(\theta) \frac{d\theta}{2\pi} +  \int_{-\infty}^{\infty} \frac{d\theta_1}{2\pi} \int_{-\infty}^{\infty} \frac{d\theta_2}{2\pi}
f_f(\theta_1) f_f(\theta_2) \fii_f(\theta_1-\theta_2) \cosh(\theta_1-\theta_2) 
\label{iterationLM}\right. + 
\\
& \cdots &\left.  + \int_{-\infty}^{\infty}  \frac{d\theta_1}{2\pi} \cdots  \int_{-\infty}^{\infty} \frac{d\theta_n}{2\pi}  f_f(\theta_1) \ldots f_f(\theta_n)
\fii_f(\theta_1-\theta_2) \fii_f(\theta_2-\theta_3) \ldots \fii_f(\theta_{n-1}-\theta_n) \cosh(\theta_1-\theta_n) + \cdots   \right]
\nonumber 
\end{eqnarray} 
One can easily recognized that this is just the LM formula (\ref{EnsembleAverageMus23}) expressed in the fermionic formulation of the theory. If we had used instead the bosonic formulation of the theory, we would have ended up instead in a similar expression 
\begin{eqnarray} 
&&\langle \Theta \rangle \,=\, 2\pi M^2 \left[ \int_{-\infty}^{\infty}  f_b(\theta) \frac{d\theta}{2\pi} +  \int_{-\infty}^{\infty} \frac{d\theta_1}{2\pi} \int_{-\infty}^{\infty} \frac{d\theta_2}{2\pi}
f_b(\theta_1) f_f(\theta_2) \fii_b(\theta_1-\theta_2) \cosh(\theta_1-\theta_2) 
\label{iterationLMbosonic}\right. + 
\\
& \cdots &\left.  + \int_{-\infty}^{\infty}  \frac{d\theta_1}{2\pi} \cdots  \int_{-\infty}^{\infty} \frac{d\theta_n}{2\pi}  f_b(\theta_1) \ldots f_b(\theta_n)
\fii_b(\theta_1-\theta_2) \fii_b(\theta_2-\theta_3) \ldots \fii_b(\theta_{n-1}-\theta_n) \cosh(\theta_1-\theta_n) + \cdots   \right]
\nonumber 
\end{eqnarray} 
which involves this time the bosonic filling $f_b(\theta)$ and the bosonic kernel $\fii_b(\theta)$. 

But, are these two expressions of $\langle \Theta\rangle$ given respectively in eq.\,(\ref{iterationLM}) and eq\,(\ref{iterationLMbosonic}) equal? They are, indeed! Notice, in fact, that the bosonic kernel $\fii_b(\theta)$ differs from the fermionic kernel $\fii_f(\theta)$ for an extra $\delta(\theta)$ term, see eq.\,(\ref{kernels}). Imagine then initially to collect maximally all these extra $\delta(\theta)$ terms from all the bosonic kernels present in eq.\,(\ref{iterationLMbosonic}): together with the first term in eq.\,(\ref{iterationLMbosonic}), 
they group together as 
\be 
\int_{-\infty}^{\infty}  \frac{d\theta}{2\pi}\, \left[f_b(\theta) - f_b^2(\theta) + f^3_b(\theta) - f_b^4(\theta) + \cdots\right] \,=\,
\int_{-\infty}^{\infty}  \frac{d\theta}{2\pi}\, \frac{f_b(\theta)}{1+  f_b(\theta)} \,=\, 
\int_{-\infty}^{\infty} \frac{d\theta}{2\pi}\,  f_f(\theta) 
\ee
where the last identity is ensured by the relation (\ref{transffilling1}). So, in virtue of an exact resummation of all the bosonic filling fractions and 
the identity (\ref{transffilling1}), we were able to get the first term of the LM formula (\ref{iterationLM}) given in the fermionic formulation. Analogously, keeping track of all the other ways of grouping together the $\delta(\theta)$ terms coming from the bosonic kernel and resumming them, we are able to reproduce all other terms of the fermionic expression (\ref{iterationLM}). 

The important result of the calculation discussed above is that the LM formula can be equivalently implemented both in the fermionic or in the bosonic formulation, as far as both the Form Factors and the filling factor are computed in one scheme or the other. 

\vspace{3mm}
\noindent
{\bf Classical limit}. We can use the bosonic formulation of the Generalized Bethe Ansatz to study the classical limit of the corresponding formulas. As we will see, some care is needed in this case. Given the classical phase-shift of eq.~(\ref{classicalphaseshift})
\be 
\varsigma_{cl}(\theta) \,=\,  - \frac{g^2 \hbar}{4} \frac{1}{\sinh\theta} \,\,\,,
\ee
one can compute the classical kernel 
\be 
\varphi_{cl}(\theta) \,=\,\frac{d}{d\theta} \varsigma_{cl}(\theta) \,=\,
\frac{g^2 \hbar}{4} \frac{\cosh\theta}{\sinh^2\theta} \,\,\,.
\label{classicalkernel}
\ee
This kernel is highly singular and it is necessary a prescription for using it in integral expressions. This is provided by the rule 
\be
\int d\theta' \varphi_{cl}(\theta-\theta') \,f(\theta') \,\equiv \,{\mathcal P} \, \int d\theta' \varsigma_{cl}(\theta-\theta') \, \frac{d f (\theta')}{d\theta'}   
\,\,\,,
\label{prescription}
\ee
where the symbol ${\mathcal P}$ stays for the Principal Value of the integral and the definition (\ref{classicalkernel}) of the kernel and its symmetry have been used to integrate by part the expression on the right-hand-side. 

To study the limit $\hbar \rightarrow 0$ of this equation, we have to remember that $R \rightarrow \hbar R$. This allows us to make an expansion in $\hbar$ in the bosonic Bethe Ansatz equations. Redefining $\hbar^{-1} \epsilon \rightarrow \epsilon$, $\hbar^{-1} \fii_{cl}(\theta) \rightarrow \fii_{cl}(\theta)$, the integral equation for the classical pseudo-energy becomes 
\begin{eqnarray}
\eps_{cl}(\theta) &\,= & \label{aprrocleps}
R\,\Lambda(\theta,\{\alpha\}) \, - \, 
\int_{-\infty}^\infty\frac{\mathrm{d}\theta'}{2\pi}\,\fii_{cl}((\theta-\theta') \,
\log\left(\hbar \eps_{cl}(\theta')\right)  
\end{eqnarray}
With the same substitution, for the Free Energy we have 
\be
R\,F_{cl} \,=\,
\frac{1}{2\pi }\int_{-\infty}^\infty\mathrm{d}\theta\,M \cosh \theta\,
\log\left(\hbar \eps_{cl}(\theta)\right) \,.
\labl{eq:GFEclasfefinale}
As shown in Appendix B, these equations allows us to find an integral expression for the lowest-energy value ${\mathcal E}_0(R)$ of the modified Mathieu equation.  In the same Appendix it is also discussed how $\hbar$ disappears from the final formula for 
${\mathcal E}_0(R)$.

\subsection{Caux-Konik formalism \label{sec:ck}}
In order to implement successfully the Generalized Bethe Ansatz one needs to provide, as driving terms of eq.~(\ref{eq:TBAf}), the values of the Lagrangian multipliers of the conserved charges. In principle, these values are self-consistently fixed by solving the eqs. 
\be 
\overline {\mathcal E}_{2n+1} \,=\, \frac{\partial F}{\partial \alpha_n} \,\,\,, 
\label{difficultproblem}
\ee
where $\overline{\mathcal E}_{2n+1}$ are the assigned values of the conserved charges. However the solution of this system of equations turns out, in general, to be quite an insurmountable problem and it would be highly desirable to have another way to determine the equilibrium properties of the system.

An important step forward the solution of this problem has been made in a paper by Caux and Konik \cite{CK}. The main idea of the Caux-Konik approach is that the knowledge of the root distribution of the initial state is enough to fix all relevant data of the Generalized Bethe Ansatz, in particular the pseudo-energy. Let see how this is done using, as example, the fermionic formulation. Similar result holds for the bosonic case. 

Let's consider as initial state $|\psi_0\rangle$ of our system a translation invariant state, for simplicity of zero momentum and zero value of all odd charges (initial states of non-zero values of the momentum and odd charges can be obtained by a simple boost). In the continuum we can expand such a state on the common eigenvectors of $H$ and all higher charges given by the rapidity basis 
\be 
|\psi_0 \rangle \,=\,|0\rangle + \int \frac{d\theta}{2\pi} K_1(\theta) |\theta\rangle + \cdots \frac{1}{k!} 
\int\frac{d\theta_1}{2\pi} \ldots \frac{d\theta_k}{2\pi} K_k(\theta_1,\ldots,\theta_k) |\theta_1,\ldots \theta_k\rangle + \cdots
\label{generalexpansioninfiniteV}
\ee
For global quenches, this expansion should necessarily have an infinite number of terms \cite{FM} and therefore $|\psi_0\rangle$ is not normalizable. In order to regularize it, let's define the theory on a finite interval $L$: in this case, the rapidities are solutions of the Bethe Equation \ref{BETHEEQUATIONS}) and the finite volume eigenstates are denoted as $| \theta_1,\ldots,\theta_n\rangle_L$. Being in correspondence with a given set of integers $\{N_j\}$ (all different and here taken to be symmetric with respect to $0$), such a set of states can be ordered in terms of an integer $p$ and therefore the finite volume expression of the original state is given by 
\be 
|\psi_0\rangle \,=\, \sum_{p} c_{p} |p\rangle \,\,\,,
\label{initialstateBethe}
\ee
where $|p\rangle \rightarrow |p; \theta_{1_p},\ldots,\theta_{n_p}\rangle$ while 
$c_p = K_{n_p}(\theta_{1_p},\ldots,\theta_{n_p})$ are the overlap coefficients, which we assume to be calculable. Notice that to recover the continuum limit for $L \rightarrow \infty$ one needs to take into account the density of state, given by the Jacobian $J(\theta_1,\ldots,\theta_n)$ \cite{PozsayTakacs}  
\be
{\cal J}_n(\theta_1,\ldots,\theta_n) \,=\,{\rm det}\, {\cal J}_{jk} 
\,\,\,\,\,\,
,
\,\,\,\,\,\,
{\cal J}_{jk} \,=\,\frac{\partial {\cal J}_j}{\partial \theta_k} \,\,\,,\label{JACOBIAN}
\ee
where the functions ${\cal J}_j$ are defined in eq.~(\ref{BETHEEQUATIONS}). 

Given a set of integers $\{N_i\}$'s, the rapidities which are solution of the Bethe Ansatz equations (\ref{BETHEEQUATIONS}) (the {\em roots}) are uniquely fixed and, for $L\rightarrow \infty$, they form a distribution $\rho^{(r)}(\theta)$. This function, in turns, determines through the {\em linear} integral equation (\ref{eq:rhos}) the total distribution $\rho(\theta)$.  Hence, if one knew both $\rho^{(r)}(\theta)$ and $\rho(\theta)$, one could have a direct access to $\epsilon(\theta)$, since this quantity is given by 
\be 
\epsilon(\theta) \,=\,- \log \frac{\rho^{(r)}(\theta)}{\rho(\theta) - \rho^{(r)}(\theta)}\,, 
\label{epsilon}
\ee
and finally to the filling fraction $f(\theta)$ given in eq.~(\ref{filling}). 
 
\begin{figure}[t]
\centering
$\begin{array}{ccc}
\includegraphics[width=0.3\textwidth]{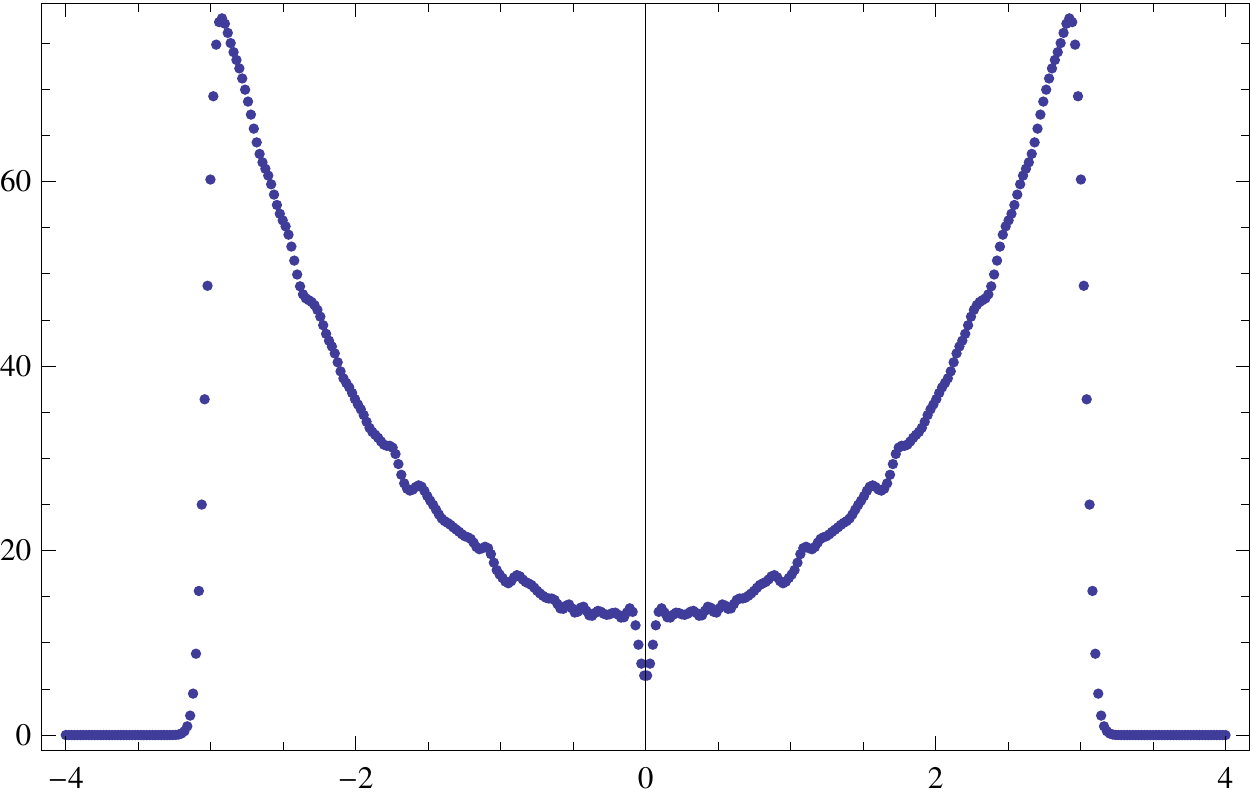} \,\,\,\,\, & \,\,\,\,\,\,
\includegraphics[width=0.3\textwidth]{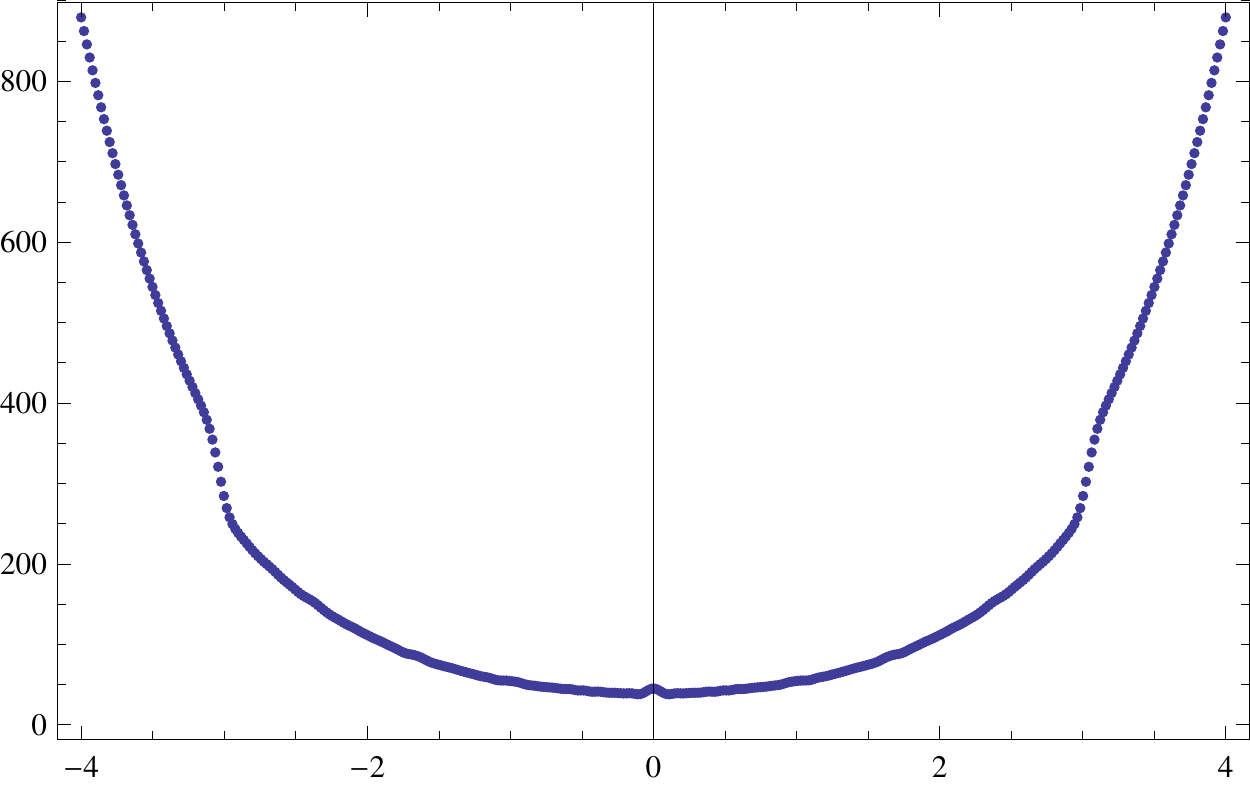} \,\,\,\,\, & \,\,\,\,\,\, 
\includegraphics[width=0.3\textwidth]{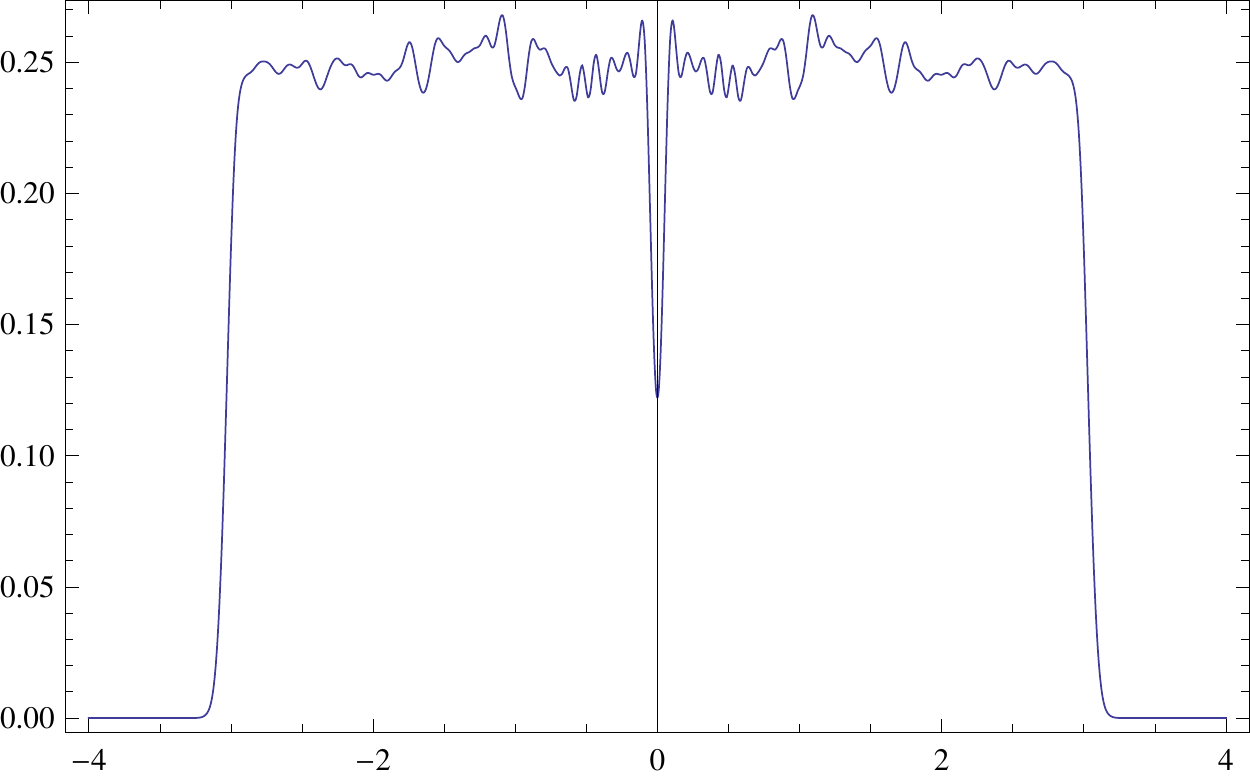} 
\\
\end{array}$
\caption{On the left, the root density for a typical initial state $| \psi_0\rangle$; on the middle, the hole density obtained by solving the integral equation (\ref{eq:rhos}); on the right, the resulting filling fraction $f_f(\theta)$.}
\label{CKnumeproc}
\end{figure}

\section{Summary of the results of Part B}\label{summarynextsteps}

In this part we have discussed in details the connected Form Factors and the filling fraction entering the LM formula (\ref{EnsembleAverageMus23})
of the GGE average. In particular, we have seen that the Form Factors admit a classical limit which coincide with the classical equation of motion of the corresponding operators. A classical limit can also be taken for the Bethe Ansatz equations which determine the filling fraction $f(\theta)$. Moreover, we have seen that the most efficient way to fix the filling fraction is through the knowledge of the density of the roots $\rho^{(r)}(\theta)$ associated to the initial state. The final form can be either expressed in the fermionic or in the bosonic language, since the two formulations are in one-to-one correspondence: this is an important property of the problem, since in the quantum case the natural formulation is the fermionic one while in the classical case is the bosonic one. However, as we have seen for the important example of the trace of the stress-energy tensor, these two formulations are not in contrast with each other and we can swap equivalently from one to the other. 

What remains now to be done is to show how to get the density of roots $\rho^{(r)}(\theta)$ for the classical integrable field theory. This is the subject of the last part of the paper which relies on the formalism of the classical Inverse Scattering Method.

\newpage

\begin{center}
{\Large {\bf PART C} }
\end{center}


\vspace{1mm}

In the Part C of the paper we will present a series of results relative to classical integrable field theories. The final goal is to extract the density of roots $\rho^{(r)}(\theta)$ of the classical field theories that is needed to determine the filling fraction $f(\theta)$ and to ultimately implement the LM formula for the classical GGE average. We will see that the classical density of roots $\rho^{(r)}(\theta)$ are associated to a classical version of the Bethe Ansatz equation. To better appreciate the meaning of this result, we find useful to present a basic review of the classical Inverse Scattering Method formalism, both in the light-cone and in the laboratory frame, since each of these two formulations has its own advantages. We will first discuss the case of infinite volume (alias the case of zero energy density) and then the case of the finite volume (and finite energy density) which is the case of our main concerns.  Along the presentation we are going to do, we will come across many important by-products, such as the identification of the action/angle variables, the explicit expressions of the local conserved charges, the numerical implementation of the transfer matrix, the spectral curve and the so-called {\em finite-gap solutions}. 

For a classical field theory at a finite energy density, notice that a straightforward application of the Inverse Scattering Method would lead to consider the {\em infinite gap solution} and subsequently to employ such a solution for computing the time average of all possible observables. Unfortunately, the infinite gap solution gives rise to quite an intractable formula and, as we will see, of very little use. Our presentation should then help in clarifying the advantage of using the classical version of the LM formula for handling the time averages of the various observables based on the classical field $\varphi(x,t)$. 
 
Guiding references of this Part are the "classical" books by Faddev and Takhtajan \cite{Faddev}, Novikov et al. \cite{Novikov}, Ablowitz and Segur 
\cite{Ablowitz} plus a series of other important papers that will be quoted at the relevant places.  

\section{The Classical Sinh-Gordon model in infinite volume: light-cone coordinates} \label{lightconeSH}

In this section we will discuss the integrable structure of the Sinh-Gordon theory in infinite volume:  although these results need to be modified once we will study the formulation of the model on a finite volume, yet they are crucial for establishing the integrable nature of the model and for identifying the proper action-angle variables. Since a certain degree of simplification occurs if one adopts the light-cone formalism, we will first study the Sinh-Gordon model in the light-cone variables and later on in the laboratory variables. For more details on various points discussed below, the reader is referred to the aforementioned books \cite{Faddev,Novikov,Ablowitz}. The Lagrangian density of the Sinh-Gordon model is 
\be 
{\mathcal L} \,=\, \frac{1}{2} (\partial_\mu \phi)^2 - \frac{m^2}{g^2} (\cosh g \phi -1) \,\,\,,
\ee
and in the light-cone variables $\tau$ and $\sigma$, defined in eq.~(\ref{lightconevariable}), the equation of motion is given by 
\be 
\phi_{\tau \sigma} \,=\,\frac{m^2}{g} \sinh g \phi \,\,\,.
\label{eqmotlight}
\ee
For an integrable model such as the Sinh-Gordon's, its equation of motion can be derived as a consistency requirement of an auxiliary linear system of differential equations for a bi-dimensional vector $\Psi = \left(\begin{array}{c} \psi_1\\\psi_2\end{array}\right)$ 
\be
\begin{array}{c} 
\partial_{\sigma}\, \Psi \,=\, U(\lambda) \, \Psi \\
\partial_{\tau} \, \Psi \,=\, V(\lambda) \, \Psi 
\end{array}
\label{systempsi}
\ee
where 
\be 
U(\lambda) \,=\, 
\left( 
\begin{array}{cc}
i \frac{\lambda}{2} &  i \frac{g}{2} \phi_\sigma \\
- i \frac{g}{2} \phi_\sigma & - i \frac{\lambda}{2} 
\end{array} 
\right) 
\,\,\,\,\,\,\,\,\,\,\,\, 
,
\,\,\,\,\,\,\,\,\,\,\,\, 
V(\lambda) \,=\, \frac{m^2}{2 i \lambda} \,  
\left( 
\begin{array}{cc}
\cosh g \phi &  - i \sinh g \phi \\
- i \sinh g \phi &  -\cosh g \phi 
\end{array} 
\right) \,\,\,.
\label{UVMatrices}
\ee
Assuming that the spectral parameter $\lambda$ does not depend upon $\sigma$ and $\tau$, the compatibility of the system of equations  (\ref{systempsi}) implies 
\be 
\frac{\partial U}{\partial \tau} - \frac{\partial V}{\partial \sigma} + [U,V] \,=\,0 \,\,\,,
\ee
which is nothing else but the equation of motion (\ref{eqmotlight}) for the field $\phi$ entering the two matrices $U(\lambda)$ and $V(\lambda)$. This is, in a nutshell, the main idea of the so-called Inverse Scattering Method \cite{Faddev,Novikov,Ablowitz}, alias to regard the equation of motion of integrable systems as compatibility conditions of a linear system of differential equations. The consequences of this idea are discussed in more details in the next Sections. 

\subsection{Scattering data and action/angle variables}
Consider for an arbitrary time $\tau$, the scattering problem defined by the first equation in (\ref{systempsi})  
\be 
\left(\begin{array}{cc} 
- i \partial_{\sigma} & - \frac{g}{2} \phi_\sigma \\
- \frac{g}{2} \phi_\sigma & i \partial_\sigma 
\end{array} 
\right) \,\left(\begin{array}{c} \psi_1\\ \psi_2\end{array}\right) 
\,\equiv \, M \left(\begin{array}{c} \psi_1\\ \psi_2\end{array}\right) \,=\, \frac{\lambda}{2} \, \left(\begin{array}{c} \psi_1\\ \psi_2\end{array}\right)\,\,\,.
\label{eqscat}
\ee
Since the matrix $M$ is hermitian, its eigenvalues $\lambda$ are real. Let's now assume that $\phi_\sigma \rightarrow 0$ when 
$\sigma \rightarrow \pm \infty$: in this case, the two solutions (known as Jost functions) can be specified by the asymptotic behaviour
\be 
\Psi(\sigma,\lambda) \,\simeq \, \Upsilon(\sigma,\lambda) 
\,\,\,\,\,\,\,\,\,\,
,
\,\,\,\,\,\,\,\,\,\, 
\sigma \rightarrow - \infty\,\,\,,
\ee
where 
\be
\Upsilon \,=\,\left(
\begin{array}{cc} 
e^{i \lambda \sigma /2} & 0 \\
0 & e^{-i \lambda \sigma /2} 
\end{array}
\right) \,\,\,. 
\ee
Moreover, it is simple to show that $\Psi(\sigma,\lambda)$ can be organized as 
\be
\Psi \,=\, 
\left(
\begin{array}{cc} 
\psi_1 & \bar \psi_1 \\
\psi_2 & \bar \psi_2 
\end{array}\right) \,\,\,, 
\ee
with $\bar\psi_1(x,\lambda) = \psi_2^*(\sigma,\lambda)$ and $\bar\psi_2(\sigma,\lambda) = \psi_1^*(\sigma,\lambda)$. The solutions 
at $\sigma = - \infty$ are connected to the ones at arbitrary $\sigma$ by means of the transfer matrix $T$ 
\be 
\Psi(\sigma,\lambda) \,=\,\Upsilon(\sigma,\lambda) \, T(\sigma,\lambda) \,\,\,. 
\label{solutionT}
\ee
$T(\sigma,\lambda)$ has a finite limit for $\sigma \rightarrow +\infty$ and its asymptotic expression can be written as  
\be 
\lim_{\sigma\rightarrow +\infty}T(\sigma,\lambda) \,\equiv\, T_{\infty}(\lambda)\,=\,\left(
\begin{array}{cc} 
a(\lambda) & b^*(\lambda) \\
b(\lambda) & a^*(\lambda) 
\end{array}
\right) \,\,\,, 
\ee
with 
\be 
|a(\lambda)|^2 - |b(\lambda)|^2 \,=\, 1\,\,\,.
\label{unitarity} 
\ee
This equation can be also written as 
\be
| r(\lambda) |^2 + | t(\lambda)|^2 \,=\, 1 \,\,\,,
\ee
where the reflection and transmission amplitudes are given by 
\be
r(\lambda) \,=\,\frac{b(\lambda)}{a(\lambda)} \,\,\,\,\,\,\,\,,\,\,\,\,\,\,
t(\lambda) \,=\,\frac{1}{a(\lambda)} \,\,\,.
\label{reflectiontransmissionamplitudes}
\ee
All information is essentially encoded in the reflection coefficient: indeed, in view of eq. (\ref{unitarity}), the modulus of $r(\lambda)$ uniquely fixes $| a(\lambda)|$ while the argument of $a(\lambda)$ can be fixed by a dispersion relation since $a(\lambda)$ is analytic in the upper half-plane of the complex variable $\lambda$, vanishes as $| \lambda | \rightarrow 0$ and moreover (in the Sinh-Gordon model) does not have any zeros in ${\rm Im}\, \lambda >0$. Therefore it can be expressed as 
\be 
a(\lambda) \,=\,| a(\lambda) | \, \exp\left[-{\cal P} \frac{1}{2 \pi i} \left\{ \lambda \int_{-\infty}^{\infty} \frac{d\lambda' \, \log | a(\lambda') |^2}{(\lambda^2 - \lambda'^2)}\right\}\right] \,\,\,, 
\label{scatteringamplitudea}
\ee
where ${\cal P}$ in front of the integral denotes its Principal Part. 

One can derive how the scattering data $a(\lambda)$ and $b(\lambda)$ evolve in time by employing the second equation of the original system (\ref{systempsi}). To this aim, consider the Jost solution $f(\sigma,\lambda)$ that behaves at $\sigma \rightarrow -\infty$ as 
\be 
f(\sigma,\lambda) \simeq \, f_1 \equiv \, \left(
\begin{array}{c} e^{i \lambda \sigma /2} \\
0 \end{array} \right) 
\,\,\,\,\,\,\,\,\,
, 
\,\,\,\,\,\,\,\,\,
\sigma \rightarrow - \infty
\ee
and at $\sigma \rightarrow + \infty$ as 
\be 
f(\sigma,\lambda) \, \simeq \, a(\lambda) \, \left(\begin{array}{c} e^{i \lambda \sigma /2} \\
0 \end{array} \right) + b(\lambda) \, \left(\begin{array}{c} 0 \\e^{-i \lambda \sigma /2} \end{array}\right) \equiv 
a(\lambda) f_1 + b(\lambda) f_2 
 \,\,\,.
\label{theotherlimit}
\ee
Let's take now $\tilde f(\sigma,\tau) = \eta(\tau)\,f(\sigma,\lambda)$ and consider the second equation of the system (\ref{systempsi}) for this solution 
\be 
\partial_\tau \tilde f \,=\, V(\lambda) \, \tilde f \,\,\,.
\ee
At $\sigma \rightarrow \pm \infty$, where the field $\phi$ and its derivative $\phi_\sigma$ vanish, we have 
\be 
V_\infty \,=\,\frac{m^2}{2 i \lambda} \,
\left( 
\begin{array}{cc} 1 & 0 \\
0 & 1 \end{array} 
\right) 
\ee
Hence, for $\sigma \rightarrow -\infty$, we get the differential equation satisfied by the function $\eta(\tau)$
\be 
\frac{d \eta (\tau)}{d \tau} \,=\, \frac{m^2 \, \eta(\tau)}{2 i \lambda} \,\,\,.
\label{solutioneta}
\ee
Consider now the other limit $\sigma \rightarrow +\infty$, where $f(\sigma,\lambda)$ behaves as in (\ref{theotherlimit}). In this case 
we have 
\begin{eqnarray}
\frac{d}{d\tau} [\eta \, a(\lambda)] \,=\,\frac{m^2}{2 i \lambda} \eta \, a(\lambda) \,\,\,,\\
\frac{d}{d\tau} [\eta \, b(\lambda)] \,=\, -\frac{m^2}{2 i \lambda} \eta \, b(\lambda) \nonumber \,\,\,.
\end{eqnarray}
Taking into account eq.~(\ref{solutioneta}), one then   
finds 
\begin{eqnarray}
\label{abevol}
\frac{d}{d \tau} a(\lambda) & \,=\, & 0  \,\,\,,\\
\frac{d}{d\tau} b(\lambda) & \,=\, & i\,\frac{m^2}{\lambda}\, b(\lambda) \nonumber \,\,\,.
\end{eqnarray}
The first equation implies that $a(\lambda)$ is a conserved quantity for any value of $\lambda$. As a consequence also its modulus is 
conserved and therefore, from eq.~(\ref{unitarity}), also the modulus of $b(\lambda)$. With this information, the second equation implies instead that the argument of $b(\lambda)$ evolves linearly as 
\be 
\rm{arg}\,b(\lambda) \,=\,\frac{m^2}{\lambda} \,\tau \,=\, \omega(\lambda) \, \tau  \,\,\,. 
\ee
Notice that $\omega(\lambda) = m^2/\lambda$ is the dispersion relation in the light-cone variables (see 
eq.\,(\ref{dispersionrelationlc})). The action-angle variables in this case are given by 
\be
P(\lambda) \,=\,\frac{2}{g^2 \pi \lambda} \log |a(\lambda)| 
\,\,\,\,\,\,\,\,\,\,
,
\,\,\,\,\,\,\,\,\,\,
Q(\lambda) \,=\,\rm{arg} \,b(\lambda) \,\,\,, 
\label{actionanglevariable}
\ee
and the Hamiltonian of the Sinh-Gordon model can be written as 
\be 
H \,=\,\int \,\omega(\lambda) \, P(\lambda) \, d\lambda\,\,\,.
\ee

\subsection{Integral equation for the transfer matrix}
Using eqs. (\ref{eqscat}) and (\ref{solutionT}), it is easy to see that the transfer matrix $T(\sigma,\lambda)$ satisfies the differential equation 
\be 
\frac{\partial}{\partial \sigma} T(\sigma,\lambda) \,=\, \tilde M \,T(\sigma,\lambda) \,\,\,,
\label{diffeqT}
\ee
where  
\be 
\tilde M \,=\, \Upsilon^{-1} (U - \partial_{\sigma} \Upsilon) \,=\,
\left(\begin{array}{cc} 
0 & i \frac{g}{2} \phi_\sigma e^{ i \lambda \sigma} \\
- i \frac{g}{2} \phi_\sigma e^{- i \lambda \sigma} & 0  
\end{array}
\right)  \,\,\,.  
\ee
This can be equivalently written as a linear integral equation 
\be 
T(\sigma,\lambda) \,=\, {\bf I} + \,\int_{-\infty}^{\sigma} d\sigma' \tilde M(\sigma',\lambda) \, T(\sigma',\lambda) \,\,\,. 
\ee
The Fredholm solution obtained by iterating this equation 
\be 
T(\sigma,\lambda) \,=\, {\bf I}  + \int_{-\infty}^\sigma \tilde M(\sigma',\lambda) + \int_{-\infty}^\sigma d\sigma' \, 
\int_{-\infty}^{\sigma'} d\sigma" \, \tilde M(\sigma',\lambda) \, \tilde M(\sigma",\lambda) + \cdots 
\ee
may be regarded as the perturbative expansion of the solution in the coupling constant $g$ and this holds in particular for the scattering coefficients $a(\lambda)$ and $b(\lambda)$ 
\begin{eqnarray} 
a(\lambda) & \,= \, & 1 - \frac{g^2}{4} \int_{-\infty}^{+\infty} dx_1\, \int_{-\infty}^{+\infty} dx_2 \, 
\theta(x_1-x_2) \, \phi_\sigma(x_1) \,e^{ i \lambda x_1} \, \phi_\sigma(x_2) \, e^{ i \lambda x_2} + \ldots
\label{perturbativeexpressionsab}\\
b(\lambda) & \,=\,& i\,\frac{g}{2} \int_{-\infty}^{+\infty} dx \,\phi_\sigma(x) \, e^{ i \lambda x} + \ldots \nonumber   
\end{eqnarray}

\subsection{Numerical calculation of the transfer matrix}
Given a field configuration $\phi(\sigma,\tau)$ and its derivative $\phi_\sigma(\sigma,\tau)$ it is easy to determine numerically the transfer matrix at any time $\tau$. As a matter of fact, there are two different procedures which give rise to the same output, up to an overall phase. 
\begin{enumerate}
\item The first procedure starts by considering the first equation of the system (\ref{systempsi})
\be 
\partial_\sigma \Psi \,=\, U(\lambda) \, \Psi \,\,\,,
\label{eqpsii}
\ee 
and integrating at the first order in the infinitesimal step $\Delta \sigma$ 
\be
\Psi(\sigma + \Delta \sigma) \,=\,W(\lambda,\Delta \sigma) \, \Psi(\sigma) \,\,\,,
\ee
with the matrix $W(\lambda,\Delta\sigma)$ given by 
\begin{eqnarray}
W(\lambda,\Delta \sigma) & \,=\,& \exp[\Delta \sigma U(\lambda)] \,=\,\\ 
& = & \left(\begin{array}{cc} 
\cosh\left(k \frac{\Delta\sigma}{2}\right) + i \frac{\lambda}{2 k} \, \sinh\left(k \frac{\Delta\sigma}{2}\right) & 
i \frac{g}{k} \,\phi_\sigma \,\sinh\left(k \frac{\Delta\sigma}{2}\right) \\
- i \frac{g}{k} \,\phi_\sigma \,\sinh\left(k \frac{\Delta\sigma}{2}\right)
& \cosh\left(k \frac{\Delta\sigma}{2}\right) - i \frac{\lambda}{ 2 k} \, \sinh\left(k \frac{\Delta\sigma}{2}\right)
\end{array} \right) \,\,\,,\nonumber
\end{eqnarray}
and
\be 
k \,\equiv\,\sqrt{g^2 \phi_\sigma^2 - \frac{\lambda}{2}} \,\,\,.
\ee
In the expression above, $\phi_\sigma = \phi_\sigma(\sigma,\tau)$ is the derivative of the field computed at the time $\tau$ by solving the equation of motion (\ref{eqmotlight}). Let's now imagine that the field and its derivative is different from zero only in an interval $I = (-\sigma_0,\sigma_0)$ in $\sigma$ which can be discretized in terms of $N$ steps $\Delta\sigma$. In this way, given the initial values of the solution at one end -$\sigma_0$ of the interval, we can find its values at the other end by means of   
\be 
\left(\begin{array}{c} 
\psi_1 \\
\psi_2 
\end{array}\right) \,=\, 
\left[\prod_{j=1}^{N+1} W(\lambda, -\sigma_0+ j \Delta\sigma) \right] \, 
\left(\begin{array}{c} 
\psi_1^0 \\
\psi_2^0 
\end{array}\right) \,\,\,.  
\ee
In other words, the asymptotic transfer matrix $T_{\infty}(\lambda)$ is given by 
\be 
T_\infty(\lambda) \,=\,\prod_{j=1}^{N+1} W(\lambda, -\sigma_0 + j \Delta\sigma) \,\,\,.
\ee
Since the expression $W(\lambda, j \Delta\sigma)$ coincides with the transfer matrix of a quantum mechanics scattering problem on a positive but very small rectangular with respect to the incident energy $E$ (centred at $j \Delta\sigma)$, the final expression of $T_\infty(\lambda)$ can be regarded as the transfer matrix of a potential that has been locally approximated by rectangular barriers, as in Figure \ref{barrier}.   

\begin{figure}[h]
\centering
$\begin{array}{c}
\includegraphics[width=0.5\textwidth]{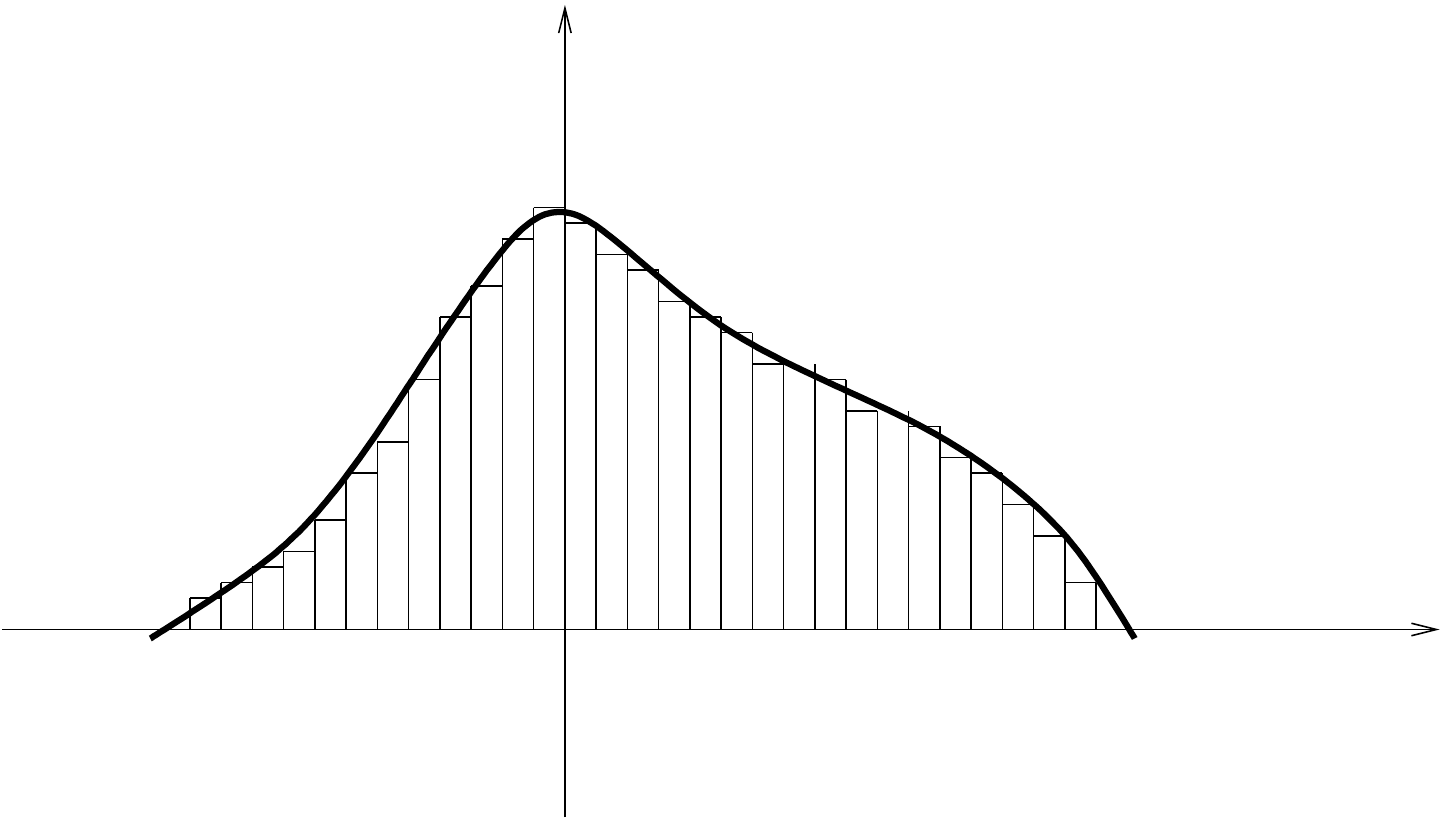} 
\end{array}$
\caption{Scattering potential approximated by a sequence of rectangular barriers.}
\label{barrier}
\end{figure}

In order to get enough accuracy, the lattice spacing $\Delta\sigma$ has to be sufficient small to catch up with the variation of the field. In typical runs of this algorithm we have implemented the numerical multiplication of around $10^4$ matrices. 
\item The second procedure starts instead directly from the differential equation (\ref{diffeqT}) satisfied by the 
transfer matrix and solving it at the first order in the infinitesimal step $\Delta\sigma$ 
\be 
T(\sigma+\Delta\sigma,\lambda) \,=\,\hat W(\sigma,\lambda) \, T(\sigma,\lambda) \,\,\,,
\ee
where 
\be 
\hat W(\sigma,\lambda) \,=\,e^{i \tilde M \Delta\sigma} \,=\,
\left(\begin{array}{cc} 
\cosh\left(\frac{\Delta\sigma}{2} g \phi_\sigma\right) & 
i e^{2 \lambda \sigma} \, \sinh\left(\frac{\Delta\sigma}{2} g \phi_\sigma\right) \\
- i e^{ 2 \lambda \sigma} \, \sinh\left(\frac{\Delta\sigma}{2} g \phi_\sigma\right) 
& \cosh\left(\frac{\Delta\sigma}{2} g \phi_\sigma\right) \end{array} \right)\,\,\,. 
\ee
Imagine once again that the support of the field and its derivative is a finite interval $I = N \Delta\sigma$, with the left end equal to $- \sigma_0$,  the asymptotic transfer matrix $T_\infty(\lambda)$ is expressed by 
the product 
\be  
T_\infty(\lambda) \,=\,\prod_{j=1}^{N+1} \hat W(\lambda, -\sigma_0 + j \Delta\sigma) \,\,\,.
\ee
\end{enumerate}

\subsection{An interesting check}
The limit $g \rightarrow 0$ provides an interesting check of both the numerical determination of $T_\infty(\lambda)$ and the action/angle variables (\ref{actionanglevariable}). In particular $P(\lambda)$,  in the limit $g\rightarrow 0$, can be expressed in terms of the perturbative expression (\ref{perturbativeexpressionsab}) of $a(\lambda)$. To this aim, let's use the mode expansion of the field in the light-cone coordinates 
\be
\phi(\sigma,\tau) \,=\,\int_0^{\infty} \frac{dp_+}{2\pi p_+} 
\left(A(p_+) \,e^{-i p_-\tau + ip_+\sigma} + A^{\dagger}(p_+) \, 
 e^{i p_-\tau - ip_+\sigma} \right) \,\,\,,
 \ee
so that 
\be
\phi_\sigma(\sigma,\tau) \,=\,i \,\int_0^{\infty} \frac{dp_+}{2\pi} 
\left(A(p_+) \,e^{-i p_-\tau + ip_+\sigma} - A^{\dagger}(p_+) \, 
 e^{i p_-\tau - ip_+\sigma} \right) \,\,\,. 
 \ee
The quantity which counts the number occupation of the modes (which is the conserved quantity in the free theory and plays the role of action variable) is
\be 
N(p_+) \,=\,\frac{1}{2 \pi p_+} |A(p_+)|^2 \,\,\,.
\label{numberoccupationlight}
\ee  
Notice that, in the light-cone variables, one has 
\be 
\int_{-\infty}^{\infty} d\sigma \,\phi_\sigma(\sigma,\tau=0) \,e^{i k \sigma} \,=\,
\left\{\begin{array}{cl} 
- i A^{\dagger}(k) & k > 0 \\
i A(k) & k < 0 
\end{array} 
\right. \,\,\,.
\ee
Hence the first terms of $a(\lambda)$ are given by 
\be 
a(\lambda) \,= \,  1 - \frac{g^2}{4} \int_{-\infty}^{+\infty} dx_1\, \int_{-\infty}^{+\infty} dx_2 \, 
\theta(x_1-x_2) \, \phi_\sigma(x_1) \,e^{ i \lambda x_1} \, \phi_\sigma(x_2) \, e^{ i \lambda x_2} + \cdots 
\ee
Using the Fourier transform of the step function 
\[
\theta(x) \,=\,\int_{\infty}^{\infty} \hat\theta(s) \,e^{i s x} \,\,\,,
\]
where 
\[
\hat\theta(s) \,=\,\frac{1}{2} \left[\delta(s) - \frac{i}{\pi} P\left(\frac{1}{s}\right)\right] \,\,\,,
\]
in the limit $g \rightarrow 0$, we get 
\be 
a(\lambda) \,\simeq\, 1 + \frac{g^2}{4} \left[ |A(\lambda)|^2 + \frac{i}{2\pi} P \,\int_0^{\infty} dp \, |A(p)|^2 \left(
\frac{1}{p+\lambda} - \frac{1}{p-\lambda}\right) \right] \,\,\,.
\ee
\begin{figure}[b]
\centering
$\begin{array}{cc}
\includegraphics[width=0.4\textwidth]{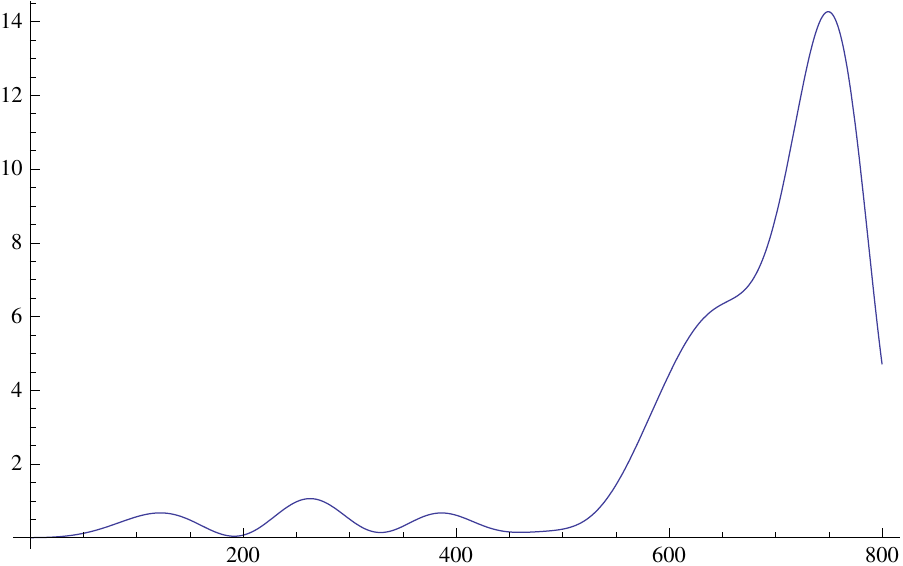}\,\,\,\,\, & \,\,\,\,\,
\includegraphics[width=0.4\textwidth]{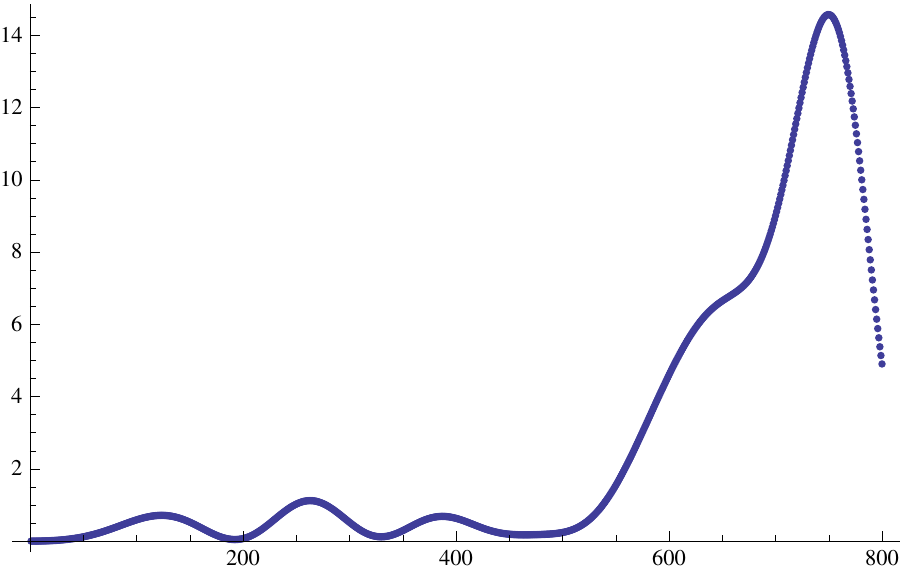}\\
\end{array}$
\caption{On the left: plot of $\frac{1}{2 \pi \lambda} |A(\lambda)|^2$ relative to a field configuration on a compact interval $I$.  
On the right: plot of $\lim_{g\rightarrow 0} \frac{1}{g^2 \pi \lambda} \log | a(\lambda) |^2$ obtained by the numerical evaluation of the transfer matrix.}
\label{checkfree}
\end{figure}
\begin{figure}[t]
\centering
$\begin{array}{c}
\includegraphics[width=0.5\textwidth]{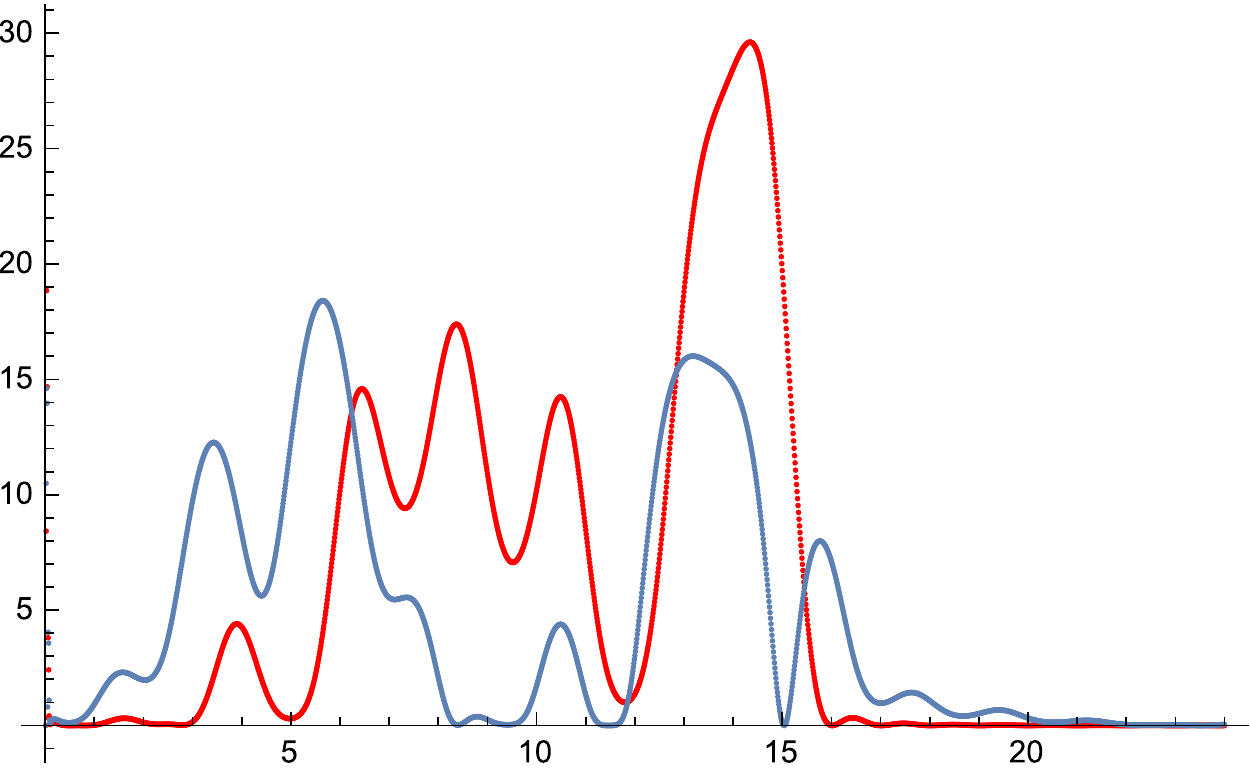} 
\\
\end{array}$
\caption{Mode occupation $P(\lambda)$ versus $\lambda$ for the interacting theory (blue curve) and the free theory (red curve).}
\label{comparisonmodeocupation}
\end{figure}
It is easy to see that the imaginary part of the expression above does not contribute to the module of $a(\lambda)$ at 
order $g^2$. Indeed we have  
\be 
| a(\lambda) |^2 \,\simeq\, 1+ \frac{g^2}{2} |A(\lambda)|^2  
\ee
and therefore 
\be 
P(\lambda) \,=\,\frac{1}{g^2 \pi \lambda} \log | a(\lambda) |^2 \,\simeq\,\frac{1}{2 \pi \lambda} |A(\lambda)|^2 \,\,\,, 
\ee
namely, in the limit $g\rightarrow 0$ the action variable $P(\lambda)$ coincides with the number occupation $N(\lambda)$ of the free theory given in eq.~(\ref{numberoccupationlight}). The numerical confirmation of this identity is presented in Figure \ref{checkfree}, where: (i) we have prepared the field $\phi(\sigma,\tau=0)$ in a random configuration with finite support on an interval $I$ and we have computed its modes by the Fourier transform; (ii) we have used this field configuration in the numerical computation of the transfer matrix in the limit $g\rightarrow 0$. As shown in Figure \ref{checkfree}, the two curves coincide.   

This check is particularly instructive as it enlightens the integrable structure of the Sinh-Gordon model in comparison to the free theory, that is also integrable. Namely, while in the free theory the conserved quantities (which are the mode occupations) are directly expressed in terms of the oscillator modes as $|A(\lambda)|^2$, in the Sinh-Gordon model the mode occupation is given by the action variable $P(\lambda)$ which consists of a non-linear and non-local expression involving all the free modes $|A(\lambda)|^2$. A comparison between these 
two quantities is shown in Figure \ref{comparisonmodeocupation}.  

\subsection{Conserved Charges}\label{ConservedChargesShG}
The light-cone formulation is particularly convenient for deriving the explicit form of the conserved charges of the Sinh-Gordon model. As a matter of fact, one can exploit the relationship between the KdV equation and a close model to the Sinh-Gordon model, alias the Sine-Gordon model \cite{Chodos}, and use later the analytic continuation that links the two models. To start with, let's briefly remind the main steps which lead to the conserved quantities of the KdV equation. 

\subsubsection{KdV system}
As well known \cite{Novikov}, the KdV equation 
\be 
u_{\tau} - 6 u u_{\sigma} + u_{\sigma\sigma\sigma} \,=\,0 \,\,\,,
\label{KdV}
\ee 
is associated to the Lax pairs 
\be 
L\,=\, -\frac{\partial^2}{\partial\sigma^2} + u(\sigma,\tau) 
\,\,\,\,\,\,\,\,
,
\,\,\,\,\,\,\,\,
M \,=\,-4 \frac{\partial^3}{\partial\sigma^3} + 3 u \frac{\partial}{\partial\sigma} + 3 \frac{\partial u}{\partial\sigma} 
\,\,\,,
\label{Laxpair}
\ee
and to the evolution equation 
\be 
L_{\tau} \,=\,[ M , L ] \,\,\,. 
\ee 
This equation ensures that the eigenvalues $\lambda^2$ of the operator $L$ 
\be 
\left[-\frac{\partial^2}{\partial\sigma^2} + u(\sigma,\tau) \right] \varphi(\sigma) \,=\,\lambda^2 \, \varphi(\sigma) \,\,\,,
\label{KdVSchr}
\ee
are time independent. $L$ is clearly a Schr\"odinger operator type where $u(\sigma,\tau)$ plays the role of the potential. Hence, also in this case one can define the scattering coefficients $a(\lambda)$ and $b(\lambda)$ and prove that $a(\lambda)$ does not evolve in time. Since for sufficiently large $| \lambda |$, the solution of the Schr\"odinger equation $\varphi(\sigma,\lambda)$ can be represented as 
\be 
\varphi(\sigma,\lambda) \,=\,\exp\left[-i \lambda \sigma + \int_{-\infty}^{\sigma} \chi(\sigma',\lambda) \,d\sigma' \right] \,\,\,,
\ee
and moreover 
\be 
\lim_{\sigma \rightarrow\infty} \varphi(\sigma,\lambda) \, e^{i \lambda \sigma} \,=\, a(\lambda) \,\,\,,
\ee
we have 
\be 
\log a(\lambda) \,=\,\int_{-\infty}^{\infty} \chi(\sigma,\lambda) \,d\sigma \,\,\,.
\ee
This expression gives rise, for all values of $\lambda$, to conserved quantities. With $\varphi(\sigma,\lambda)$ solution of the Schr\"odinger equation, $\chi(\sigma,\lambda)$ satisfies the Riccati equation 
\be 
\frac{d \chi}{d\sigma} + \chi^2 - u - 2 i \lambda \chi \,=\,0 \,\,\,,
\label{Riccati}
\ee
and admits the asymptotic expansion 
\be 
\chi(\sigma,\lambda) \,=\,\sum_{n=1}^{\infty} \frac{\chi_n(\sigma)}{(2 i \lambda)^n} \,\,\,.
\ee  
Substituting into (\ref{Riccati}), we get the recursive equations 
\begin{eqnarray}
\chi_{n+1}(\sigma) & = & \frac{d}{d\sigma}\chi_n(\sigma) + \sum_{k=1}^{n-1} \chi_n(\sigma) \chi_{n-k}(\sigma) \\
\chi_1(\sigma) & = & - u(x) \nonumber \,\,\,.
\label{recursivecons}
\end{eqnarray}
While all even $\chi_{2n}(\sigma)$ can be proved to be total derivatives, the odd terms $\chi_{2n+1}(\sigma)$ give rise instead to genuine conserved charges 
\be
{\mathcal Q}_{2n+1} \,=\,\int_{-\infty}^{\infty} \chi_{2n+1}(\sigma) \, d\sigma \,\,\,,
\ee 
and they satisfy the continuity equations 
\be
\partial_\tau \chi_{2n+1} \,=\,\partial_\sigma \theta_{2n+1} \,\,\,,
\label{opportunedensities}
\ee
with opportune densities $\theta_{2n+1}$. The first expressions of this infinite chain of conserved densities (where we drop total derivative terms) are 
\begin{eqnarray}
\chi_1(\sigma) & = & - u(\sigma) \nonumber \\
\chi_3(\sigma) & = & u^2(\sigma) \label{densitychi}\\
\chi_5(\sigma) & = & - (u^2_\sigma(\sigma) + 2 u^3(\sigma)) \nonumber \\
\chi_7(\sigma) & = & (u^2_{\sigma\sigma} - 5 u^3 u_{\sigma\sigma} + 5 u^4 ) \nonumber 
\end{eqnarray}

\subsubsection{Sine-Gordon vs KdV} 
As the Sinh-Gordon model, the equation of motion of the Sine-Gordon model in the light-cone coordinates 
\be 
\phi_{\tau\sigma} \,=\,\frac{m^2}{g} \,\sin g \phi \,\,\,,
\label{eqmotionSGlc}
\ee
can be obtained as compatibility condition of first order system of equations with matrices 
\be 
U_{SG}(\lambda) \,=\, 
\left( 
\begin{array}{cc}
i \frac{\lambda}{2} &  i \frac{g}{2} \phi_\sigma \\
i \frac{g}{2} \phi_\sigma & - i \frac{\lambda}{2}
\end{array} 
\right) 
\,\,\,\,\,\,\,\,\,\,\,\, 
,
\,\,\,\,\,\,\,\,\,\,\,\, 
V_{SG}(\lambda) \,=\, \frac{m^2}{2 i \lambda} \,  
\left( 
\begin{array}{cc}
\cos g \phi &  - i \sin g \phi \\
 i \sin g \phi &  -\cos g \phi 
\end{array} 
\right) \,\,\,.
\ee
Consider now the eigenvalue problem 
\be 
\partial_\sigma \Psi \,=\,U_{SG}(\lambda) \, \Psi \,\,\,,
\ee
that, written in components, reads 
\be
\begin{array}{l}
\partial_\sigma \psi_1 \,=\, i \frac{\lambda}{2} \,\psi_1 + i \frac{g}{2} \phi_\sigma \, \psi_2  \,\,\,,\\
\partial_\sigma \psi_2 \,=\, -i \frac{\lambda}{2} \,\psi_2 + i \frac{g}{2} \phi_\sigma \, \psi_1 \,\,\,.
\end{array}
\ee
Expressed in terms of $\psi_+ \equiv (\psi_1 + \psi_2)$ and $\psi_- \equiv (\psi_1 - \psi_2)$, we have 
\be 
\begin{array}{l}
\partial_\sigma \psi_+ \,=\,i \frac{\lambda}{2}\, \psi_- + i \frac{g}{2} \psi_+ \,\,\,,\\
\partial_\sigma \psi_- \,=\,i \frac{\lambda}{2}\, \psi_+ - i \frac{g}{2} \psi_- \,\,\,. 
\end{array}
\ee
Eliminating $\psi_-$, one gets a second order differential equation for $\psi_+$ 
\be 
-\partial_\sigma^2 \psi_+ + \left[i \frac{g}{2} \phi_\sigma - \left(\frac{g}{2}\right)^2 \phi_\sigma^2 \right] \psi_+ \,=\,\frac{\lambda^2}{4} \,\psi_+ \,\,\,.
\ee
Comparing now with (\ref{KdVSchr}), we see that the role of the KdV function $u(\sigma,\tau)$ is played by the expression 
\be 
u(\sigma,\tau) \rightarrow  \left[i \frac{g}{2} \phi_\sigma - \left(\frac{g}{2}\right)^2 \phi_\sigma^2 \right]\,\,\,,
\ee
and therefore, making this substitution in (\ref{densitychi}), we can borrow the expression of the conserved charges of KdV to get the corresponding conserved charges of the Sine-Gordon model \cite{Chodos}. The conserved charges of the Sinh-Gordon model can then be obtained by analytic continuation of the coupling constant $g \rightarrow i g$. 

Two comments are in order. The first concerns the normalization of the charges: in the way they have been obtained, their normalization seems to be fixed. However, being in involution $\left\{{\mathcal Q}_s,{\mathcal Q}\right\}_s = 0$, their relative normalization can be fixed differently, as in fact we will do later. The second comment concerns the fact that this method is quite efficient in getting the densities which are conserved in the light-cone coordinates but unfortunately it leaves out the determination of the corresponding densities $\theta_{2n+1}$ entering the continuity equations (\ref{opportunedensities}). These densities are needed if one wants to write down the conserved quantities in the laboratory frame, as evident from eqs. (\ref{lc1}) and (\ref{lc2}). In the case of Sin(h)-Gordon, these densities can be explicitly computed by making use of the equation of motion: notice, in fact, that the equation of motion gives rise to the following chain of identities 
\begin{eqnarray} 
\phi_{\tau\sigma} & \,=\,& \frac{m^2}{g} \sin(g \phi) \,\,\,, \nonumber \\
\phi_{\tau\sigma\sigma} & \,=\,& m^2 \phi_\sigma \cos(g \phi) \,\,\,, 
\label{chainofidentities} \\
\phi_{\tau\sigma\sigma\sigma} & \,=\,& m^2 \left[\phi_{\sigma\sigma} \cos(g \phi) - g \phi_\sigma^2 \sin(g \phi) \right] \,\,\,, \nonumber \\
\cdots & \,=\, \cdots 
\end{eqnarray}
(and analogously for the Sinh-Gordon model) which allows us to identify the corresponding densities $\theta_{2n+1}$ of these models.  Few examples will clarify this point. 
\begin{itemize} 
\item Let's consider the chiral component of the first conserved current of the Sine-Gordon model, here called $J_1$, normalized as
\be 
J_1 \,=\,\frac{1}{2} \phi_\sigma^2 \,\,\,.
\ee 
It satisfies $\partial_{\tau} J_1 = \partial_\sigma \tilde J_1$ and we would like to determine $J_1$. Taking the derivative with respect to $\tau$ of $J_1$ and using the eq. of motion (\ref{eqmotionSGlc}), we have 
\be 
\partial_\tau J_1 \,=\,\phi_\sigma \phi_{\tau\sigma} \,=\,\frac{m^2}{g}\, \phi_{\sigma} \sin (g \phi) \,=\,-\frac{m^2}{g^2} \partial_{\sigma} 
\cos (g \phi) \,\,\,,
\ee
so that the pair components of the conserved current are 
\be
J_1 \,=\,\frac{1}{2} \phi_\sigma^2 
\,\,\,\,\,\,\,\,\,\,
, 
\,\,\,\,\,\,\,\,\,\,
\tilde J_1 \,=\, - m^2/g^2 \cos (g \phi) \,\,\,,
\ee
with 
\[
\partial_\tau J_1 \,=\,\partial_\sigma \tilde J_1 \,\,\,.
\]
\item 
Consider now the chiral component of the second conserved current, denoted by $J_3$ and here normalized as 
\be 
J_3 \,=\, \frac{1}{4} \phi_\sigma^4 - \frac{1}{g^2} \phi_{\sigma\sigma}^2 \,\,\,, 
\ee
of which we want to determine the corresponding component $\tilde J_3$. Taking the derivative w.r.t. $\tau$ of this quantity and 
using (\ref{chainofidentities}), we get 
\begin{eqnarray}
\partial_\tau J_3 &\,=\,& \phi_\sigma^3 \phi_{\sigma\tau} -\frac{2}{g^2} \phi_{\sigma\sigma} \phi_{\sigma\sigma\tau} \,=\,
m^2 \left[\frac{\phi_\sigma^3}{g} \sin(g \phi) - \frac{2}{g^2} \phi_{\sigma\sigma} \phi_\sigma \cos(g \phi) \right]\,=\,\\
& = & - \partial_\sigma \,m^2 \left[\frac{1}{g^2} \phi_\sigma^2 \cos(g \phi) \right] \,\,\,,
\end{eqnarray} 
so that the pair components of this conserved current are 
\be
J_3 \,=\, \frac{1}{4} \phi_\sigma^4 - \frac{1}{g^2} \phi_{\sigma\sigma}^2 
\,\,\,\,\,\,\,\,\,\,
, 
\,\,\,\,\,\,\,\,\,\,
\tilde J_3 \,=\,- m^2 \left[\frac{1}{g^2} \phi_\sigma^2 \cos(g \phi) \right] \,\,\,,
\ee
with 
\[
\partial_\tau J_3 \,=\,\partial_\sigma \tilde J_3 \,\,\,.
\] 
\item Applying this procedure to the chiral component of the higher conserved currents one can get all the pairs $(J_{2n+1}, \tilde J_{2n+1})$ and determine the corresponding conserved densities in the laboratory frame. The first few of these expressions for the Sinh-Gordon model are collected in Appendix D. 
\end{itemize}

\section{The Classical Sinh-Gordon model in infinite volume: laboratory coordinates}\label{ShLabframe}

\subsection{A different gauge}
It is possible to obtain the formulation of the Sinh-Gordon equation as the consistency equation of a linear system directly taking the change of variables from the light-cone coordinates to the laboratory frame. To simplify the notation, here, we also apply the following gauge transformation
\begin{equation}
 \mathcal{G} = \frac{1}{\sqrt{2}}\left(
\begin{array}{cc}
 -i e^{\frac{1}{2} g \phi} & -e^{\frac{1}{2} g \phi } \\
 e^{-\frac{1}{2} g \phi }& i e^{-\frac{1}{2} g \phi} \\
\end{array}
\right)
\end{equation}
and set $\Phi = \mathcal{G} \Psi$ in \eqref{systempsi}. 
In this way, we arrive to the linear differential system
\begin{subequations}
\label{linearlab}
 \begin{align}
\partial_{x}\, \Phi \,&=\, U_{\ell} \, \Phi \label{UPhiEq}\\
\partial_{t} \, \Phi \,&=\, V_{\ell} \, \Phi 
 \end{align}
\end{subequations}
where the differential operators $U_{\ell}$ and $ V_{\ell}$ are obtained by the gauge transformations:
\begin{align}
 U_{\ell} &= \partial_x G G^{-1} + \frac 12 G (U + V) G^{-1} = \left(
 \begin{array}{cc}
 \frac{g (\phi_x + \phi_t)}{4} & \frac{\lambda ^2-e^{g \phi } m^2}{4 \lambda } \\
 -\frac{\lambda ^2-e^{-g \phi } m^2}{4 \lambda } & -\frac{g (\phi_x + \phi_t)}{4} \\
\end{array}
 \right)\\
 V_{\ell} &= \partial_t G G^{-1} +  \frac 12 G (V - U) G^{-1} = \left(\begin{array}{cc}
 \frac{g (\phi_x + \phi_t)}{4} & \frac{\lambda ^2+ e^{g \phi } m^2}{4 \lambda } \\
 -\frac{\lambda ^2+e^{-g \phi } m^2}{4 \lambda } & -\frac{g (\phi_x + \phi_t)}{4} \\
\end{array}
\right)
\end{align}
where in the last expression, we also substituted $\lambda \to -m^2/\lambda$. This formulation coincides with the one adopted in \cite{Forest}. The advantage of this gauge choice is the manifest symmetry between the two operators $U_{\ell}$ and $V_{\ell}$. Then, the consistency constraint leads to the Sinh-Gordon equation
\be 
\frac{\partial U_{\ell}}{\partial t } - \frac{\partial V_{\ell}}{\partial x} + 
[ \, U_\ell \,, \, V_{\ell} \,] \,=\,0 \qquad \Rightarrow \qquad \phi_{tt} - \phi_{xx} = - \frac{m^2}{g} \sinh g \phi\;.
\label{zerocurvature}
\ee
It is useful to proceed directly to the derivation of the conserved charges within this gauge choice.

\subsection{Derivation of conserved charges \label{sec:chargeswhole}}

We will now build explicitly the conserved charges and show how they are related to the action/angle variables.  The main idea is to show that $a(\lambda)$ is a conserved quantity for arbitrary value of the spectra parameter $\lambda$ and using it as a generating function for the local conserved charges. We will actually see that two different points of view lead to an expansion in local charges around two values of $\lambda$, alias 
$\lambda = 0$ and $\lambda= \infty$. We start introducing again the \textit{Jost solutions} of \eqref{UPhiEq}
\begin{align}\label{jost}
\mathbf{f}_{\pm} &\stackrel{x\to+\infty}{\longrightarrow} \psi_{\pm}\\ 
\mathbf{g}_{\pm} &\stackrel{x\to-\infty}{\longrightarrow} \psi_{\pm} \; ,
\end{align}
where we $\psi_{\pm}$ is a basis of solutions for $\lim_{x\to \infty} U_{\ell}(x)$. Explicitly:
\be\label{basisasymptotic}
\psi_{\pm} = \left(\begin{array}{c} 1 \\ \pm i \end{array}\right) e^{\pm i k x}\;,\quad
\lambda = m e^{\theta}\;, \quad k = \frac{m  \sinh\theta}{2} = \frac{1}{4}(\lambda - m^2 \lambda^{-1})\;.
\ee
Since $U_{\ell}$ is a real matrix, we have 
\begin{equation}
\label{complexjost}
\mathbf{f}_{+} = \mathbf{f}_{-}^\ast\;, 
\qquad
\mathbf{g}_{+} = \mathbf{g}_-^\ast\;.
\end{equation}
In general, once we fix a point $x_0$ and a vector $\mathbf{f}_0$, we can obtain a solution of \eqref{UPhiEq} that takes the value $\mathbf{f}_0$ at $x=x_0$ as 
\be
\label{xorder}
\mathbf{f}(x, \lambda) = T(x_0, x, \lambda) \mathbf{f}_0 \,\,\,, 
\ee
where $T(x_0, x, \lambda)$ means the $x$-ordered exponential of the operator $U_\lambda$ defined by the equation
\be
\label{xexpdef}
\frac{dT(x_0,x, \lambda)}{dx} = U_{\ell}(x) T(x_0,x, \lambda) \,\,\,\,\,\,\,\, , \,\,\,\,\,\,\,\,\, T(x,x, \lambda)= 1\,\,\,.
\ee
Taking $x_0 \to \pm \infty$ and $\mathbf{f}_0 \to \psi_{\pm}$ we obtain
\begin{align}
\label{xorderexp}
\twovecT{\mathbf{f}_+}{\mathbf{f}_-} &= \lim_{y_0 \to \infty} T(y_0,x, \lambda) \twovecT{\psi_+ (y_0)}{\psi_- (y_0)}\\
\twovecT{\mathbf{g}_+}{\mathbf{g}_-} &= \lim_{x_0 \to -\infty} T(x_0,x, \lambda) \twovecT{\psi_+ (x_0)}{\psi_- (x_0)}\;. 
\end{align}
As the transfer matrix can be written as the change of basis between the Jost solutions, we have
\be
\label{transferintegral}
T(\lambda) = 
\left(
\begin{array}{cc} 
a(\lambda) & b^*(\lambda) \\
b(\lambda) & a^*(\lambda) 
\end{array}
\right) =
\twovecT{\mathbf{f}_+}{\mathbf{f}_-}^{-1}\twovecT{\mathbf{g}_+}{\mathbf{g}_-} = 
\lim_{\substack{x_0 \to -\infty\\y_0 \to \infty}}\Psi(y_0)^{-1} T(x_0,y_0, \lambda) \Psi(x_0)\; , 
\ee
where the first equality is implied by eq.~\eqref{complexjost}. First we need an expression for $T(y,x, \lambda)$ and, in order to solve \eqref{xexpdef}, it is useful to recast it in the form
\be
\label{ansatzT}
T(y,x, \lambda) = (1 + W(x)) e^{Z(x)} C(y)\,\,\,,
\ee
where the matrix $W(x)$ is off-diagonal, $Z(x)$ is diagonal and $C(y)$ is fixed by the condition at $T(x, x,\lambda) = 1$. Putting \eqref{ansatzT} inside \eqref{xexpdef} and separating the diagonal and off-diagonal part of $U_\ell$, one obtains 
\begin{align}
\label{eqT}
W'(x) + W(x) Z'(x) &= U_\ell^{(O)}(x)  + U_\ell^{(D)}(x) W(x)\\
Z'(x) &=  U_\ell^{(D)}(x)  + U_\ell^{(O)}(x) W(x) \label{zfirst}
\end{align}
that after eliminating $Z'(x)$ gives 
\be\label{Weq}
W'(x) + [W(x),U^{(D)}_\ell(x)] + W(x) U^{(O)}_\ell(x) W(x) = U^{(O)}_\ell(x) \,\,\,.
\ee
We write explicitly these matrices as
\be\label{Wform}
W(x) = \left(\begin{array}{cc} 0 & -w_-(x) \\ w_+(x) & 0 \end{array} \right), \qquad 
Z(x) = \left(\begin{array}{cc} z_+(x)  & 0 \\ 0 & z_-(x)\end{array} \right)
\ee
and \eqref{Weq} becomes a Riccati differential equation 
\be\label{weq}
w_{\pm}(x)^2 \left(\frac{m^2 e^{\pm g \phi (x)}}{\lambda }-\lambda \right)+\frac{m^2 e^{\mp g \phi (x)}}{\lambda }=\pm 2 g w_{\pm}(x) \Omega (x)+\lambda +4 w'_{\pm}(x) \; ,
\ee
with $\Omega = \phi_x + \phi_t$. With the power expansion $w_{\pm}(x) = \sum_{k=0}^\infty w_{k,\pm}(x) \lambda^{-k}$, we get the recursion
\begin{multline}
\label{wrec}
w_{n+1, \pm}  = \frac{1}{2 w_{0,\pm}} \bigl(-4 w_{n,\pm}' \mp 2 g w_{n,\pm}\Omega + m^2 e^{\mp g \phi} \delta_{n,1} +
\\ \hspace{-3mm} + m^2 e^{\pm g \phi} \sum_{k=0}^{n-1} w_{n-1-k,\pm} w_{k,\pm} - \sum_{k=1}^n w_{n+1-k,\pm} w_{k,\pm}\bigr)
\end{multline}
with $w_{0,\pm} = \mp i$ and, for simplicity, we omit the dependence on $x$. Employing \eqref{zfirst}, we get the expansion
\be\label{zexp}
\frac{dz_{\pm}(x)}{dx} = 
\mp \frac{\lambda i}{4}   + \frac{1}{4}\sum_{k=1}^\infty (w_{k+1,\pm}(x) - e^{\pm g \phi} m^2 w_{k-1,\pm}(x) )\lambda^{-k} \;.
\ee
Now, it is easy to prove by induction in \eqref{wrec} that $w_{k,\pm}(x) \stackrel{x\to \infty}{\to} 0$, for $k\geq1$. Therefore we have
\be\label{limitT}
T(\lambda) = \left(\begin{array}{cc} 
                    e^{i p(\lambda)} & 0 \\ 0 & e^{- i p(\lambda)}  
                   \end{array}
\right)\,\,\,, 
\ee
where we used that $\det T(\lambda) = 1$ and we set
\be\label{pdef}
\log a(\lambda) = i p(\lambda) \equiv \sum_{k=1}^\infty \lambda^{-k} I_k \equiv 
\sum_{k=1}^\infty \frac{\lambda^{-k}}{4} \int_{-\infty}^\infty dx \;  w_{k+1,+}(x) - 
(e^{g \phi}-\delta_{k,1}) m^2 w_{k-1,+}(x) \,\,\,.
\ee
As, $a(\lambda)$ is conserved, also the $I_k$ are conserved quantities. In particular we have
\be\label{I1}
I_1 = \frac{i g^2}{4}\int_{-\infty}^\infty dx\left[ \frac{ (\phi_x + \phi_t)^2}{2}+ 
\frac{m^2}{g^2} (\cosh g \phi -1)\right] \;. 
\ee
In a similar way, one can repeat the expansion for small $\lambda \simeq 0$, which leads to another series of conserved
charges, analogous to eq.~\eqref{pdef}
\be\label{ptildedef}
\log a(\lambda) = i p(\lambda) \equiv \sum_{k=1}^\infty \lambda^{k} \tilde I_k \,\,\,.
\ee
In particular, one has for the first one
\be\label{I1tilde}
\tilde I_1 = - \frac{i g^2}{4m^2} \int_{-\infty}^\infty dx \left(\frac{\left(\phi_t-\phi_x \right)^2}{2}
   +\frac{m^2}{g^2} (\cosh g \phi-1)\right) \; .
\ee
Thanks to \eqref{pdef} and \eqref{ptildedef}, it is easy to connect the expression of  the charges to the action/angle variables. Indeed, from the dispersion relation of $a(\lambda)$, we have
\begin{align}
\log a (\lambda)  &= -\frac{i}{\pi}\int_{-\infty}^\infty d\lambda \frac{\log | a(\lambda')|}{\lambda' - \lambda - i \epsilon} \stackrel{\lambda \to \infty}{=} \frac{i}{\pi}\sum_{n=0}^\infty \lambda^{-n - 1} \int_{-\infty}^\infty d\lambda (\lambda')^n\log | a(\lambda')|\,\,\,,\\
\log a (\lambda)  &\stackrel{\lambda \to 0}{=} -\frac{i}{\pi}\sum_{n=0}^\infty \lambda^{n} \int_{-\infty}^\infty d\lambda (\lambda')^{-n-1}\log | a(\lambda')|\;.
 \label{disprellargesmall}
\end{align}
Clearly, as $a(\lambda) = a^{\ast}(-\lambda)$, only the odd $n$ give non-trivial charges as already discussed in Sec.~\ref{ConservedChargesShG}. 
Setting $I_{-n} = \tilde I_{-n}$, we can write compactly
\begin{equation}
 \label{chargedispersion}
I_{2n+1} = \sgn(2n+1)\frac{2 i}{\pi}\int_0^\infty (\lambda')^{2 n} \log | a(\lambda) | \,\,\,. 
\end{equation}
Then, the charges $I_{\pm n}$ and $\tilde I_n$ can be combined to obtain the even/odd charges
\begin{equation}
 E_n = \frac{2i}{g^2}  \left[ m^{2n + 2} I_{-2n -1} - \frac{I_{2n+1}}{m^{2n}} \right] \;, \qquad 
 O_n = \frac{2i}{g^2} \left[ \frac{I_{2n+1}}{m^{2n}} + m^{2n + 2} I_{-2n -1}\right]  \,\,\,.
\end{equation}
Comparing with the explicit expressions in \eqref{I1} and \eqref{I1tilde}, we see that, as expected,
the Hamiltonian and the total momentum are the first conserved quantities
\begin{subequations}
\label{firstcharges}
\begin{align}
 \mathcal{H} &\equiv E_0 = \int_{-\infty}^\infty dx \;  \frac{\Pi (x)^2 + \phi_x(x)^2}{2} + \frac{m^2 (\cosh (g \phi (x))-1)}{g^2}\,\,\,, \label{allinterval1}\\
 \mathcal{P} &\equiv O_0 = - \int_{-\infty}^\infty dx \;  \Pi(x) \phi_x(x) \;, \label{allinterval2}
\end{align}
\end{subequations}
and we deduce the expressions for $n\geq 0$
\begin{equation}
\label{chargesEO}
 E_n = \int_{-\infty}^\infty d\theta \; P(\theta) m\cosh (2n+1)\theta \;, \qquad
 O_n = \int_{-\infty}^\infty d\theta \; P(\theta) m\sinh (2n+1)\theta 
\end{equation}
where we changed variable from the spectral parameter to the rapidity with $\lambda = m e^{\theta}$
and we introduced the action variable 
\begin{equation}
P(\theta) \,=\, \frac{8}{\pi g^2} \ln |a (\lambda(\theta))| \,\,\,.
\label{actionframe}
\end{equation}
In closing this section let's comment that, as in the light-cone coordinates, it is also easy to set up a numerical procedure to 
compute the transfer matrix $T(x,\lambda)$ in the laboratory frame. One has simply to integrate numerically the 
differential equation (\ref{xexpdef}) and thus determine numerically the transfer matrix $T_\infty(\lambda)$ for the Sinh-Gordon 
model when the field and all its derivatives vanish at $|x| \rightarrow \infty$: if $I = (-x_0,x_0)$ is the finite support of the field 
configurations, we have 
\be 
T_\infty(\lambda) \,=\,\prod_{j=1}^{N+1} \hat W(\lambda, -x_0 + j \Delta x) \,\,\,.
\label{formulanumericatransfer}
\ee
where $
\hat W(\lambda,\Delta x) \,=\,e^{ U_l\,\Delta x}  
$. 

\section{Life on a circle} \label{lifeonacircle}

The problem we are interested to concerns the equilibration of a classical field theory at a {\em finite density energy}. A way to implement this condition is to define the theory on a finite length $L$ and then send $L\rightarrow \infty$ with the ratio $E/L$ kept fixed in this limit. Since our models are massive, the boundary condition should be essentially irrelevant once we send $L \rightarrow \infty$ and for this reason we will choose to work with periodic boundary conditions. The Inverse Scattering Method becomes in this case more involved but also much more interesting since it puts nicely together many of the previous results. In the following we will adopt the laboratory frame variables. In addition to the aforementioned books \cite{Faddev,Novikov,Ablowitz}, for the aims of this section it is also useful to consult the references \cite{BulloughPillingTimonenQCShG,Forest,Flaska}.



\vspace{0.3cm}
\noindent 
\subsection{Periodic Scattering Problem}
Let the field $\phi(x,t)$ of the Sinh-Gordon model and all its derivatives be periodic functions in $x$ with period $L$ and let's choose the interval $I = (0, L)$.  In such a case, the scattering problem posed by eq.~(\ref{UPhiEq}) 
\be 
\partial_{x}\, \Phi \,=\, U_{\ell} \, \Phi \,\,\,,
\label{scatteringproblemlab2}
\ee
becomes the quantum mechanical problem of a particle in a periodic potential (for the properties of this problem see, for instance, \cite{Merzbacher}). 
Fixing a point $x_0$ in the interval $I$, we can specify a complete basis $(\zeta_+(x,x_0,k), \zeta_-(x,x_0,k)$ for the solution of (\ref{scatteringproblemlab2}) by the following initial conditions 
\be 
\label{bcFloquet}
\zeta_+(x=x_0,x_0,k) \,=\, \left(\begin{array}{c} 1\\0\end{array}\right) 
\;,
\qquad \zeta_-(x=x_0,x_0,k) \,=\,\left(\begin{array}{c}0\\1\end{array}\right)\,\,\,.
\ee
A simple consequence of the periodicity of the potential is that $\tilde\zeta_{\pm}(x, x_0, \lambda) \equiv \zeta_{\pm}(x + L, x_0, \lambda)$ 
will be another basis of solutions. A change of basis between the two will exist 
\begin{equation}
\label{zetazetatilde}
 \left\{\begin{array}{ll}
         \tilde \zeta_+(x, x_0, \lambda) &= t_{11} \zeta_+(x,x_0, \lambda) +  t_{12} \zeta_+(x,x_0, \lambda) \;,\\
         \tilde \zeta_-(x, x_0, \lambda) &= t_{21} \zeta_+(x,x_0, \lambda) +  t_{22} \zeta_-(x,x_0, \lambda) \;, 
        \end{array}
\right.
\end{equation}
and define a \textit{periodic transfer matrix} $\mathcal{T}(\lambda)$
\be
\label{periodictf}
\twovec{\zeta_+(x + M L, x_0, \lambda)}{ \zeta_-(x + M L, x_0, \lambda)} = \TT^M(\lambda) 
\twovec{\zeta_+(x, x_0, \lambda)}{\zeta_-(x, x_0, \lambda)} \;.
\ee
By definition a point $\lambda$ will be in the spectrum if the corresponding solution is bounded on the real line; this will require that both the two eigenvalues $\rho_+$ and $ \rho_-$ of $\TT(\lambda)$ have unit modulus: $|\rho_{\pm}| = 1$. Since $\rho_+ \rho_- = \det \TT(\lambda) = 1$, the eigenvalues are given by
\be 
\rho_\pm \,= \,\frac{1}{2}\left[\Delta(k) \,\pm \, \sqrt{\Delta^2(k) -4}\right] \,\,\,, 
\label{transeigenvalues}
\ee
where $\Delta(\lambda) \equiv \Tr\TT(\lambda)$ is called the discriminant. Then, the unit modulus conditions becomes 
\be\label{conditionDiscriminant}
\Delta(\lambda) \in \mathbb{R} \; , \qquad |\Delta(\lambda)| \leq 2 \;.
\ee
It is useful to re-parametrize $\Delta(\lambda) \to \Delta(k)$ using the expression for $k(\lambda)$ in eq.~\eqref{basisasymptotic}. Then, we have the usual situations of the band structure:
\begin{enumerate}
\item If $|\Delta(k)| > 2$, the two eigenvalues are both real and positive, with $\rho_+ > 1$ while $\rho_- < 1$. Hence, iterating eq. (\ref{periodictf}), $\zeta_{\pm}$ will necessarily become exponentially large at $x \to + \infty$ or $x \to - \infty$. These situations are not admissible as spectral data and therefore the regions in $k$ where $\Delta(k) > 2$ are outside the spectrum and constitutes the so-called {\em gaps} of the spectrum. 
\item If $|\Delta(k)| < 2$, both eigenvalues are complex number of modulus $1$, with equal and opposite phases. In this case, 
for arbitrary $M$, in eq.~\eqref{periodictf}, the functions $\zeta_{\pm}(x, x_0, \lambda)$ will remain bounded as the eigenvectors of $\TT(\lambda)$ 
get multiplied by a phase. 
\item When $|\Delta(k_n)| =2$ and the roots $k_n$ of this equation are not degenerate (i.e. we have simultaneously $\Delta'(k_n) \neq 0$), 
the associated eigenvectors $\psi_\pm(x)$ are respectively periodic ($\Delta = 2$) or anti-periodic ($\Delta = -2$) functions. 
\end{enumerate}
Note that $\Delta(k)$ can be expressed in terms of the basis $\left(\zeta_+,\zeta_-\right)$ as 
\be 
\Delta(k) \,=\, \zeta_{+,1}(x_0+L,x_0,k) + \zeta_{-,2}(x_0 + L, x_0,k) \,\,\,. 
\label{alternativeDelta}
\ee
To verify this expression, it is enough to set $x=x_0$ in \eqref{zetazetatilde} and use the initial conditions (\ref{bcFloquet})
\begin{eqnarray}
\zeta_+(x_0 + L, x_0, k) & \,=\, & t_{11} \left(\begin{array}{c} 1\\0 \end{array}\right) + t_{12} \left(\begin{array}{c} 0\\1 \end{array}\right) \\
\zeta_-(x_0 + L, x_0, k) & \,=\, & t_{21} \left(\begin{array}{c} 1\\0 \end{array}\right) + t_{22} \left(\begin{array}{c} 0\\1 \end{array}\right) 
\end{eqnarray}
and therefore 
\be
\label{TTzeta}
\TT(\lambda) \,=\, \left(\begin{array}{lr}
\zeta_{+,1}(x_0 + L, x_0, \lambda) &  \zeta_{+,2}(x_0 + L, x_0, \lambda) \\
\zeta_{-,1}(x_0 + L, x_0, \lambda) &  \zeta_{-,2}(x_0 + L, x_0, \lambda) 
\end{array}
\right) \,\,\,. 
\ee 
Taking now the trace of $\TT(\lambda)$ we arrive to the expression (\ref{alternativeDelta}).
\begin{figure}[t]
\centering
$\begin{array}{c}
\includegraphics[width=0.4\textwidth]{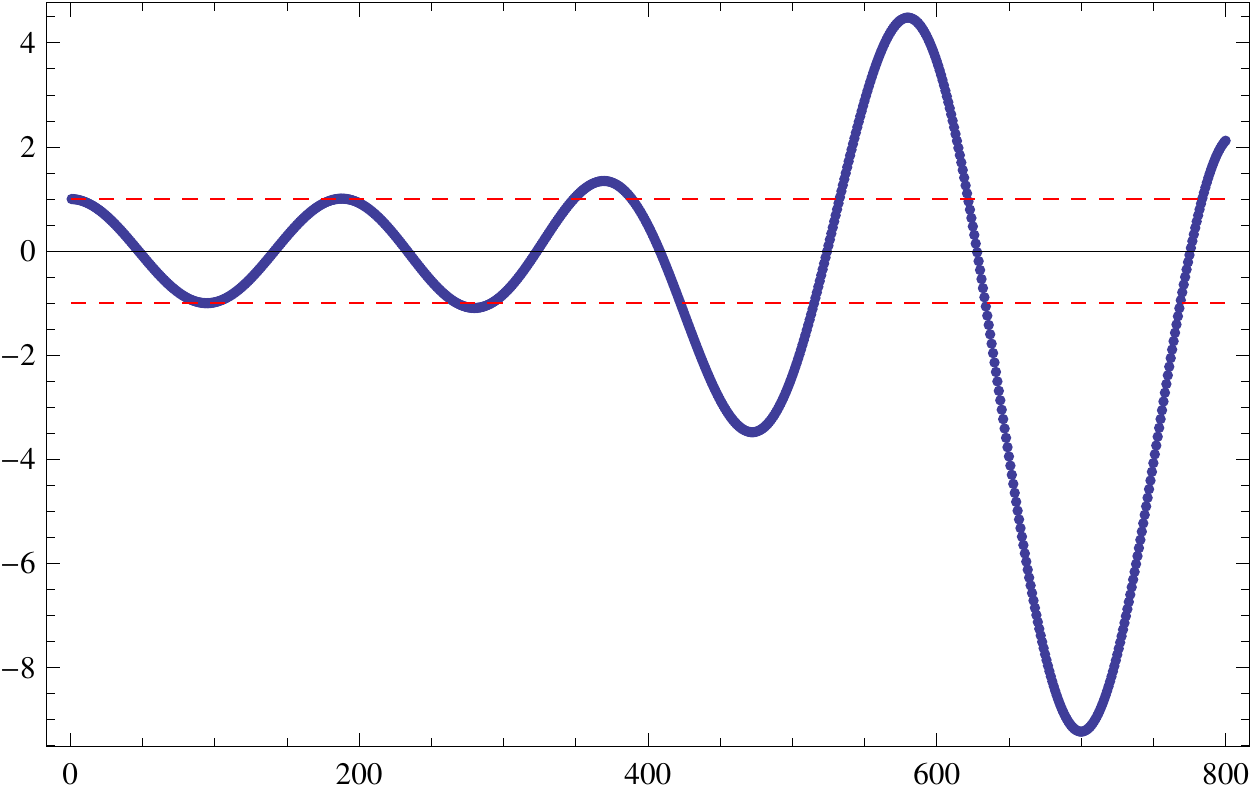}
\end{array}$
\caption{Plot of $\Delta(k)/2$ versus $k$ (in unit 1/100) where are visible the intersections with the lines $\pm 1$ .}
\label{spectralfig}
\end{figure}
A typical plot of $\Delta(k)$ is given in Figure \ref{spectralfig}, where the simple roots $k_n$ are explicitly visible. The knowledge of the roots $k_n$ contains all the informations about the function $\Delta(k)$ and therefore determines the spectrum. In Sec. \ref{sec:finitegap}, we will show how these roots lead to the potential at all times and thus to an analytic expression for the solution of the Sinh-Gordon equation at all times. We will now clarify some properties of $\Delta(k)$ and its relation with the whole-line scattering data.

\vspace{0.3cm}
\subsection{Relation to the whole-line problem}

An important step is represented by the relation between the whole line and the periodic case. Given a periodic field $\phi(x,t)$ which solves the Sinh-Gordon equation in \eqref{zerocurvature}, we can always associate a compact support field by restricting the periodic one to any fundamental period, i.e. 
\begin{equation}
\label{truncatedphi}
\phi_{L}(x,t) = \begin{cases} 
                         \phi(x,t)  	& x \in [x_0,x_0 + L]\\
                         0		& \mbox{otherwise}
                        \end{cases} 
\end{equation}
As the field $\phi_L(x,t)$ vanishes identically outside $[x_0,x_0 + L]$, the solution to \eqref{UPhiEq} outside $[0,L]$ must be a combination of the asymptotic solutions $\psi_{\pm}$ given in eq.\,\eqref{basisasymptotic}. In particular, from eq.\,\eqref{bcFloquet}, we deduce
\begin{align}
 \zeta_{+}(x, x_0, \lambda) &\simeq \frac{1}{2} \psi_- e^{i k(\lambda) x_0} + \frac{1}{2} \psi_+ e^{- i k(\lambda) x_0} \;, \quad x \simeq x_0\\
 \zeta_{-}(x, x_0, \lambda) &\simeq \frac{i}{2} \psi_- e^{i k(\lambda) x_0} - \frac{i}{2} \psi_+ e^{- i k(\lambda) x_0} \;, \quad x \simeq x_0
 \end{align}
The behaviour for $x \simeq x_0 + L$ can then be obtained using the transfer matrix $T(\lambda)$ in eq.~\eqref{transferintegral}
associated to $\phi_L(x,t)$. Using these results in eq.~\eqref{TTzeta}, it follows
\be\label{periodictowl}
\Delta(\lambda) = a(\lambda) e^{- i k(\lambda) L} + a^\star(\lambda) e^{i k(\lambda) L} \,\,\,, 
\ee
where $a(\lambda)^{-1}$ is the transmission coefficient for the potential $\phi_L(x,t)$. For real $\lambda$, this can be also written as 
\begin{equation}
\label{periodictowlcos}
\Delta(\lambda) = 2|a(\lambda)| \cos\left[ k(\lambda) L - \arg a(\lambda)\right] \,\,\,,
\end{equation}
an expression whose meaning will become clear soon. 

\subsection{Finite and Infinite Gap Solution \label{sec:finitegap}}
Let's briefly discuss how to obtain an analytic expression of the field at all time $t$ using directly the knowledge of the spectrum of 
eq.~\eqref{scatteringproblemlab2}, a method known as {\em finite gap solution} \cite{Novikov,Ablowitz,Forest}.  In order to simplify the notation, we set here $m = 1/4$. 

In absence of an external potential, the transmission coefficient is simply $1$ and the discriminant oscillates always inside the range $[-2,2]$.
\begin{figure}[b]
\centering
\includegraphics[width=0.4\columnwidth]{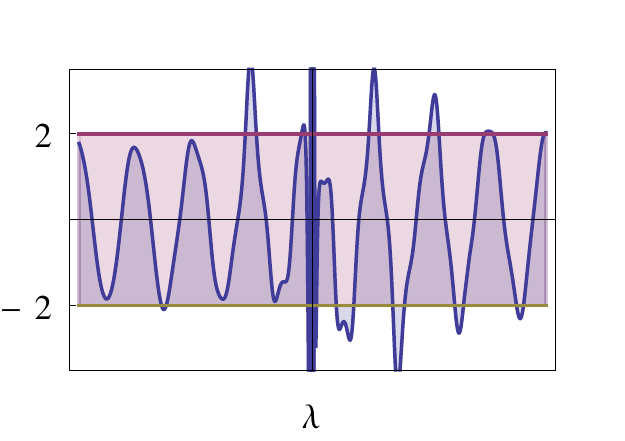}
\includegraphics[width=0.4\columnwidth]{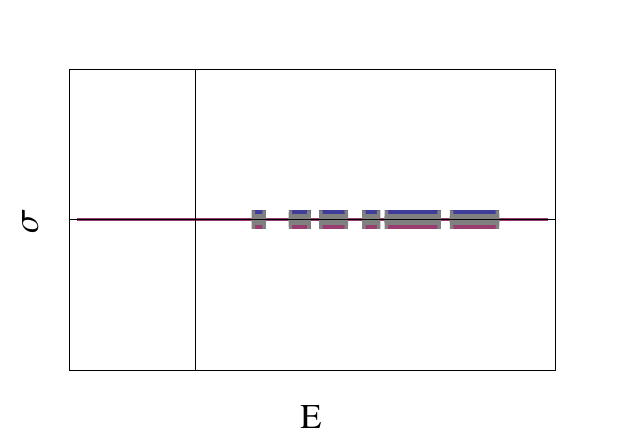}
\caption{(Left) Typical plot of the discriminant $\Delta(\lambda)$ vs $\lambda$. The points where the curve meets $\pm 2$ identify the simple spectrum that completely characterizes the full function. (Right) A cartoon of the spectrum in the $E$-plane ($E = \lambda^2$), showing bands separated by gaps inside $E>0$.} 
\label{spectralCurve}
\end{figure}
In a more interesting examples, the curves of the discriminant, \textit{the spectral curve}, crosses the two lines $\Delta = \pm 2$ in several points corresponding to the boundaries of the region of spectrum inside the real line, see Figure \ref{spectralCurve}. This set of points is called \textit{simple spectrum}. From eq.~\eqref{transeigenvalues}, the eigenvalue of $\TT(\lambda)$ are expressed in terms of
\be\label{elliptic}
y(E) = \sqrt{\Delta(E)^2 - 4}
\ee
with $E = \sqrt{\lambda}$ and it is possible to choose the branch-cuts of this function such that they coincide with the spectrum of \eqref{linearlab}. The analyticity properties of this function ensure that it is completely defined by the simple spectrum \cite{Novikov,Ablowitz,Forest}. It is possible to explicitly determine the most general solution of \eqref{zerocurvature} with a given (simple) spectrum, at least in the simplifying hypothesis that the number of elements inside the simple spectrum is finite, say $2n$. In practice, we have that $y(E)^2$ is a polynomial and the cuts-structure of \eqref{elliptic} defines a two-sheets Riemann surfaces of genus $n-1$. This fact has remarkable consequences because it puts into the game a powerful machinery coming from algebraic geometry. Let's briefly sketch the recipe and we remind to \cite{Forest} for the details. 

By setting $y(E)^2 = E\prod_{i=1}^{2n} (E-E_i)$, the spectrum and the branch cuts will lie in the intervals $[E_{2k}, E_{2k+1}]$, for $k=0,\ldots,n$ and $E_0=0, E_{2n+1} = \infty$. These intervals identify a Riemann surface and it can be shown that the relevant quantities are contour integral of $E^j/y(E)$, where $j=0,\ldots, n-1$. The crucial property is that the set of independent contours is actually finite dimensional. Picking two different basis, they are 
said to be canonical if each contour crosses only one contour of the other basis. In this cases, the matrix of change of basis is called \textit{period-matrix}. To be more concrete, once defined the two $n\times n$ matrices ($i,j = 1,\ldots,n$)
\be\label{cycles}
\alpha_{ij} = 2\int_{E_{2j-1}}^{E_{2j}} \frac{E^{i-1} dE}{y(E)} , \quad
\beta_{ij} = 2\int_{E_{2(j-1)}}^{E_{2j-1}} \frac{E^{i-1} dE}{y(E)}\,\,\,,
\ee
we can obtain the period matrix as
\be\label{periodmatrix}
B = -\alpha^{-1} \tilde \beta \,\,\,, 
\ee
where $\tilde\beta_{ij} = \sum_{k\leq j} \beta_{ik}$ is introduced to make the two basis canonical. In terms of these quantities the general solution of the field $\phi(x,t)$ living on the circle takes the form 
\be\label{siegelrepr}
\phi(x,t) \equiv 2 \log \left(\frac{\Theta(\mathbf{l}(x,t)+\frac{1}{2}|B)}{\Theta(\mathbf{l}(x,t)|B)}\right)\,\,\,, 
\ee
where $\Theta(\mathbf{v}|M)$ is the Riemann-Siegel function, defined for a $n$-vector $\mathbf{v}$ and a $n\times n$ matrix $M$
\be
\label{siegeldef}
\Theta(\mathbf{v}| M) = \sum_{\mathbf{w} \in \mathbb{Z}^n} \exp ( \pi i w^t M w + 2 \pi i w^t v )  \;,  
\ee
and the vector $\mathbf{l}(x,t)$  has a simple time-evolutions 
\be
\label{lvec}
l_j (x,t) = 2[ (\alpha^{-1}_{1j} + \alpha^{-1}_{1j}\mathcal{K}) x  + (\alpha^{-1}_{1j} - \alpha^{-1}_{nj}\mathcal{K})t] + l_j^{(0)}
\ee
with $\mathcal{K} = \frac{1}{16 \sqrt{E_1\ldots E_n}}$ and $l_j^{(0)}$ arbitrary constants determining the initial conditions. The Riemann-Siegel function ensures (quasi)-periodicity in the different components: as expected from an integrable system, the phase space can be described as a multidimensional torus and the dynamics on each component of the torus reduces to a simple rotation. 

The formula (\ref{siegelrepr}) provides the solution of the equation of motion for the field of the Sinh-Gordon model. Roughly speaking, the number $n$ of the gaps is associated to the number of excitations of the field present on the circle: if the (numerical) computation of the formula (\ref{siegelrepr}) is already difficult for small values of $n$ (such as $n =2$ or $n=3$), it shall be stressed that it becomes absolutely prohibitive for large values of $n$, in particular for the infinite-gap limit $n \rightarrow \infty$ associated to the {\em finite-energy} case that we are interested in. In other words, the classical Inverse Scattering Method provides eq.\,(\ref{siegelrepr}) as a close and elegant formula for the time-evolution of the field but this formula is of very little practical use, in particular to compute the time average of the various functions $F[\phi(x,t)]$ based on the field $\phi(x,t)$ in the finite-energy case.  It is in this respect that the formulation based on the classical limit of the LM formula comes in our aid, providing a simplest approach to compute asymptotic time averages for classical integrable models. 

\noindent

\section{Identification of the action variables and root density}\label{TraceTM}

As explained in Sec.~\ref{sec:ck}, given a finite density initial condition, the essential ingredient in the Caux-Konik construction for the calculation of the stationary expectation values is to identify the root density $\rho^{(r)}(\theta)$ of the initial state. We discuss in this section how such a root density can be explicitly calculated and computed numerically. To do so it is worth start discussing how the conserved charges of the periodic problem are related to the ones in the whole-line case presented in Sec.~\ref{sec:chargeswhole}.

\subsection{Trace of the Transfer Matrix}
The trace of the transfer matrix for the periodic problem in \eqref{zetazetatilde} can be expressed as:
\begin{equation}
\label{deltaeq}
\Delta(\lambda) \equiv \Tr \TT(\lambda) = \Tr [\zeta(x_0)^{-1} T(x_0, x_0 + L, \lambda)\zeta(x_0) ]=
\Tr [T(x_0, x_0 + L, \lambda) ]
\end{equation}
where $T(y,x, \lambda)$ was defined in eq.~\eqref{xexpdef} and $\zeta(x_0) = (\zeta_+, \zeta_-)$
contains the solutions introduced in eq.~\eqref{bcFloquet}. It is easy to show that $\Delta(\lambda)$ is a conserved quantity. 
Indeed, the compatibility equation (\ref{zerocurvature}) of the linear system can be equivalently 
seen as the zero-curvature condition of the differential operators \cite{Faddev}
\begin{equation}
L_+ = \frac{d}{dx} - U_{\ell}(\lambda) \; ,
\qquad 
L_- = \frac{d}{dt} - V_{\ell}(\lambda)\; . 
\end{equation}
This implies that for any closed contour ${\mathcal C}$ in the space-time plane the Wilson loop given by the path-ordered integral
\be
W({\mathcal C}) \,=\, {\mathcal P} \,\exp\left[\int_{\mathcal C} \, U_\ell \, dx +  V_\ell \, dt \right] \,\,\,
\ee
is just the unit matrix. Taking the contour as in Figure \ref{contourtrans} and separating the space and time contributions
\begin{figure}[b]
\centering
$\begin{array}{c}
\includegraphics[width=0.45\textwidth]{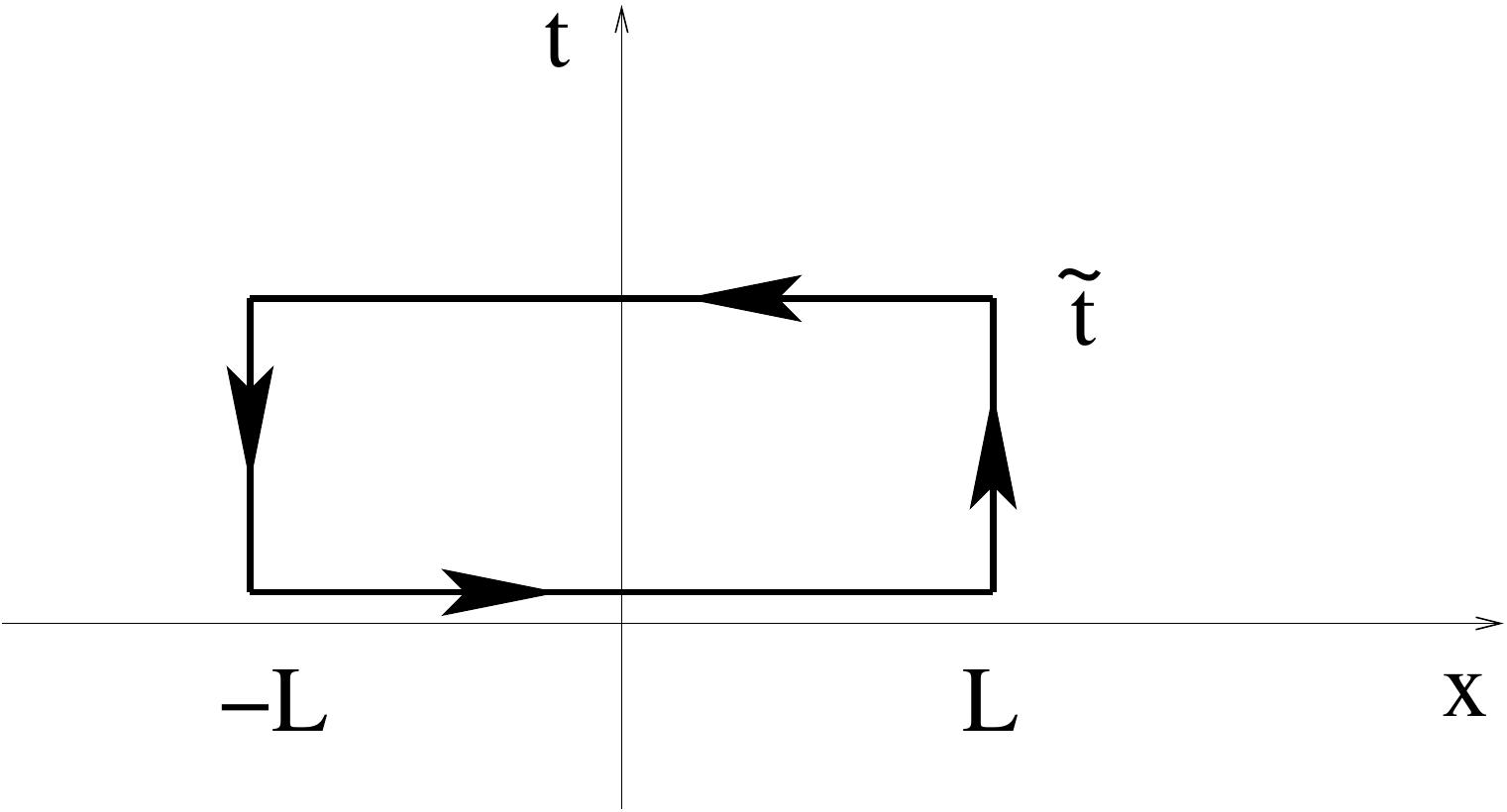}
\end{array}$
\caption{Contour on which is evaluated the Wilson loop $W({\mathcal C)}$.}
\label{contourtrans}
\end{figure}
\be
\TT_{t}(\lambda) \,=\,{\mathcal P} \, \exp\left[\int_{x_0}^{x_0 + L} U_{\ell}(x,t) \, dx \right]
\;, 
\qquad 
\mathcal{S}_x(\lambda) \,=\,
{\mathcal P} \, \exp\left[\int_{t}^{\bar t} V_{\ell}(x,t) \, dt \right] \,\,\,,
\ee
we have 
\be 
\TT_{\bar t}(\lambda) \,=\, \mathcal{S}^{-1}_{x_0} (\lambda)\, \TT_{t}(\lambda) \, \mathcal{S}_{x_0 + L}(\lambda) \,\,\,,
\ee
and because of the periodicity of the field $\phi(x,t)$: $\Tr \TT_{t}(\lambda) = \Tr \TT_{\bar t}(\lambda) = \Delta(\lambda)= t_{11}(\lambda) + t_{22}(\lambda)$ is a conserved quantity. We can interpret $t_{11}(\lambda) = t_{22}^\ast(\lambda) = a^{(L)}(\lambda)$ in analogy to the (inverse) transmission coefficient $a(\lambda)$ of the field with compact support on the whole-line introduced in eq.~\eqref{transferintegral}.
Note that, while for the infinite volume case, it was $a(\lambda)$ to be conserved (and therefore $\Re [a(\lambda)]$ as well), for the periodic case with 
finite $L$ only $\Tr \TT(\lambda) = 2 \Re [a^{(L)}(\lambda)] $ is conserved.

As $\Delta(\lambda)$ is conserved, it is a good candidate as a generating function for the local conserved charges as it was $\ln a(\lambda)$
for the whole line. In particular, one can repeat the procedure described in Sec.~\ref{sec:chargeswhole} and use the representation in eq.~\eqref{ansatzT} for
$T(y, x, \lambda)$ in eq.~\eqref{deltaeq}. As $ W(x_0) = W(x_0 + L)$, 
we arrive at 
\begin{equation}
 \Delta(k) = \Tr e^{Z(x_0 + L) - Z(x_0)}\,\,\,,
\end{equation}
where $Z(x)$ was defined in eq.~\eqref{Wform}. From Eqs.\eqref{zexp}, we can set
\begin{equation}
 p_{\mbox{\tiny per}}(\lambda) = \arccos\left(\frac{\Delta(\lambda)}{2}\right) + k(\lambda) L \;.
\end{equation}
Then, the function  $p_{\mbox{\tiny per}}(\lambda)$, as the one introduced in eq.~\eqref{pdef}, generates two sets of charges expanding around $\lambda = 0$ and $\lambda = \infty$. These charges have the same densities as the one defined for instance in eqs. (\ref{allinterval1} - \ref{allinterval2}) 
but with an integral restricted this time to the fundamental domain $[x_0, x_0 + L]$, as  
\begin{subequations}
\label{firstchargesperiodic}
\begin{align}
 \mathcal{H} &\equiv E_0 = \int_{x_0}^{x_0 + L} dx \;  \frac{\Pi (x)^2 + \phi_x(x)^2}{2} + \frac{m^2 (\cosh (g \phi (x))-1)}{g^2} \\
 \mathcal{P} &\equiv O_0 = - \int_{x_0}^{x_0 + L} dx \;  \Pi(x) \phi_x(x) \;.
\end{align}
\end{subequations}

\vspace{1mm}
\noindent
{\bf Relation between periodic and whole-line conserved charges.}
The construction, presented above, seems to suggest that the conserved charges in the periodic case can be simply be obtained 
considering the scattering problem for the truncated field $\phi_L(x,t)$ introduced in eq.~\eqref{truncatedphi}. However, the truncation in \eqref{truncatedphi} introduces a singularity at both boundaries $x = x_0, x_0 + L$. As the conserved charges also involve space derivatives of the field $\phi_L(x,t)$, it is clear that the whole-line charges (e.g. eq.~\eqref{firstcharges}) and the periodic one (e.g. eq.~\eqref{firstchargesperiodic}) will be different. However, notice that the difference comes only from boundary terms, while for the extensive bulk of width $L$, the charges are exactly the same. Hence, it follows that in the limit of large $L$, the two expressions can be identified. Therefore, we can employ the expressions in 
eq.~\eqref{chargesEO} and a simple comparison with eq.~\eqref{check} shows that we have to take
\begin{equation}
\label{rootsfroma}
 \rho^{(r)}(\theta) = \frac{P(\theta)}{L} = \frac{8}{\pi g^2 L} \ln |a(\theta)|
\end{equation}
where the action variable $P(\theta)$ was introduced in eq.~\eqref{actionframe}. We will give in the next section another verification that this is indeed the correct choice.

\subsection{Classical Bethe Ansatz} 
In this section we are going to show that the roots $k_n$ of the equation 
\be 
|\Delta(k_n)| \,=\, 2 \,\,\,, 
\label{rootequations}
\ee
which determines the spectrum for the periodic scattering problem (\ref{scatteringproblemlab2}), can be interpreted as solutions of a classical Bethe Ansatz equation, once the identification in eq.~\eqref{rootsfroma} has been used. Starting from eq.~\eqref{periodictowlcos}, in fact the root equation (\ref{rootequations}) can be expressed as 
\be 
k_n L - \rm{arg}\,a(k_n)  \,=\, \pi n + \cos^{-1}(| a(k_n)|^{-1}) \,\,\,.
\label{primostep}
\ee
It is now easy to show that, in the limit $L \rightarrow \infty$, these are nothing else but the bosonic Bethe Ansatz equations (\ref{BETHEEQUATIONS}) in their classical limit: using for $k$ and $\lambda$ the parametrization in terms of the rapidity in eq.~\eqref{basisasymptotic}, thanks to the dispersion relation in eq.~\eqref{scatteringamplitudea} the argument of $a(\lambda)$ can be expressed as 
\be
\label{dispersionphaseshift}
\arg a(\lambda) = \frac{2\lambda}{\pi } {\cal P} \int_{0}^{\infty} \frac{d\lambda' \, \log | a(\lambda') |}{(\lambda^2 - \lambda'^2)}= 
\frac{1}{\pi } {\cal P} \int_{-\infty}^{\infty} \frac{d\theta'\, \log | a(\theta') |}{\sin(\theta - \theta')} = 
- \frac{L}{2} {\cal P} \int_{-\infty}^{\infty} d\theta'\, \rho^{(r)}(\theta') \varsigma_{cl}(\theta - \theta') \,\,\,,
\ee
where the classical phase-shift $\varsigma(\theta)$ was introduced in eq.~\eqref{classicalphaseshift}. Then, eq. (\ref{primostep}) can be written as 
\be 
m \sinh\theta_n + \int \,d\theta'\, 
\varsigma_{cl}(\theta-\theta')\,\rho^{(r)}(\theta') \,=\,\frac{2\pi n}{L} + \frac{2}{L} \, \cos^{-1}(| a(k_n)|^{-1})
\,\,\,.
\label{secondstep}
\ee
Notice that $| a(k_n)|^{-1}$ is the modulus of the transmission amplitude $|t(k)|$ of the truncated potential $\phi_L(x,t)$: sending $L \rightarrow \infty$ we have $| t(k)| \rightarrow 1$ and therefore the last term in the equation above simply modify the integer $n$ since 
$
 \cos^{-1}(| a(k_n)|^{-1}) \rightarrow 2 \pi n' 
$. 
Therefore, the final form of eq. (\ref{secondstep}) is given by 
\be
m \sinh\theta_n + \int \,d\theta'\, 
\varsigma_{cl}(\theta-\theta')\,\rho^{(r)}(\theta') \,=\,\frac{2\pi n}{L} \,\,\,, 
\label{laststep}
\ee
which indeed coincides with the classical limit of the Bethe Ansatz equations (\ref{BETHEEQUATIONS}). 

\subsection{From the root density to the filling fraction}
The roots $k_n$ of the equation $ \Delta(k_n) \,=\, \pm 2$ are then nothing else but the root solutions of the classical Bethe Ansatz equations. 
Their actual positions ultimately depend upon the initial values at $t=0$ of the field $\phi(x,t)$ and its time derivative $\phi_t(x,t)$: in the limit of large $L$, the behaviour of  $\Delta(k)$ becomes highly oscillatory and the intersections $\Delta(k_n) = \pm 2$ tend to the continuous root density distribution $\rho^{(r)}(k)$. According to eq.\,(\ref{rootsfroma}}), this root density distribution $\rho^{(r)}(k)$ can be extracted directly from the coefficient $a(\theta)$ of the transfer matrix $T(\lambda)$. On the other hand, the transfer matrix $T(\lambda)$ can be numerically evaluated from the initial configuration of the field $\phi(x,t)$ following the procedure discussed in Section \ref{sec:chargeswhole}, and in particular applying the formula (\ref{formulanumericatransfer}). 
  
Since the classical phase-shift obtained in eq.~\eqref{laststep} is consistent with the classical limit of the bosonic phase-shift given in eq.~\eqref{phaseshiftbosonic}, we can use in this case the bosonic version of the Bethe Ansatz equation (\ref{eq:rhos}) (in the classical limit) to get the bosonic density of state $\rho(\theta)$ and then, from the formula (\ref{filling}) we can recover the sought filling fraction $f(\theta)$ entering the LM formula (\ref{EnsembleAverageMus23}).

\section{Conclusions}\label{conclusions}
In this paper we have addressed the problem of computing the time averages of a classical integrable model (as for instance the Sinh-Gordon model): we have shown that rather than using the almost intractable formula coming from the infinite-gap solution of the classical Inverse Scattering Method, it is instead possible to employ (the classical limit of) the Leclair-Mussardo formula (\ref{EnsembleAverageMus23}) that rules the asymptotic values of various observables in a quantum quench. The classical limit of this formula is obtained by restoring the $\hbar$ dependence in the quantum field theory and then sending $\hbar \rightarrow 0$ in the relevant quantities. 

The main points of the procedure presented in this paper consists of properly interpreting the expressions associated to the classical limit of the two basic quantities entering the quantum Leclair-Mussardo formula, namely the Form Factors and the filling fraction of the states. Concerning the former quantities, we have shown that the classical limit of the Form Factors indeed exists and moreover that the resulting functions can be directly associated to the solution of the classical equation of motion. These functions are naturally associated to a bosonic formulation of the Form Factors. About the latter quantities, they are also naturally associated to a bosonic formulation of the Bethe Ansatz. Moreover, we have shown that the classical root density $\rho^{(r)}(\theta)$ can be directly extracted from the initial conditions of the classical field $\phi(x,t)$ by employing the numerical determination of the transfer matrix of the classical Inverse Scattering Method: once this root density $\rho^{(r)}(\theta)$ is known, one can then apply a classical version of the Caux-Konik procedure to extract the corresponding total density $\rho(\theta)$ and finally the filling fraction $f(\theta)$. 

In this paper we focus the attention on the Sinh-Gordon model for its relative simplicity but we expect that the application to other classical relativistic integrable models (such as the Sine-Gordon or the Bullough-Dodd models) is just question of simple details, but it is a question which is nevertheless worth exploring. Equally interesting is the problem to study the classical limit of the formula recently proposed by Smirnov and Nigro \cite{Negro} and opportunely reformulated for the quantum quench process by Bertini et al. \cite{Piroli} for the time average of the exponential operators of the 
Sinh-Gordon model: since in the quantum case this formula provides an exact resummation of all terms present in the Leclair-Mussardo formula, we expect that if its classical limit exists, it should be equivalent to the exact values of the time averages of the exponential operators in the classical integrable Sinh-Gordon model.

\vspace{5mm}

\noindent {\bf Acknowledgements:}
We would like to thank Paulo de Assis for his collaboration on the early stage of this project. We are grateful to Boris Dubrovin for very useful discussions on classical integrable models. On quantum integrable systems out of equilibrium we enjoyed very useful discussions and collaborations along the years with many colleagues, in particular we would like to thank Alvise Bastianello, Denis Bernard, J-S. Caux, John Cardy, Pasquale Calabrese, Axel Cortes Cubero, Benjamin Doyon, Fabian Essler, Maurizio Fagotti, Andrea Gambassi, Robert Konik, Milosz Panfil, Lorenzo Piroli, Marcos Rigol, Alberto Rosso, Raoul Santachiara, Alessandro Silva, Spyros Sotiriadis, Jacopo Viti. The final part of the project was carried out in Cambridge, at the Isaac Newton Institute, where both authors were taking part to the program {\em Mathematical Aspects of Quantum Integrable Models in and out of Equilibrium}, under the grant {\em Isaac Newton Institute for Mathematical Sciences EP/K032208/1}.

\newpage

\appendix
\section{Fourier analysis}


In $(1+1)$ dimensions and in the infinite volume, the free real scalar field of mass $m$ admits the following Fourier decomposition
\be 
\phi(x,t)\,=\,\int_{-\infty}^{\infty} \frac{dk}{2 \pi} \frac{1}{\sqrt{2 \omega(k)}} \,
\left[A(k) \,e^{-i \omega t + i k x} + A^{\dagger}(k) \,e^{i \omega t - i k x} \right] \,\,\,,
\label{modeexpansionfreetheory}
\ee
where $\omega(k) = \sqrt{m^2 + k^2}$. The modes $A(k)$ are given by the field $\phi(x,t)$ and its time-derivative 
$\frac{\partial \phi(x,t)}{\partial t} \equiv \dot{\phi}(x,t)$ at $t=0$ as 
\begin{eqnarray}
A(k) &\,=\,& \frac{i}{\sqrt{2 \omega(k)}} \int_{-\infty}^{\infty} dx \,e^{- i k x} \left[\dot{\phi}(x,0) - i \omega(k) \phi(x,0)\right]
\,\,\,, \label{modeexpansion}\\
A^{\dagger}(k) &\,=\,& -\frac{i}{\sqrt{2 \omega(k)}} \int_{-\infty}^{\infty} dx \,e^{ i k x} \left[\dot{\phi}(x,0) + i \omega(k) \phi(x,0)\right] \nonumber
\end{eqnarray} 
For the free theory with Hamiltonian and momentum 
\begin{eqnarray*}
H & \,=\,& \frac{1}{2} \,\int \,  \left[(\phi_t)^2 + (\phi_x)^2 + m^2 \phi \right] \, dx \,\,\,,\\
P & \,=\,& - \int \, \phi_t \, \phi_x \, dx \,\,\,, 
\end{eqnarray*}
substituting the mode expansion we have
\begin{eqnarray} 
H & \,=\,& \int \, \frac{dk}{2 \pi} \,\omega(k) \, |A(k)|^2  \,\,\,,\\
P & \,=\, & \int \, \frac{dk}{2 \pi} \, k \, |A(k)|^2 \,\,\,.\nonumber 
\end{eqnarray} 
Hence, for the free theory in the laboratory frame the action variable which gives the mode occupation is 
\be 
P(k) \,=\,\frac{1}{2\pi} \,|A(k)|^2 \,\,\,. 
\ee
We will also make often use of the rapidity variable $\theta$ entering the dispersion relations
\be
E\,=\,m \cosh\theta 
\,\,\,\,\,\,\,\,\,\,
,
\,\,\,\,\,\,\,\,\,\,
P\,=\,m \sinh\theta \,\,\,, 
\ee
and will expand the field as 
\be 
\phi(x,t)\,=\,\frac{1}{\sqrt 2}\, \int_{-\infty}^{\infty} \frac{d\theta}{2 \pi} \,
\left[A(\theta) \,e^{-i m t \cosh\theta  + i m x \sinh\theta } + A^{\dagger}(\theta) \,e^{i m t \cosh\theta  - i m x \sinh\theta } \right] \,\,\,. 
\label{rapidityexpansion}
\ee
Although for simplicity we use, in both $\theta$ and $k$ variables, the same symbols for the creation/annihilation operators, notice however that these operators have a different normalization of the two schemes. The corresponding formulas for the Hamiltonian and momentum of free 
theory in $\theta$ variable are 
\begin{eqnarray}
H & \,=\,& \int \, \frac{d\theta}{2 \pi} \, m \cosh\theta \, |A(\theta)|^2  \,\,\,,\\
P & \,=\, & \int \, \frac{dk}{2 \pi} \, m \sinh\theta \, |A(\theta)|^2 \,\,\,.\nonumber 
\end{eqnarray} 
On a circle of radius $L$, with periodic boundary conditions, the expansion of the free field is given in terms of the Fourier 
series 
\be 
\phi(x,t)\,=\, \sum_{n=-\infty}^{\infty} \frac{1}{\sqrt{2 L \omega_n}} \,
\left[A_n \,e^{-i \omega_n t + i 2 \pi n x/L} + A^{\dagger}_n \,e^{i \omega_n t - i 2 \pi n x/L} \right] \,\,\,,
\ee
where 
\be
\omega_n = \sqrt{m^2 + (2 \pi n/L)^2}
\label{omegaFS}
\ee
and 
\be
A_n \,=\,\frac{i}{\sqrt{2 L \omega_n}} \int_{0}^{L} dx \,e^{- 2 \pi i n x/L} \left[\dot{\phi}(x,0) - i \omega_n \phi(x,0)\right]
\,\,\,. 
\ee
with $A_n^{\dagger}$ given by the complex conjugate of the formula above.

\begin{figure}[b]
\centering
$\begin{array}{cc}
\includegraphics[width=0.4\textwidth]{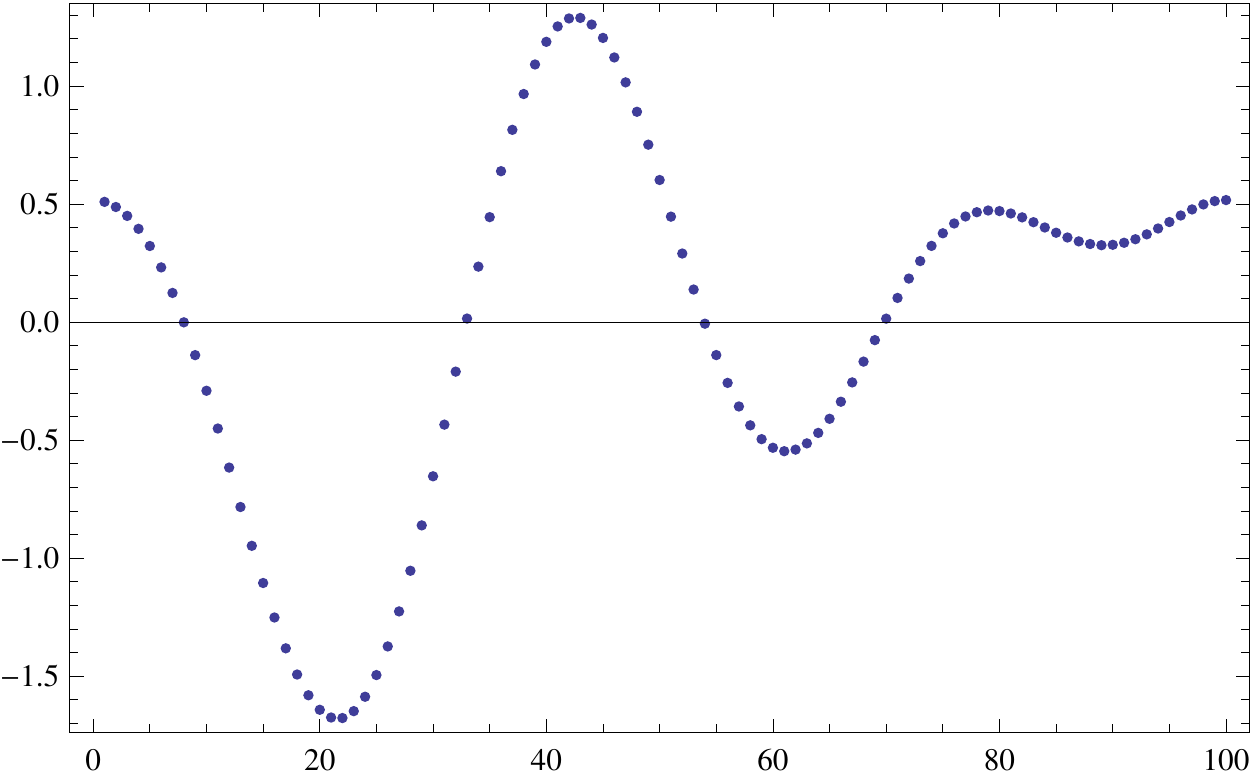} &
\includegraphics[width=0.4\textwidth]{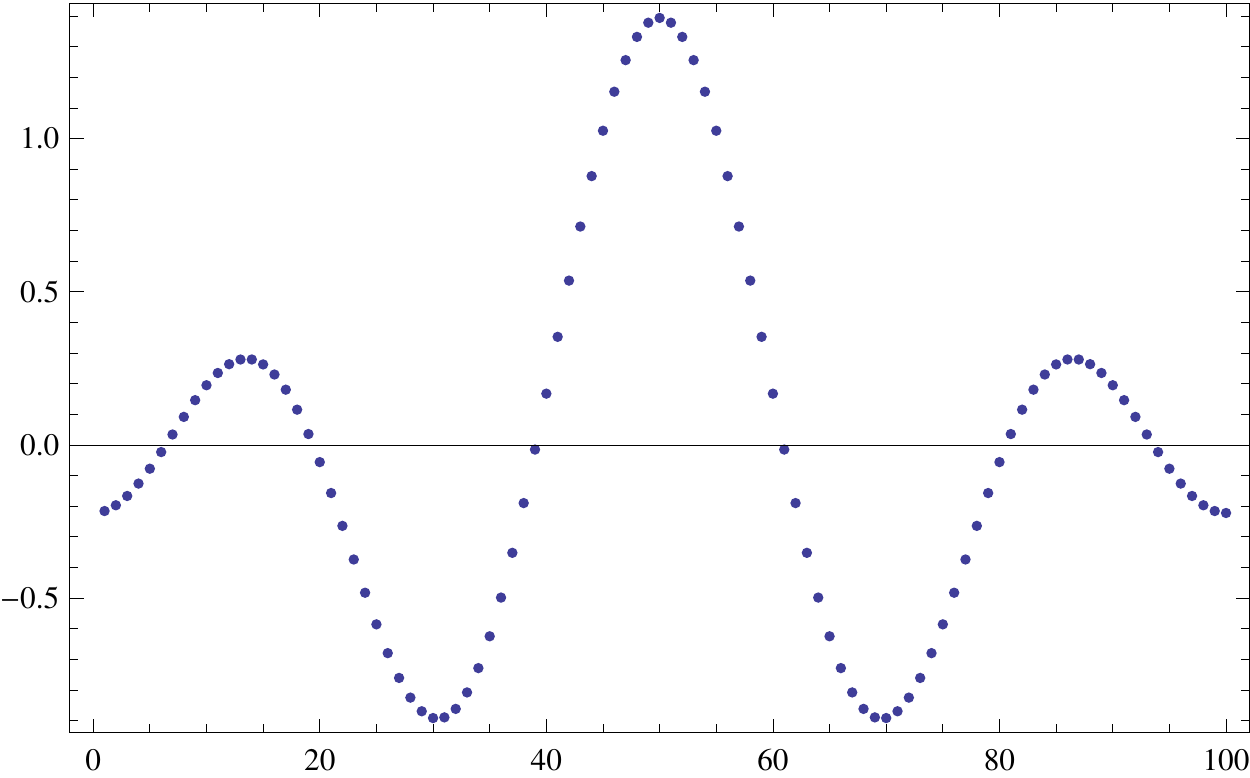}\\
\end{array}$
\caption{$\phi(x,0))$ (on the left) and $\dot\phi(x,0)$ (on the right) discretized on a finite interval of length $L$.}
\label{discretization}
\end{figure}

If the interval $L$ is discretized in $N$ steps of lattice spacing $a$, with $L = N a$, the field $\phi(x,t)$ and its time derivative 
$\dot{\phi}(x,t)$ will be represented at any time $t$ by the set of $2N$ values $\phi(m a,t) \equiv \phi_m(t)$ and 
$\dot{\phi}(n a,t) \equiv \dot{\phi}_n(t)$ (see Figure \ref{discretization}): these values can be associated to the 
Discrete Fourier Series of $A_n$ and $A^{\dagger}_n$, obtained by restricting the possible values of momentum 
to the finite set 
\be 
k_n \,=\,\frac{2 \pi n}{N} 
\,\,\,\,\,\,\,\,\,\,
,
\,\,\,\,\,\,\,\,\,\,
n= 0,1,2,\cdots, N-1 \,\,\,.
\label{firstset}
\ee  

In this case the direct and inverse formulas are given by 
\be
\phi_m(t) \,=\,\,\sum_{n =0}^{N-1} \frac{1}{\sqrt{2 N \omega_n}} \,
\left[A_n \,e^{-i \omega_n t + i 2 \pi m n /N} + A^{\dagger}_n \,e^{i \omega_n t - i 2 \pi m n/N} \right] \,\,\,,
\ee
\be
A_n \,=\,\frac{i}{\sqrt{2 N \omega_n}} \sum_{m=1}^N  \,e^{- 2 \pi i m n /N} \left[\dot{\phi}_m(0) - i \omega_n \phi_m(0)\right]
\,\,\,. 
\ee 
Moreover, in this case the dispersion relation is expressed as  
\be 
\omega_n \,=\,\sqrt{m^2 + \left(\frac{\sin \frac{1}{2}k_n}{\frac{1}{2} a} \right)^2} \,\,\,.
\ee 
and it reduces to (\ref{omegaFS}) when $a \rightarrow 0$. We can make the formulas more symmetric by considering an even integer value for $N$ and by shifting the index $n$ in (\ref{firstset}), so that the Brillouin zone is described by the momenta 
\be 
k_n \,=\,\frac{2 \pi n}{N} 
\,\,\,\,\,\,\,\,\,\,
,
\,\,\,\,\,\,\,\,\,\,
n= -\frac{N}{2}, \cdots,  \frac{N}{2} \,\,\,.
\label{firstset1}
\ee  
Since we are mostly concerning with field configurations at finite density, in the paper we will often define the theory on a finite interval $L$ and on a lattice of $N$ points with $L = N a$. Keeping finite the ultraviolet cut-off $a$, in the thermodynamic limit we have 
\be 
\sum_{n=-N/2}^{N/2} f(k_n) \longrightarrow \frac{L}{2\pi} \int_{-\pi/a}^{\pi/a} dk \, f(k) 
\ee
 

\section{Ground state of modified Mathieu equation}

In this Appendix we use the Transfer Matrix formalism and the Thermodynamics Bethe Ansatz equation to express the 
ground state energy of the modified Mathieu equation in terms of a solution of an integral equation. Let's consider 
the thermal partition function of the classical Sinh-Gordon model 
\be 
Z(\beta) \,=\,\int {\cal D}\pi \,{\cal D}\phi \exp\left[-\beta \left(\frac{1}{2} \pi^2 + \frac{1}{2}\left(\frac{d\phi}{dx}\right)^2 + 
\frac{m^2}{g^2} (\cosh (g\phi) -1) \right)\right] \,=\,Z_{\pi}(\beta) Z_{\phi}(\beta)\,\,\,.
\ee
Discretizing the path integral, we have 
\be 
Z_{\pi}(\beta) \,=\,\frac{1}{h^{N/2}} \, \int \prod_{i=1}^N \left(d\pi_i \right)\, e^{-\frac{\beta}{2a} \sum_{i=1}^N \pi_i^2}
 =  \left(\frac{2 \pi a}{h \beta}\right)^{N/2} \,\,\,.
\ee 
For the term which depends on $\phi$ we have 
\begin{eqnarray}
Z_\phi(\beta) & = & \frac{1}{h^{N/2}}\,\int \prod_{i=1}^N d\varphi_i \,e^{-\beta\,a\,\sum_{j=1}^N \left[
\frac{(\varphi_j - \varphi_{j-1})^2}{2 a^2} + \frac{m^2}{g^2} (\cosh(g\phi_j) -1) \right] } 
\end{eqnarray}
Let's define 
\be
f(x_{i+1},x_i) \equiv a \left[\frac{(x_{i+1}-x_i)^2}{2 a^2} + \frac{m^2}{g^2} (\cosh(g x_{i+1})-1)\right]
\,\,\,.
\ee 
Introducing a complete set of functions $\psi_m(x)$ 
\be 
\delta(x_{i+1} - x_i) \,=\,\sum_{m=0}^{\infty} \psi_m(x_{i+1}) \, \psi(x_i) \,\,\,,
\ee
eigenfunctions of the Transfer Matrix operator 
\be
\int_{-\infty}^{\infty} dx e^{-\beta f(y,x)} \psi_m(x) \,=\,e^{- a {\cal E}_m(\beta)} \, \psi_m(y) 
\label{integralequationtransfermatrix}
\ee
the path integral above can be written as 
\be
Z_\phi(\beta) \,=\,\sum_{m=0}^{\infty} e^{- N a {\cal E}_m(\beta)} \,\,\,,
\ee
that, in the limit $L = N a \rightarrow \infty$, becomes 
\be 
Z_\phi(\beta) \,=\,e^{- L E_0(\beta)} \,\,\,,
\ee
When $a \rightarrow 0$, the integral equation (\ref{integralequationtransfermatrix}) 
can be converted into the Schr\"odinger equation 
\be
\left[-\frac{1}{2\beta} \frac{d^2}{dx^2} + \frac{\beta m^2}{g^2} (\cosh(g x)-1) + V_0\right] \psi_m(x) \,=\, {\cal E}_m(\beta)\,\psi(x) 
\label{Sch}
\ee
where $V_0$ is a constant given by 
\be 
V_0 \,=\,\frac{1}{2 a} \log\left(\frac{\beta}{2 \pi h a}\right)\,\,\,.
\ee
It is easy to see that $V_0$, added to $-L^{-1} \log Z_{\pi}$, gives rise to the $a\rightarrow 0$ limit of the Free Energy per unit length of the free theory (see eq.~(\ref{freefreeenergyazero}) 
\be 
V_0 - L^{-1} \log Z_{\pi} \,=\,a^{-1} \,\log\left(\frac{\beta \hbar}{a}\right) \,=\, L^{-1} \beta F_{free}(\beta) \,\,\,.
\ee
Subtracting $V_0$ and making a canonical transformation on the coordinate and momentum, $q \rightarrow \sqrt{\beta} q$ and 
$p \rightarrow p/\sqrt{\beta}$, the Schr\"odinger equation (\ref{Sch}) can be written as 
\be
\left[-\frac{1}{2} \frac{d^2}{dx^2} + \frac{\beta m^2}{ g^2} (\cosh(\frac{g x}{\sqrt{\beta}})-1) \right] \psi_m(x) = {\cal E}_m(\beta)\,\psi(x) 
\label{Sch2}
\ee 
i.e. a modified Mathieu equation. So, the Transfer Matrix method leads to the following expression for the Free Energy 
of the classical Sinh-Gordon model 
\be
\beta F_{cl}(\beta) \,=\,\beta F_{free}(\beta) + {\mathcal E}_0(\beta) \,\,\,,
\label{finalexprefreetotale}
\ee
where ${\cal E}_0(\beta)$ is the ground state energy of the modified Mathieu equation (\ref{Sch2}).

On the other hand, the Free Energy of the classical Sinh-Gordon model is given by the classical Thermodynamics Bethe Ansatz equations given in the text: 
\begin{eqnarray} 
\beta F_{cl}(\beta) &\,=\,& \int_{-\infty}^{\infty} m \cosh\theta \,\log(\hbar \epsilon_{cl}(\theta)\,\frac{d\theta}{2\pi} \,\,\,,\\
\epsilon_{cl}(\theta) & \,=\,& \beta \cosh\theta + \int_{-\infty}^{\infty} \fii_{cl}(\theta-\theta') \,\log(\hbar \epsilon_{cl}(\theta')) \frac{d\theta'}{2\pi} 
\,\,\,.
\nonumber 
\end{eqnarray}
Substituting $\epsilon_{cl}(\theta)$ into the first equation and using eq.~(\ref{nicecheckfreeenergy}), 
the first iterative term is identified with $\beta F_{free}(\beta)$. The remaining terms provides then an expression of 
${\mathcal E}_0(\beta)$ of the Mathieu equation. 

It is worth noticing how $\hbar$ disappears from the final expression of ${\mathcal E}_0(\beta)$. First of all, notice that $\hbar$ does not enter the solution of $\epsilon_{cl}(\theta)$ because, splitting the logarithmic term as  $\log(\hbar \epsilon_{cl}(\theta')) = \log \hbar + \log(\epsilon_{cl}(\theta'))$ and using the integration rule (\ref{prescription}), we have 
$$ 
\int_{-\infty}^{\infty} \fii_{cl}(\theta-\theta') \,\log(\hbar) \,=\,0 \,\,\,.
$$
Secondly, using eq.~(\ref{finalexprefreetotale}) and the expression (\ref{nicecheckfreeenergy}) 
for the Free Energy of the free theory, we can express ${\cal E}_0(\beta)$ as 
\be 
{\cal E}_0(\beta) \,=\, \int_{-\infty}^{\infty} m \cosh\theta \,\log\left(\frac{\epsilon_{cl}(\theta)}{\cosh\theta}\right)\,\frac{d\theta}{2\pi} 
\,\,\,,
\label{MATHEIUENERGY}
\ee
which does not contain any longer $\hbar$. 

Notice that in principle one can also obtain the energies ${\mathcal E}_m$ ($m > 0$) of the excited states of the Mathieu equation using the analytic continuation of the Thermodynamics Bethe Ansatz equations, along the line of ref. \cite{DoreyTateo}.

\section{Light-cone dynamics}

In this Appendix we collect the relevant formulas for the light-cone dynamics. Let's define 
\be
\begin{array}{l}
\tau\,=\,x + t \\
\sigma \,=\, x - t 
\end{array}
\,\,\,\,\,\,\,
, 
\,\,\,\,\,\,\,
\begin{array}{lc}
x \,=\,& \frac{1}{2}(\tau + \sigma) \\
t \,=\, & \frac{1}{2}(\tau - \sigma) 
\end{array}
\ee
so that 
\be
\begin{array}{l}
\partial_\tau \,=\,\partial_t + \partial_t \\
\partial_\sigma \,=\, \partial_x - \partial_t 
\end{array}
\,\,\,\,\,\,\,
, 
\,\,\,\,\,\,\,
\begin{array}{lc}
\partial_x \,=\,& \frac{1}{2}(\partial_\tau + \partial_\sigma) \\
\partial_t \,=\, &\frac{1}{2}(\partial_\tau - \partial_\sigma) 
\end{array}
\ee
The Laplacian operator becomes 
\be 
\Box \,=\,\partial_t^2 - \partial_x^2 \,=\,-\partial_\tau \,\partial_\sigma \,\,\,
\ee 
while the Lagrangian density reads 
\be
{\mathcal L} \,=\,\frac{1}{2} (\partial_\mu \phi)^2 - V \,=\, -\frac{1}{2} \phi_\tau \, \phi_\sigma -V \,\,\,.
\ee
The corresponding equation of motion is 
\be 
\phi_{\tau\sigma} \,=\,\frac{\delta V}{\delta \phi} \,\,\,.
\ee
Considering $\tau$ as time variable and $\sigma$ as space variable, the corresponding conjugate variables are the 
light-cone Hamiltonian ${\mathcal H}$ and momentum ${\mathcal P}$, given by 
\be 
{\mathcal H} \,=\,  V  
\,\,\,\,\,\,\,
, 
\,\,\,\,\,\,\,
{\mathcal P} \,=\, \frac{1}{2}\,(\phi_\sigma)^2\,\,\, .
\ee
The Fourier expansion of the field is given by 
\be
\phi(\sigma,\tau) \,=\,\int_0^{\infty} \frac{dp_+}{2\pi p_+} 
\left(A(p_+) \,e^{-i p_-\tau + ip_+\sigma} + A^{\dagger}(p_+) \, 
 e^{i p_-\tau - ip_+\sigma} \right) \,\,\,,
\ee
with the dispersion relation 
\be
p_- \,=\,\frac{m^2}{p_+} \,\,\,.
\label{dispersionrelationlc}
\ee
For the free theory, with $V = \frac{m^2}{2} \phi^2$, the Hamiltonian ${\mathcal H}$ and the momentum ${\mathcal P}$ can be 
expressed as 
\begin{eqnarray}
{\mathcal H} & \,=\, & m^2 \int_0^{\infty} \frac{dp_+}{2 \pi p_+^2} |A(p_+)|^2 \,\,\,, \nonumber \\
{\mathcal P} & \,=\, & \int_0^{\infty} \frac{dp_+}{2 \pi p_+} p_+ \,|A(p_+)|^2 \,\,\,. \nonumber \\
\end{eqnarray}
Going back to the laboratory frame, for the corresponding Hamiltonian $H$ and momentum $P$ we have 
\begin{eqnarray}
H & \,=\,& \frac{{\mathcal H} + {\mathcal P}}{2} \,=\,  \int_{0}^{\infty} \frac{dp_+}{2 \pi p_+} |A(p_+)|^2 
\,\left(\frac{m^2}{2 p_+} + \frac{p_+}{2} \right)  \nonumber \\
P & \,=\,& \frac{{\mathcal H} - {\mathcal P}}{2} \,=\,  \int_{0}^{\infty} \frac{dp_+}{2 \pi p_+} |A(p_+)|^2 
\,\left(\frac{m^2}{2 p_+} - \frac{p_+}{2} \right)  \nonumber \\
\end{eqnarray}
Let us now pose 
\be 
k \,\equiv \, \frac{m^2}{2 p_+} - \frac{p_+}{2} \,\,\,,
\ee
and solve for $p_+$. Notice that $k(p_+)$ is a monotonic function for $p_+ >0$ and takes values in 
$(-\infty,\infty)$: it can be inverted as 
\be
p_+ \,=\, \sqrt{k^2 + m^2} - k  \,\,\,, 
\ee
so that we have 
\be 
\frac{m^2}{2 p_+} + p_+ \,=\,\sqrt{k^2 + m^2} \,\,\,. 
\ee
Introducing the rapidity variable $\theta$, with $k = m \sinh\theta$, one can see that $p_+ = m e^\theta$ 
and therefore $H$ and $P$ take the usual form of the integral on the modes  
\begin{eqnarray}
H & \,= \,& \int_{-\infty}^{\infty} \cosh\theta \,|A(\theta)|^2\, d\theta \,\,\,,\\
P & \,=\, & \int_{-\infty}^{\infty} \sinh\theta\, |A(\theta)|^2 \, d\theta \,\,\,. \nonumber 
\end{eqnarray}

\section{Conserved currents}\label{CCShGexplicit}

In this Appendix we report the conserved densities of the Sinh-Gordon model in the laboratory frame, making use of the 
corresponding quantities obtained in the light-cone coordinates. 
For a continuity equation written in light-cone variables,
\begin{equation}
\partial_{\mp} \mathit{j}_{\pm}+\partial_{\pm} \mathcal{J}_{\mp}=0,
\end{equation}
we have associated conservation laws in space and time variables,
\begin{equation}
\partial_0 J_0 + \partial_1 J_1=0,
\end{equation}
where $x_0=t, x_1=x$, and
\begin{eqnarray} 
J_0(x,t) = \rho(x,t) =  \left(\mathit{j}_++\mathcal{J}_-\right)+ \left(\mathit{j}_-+\mathcal{J}_+\right), \\
J_1(x,t) = j(x,t) =  \left(\mathit{j}_+-\mathcal{J}_-\right)+ \left(\mathcal{J}_+-\mathit{j}_-\right).
\end{eqnarray}
The conserved quantities for such models are well known, with the first ones given by:
\begin{eqnarray}
\rho_1(x,t) &=& \frac{1}{2}\left( \phi_t^2 + \phi_x^2 \right) + \frac{\mu^2}{\gamma^2} \left( \cosh (\gamma \phi) -1 \right) = \mathcal{H}(x,t) ,\\[8pt]
\nonumber
\rho_2(x,t)  &=& 2 (\phi_{xt}^2 + \phi_{xx}^2) + \frac{3}{4} \phi_t^2 \phi_x^2 \gamma^2 + \frac{1}{8} (\phi_t^4 + \phi_x^4) \gamma^2 + \\ 
\nonumber
&+& \frac{\mu^2}{2} (\phi_t^2 + \phi_x^2)  \cosh (\gamma \phi) +  
\frac{\mu^2}{2 \gamma^2}  \sinh (\gamma \phi ) (\mu^2 \sinh(\gamma\phi) - 4 \gamma \phi_{xx})\\[8pt]
\nonumber
\rho_3(x,t)  &=& 8 (\phi_{xxt}^2 + \phi_{xxx}^2) +  5 ((\phi_t^2 + \phi_x^2) \phi_{xt}^2 + 4 \phi_t \phi_x \phi_{xt} \phi_{xx} + (\phi_t^2 + \phi_x^2) \phi_{xx}^2) \gamma^2 + \\
\nonumber
&+& \frac{1}{16} (\phi_t^2 + \phi_x^2) (\phi_t^4 + 14 \phi_t^2 \phi_x^2 + \phi_x^4) \gamma^4 
+ \frac{\mu^4}{4 \gamma^2}\sinh(2 \gamma\phi) (\mu^2 \sinh(\gamma\phi) - 4 \phi_{xx} \gamma) +\\
\nonumber
&+&  \frac{1}{8} \left(16 (\phi_{xt}^2 + \phi_{xx}^2 - 2 \phi_t \phi_{xxt} - 6 \phi_x \phi_{xxx}) 
+  3 (\phi_t^4 + 6 \phi_t^2 \phi_x^2 + \phi_x^4) \gamma^2 \right) \mu^2 \cosh(\gamma\phi) + \\
\nonumber
&+& \frac{1}{4} (\phi_t^2 + \phi_x^2) \mu^4 \sinh(\gamma\phi)^2 + \frac{1}{2} (\phi_t^2 + 9 \phi_x^2) \mu^4 \cosh(\gamma\phi)^2 +\\
&-& 3 (2 \phi_t \phi_x \phi_{xt} + (\phi_t^2 + \phi_x^2) \phi_{xx}) \gamma \mu^2 \sinh(\gamma\phi) \nonumber 
\end{eqnarray}
\begin{eqnarray*}
\rho_4(x,t) &=&
4 \mu ^8 \sinh ^4 (\gamma  \phi ) +  4 \gamma ^2 \mu ^6 \left (7 \phi_t^2 +   31 \phi_x^2 \right) \sinh (\gamma  \phi ) \sinh (2 \gamma  \phi ) + \\
&+& 16 \gamma ^2 \mu ^6 \left (\phi_t^2 +      9 \phi_x^2 \right) \cosh ^3 (\gamma  \phi ) +  \\
&+& 16 \gamma ^3 \mu ^4 \sinh (2 \gamma  \phi ) \left (7 \phi_t^2 \phi_ {xx} + 2 \phi_t \phi_x \phi_ {xt} + 43 \phi_x^2 \phi_ {xx} \right) +  \\
&+& 4 \mu ^4 \sinh (2 \gamma  \phi ) \left(\mu ^4 \sinh (2 \gamma  \phi ) 
- 4 \gamma \mu ^2 \left(\gamma  \left (\phi_t^2 +  7 \phi_x^2 \right) \sinh (\gamma  \phi ) + \phi_ {xx} \cosh (\gamma  \phi ) \right) +  32 \gamma   \phi_ {xxxx} \right) + \\
&-& 8 \gamma ^3 \mu ^2 \sinh (\gamma  \phi ) \left(32 \phi_ {xxxx} \left (\phi_t^2 + 7 \phi_x^2 \right) 
+ 15 \gamma ^2 \left (4 \phi_t \phi_x \phi_ {xt} \left(\phi_t^2 + \phi_x^2 \right) + \phi_ {xx} \left (\phi_t^4 + 6 \phi_t^2  \phi_x^2 + \phi_x^4 \right) \right) + \right.\\
&-& \left. 16 \left (6 \phi_ {xt} (\phi_t \phi_ {xxx} + \phi_x \phi_ {xxt}) + 6 \phi_ {xx} (\phi_t \phi_ {xxt} + \phi_x \phi_ {xxx}) - 8 \phi_t \phi_x 
    \phi_ {xxxt} + 
    15 \phi_ {xt}^2 \phi_ {xx} +  5 \phi_ {xx}^3 \right) \right) + \\
&+& 6 \gamma ^2\mu ^4 \sinh ^2 (\gamma  \phi ) \left(\gamma ^2 \left (\phi_t^4 + 70 \phi_t^2 \phi_x^2 + 129 \phi_x^4 \right) - 
    16 \left (4 \phi_t \phi_ {xxt} + 4 \phi_x \phi_ {xxx} + 3 \phi_ {xt}^2 + 3 \phi_ {xx}^2 \right) \right) +  \\
&+& 8 \gamma ^2 \mu ^4 \cosh ^2 (\gamma  \phi ) \left(\gamma ^2 \left(3 \phi_t^4 + 106 \phi_t^2 \phi_x^2 + 51 \phi_x^4 \right) - 16 \left(\phi_t \phi_ {xxt} + 3 \phi_x \phi_ {xxx} 
    - 8\phi_ {xx}^2 \right) + 32 \phi_ {xt}^2 \right) +  \\
&+& 2 \gamma ^2 \mu ^2 \cosh (\gamma  \phi ) \left(5 \gamma ^4 \left(\phi_t^2 + \phi_x^2 \right) \left (\phi_t^4 + 14 \phi_t^2 \phi_x^2 + \phi_x^4 \right) 
- 16 \gamma ^2 \left (10 \phi_t^3 \phi_ {xxt} + \phi_t^2 \left (58 \phi_x \phi_ {xxx} + \right. \right. \right.\\
& + & \left. \left. \left. 
5 \left (\phi_ {xt}^2 + \phi_ {xx}^2 \right) \right) + 2 \phi_t \phi_x (43 \phi_x \phi_ {xxt} + 10 \phi_ {xt}
               \phi_ {xx}) + \phi_x^2 \left (38 \phi_x \phi_ {xxx} + \right. \right. \right. \\
&+& \left. \left.     \left.   5 \left (\phi_ {xt}^2 + \phi_ {xx}^2 \right) \right) \right) + 128 \left (-4 \phi_ {xt}
            \phi_ {xxxt} - 8 \phi_ {xx} \phi_ {xxxx} + \phi_ {xxt}^2 + \phi_ {xxx}^2 \right) \right) +  \\
&+& 32 \gamma  \mu ^6 \phi_ {xx} \sinh ^3 (\gamma  \phi) 
            + \frac {1} {4} \left(-1792 \gamma ^4 \left (-2 \phi_{xxt}^2 \left (\phi_t^2 + \phi_x^2 \right) - 2 \phi_ {xxx}^2 \left(\phi_t^2 + \phi_x^2 \right) + \right. \right.\\
& - & \left. \left.            
       8 \phi_t \phi_x \phi_ {xxt} \phi_ {xxx} + \phi_ {xt}^4 + 6 \phi_ {xt}^2 \phi_ {xx}^2 + \phi_ {xx}^4 \right) + 1120 \gamma ^6 \left(8 \phi_t \phi_x \phi_ {xt} \phi_ {xx}
            \left(\phi_t^2 + \phi_x^2 \right) + \right. \right.\\
& + & \left. \left.              
             \phi_ {xt}^2 \left(\phi_t^4 + 6 \phi_t^2 \phi_x^2 + \phi_x^4 \right) + \phi_ {xx}^2
            \left(\phi_t^4 + 6 \phi_t^2 \phi_x^2 + \phi_x^4 \right) \right) + \right. \\ 
            & + & \left. 
            5 \gamma ^8 \left(\phi_t^8 + 28 \phi_t^6 \phi_x^2 + 70 \phi_t^4 \phi_x^4 + 28 \phi_t^2 \phi_x^6 + \phi_x^8 \right) + 
    4096 \gamma ^2 \left (\phi_ {xxxt}^2 + \phi_ {xxxx}^2 \right) \right)
\end{eqnarray*}
The corresponding first conserved currents read
\begin{eqnarray}
j_1(x,t) &=& \phi_t \phi_x = \mathcal{P}(x,t), \\
j_2(x,t) &=& (4 \phi_{xt} (\phi_{tt} + \phi_{xx}) + \phi_t \phi_x ((\phi_t^2 + \phi_x^2) \gamma^2 - 2 \mu^2 \cosh(\gamma \phi)),\\
\cdots 
\end{eqnarray}

In the free coupling limit, for the first representatives we have 
\begin{eqnarray}
\rho_1^{(0)} &=& \frac{1}{2} \left( \phi_t^2+\phi_x^2 + \mu ^2 \phi ^2 \right), \\
\rho_2^{(0)} &=& 2 \left( \phi_{xt}^2+\phi_{xx}^2 -\mu ^2 \phi \phi_{xx} \right) + \mu^2 \rho_1^{(0)}, \\
\rho_3^{(0)} &=& 8 (\phi_{xxt}^2 + \phi_{xxx}^2) +  5 ((\phi_t^2 + \phi_x^2) \phi_{xt}^2 + 4 \phi_t \phi_x \phi_{xt} \phi_{xx} 
+ \frac{\mu^4}{2} \phi (\mu^2 \phi - 4 \phi_{xx} )  \\ 
&+&   2 (\phi_{xt}^2 + \phi_{xx}^2 - 2 \phi_t \phi_{xxt} - 6 \phi_x \phi_{xxx}) 
+ \frac{1}{2} (\phi_t^2 + 9 \phi_x^2) \mu^4  \nonumber
\end{eqnarray}


\newpage 

\begin{thebibliography}{99}

\bibitem{Weiss1} T. Kinoshita, T. Wenger, D. S. Weiss, Nature 440, 900 (2006); M. Greiner, O. Mandel, T. W. H\"ansch, and I. Bloch, Nature 419 51 (2002); S. Hofferberth, I. Lesanovsky, B. Fischer, T. Schumm, and J. Schmiedmayer, Nature 449, 324 (2007); S. Trotzky Y.A. Chen, A. Flesch, I. P. McCulloch, U. Schollw\"ock, J. Eisert, I. Bloch, Nature Phys. 8, 325 (2012).
\bibitem{Weiss2} F. Meinert, M. J. Mark, E. Kirilov, K. Lauber, P. Weinmann, A. J. Daley, H.-C. N\"agerl, Phys. Rev. Lett. 111, 053003
(2013); T. Langen, R. Geiger, M. Kuhnert, B. Rauer, and J. Schmiedmayer, Nature Phys. 9, 640 (2013); M. Cheneau, P. Barmettler, D. Poletti, M. Endres, P. Schauss, T. Fukuhara, C. Gross, I. Bloch, Nature 481, 484 (2012). 
\bibitem{Weiss3} T. Langen, S. Erne, R. Geiger, B. Rauer, T. Schweigler, M. Kuhnert, W. Rohringer, I. E. Mazets, T. Gasenzer, and J. Schmiedmayer, Science 348, 207 (2015).

\bibitem{CC1} P. Calabrese and  J. Cardy, Phys. Rev. Lett. {\bf 96}, 136801 (2006);
P. Calabrese and  J. Cardy, J. Stat. Mech. P06008 (2007). 
\bibitem{CC2} P. Calabrese and  J. Cardy, this volume.
\bibitem{inho1} D. Bernard and B. Doyon, J. Phys. A {\bf45}, 362001 (2012); Ann. Henri Poincaré {\bf 16}, 113 (2015); this volume.
\bibitem{inho2} 
C.  Karrasch,  J.  H.  Bardarson,  and  J.  E.  Moore,  Phys. Rev. Lett. {\bf 108}, 227206 (2012); 
New J. Phys. {\bf 15}, 083031 (2013).
C. Karrasch, R. Ilan, and J. E. Moore, Phys. Rev. B {\bf 88}, 195129 (2013).
R. Vasseur, C. Karrasch, and J. E. Moore, Phys. Rev. Lett. {\bf 115}, 267201 (2015).
\bibitem{inho3} A. De Luca, J. Viti, D. Bernard, and B. Doyon, Phys. Rev. B {\bf 88}, 134301 (2013).
A. De Luca, J. Viti, L. Mazza, and  D. Rossini, Phys. Rev. B {\bf 90}, 161101 (2014).
D. Bernard, B. Doyon, and J. Viti, J. Phys. A {\bf 48}, 05FT01 (2015).
A. Biella, A. De Luca, J, Viti, D. Rossini, L. Mazza, R. Fazio, arXiv:1602.05357
\bibitem{inho4} R. Vasseur and J. E. Moore, this volume. 
\bibitem{quenches1}
A. Polkovnikov, K. Sengupta, A. Silva and M. Vengalattore,
Rev. Mod. Phys. \textbf{83}, 863-883  (2011); 
J. M. Deutsch, Phys. Rev. A {\bf 43}, 2046 (1991); M. Srednicki,
Phys. Rev. E {\bf 50}, 888 (1994); M.~Greiner, O.~Mandel, T.~W.~H\"ansch, and I.~Bloch,
Nature {\bf 419} 51 (2002).
\bibitem{quenches2} S. Hofferberth, I. Lesanovsky, B. Fischer, T. Schumm, and J. Schmiedmayer,
Nature {\bf 449}, 324 (2007); A. Iucci and M. A. Cazalilla, Phys. Rev. A {\bf 80}, 063619 (2009); 
M. C. Ba\~nuls, J. I. Cirac, and M. B. Hastings,
Phys. Rev. Lett. {\bf 106}, 050405 (2011); M. Rigol and M. Srednicki,  Phys. Rev. Lett. 108, 110601 (2012); D. Iyer and N. Andrei, Phys. Rev.
Lett. 109, 115304 (2012). 
\bibitem{quenches3}
 D. Rossini, A. Silva, G. Mussardo, G. Santoro, Phys.
Rev. Lett. 102, 127204 (2009); D. Rossini, S. Suzuki, G. Mussardo, G. E. Santoro, A. Silva, Phys. Rev. B 82, 144302 (2010); G. Biroli, C. Kollath, and A.M. L\"auchli, Phys. Rev. Lett. {\bf 105},
250401 (2010); G. P. Brandino, A. De Luca, R.M. Konik, G. Mussardo, 
Phys. Rev. B 85, 214435; C. Gogolin, M. P. M\"uller, and J. Eisert,
Phys. Rev. Lett. {\bf 106}, 040401 (2011); M. Kormos, M. Collura, P. Calabrese, Phys. Rev. A 89, 013609 (2014); 
S. Sotiriadis, P. Calabrese and J. Cardy, EPL \textbf{87} 20002 (2009).
\bibitem{quenches4} 
P. Calabrese, F.H.L. Essler, and M. Fagotti,
Phys. Rev. Lett. {\bf 106}, 227203 (2011);
P. Calabrese, F.H.L. Essler, and M. Fagotti, J. Stat. Mech. P07016
(2012); P. Calabrese, F.H.L. Essler, and M. Fagotti,
J. Stat. Mech. P07022 (2012); J. Sirker, N.P. Konstantinidis, and N. Sedlmayr, Phys. Rev. A 89, 042104 (2014); 
M. Collura, P. Calabrese, F. H. L. Essler, Phys. Rev. B 92, 125131 (2015). 

\bibitem{Rigol}
M. Rigol, V. Dunjko, V. Yurovsky, and M. Olshanii, Phys. Rev. Lett. 98, 050405 (2007); 
M.~Rigol, V.~Dunjko, M.~Olshanii, 
Nature {\bf 452}, 854 (2008). 
\bibitem{2011_Pozsgay_JSTAT_P01011}
B.~Pozsgay, J. Stat. Mech.: Th. Exp. P01011 (2011).
\bibitem{MC} J. Mossel, J.S. Caux, J. Phys. A:  45, 255001 (2012). 
\bibitem{CK} J.S. Caux, R. Konik, Phys. Rev. Lett. 109, 175301 (2012). 
\bibitem{CE} J.S. Caux and F. Essler, Phys. Rev. Lett. 110, 257203 (2013). 
\bibitem{Cauxreview} J.S. Caux, this volume. 
\bibitem{FEreview} F.H.L. Essler and M. Fagotti, this volume. 
\bibitem{CEF}  P. Calabrese, F.H.L. Essler, and M. Fagotti, Phys. Rev. Lett. 106, 227203 (2011); 
J. Stat. Mech. P07016 (2012); J. Stat. Mech. P07022 (2012); F.H.L. Essler, M. Fagotti, Phys. Rev. B 87, 245107 (2013).
\bibitem{Sotiriadis} S. Sotiriadis, D. Fioretto, and G. Mussardo, J. Stat. Mech. (2012) P02017; 
S. Sotiriadis, G. Takacs, G. Mussardo, Phys. Lett. B 734 52 (2014); S. Sotiriadis, P. Calabrese, 
J. Stat. Mech. (2014) P07024. 
\bibitem{prethermalization} 
B. Bertini, F. H.L. Essler, S. Groha, N. J. Robinson, 
Phys. Rev. Lett. 115, 180601 (2015); J. Marino, A. Silva, 
Phys. Rev. B 89, 024303 (2014); 
T. Kitagawa, A. Imambekov, J. Schmiedmayer, E. Demler
New J. Phys. 13 (2011) 073018; 
B. Bertini, M. Fagotti, J. Stat. Mech. (2015) P07012.
\bibitem{BertiniSchurichEssler} B. Bertini, D. Schuricht, F. H. L. Essler, 
J. Stat. Mech. (2014) P10035
\bibitem{FM} D. Fioretto, G. Mussardo, New J. Phys. 12, 055015 (2010).
\bibitem{GMPRL13} G. Mussardo, Phys. Rev. Lett. {\bf 111}, 100401 (2013).
\bibitem{GGE_problems}
B. Wouters, M. Brockmann, J. De Nardis, D. Fioretto, M. Rigol and J.-S. Caux,
Phys. Rev. Lett. {\bf 113}, 117202 (2014);
B. Pozsgay, M. Mesty\'{a}n, M. A. Werner, M. Kormos, G. Zar\'{a}nd and
G. Tak\'{a}cs, Phys. Rev. Lett. {\bf 113}, 117203 (2014);
G. Goldstein and N. Andrei, arXiv:1405.4224.
\bibitem{completeGGE} E. Ilievski, J. De Nardis, B. Wouters, J.S. Caux, F. H. L. Essler, T. Prosen, 
Phys. Rev. Lett. 115, 157201 (2015). 
\bibitem{EMP} F. H.L. Essler, G. Mussardo, M. Panfil, Phys. Rev. A 91, 051602 (2015). 



\bibitem{Faddev}  L.D. Faddev, L.A. Takhtajan, \textit{Hamiltonian Methods}, Springer, Berlin (2000). 
\bibitem{Novikov} S. Novikov, S.V. Manakov, L.P. Pitaevskii, V.E. Zakharov, \textit{Theory of Solitons}, Plenum Publ. Corporation, New York (1984). 
\bibitem{Ablowitz} M.J. Ablowitz and H. Segur, \textit{Solitons and the Inverse Scattering Transform}, SIAM Philadelphia 1961. 




\bibitem{LM}
A. LeClair, G. Mussardo,
Nucl. Phys. B 552, 624 (1999).




\bibitem{KW}
M. Karowski and P. Weisz,
Nucl. Phys. B \textbf{139}, 455 (1978).


\bibitem{Smirnov} F. A. Smirnov, Form Factors in Completely Integrable Models
of Quantum Field Theory (World Scientific, Singapore, 1992).
\bibitem{Luky} V. Brazhnikov, S. Lukyanov, Nucl.Phys. B512 (1998) 616-636. 
\bibitem{KM}
A. Koubek and G. Mussardo
Phys. Lett B \textbf{311}, 193-201 (1993); A. Fring, G. Mussardo and P. Simonetti,
Nucl. Phys. B \textbf{393}, 413-441 (1993).

\bibitem{zamzam} A.B. Zamolodchikov and Al.B. Zamolodchikov, Annals of Physics 120, 253 (1979). 
\bibitem{GMUSSARDO} G. Mussardo, Statistical Field Theory. Oxford Univ. Press, Oxford (2010). 
\bibitem{BDD}
D. Boyanovsky, C. Destri and H.J. de Vega,
Phys. Rev. D \textbf{69}, 045003  (2004).
\bibitem{FPU}
E. Fermi, J. Pasta, and S. Ulam,
\textit{Studies of Nonlinear Problems},
Reprint from the American Mathematical Monthly, Vol. 74, No. 1 (1967).
\bibitem{FPUreport} {\em The Fermi-Pasta-Ulam Problem. A status report}, Editor G. Gallavotti, Springer (Berlin 2008). 
\bibitem{fucito1982approach} F. Fucito, F. Marchesoni, E. Marinari, G. Parisi, L. Peliti, S. Ruffo 
and A. Vulpiani, J. Physique 43 (1982) 707-713. 
\bibitem{bassetti1984complex} B.Bassetti,  P. Butera, M. Raciti, M. Sparpaglione, Phys. Rev. A 30, 1033 (1984). 

\bibitem{Scalapino} D.J. Scalapino, M. Sears, R.A. Ferrell, 
Phys. Rev. B \textbf{6}, 3409 (1972).  


  






\bibitem{PozsayTakacs} B. Pozsgay and G. Takacs,  Nucl. Phys. B \textbf{788} (2008) 209. 


\bibitem{YY}
C.N.~Yang, C.P.~Yang,
J. Math. Phys. {\bf 10}, 1115 (1969). 
\bibitem{ZamTBA} Al.B. Zamolodchikov, Nucl. Phys. B 342 (1990), 695. 
\bibitem{Wadati} M. Wadati, Journ. Phys. Society of japan, 54 (1985), 3727. 
\bibitem{KMT} M.~Kormos, G.~Mussardo, A.~Trombettoni, Phys. Rev. A {\bf 81}, 043606 (2010). 





\bibitem{BulloughPillingTimonenQCShG}
R.K. Bullough, D.J. Pilling and J. Timonen,
J. Phys. A \textbf{19} L955-L960 (1986).
\bibitem{Forest} M. G. Forest, D. W. McLaughlin, J. Math. Phys. 23, 1248 (1982). 
\bibitem{Flaska} H. Flaschka and D.W. Mclaughlin, Progress of Theoretical Physics 55, 438 (1976).    
\bibitem{Chodos} A. Chodos, Phys. Rev. B \textbf{21}, 2818 (1980). 

\bibitem{Brodsky} S. J. Brodsky, P. Hoyer, 
Phys.Rev. D \textit{83} (2011) 045026 

\bibitem{Amit} D. Amit, {\em Field Theory, the Renormalization Group and Critical Phenomena}, 
Revised Second Edition. Singapore, World Scientific 1984.




\bibitem{Rajaraman} R. Rajaraman, {\em Solitons and Instantons}, Elsevier (Amsterdam), 1982. 



\bibitem{ari} A. E. Arinshtein, V. A. Fateev, and A. B. Zamolodchikov, Phys.
Lett. B \textbf{ 87}, 389 (1979).


\bibitem{balog} J. Balog, Nucl. Phys. B \textbf{419}, 480 (1994).





\bibitem{sudarshan} J. R. Klauder and E.C.G Sudarshan, {\em Fundamental of Quantum Optics}, 
W.A. Benjamin (New York), 1968. 











\bibitem{Merzbacher} E. Merzbacher, \textit{Quantum Mechanics}, John Wiley \& Sons, New York 1970.
\bibitem{DoreyTateo} P. Dorey and R. Tateo, Nucl.Phys. B \textit{482} (1996) 639.  


\bibitem{Negro} S. Negro and F. Smirnov, Nucl. Phys. B 875, 166 (2013); S. Negro, Int. J. Mod. Phys. A 29, 1450111 (2014).

\bibitem{Piroli} B. Bertini, L. Piroli and P. Calabrese, {\em Quantum quenches in the sinh-Gordon model: steady state and one point correlation
functions}, arXiv: 1602.08269. 
\end{thebibliography}
\end{document}